\documentclass[12pt]{report}

\usepackage{axodraw}

\makeatletter
\if@twoside \oddsidemargin -17pt \evensidemargin 00pt
\else \oddsidemargin 00pt \evensidemargin 00pt
\fi
\expandafter\ifx\csname mathrm\endcsname\relax\def\mathrm#1{{\rm #1}}\fi

\renewcommand{\theequation}{\thesection.\arabic{equation}}
\newcounter{saveeqn}

\@addtoreset{equation}{section}

\makeatother

\def\co{\relax}
\def\co{,}
\def\nl{\nonumber\\}
\def\nlc{\co\nonumber\\}
\def\nln{\nonumber\\*[-1ex]\phantom{\fbox{\rule{0em}{2ex}}}}

\def\beq{\begin{equation}}
\def\eeq{\end{equation}}
\def\beqar{\begin{eqnarray}}
\def\eeqar{\end{eqnarray}}
\def\barr#1{\begin{array}{#1}}
\def\earr{\end{array}}
\def\bfi{\begin{figure}}
\def\efi{\end{figure}}
\def\btab{\begin{table}}
\def\etab{\end{table}}
\def\bce{\begin{center}}
\def\ece{\end{center}}
\def\nn{\nonumber}
\def\disp{\displaystyle}
\def\text{\textstyle}
\def\arraystretch{1.4}

\def\al{\alpha}
\def\be{\beta}
\def\ga{\gamma}
\def\de{\delta}
\def\veps{\varepsilon}
\def\la{\lambda}
\def\si{\sigma}
\def\Ga{\Gamma}
\def\De{\Delta}
\def\La{\Lambda}

\def\refeq#1{\mbox{(\ref{#1})}}
\def\refeqs#1{\mbox{(\ref{#1})}}
\def\refeqf#1{\mbox{(\ref{#1})}}
\def\reffi#1{\mbox{Fig.~\ref{#1}}}
\def\reffis#1{\mbox{Figs.~\ref{#1}}}
\def\refta#1{\mbox{Table~\ref{#1}}}
\def\reftas#1{\mbox{Tables~\ref{#1}}}
\def\refch#1{\mbox{Chapter~\ref{#1}}}
\def\refse#1{\mbox{Section~\ref{#1}}}
\def\refses#1{\mbox{Sections~\ref{#1}}}

\def\refapp#1{\mbox{Appendix~\ref{#1}}}
\def\citere#1{\mbox{Ref.~\cite{#1}}}
\def\citeres#1{\mbox{Refs.~\cite{#1}}}

\def\solid{\raise.9mm\hbox{\protect\rule{1.1cm}{.2mm}}}
\def\dash{\raise.9mm\hbox{\protect\rule{2mm}{.2mm}}\hspace*{1mm}}

\newcommand{\GeV}{\unskip\,\mathrm{GeV}}
\newcommand{\MeV}{\unskip\,\mathrm{MeV}}
\newcommand{\TeV}{\unskip\,\mathrm{TeV}}
\newcommand{\fba}{\unskip\,\mathrm{fb}}

\def\mathswitchr#1{\relax\ifmmode{\mathrm{#1}}\else$\mathrm{#1}$\fi}

\newcommand{\PV}{\mathswitchr V}
\newcommand{\PW}{\mathswitchr W}
\newcommand{\PZ}{\mathswitchr Z}

\newcommand{\Pg}{\mathswitchr g}
\newcommand{\PH}{\mathswitchr H}
\newcommand{\Pe}{\mathswitchr e}
\newcommand{\Pne}{\mathswitch \nu_{\mathrm{e}}}
\newcommand{\Pnebar}{\mathswitch \bar\nu_{\mathrm{e}}}
\newcommand{\Pane}{\mathswitch \bar\nu_{\mathrm{e}}}

\newcommand{\Pd}{\mathswitchr d}
\newcommand{\Pdbar}{\bar{\mathswitchr d}}

\newcommand{\Pu}{\mathswitchr u}
\newcommand{\Pubar}{\bar{\mathswitchr u}}
\newcommand{\Ps}{\mathswitchr s}
\newcommand{\Psbar}{\bar{\mathswitchr s}}
\newcommand{\Pb}{\mathswitchr b}

\newcommand{\Pc}{\mathswitchr c}
\newcommand{\Pcbar}{\bar{\mathswitchr c}}
\newcommand{\Pt}{\mathswitchr t}

\newcommand{\Pep}{\mathswitchr {e^+}}
\newcommand{\Pem}{\mathswitchr {e^-}}

\newcommand{\PWp}{\mathswitchr {W^+}}
\newcommand{\PWm}{\mathswitchr {W^-}}
\newcommand{\PWpm}{\mathswitchr {W^\pm}}

\def\mathswitch#1{\relax\ifmmode#1\else$#1$\fi}

\newcommand{\MV}{\mathswitch {M_\PV}}
\newcommand{\MW}{\mathswitch {M_\PW}}

\newcommand{\MZ}{\mathswitch {M_\PZ}}
\newcommand{\MH}{\mathswitch {M_\PH}}
\newcommand{\Me}{\mathswitch {m_\Pe}}
\newcommand{\Mmy}{\mathswitch {m_\mu}}
\newcommand{\Mta}{\mathswitch {m_\tau}}
\newcommand{\Md}{\mathswitch {m_\Pd}}
\newcommand{\Mu}{\mathswitch {m_\Pu}}
\newcommand{\Ms}{\mathswitch {m_\Ps}}
\newcommand{\Mc}{\mathswitch {m_\Pc}}
\newcommand{\Mb}{\mathswitch {m_\Pb}}
\newcommand{\Mt}{\mathswitch {m_\Pt}}
\newcommand{\GW}{\mathswitch {\Gamma_\PW}}
\newcommand{\GZ}{\Gamma_{\PZ}}
\newcommand{\GV}{\Gamma_{\PV}}

\newcommand{\scrs}{\scriptscriptstyle}
\newcommand{\sw}{\mathswitch {s_{\scrs\PW}}}
\newcommand{\cw}{\mathswitch {c_{\scrs\PW}}}

\newcommand{\GF}{\mathswitch {G_\mu}}

\def\ie{i.e.\ }
\def\eg{e.g.\ }
\def\cf{cf.\ }

\newcommand{\ps}{p\hspace{-0.42em}/}%\hspace{0.1em}}
\newcommand{\qs}{q\hspace{-0.5em}/\hspace{0.1em}}

\renewcommand{\O}{{\cal O}}
\newcommand{\C}{{\cal C}}
\newcommand{\Oa}{\mathswitch{{\cal{O}}(\alpha)}}

\newcommand{\ri}{{\mathrm{i}}}
\newcommand{\ieps}{\ri\epsilon}
\newcommand{\rd}{{\mathrm{d}}}

\newcommand{\M}{{\cal {M}}}
\newcommand{\V}{{\cal {V}}}
\renewcommand{\L}{{\cal L}}
\newcommand{\F}{{\cal {F}}}
\newcommand{\bq}{{\bf q}}

\newcommand{\born}{{\mathrm{Born}}}
\newcommand{\nf}{{\mathrm{nf}}}
\newcommand{\U}{\mathrm{U}}
\newcommand{\SU}{\mathrm{SU}}

\def\Li{\mathop{\mathrm{Li}_2}\nolimits}
\def\cLi{\mathop{{\cal L}i_2}\nolimits}
\def\Re{\mathop{\mathrm{Re}}\nolimits}
\def\Im{\mathop{\mathrm{Im}}\nolimits}

\def\arc{\mathop{\mathrm{arc}}\nolimits}

\def\draftdate{\relax}
\def\mda{\relax}
\def\mua{\relax}
\def\mla{\relax}
\def\mua{\marginpar[\boldmath\hfil$\uparrow$]%
                   {\boldmath$\uparrow$\hfil}%
                    \typeout{marginpar: $\uparrow$}\ignorespaces}
\def\mda{\marginpar[\boldmath\hfil$\downarrow$]%
                   {\boldmath$\downarrow$\hfil}%
                    \typeout{marginpar: $\downarrow$}\ignorespaces}
\def\mla{\marginpar[\boldmath\hfil$\rightarrow$]%
                   {\boldmath$\leftarrow $\hfil}%
                    \typeout{marginpar: $\leftrightarrow$}\ignorespaces}
\def\Mua{\marginpar[\boldmath\hfil$\Uparrow$]%
                   {\boldmath$\Uparrow$\hfil}%
                    \typeout{marginpar: $\uparrow$}\ignorespaces}
\def\Mda{\marginpar[\boldmath\hfil$\Downarrow$]%
                   {\boldmath$\Downarrow$\hfil}%
                    \typeout{marginpar: $\downarrow$}\ignorespaces}
\def\Mla{\marginpar[\boldmath\hfil$\Rightarrow$]%
                   {\boldmath$\Leftarrow $\hfil}%
                    \typeout{marginpar: $\leftrightarrow$}\ignorespaces}
\def\draft{
\def\thtystars{******************************}
\def\sixtystars{\thtystars\thtystars}
\typeout{}
\typeout{\sixtystars**}
\typeout{* Draft mode!
         For final version remove \protect\draft\space in source file *}
\typeout{\sixtystars**}
\typeout{}
\def\draftdate{\today}
\def\mua{\marginpar[\boldmath\hfil$\uparrow$]%
                   {\boldmath$\uparrow$\hfil}%
                    \typeout{marginpar: $\uparrow$}\ignorespaces}
\def\mda{\marginpar[\boldmath\hfil$\downarrow$]%
                   {\boldmath$\downarrow$\hfil}%
                    \typeout{marginpar: $\downarrow$}\ignorespaces}
\def\mla{\marginpar[\boldmath\hfil$\rightarrow$]%
                   {\boldmath$\leftarrow $\hfil}%
                    \typeout{marginpar: $\leftrightarrow$}\ignorespaces}
\def\Mua{\marginpar[\boldmath\hfil$\Uparrow$]%
                   {\boldmath$\Uparrow$\hfil}%
                    \typeout{marginpar: $\Uparrow$}\ignorespaces}
\def\Mda{\marginpar[\boldmath\hfil$\Downarrow$]%
                   {\boldmath$\Downarrow$\hfil}%
                    \typeout{marginpar: $\Downarrow$}\ignorespaces}
\def\Mla{\marginpar[\boldmath\hfil$\Rightarrow$]%
                   {\boldmath$\Leftarrow $\hfil}%
                    \typeout{marginpar: $\Leftrightarrow$}\ignorespaces}
\overfullrule 5pt
\oddsidemargin -15mm
\marginparwidth 29mm
}

\makeatletter

\def\eqnarray{\stepcounter{equation}\let\@currentlabel=\theequation
\global\@eqnswtrue
\global\@eqcnt\z@\tabskip\@centering\let\\=\@eqncr
$$\halign to \displaywidth\bgroup\hskip\@centering
  $\displaystyle\tabskip\z@{##}$\@eqnsel&\global\@eqcnt\@ne
  \hskip 2\arraycolsep \hfil${##}$\hfil
  &\global\@eqcnt\tw@ \hskip 2\arraycolsep $\displaystyle\tabskip\z@{##}$\hfil
   \tabskip\@centering&\llap{##}\tabskip\z@\cr}
\def\appendix{\par
 \setcounter{section}{0} \setcounter{subsection}{0}
 \def\thesection{\Alph{section}}}

\makeatother

\newcommand{\lsim}
{\;\raisebox{-.3em}{$\stackrel{\displaystyle <}{\sim}$}\;}
\newcommand{\gsim}
{\;\raisebox{-.3em}{$\stackrel{\displaystyle >}{\sim}$}\;}

\newcommand{\eeWW}{{\Pe^+ \Pe^-\to \PW^+ \PW^-}}
\newcommand{\Wpff}{{\PW^+ \to f_1\bar f_2}}
\newcommand{\Wmff}{{\PW^- \to f_3\bar f_4}}

\newcommand{\Kp}{K_+}
\newcommand{\Km}{K_-^*}
\newcommand{\betaM}{\beta}
\newcommand{\betap}{\bar{\beta}}
\newcommand{\betaW}{\beta_\PW}

\newcommand{\xW}{x_\PW}
\newcommand{\Cbr}{{\cal C}}
\newcommand{\Dbr}{{\cal D}}
\newcommand{\Ebr}{{\cal E}}
\newcommand{\Ybr}{Y'}
\newcommand{\Ibr}{{\cal I}}
\newcommand{\CM}{\mathrm{CM}}
\newcommand{\real}{{\mathrm{real}}}
\newcommand{\virt}{{\mathrm{virt}}}
\newcommand{\ffp}{\mathswitch{\mathrm{f\/f}'}}
\newcommand{\mfp}{\mathswitch{\mathrm{mf}'}}
\newcommand{\mmp}{\mathswitch{\mathrm{mm}'}}
\newcommand{\mf}{\mathswitch{\mathrm{mf}}}
\newcommand{\mm}{\mathswitch{\mathrm{mm}}}
\newcommand{\im}{\mathswitch{\mathrm{im}}}
\newcommand{\ifp}{\mathswitch{\mathrm{if}}}

\newcommand{\eeffff}{\Pep\Pem\to 4f}
\newcommand{\eeffffg}{\eeffff\ga}
\renewcommand{\O}{{\cal O}}
\newcommand{\spab}{\langle p_a p_b \rangle}
\newcommand{\spac}{\langle p_a p_c \rangle}
\newcommand{\spad}{\langle p_a p_d \rangle}
\newcommand{\spae}{\langle p_a p_e \rangle}

\newcommand{\spbd}{\langle p_b p_d \rangle}
\newcommand{\spbe}{\langle p_b p_e \rangle}
\newcommand{\spbf}{\langle p_b p_f \rangle}
\newcommand{\spcd}{\langle p_c p_d \rangle}
\newcommand{\spce}{\langle p_c p_e \rangle}

\newcommand{\spdf}{\langle p_d p_f \rangle}
\newcommand{\spef}{\langle p_e p_f \rangle}
\newcommand{\spak}{\langle p_a k \rangle}
\newcommand{\spbk}{\langle p_b k \rangle}
\newcommand{\spck}{\langle p_c k \rangle}
\newcommand{\spdk}{\langle p_d k \rangle}
\newcommand{\spek}{\langle p_e k \rangle}
\newcommand{\spfk}{\langle p_f k \rangle}

\newcommand{\cspad}{\spad^*}

\newcommand{\cspbd}{\spbd^*}

\newcommand{\cspbf}{\spbf^*}
\newcommand{\cspcd}{\spcd^*}

\newcommand{\cspdf}{\spdf^*}

\newcommand{\cspbk}{\spbk^*}

\newcommand{\cspdk}{\spdk^*}

\newcommand{\cspfk}{\spfk^*}

\newcommand{\tp}{\tilde{p}}
\newcommand{\tm}{\tilde{m}}

\newcommand{\tK}{\tilde{K}}
\newcommand{\ns}{n\hspace{-0.52em}/\hspace{0.1em}}
\newcommand{\zik}{z_{ik}}
\newcommand{\yik}{y_{ik}}
\newcommand{\zia}{z_{ia}}
\newcommand{\xia}{x_{ia}}
\newcommand{\xab}{x_{ab}}
\newcommand{\vab}{v_{ab}}

\begin{document}
\thispagestyle{empty}
Diss. ETH No. 13363
\hfill \draftdate
\vspace*{3cm}
\begin{center}
{\Large \bf Precise Predictions for Four-Fermion Production}\\[.4em]
{\Large \bf in Electron-Positron Annihilation}\\
\vspace*{2cm}
A dissertation submitted to the \\
SWISS FEDERAL INSTITUTE OF TECHNOLOGY ZURICH \\
(ETH Z\"urich) \\
\vspace*{1.5cm}
for the degree of \\
Doctor of Natural Sciences \\
\vspace*{1.5cm}
presented by \\
{\bf Markus Roth} \\
Dipl.\  Phys.\ Univ.\ W\"urzburg \\
born March 8, 1969 \\
German citizen \\
\vspace*{1.5cm}
accepted on the recommendation of \\
Prof.\ Zoltan Kunszt, examiner \\
PD Dr.\ Ansgar Denner, co-examiner \\
Prof. J\"urg Fr\"ohlich, co-examiner \\
\vspace*{1.5cm}
1999
\end{center}
\newpage
\mbox{}
\thispagestyle{empty}
\newpage
\setcounter{page}{0}
\pagenumbering{Roman}
\tableofcontents

\chapter*{Abstract}
\addcontentsline{toc}{chapter}{Abstract}

At present, the $\PW$ boson is investigated experimentally at LEP2
via its pair production.
In order to achieve precise theoretical predictions  
for the measurement of the W-boson mass and the 
non-abelian triple-gauge-boson couplings,
the inclusion of radiative corrections is required. 
Since the W bosons decay very rapidly into light fermion pairs, 
the actual processes under investigation are
$\Pe^+ \Pe^- \to 4\,\mathrm{fermions}$.

The full ${\cal O}(\alpha )$ corrections to these processes are not 
available at present.
Since the main contributions originate from diagrams with two 
resonant W-boson propagators,
an expansion of the amplitude around the poles of the two resonant 
W bosons is a reasonable approach, which is also gauge-invariant.
In the double-pole approximation, the contributions are classified in
factorizable and non-factorizable corrections. 
The amplitudes of the factorizable corrections are 
composed of those for the on-shell W-pair production and
the on-shell W-decays  multiplied by the two propagators of the 
resonant W bosons.
All other corrections are called non-factorizable, because 
they do not factorize into a simple product of 
W-pair production and W decays.

As a first step of this work, the non-factorizable corrections of 
the processes $\Pe^+ \Pe^- \to 4\,\mathrm{fermions}$ are calculated
in double-pole approximation.
The non-factorizable corrections are implemented into an existing
Monte Carlo program and various distributions are studied. 
It turns out that the non-factorizable corrections 
are negligible with respect to the experimental accuracy of LEP2; 
however, they should become relevant for a future linear collider with 
higher luminosity.

A further building block of the 
radiative corrections to four-fermion production are 
the bremsstrahlung processes $\Pe^+ \Pe^- \to 4\,\mathrm{fermions}+\gamma$. 
These processes are of physics interest in their own right. 
For instance, the radiative processes can be used to 
obtain information on the  
quartic-gauge-boson couplings $\gamma \gamma \PW \PW$, 
$\gamma \PZ \PW \PW$, and 
$\gamma \gamma \PZ \PZ$, which are part of the tree-level amplitude.
The tree-level helicity amplitudes for the  
processes $\Pe^+ \Pe^- \to 4\,\mathrm{fermions}$ and 
$\Pe^+ \Pe^- \to 4\,\mathrm{fermions}+\gamma$
for all possible final-state fermions are calculated. 
A multi-channel Monte Carlo program
for both classes of processes is constructed. 
This is a particularly difficult task owing to the
very complex peaking structure of the differential cross section.

In order to include radiative corrections into the Monte Carlo program,
the infrared and collinear singularities must be extracted from the
bremsstrahlung process.
This is done by applying the dipole-subtraction method to 
four-fermion production.
This subtraction method has already been worked out in massless QCD for 
dimensional regularization.
In the Electroweak Standard Model it is more convenient to
regularize the amplitude with an infinitesimal photon mass and 
small fermion masses. 
Hence, the subtraction method is reformulated for 
mass regularization.

Finally, the doubly-resonant virtual corrections are implemented
into the Monte Carlo program for the processes
$\Pe^+ \Pe^- \to 4\,\mathrm{fermions}(+\ga)$. 
Therefore, the results of the 
non-factorizable corrections and the already existing
results of the on-shell W-pair production and on-shell W-decay are used.
For the real corrections the complete bremsstrahlung process 
$\Pe^+ \Pe^- \to 4\,\mathrm{fermions}+\gamma$ is taken into account.
All results are combined in a four-fermion generator,
which is the first Monte Carlo generator including 
the complete $\Oa$ corrections to 
the processes $\Pep \Pem \to \PW^+ \PW^- \to 4\,{\mathrm{fermions}}$
in double-pole approximation.
This generator is used to produce numerical results for 
the total cross section, angular, and invariant-mass distributions.

\chapter*{Zusammenfassung}
\addcontentsline{toc}{chapter}{Zusammenfassung}

Am Beschleuniger LEP2 wird zur Zeit die
$\PW$-Paar-Produktion experimentell untersucht.
Um genaue theoretische Vorhersagen f\"ur die Bestimmung der 
\PW-Boson-Masse und die Untersuchung der Drei-Eichboson-Kopplungen
zu erhalten, m\"ussen Strahlungskorrekturen 
ber\"ucksichtigt werden.
Weil die \PW-Bosonen sehr schnell in je zwei leichte Fermionen
zerfallen, werden in Wirklichkeit die Prozesse 
$\Pep \Pem \to 4\,\mathrm{Fermionen}$ untersucht.

Die vollst\"andigen Strahlungskorrekturen in $\Oa$ sind zu diesen Prozessen
noch nicht bekannt. 
Der Hauptbeitrag stammt von Diagrammen mit zwei resonanten 
\PW-Boson-Propagatoren.
Deshalb besteht ein naheliegender und auch eichinvarianter Ansatz darin,
die Amplitude um die Pole der beiden \PW-Bosonen zu entwickeln. 
In dieser Doppelpol\-n\"aher\-ung k\"onnen die Strahlungskorrekturen 
in faktorisierbare und nicht-faktorisierbare Korrekturen
klassifiziert werden.
Die Amplitude der faktorisierbaren Korrekturen setzt sich aus den Amplituden 
der \PW-Paar-Produktion, den beiden Amplituden der \PW-Boson-Zerf\"alle 
und den zwei Propagatoren der resonanten \PW-Bosonen zusammen.
Alle \"ubrigen Korrekturen werden mit nicht-faktorisierbar bezeichnet,
weil sie nicht aus einem einfachen Produkt von Beitr\"agen zur 
Produktion und zu den Zerf\"allen geschrieben werden k\"onnen.

In einem ersten Schritt wurden in der vorliegenden Arbeit 
die nicht-faktorisierbaren 
Korrekturen zu den Prozessen $\Pep \Pem \to 4\,\mathrm{Fermionen}$
in Doppelpoln\"aherung berechnet.
Sie wurden in ein existierendes
Monte-Carlo-Pro\-gramm eingebaut, und verschiedene Ver\-teil\-ung\-en 
wurden studiert.
Dabei zeigte sich, dass diese Korrekturen
vernachl\"assigbar gegen\-\"uber dem experimentellen Fehler
von LEP2 sind.
Jedoch werden sie f\"ur einen zuk\"unftigen Li\-ne\-ar\-beschleuniger 
mit h\"oherer Luminosit\"at voraussichtlich wichtig.

Ein weiterer notwendiger Bestandteil der Strahlungskorrekturen
zur Vier-Fermion-Pro\-duk\-tion sind die Bremsstrahlungsprozesse
$\Pep \Pem \to 4\,\mathrm{Fermionen} + \ga$. 
Mit diesen Prozessen k\"onnen auch die Vier-Eichboson-Kopplungen 
$\ga \ga \PW \PW$, $\ga \PZ \PW \PW$ und $\ga \ga \PZ \PZ$
studiert werden, die auf Born-Niveau enthalten sind.
Die Helizit\"ats-Amplituden zu den Prozessen
$\Pep \Pem \to 4\,$Fermionen und 
$\Pep \Pem \to 4\,$Fer\-mion\-en$ + \ga$
f\"ur alle Endzust\"ande wurden berechnet, und ein 
Monte-Carlo-Programm f\"ur beide Klassen von Prozessen geschrieben.
Die Schwierigkeit lag dabei in dem sehr komplexen 
und stark variierenden Verhalten des differentiellen 
Wirkungsquerschnittes.

Um Strahlungskorrekturen mit Hilfe eines Monte-Carlo-Programms zu berechnen,
m\"us\-sen die infraroten und kollinearen Singularit\"aten 
vom Bremsstrahlungsprozess abge\-spalten werden. 
Dazu wurde die Dipol-Subtraktionsmethode auf die Vier-Fermion-Produktion
angewandt.
Diese Subtraktionsmethode existierte in der Literatur f\"ur masselose QCD
und dimensionale Regularisierung.
F\"ur das Elektroschwache Standardmodell werden normalerweise
die Singularit\"aten mit einer infinitesimalen Photonmasse und 
kleinen Fermionmassen regularisiert.
Daher wurde die Subtraktionsmethode f\"ur die Massenregularisierung
umgeschrieben.

Alle doppeltresonanten virtuellen Korrekturen wurden in das vorher
erw\"ahnte Monte-Carlo-Programm f\"ur die Prozesse 
$\Pep \Pem \to 4\,\mathrm{Fermionen}(+\ga)$ eingebaut. 
Dazu wurden die nicht-faktorisierbaren Korrekturen und die schon
existierenden Ergebnisse f\"ur die W-Paar-Produktion 
und W-Zerf\"alle verwendet.
F\"ur die reellen Korrekturen wurde der komplette Bremsstrahlungsprozess 
$\Pep \Pem \to 4\,$Fermionen$ + \ga$ ber\"ucksichtigt.
Dieses Programm ist der erste Monte-Carlo-Generator, der alle 
Strahlungskorrekturen in $\Oa$ zur Vier-Fermion-Produk\-tion
in Doppelpoln\"aherung beinhaltet.
Mit ihm wurden numerische Ergebnisse f\"ur den totalen Wirkungsquerschnitt,
Winkelverteilungen und invariante-Massen-Verteilungen erzeugt.

\chapter{Introduction}
\pagenumbering{arabic}

The Glashow--Salam--Weinberg model \cite{GSW}, 
known as the Electroweak Standard Model (SM),
is very successful in describing electroweak phenomena.
Since the SM is a spontaneously broken gauge theory,
it is renormalizable \cite{tH71} and hence observables can
be, in principle, calculated to any finite order in perturbation theory.

An important feature of the SM lies in 
the appearance of elementary gauge-boson-self-interactions
resulting from the non-abelian structure of the gauge group.
The $\ga\PW\PW$ and $\PZ\PW\PW$ vertices can be studied in detail 
at the $\Pep \Pem$ collider LEP2 \cite{LEP2TGC97}.
Beside the investigation of the triple-gauge-boson-couplings,
LEP2 also allows for a precise determination of the $\PW$-boson mass.
Two methods are used \cite{LEP2MW97,Be96}:
the measurement of the total cross section near threshold 
and the reconstruction method, where 
the Breit-Wigner resonance shape is reconstructed 
from the decay products of the $\PW$ bosons.

LEP2 is operating above the $\PW$-pair production threshold and
produces about $10^4$ $\PW$~pairs. 
Hence, the typical experimental accuracy is of the order of
one to a few per cent. 
The accuracy of the $\PW$-boson mass measurement is expected 
to be $\lsim 50 \MeV$ at LEP2 \cite{LEP2MW97} and about $15 \MeV$ 
for a future linear collider \cite{NLC}.
This experimental accuracy should be matched or better exceeded by the
precision of the theoretical predictions.  
Since $\PW$~bosons decay very rapidly into light fermions, 
the actual reaction under investigation 
is $\Pep \Pem \to \PW^+ \PW^- \to 4\,\mathrm{fermions}$. 

\bfi
\begin{center}
\setlength{\unitlength}{1pt}
\begin{picture}(420,150)(0,-20)
\put(20,-8){
\begin{picture}(150,100)(0,0)
\ArrowLine(35,70)( 5, 80)
\ArrowLine( 5,20)(35, 30)
\ArrowLine(35,30)(35,70)
\Photon(35,30)(90,20){2}{6}
\Photon(35,70)(90,80){-2}{6}
\Vertex(35,70){2.0}
\Vertex(35,30){2.0}
\Vertex(90,80){2.0}
\Vertex(90,20){2.0}
\ArrowLine(90,80)(120, 95)
\ArrowLine(120,65)(90,80)
\ArrowLine(120, 5)( 90,20)
\ArrowLine( 90,20)(120,35)
\put(55,82){$\PW$}
\put(55,10){$\PW$}
\put( 20,50){$\nu_\Pe$}
\put(-15,75){$\mathrm e^+$}
\put(-15,15){$\mathrm e^-$}
\put(125,90){$f_1$}
\put(125,65){$\bar f_2$}
\put(125,30){$f_3$}
\put(125, 5){$\bar f_4$}
\end{picture}
}
\put(245,-8){
\begin{picture}(150,100)(0,0)
\ArrowLine( 15,50)(-20, 75)
\ArrowLine(-20,20)( 15, 50)
\Photon(15,50)(60,50){2}{5}
\Photon(60,50)(90,20){-2}{5}
\Photon(60,50)(90,80){2}{5}
\Vertex(15,50){2.0}
\Vertex(60,50){2.0}
\Vertex(90,80){2.0}
\Vertex(90,20){2.0}
\ArrowLine(90,80)(120, 95)
\ArrowLine(120,65)(90,80)
\ArrowLine(120, 5)( 90,20)
\ArrowLine( 90,20)(120,35)
\put(30,58){$\PZ,\ga$}
\put(62,70){$\PW$}
\put(62,18){$\PW$}
\put(-35,75){$\mathrm e^+$}
\put(-35,15){$\mathrm e^-$}
\put(125,90){$f_1$}
\put(125,65){$\bar f_2$}
\put(125,30){$f_3$}
\put(125, 5){$\bar f_4$}
\end{picture}
}
\end{picture}
\end{center}
\caption{Diagrams with two resonant 
$\mathrm W$-boson propagators contributing to 
$\eeffff$}
\label{fi:CC03}
\efi

In the LEP2 energy region, the lowest-order cross
section is dominated by the diagrams that involve two resonant
$\PW$~bosons, as shown in \reffi{fi:CC03}. 
All other lowest-order diagrams are typically suppressed
by a factor $\Ga_\PW/\MW\approx 2.5 \%$, but may be enhanced in certain 
phase-space regions. 
Since all these contributions are required at the one-per-cent level,
the complete lowest-order matrix element has to be taken into account.  

Furthermore, the implementation of the finite width of unstable 
particles, such as the $\PW$ bosons, has to be done properly. 
A finite width is necessary in the phase-space region where 
the unstable particle becomes resonant, \ie nearly on shell,
otherwise the cross section has a non-integrable unphysical singularity.
The finite width is naturally introduced via Dyson summation,
where the width arises from the imaginary parts of the resummed  
self-energy diagrams: 
\begin{eqnarray}
\label{eq:Dyson}
\frac{1}{k^2-M_0^2} 
\sum\limits_{n=0}^{\infty} \left[\frac{-\Sigma_\PV(k^2)}{k^2-M_0^2}\right]^n
&=&\frac{1}{k^2-M_0^2+\Sigma_\PV(k^2)},
\end{eqnarray}
where $M_0$ symbolizes the bare mass of the unstable particle,  
and $\Sigma_\PV$ denotes the one-particle-irreducible self-energy.

However, since only a part of higher-order corrections are 
included in this way, the whole result is gauge-dependent,
and wrong results can be obtained in certain phase-space regions 
\cite{Ba95,Ku95,Ar95,Be97c}.
The reason is that Ward identities are violated and, hence, 
gauge breaking terms can be amplified in the presence of small scales, or 
unitarity cancellations do not take place properly. 
Ward identities are, in particular, crucial for processes with
nearly on-shell virtual photons or for the production of longitudinal 
polarized gauge bosons at high energies.

For the calculation of the tree-level processes $\eeffff$ and $\eeffffg$,
three schemes for the implementation of the finite $\PW$-boson width 
are compared in \refch{ch:treelevel}:
the constant-width scheme, where the imaginary part
of the self-energy is replaced by a constant,
the running-width scheme with the naive running
$\Gamma (k^2)= \GV \theta(k^2) k^2 /\MV^2$,
and the complex-mass scheme, where the boson masses 
are replaced by complex masses in all couplings and propagators and, 
in particular, in the definition of the weak mixing angle.
In general, the first two schemes violate Ward identities \cite{Ar95,Be97c}, 
while the third scheme fulfils all Ward identities.

A gauge-invariant approach for the introduction of the 
finite width is the {\it pole expansion} \cite{polescheme,Ae94}
outlined in the following.
The complete matrix element for a process, with 
an unstable particle in the intermediate state, can be written as 
\begin{eqnarray}
\label{eq:pole}
\M &=&\frac{r}{k^2-M^2}+n,
\end{eqnarray}
where the residue is denoted by $r$, and the mass of the unstable particle by 
$M$. 
The symbol $n$ summarizes all terms that are regular at $k^2=M^2$.
Both the location of the complex pole, which corresponds to the mass of the 
unstable particle, 
and the residue are gauge-invariant quantities.
For stable particles the mass lies on the real axis,
for unstable particles in the complex plane.

In perturbation theory, the complex mass $M$ is determined 
by the location of the pole after Dyson summation \refeq{eq:Dyson}:
\begin{eqnarray}
\label{eq:complexmass}
M^2=M_0^2-\Sigma_\PV (M^2)&=&\MV^2-\ri \MV \GV,
\end{eqnarray}
where $\GV$ and $\MV$ denote the finite width and the real mass of the 
unstable particle, respectively.
After expanding the self-energy about $M^2$, the inverse propagator reads
\begin{eqnarray}
\nn
k^2-M_0^2+\Sigma_\PV(k^2)
&=&[1+\Sigma^\prime_\PV(M^2)](k^2-M^2)+\O \left((k^2-M^2)^2\right),
\end{eqnarray}
and the matrix element can be rewritten in the form \cite{Ae94}: 
\begin{eqnarray}
\label{eq:poleexpansion}
\nn
\M&=&\frac{R(k^2,\theta)}{k^2-M_0^2+\Sigma_\PV(k^2)}+N(k^2,\theta)
=\frac{R(M^2,\theta)}{[1+\Sigma_\PV^\prime (M^2)](k^2-M^2)}\\
&&{}
+\left[ \frac{R(k^2,\theta)}{k^2-M_0^2+\Sigma_\PV(k^2)}
         -\frac{R(M^2,\theta)}{[1+\Sigma_\PV^\prime (M^2)](k^2-M^2)}\right]
+N(k^2,\theta),
\end{eqnarray}
where $N$ includes the non-resonant diagrams, and 
$\theta$ summarizes all kinematic variables, except for 
the invariant mass $k^2$ of the unstable particle.
Since the complete amplitude is gauge-invariant, the single terms 
of the Laurent expansion about the squares of the complex mass 
are also gauge-invariant.
The first term on the right-hand side of \refeq{eq:poleexpansion} 
corresponds to the leading term in a Laurent expansion about 
$k^2=M^2$, and dominates the cross section in the resonance region.
The remaining terms are finite in the limit $k^2\to M^2$.
Therefore, a reasonable and gauge-invariant approximation is to neglect 
the non-leading terms and to keep only the resonant term.
This simplifies the calculation considerably, since all obviously
non-resonant diagrams can be left out from the calculation 
from the beginning.

In \refch{ch:nfc} and \ref{ch:radcorr}, the pole scheme is applied 
to the radiative corrections of four-fermion production.
Radiative corrections are, in general, required in the 
theoretical predictions for four-fermion production,
in order to match the accuracy of LEP2 of about one per cent.
The full $\Oa$ calculation involves $10^3$-$10^4$ diagrams and is, therefore, 
extremely complicated. 
Since the non-doubly-resonant radiative corrections are 
of the order of  $\al \GW \ln (\dots)/(\pi \MW)\approx 0.1 \%$,
the restriction to the doubly-resonant corrections is a reasonable approach.

After the introduction of the finite $\PW$-boson width, 
the matrix element reads
\begin{eqnarray}
\M&=&\frac{R_{+-}(k_+^2,k_-^2,\theta)}{(k_+^2-M^2)(k_-^2-M^2)}
+\frac{R_+(k_+^2,k_-^2,\theta)}{k_+^2-M^2}
+\frac{R_-(k_+^2,k_-^2,\theta)}{k_-^2-M^2}
+N(k_+^2,k_-^2,\theta),\quad
\end{eqnarray}
where $k_\pm$ are the momenta of the resonant virtual $\PW^\pm$ bosons and
the factor $[1+\Sigma_\PV^\prime (M^2)]$ is already included in the 
definition of $R_{+-}$, $R_+$, and $R_-$%
\footnote{We use the the on-shell renormalization scheme of \citere{De93}, 
where $\Re\{\Sigma_\PV^\prime (\MV^2)\}=0$. 
In this renormalization scheme, the deviation of the self-energy 
at the complex pole $\Sigma_\PV^\prime(M^2)$ yields only $\O(\al^2)$ 
corrections. Hence, the factor $[1+\Sigma_\PV^\prime (M^2)]$ is neglected in 
the following chapters.}.
In double-pole approximation, the matrix element is expanded about
the squares of the $\PW$-boson masses and all non-doubly-resonant terms are 
neglected:
\begin{eqnarray}
\label{eq:ee4fpoleexpansion}
\M^{\mathrm{DPA}}&=&
\frac{R_{+-}(M^2,M^2,\theta)}{(k_+^2-M^2)(k_-^2-M^2)}.
\end{eqnarray}

The double-pole approximation simplifies the calculation considerably, 
since only dia\-grams with two resonant $\PW$-boson propagators 
have to be calculated. 
Furthermore, a large part of the doubly-resonant radiative corrections
are included in the factorizable corrections. 
The factorizable corrections are composed of the 
on-shell $\PW$-pair production, the two resonant $\PW$-boson propagators,
and the two on-shell $\PW$-boson decays.
The remaining radiative corrections are called non-factorizable and
are explicitly calculated and discussed in \refch{ch:nfc}.

The double-pole approximation is a good approximation if the doubly-resonant
contributions dominate the cross section. 
If these are suppressed, as close to the $\PW$-pair production threshold,
the other contributions become important.
This is also true if non-doubly-resonant terms are enhanced, 
as \eg by nearly on-shell photons.
These terms can be suppressed by applying appropriate cuts on the phase space.

Note that the pole expansion only works if the on-shell limit exists.
Since the non-factorizable corrections, given in \refse{seanres},
involve on-shell-divergent terms, like $\ln (k_\pm^2-M^2)$, the 
pole expansion of these corrections has not such a simple form as in 
\refeq{eq:ee4fpoleexpansion}.
Thus, the non-factorizing corrections are calculated for 
off-shell $\PW$ bosons, while the limit $k_\pm^2\to M^2$ is performed
whenever possible. 

With the results of \refch{ch:treelevel} and \ref{ch:nfc},
a Monte Carlo generator is constructed
in \refch{ch:radcorr} that includes all doubly-resonant $\Oa$ radiative 
corrections to four-fermion production.
More precise, this generator includes the complete tree-level matrix element,
the virtual corrections in double-pole approximation, and the 
complete bremsstrahlung process.
The cancellation of the soft and collinear singularities 
is achieved within the subtraction method as discussed in 
\refse{se:subtraction}.
Although the real non-factorizable corrections are calculated in
\refch{ch:nfc} in double-pole approximation, the complete
bremsstrahlung process is taken into account in \refch{ch:radcorr}.
In this way, the problem of overlapping resonances is avoided 
(see \refse{se:overlappingresonances}).
The Monte Carlo program of \refch{ch:radcorr} is the first generator
that includes the complete $\Oa$ corrections to the processes
$\Pep \Pem \to \PW^+ \PW^- \to 4\,{\mathrm{fermions}}$
in double-pole approximation.

\chapter{\boldmath Tree-level processes 
$\mathrm e^+ \mathrm e^- \to 4 f (\,+\,\gamma)$}
\label{ch:treelevel}

While the most important process at LEP2 
for the studies of the gauge sector in the Electroweak Standard Model
is certainly $\Pep\Pem\to\PWp\PWm\to4f$, many
other reactions have now become accessible. Besides the  
4-fermion-production processes, including single \PW-boson production,
single \PZ-boson production, or \PZ-boson-pair production, LEP2 
and especially a future linear collider allow us
to investigate another class of processes, namely $\eeffffg$.

The physical interest in the processes $\eeffffg$ is
twofold. First of all, they are an important building block for the
radiative corrections to $\Pep\Pem\to4f$, and their effect must be
taken into account in order to get precise predictions for the
observables that are used for the measurement of the \PW-boson mass
and the triple-gauge-boson couplings. 
On the other hand, those processes themselves involve interesting
physics. They include, in particular, triple-gauge-boson-production
processes such as $\PWp\PWm\ga$, $\PZ\PZ\ga$, or $\PZ\ga\ga$ production
and can therefore be used to obtain information on the quartic
gauge-boson couplings $\ga\ga WW$, $\ga ZWW$, and $\ga\ga ZZ$.
While only a few events of this kind are expected at LEP2,
these studies can be performed in more detail at future linear
$\Pep\Pem$ colliders \cite{Be92}. 

Some results for $\eeffffg$ with an observable photon already exist in
the literature. In \citeres{Ae91,Ae91a} the contributions to the
matrix elements involving two resonant \PW~bosons have been calculated
and implemented into a Monte Carlo generator.  This generator has been
extended to include collinear bremsstrahlung \cite{vO94} and used to
discuss the effect of hard photons at LEP2 \cite{vO96}. The complete
cross section for the process $\Pep\Pem\to\Pu\,\Pdbar\,\Pem\Pnebar\ga$ has
been discussed in \citere{Fu94}.  In \citere{Ca97}, the complete
matrix elements for the processes $\eeffffg$ have been calculated
using an iterative numerical algorithm without referring to Feynman
diagrams. We are, however, interested in explicit analytical results
on the amplitudes for various reasons.  In particular, we want to have
full control over the implementation of the finite width of the
virtual vector bosons and to select single diagrams, such as the
doubly-resonant ones.
No results for $\eeffffg$ with $\Pep\Pem$ pairs in the final
state have been published in the past.
The results of this section are published in \citere{De99} and
agree very well with the recent calculations of \citere{Je99}, 
where finite-mass effects due to nearly collinear photon emission 
are discussed for the process 
$\Pep \Pem \to \Pu \, \Pdbar \, \mu^- \bar\nu_\mu \ga$. 

In order to perform the calculation as efficient as possible we have
reduced all processes to a small number of generic contributions. For
$\eeffff$, the calculation is similar to the one in \citere{Be94},
and the generic contributions correspond to individual Feynman diagrams.
In the case of $\eeffffg$ we have combined groups of diagrams 
in such a way that the
resulting generic contributions can be classified in the same way as
those for $\eeffff$. As a consequence, the generic contributions are
individually gauge-invariant with respect to the external photon. The
number and the complexity of diagrams in the generic contributions for
$\eeffffg$ 
has been reduced by using a non-linear gauge-fixing condition
for the W-boson field
\cite{nlgauge}. In this way, many cancellations between diagrams are
avoided, without any 
further algebraic manipulations. Finally, for the helicity amplitudes
corresponding to the generic contributions concise results 
have been obtained by using the 
Weyl--van~der~Waerden formalism (see
\citere{wvdw} and references therein).

After the matrix elements have been calculated, the finite widths 
of the resonant particles have to be introduced. We have done this in 
different
ways and compared the different treatments for $\eeffff$ and
$\eeffffg$. In particular, we have discussed a ``complex-mass
scheme'', which preserves all Ward identities and is still rather
simple to apply. 

The matrix elements to $\eeffff$ and $\eeffffg$
exhibit a complex peaking behaviour owing to propagators
of massless particles and Breit--Wigner resonances,
so that the integration over the
8- and 11-dimensional phase 
spaces, respectively, is not straightforward.
In order to obtain numerically stable results,
we adopt the multi-channel integration
method \cite{Be94,Multichannel} and 
reduce  the Monte Carlo error by the adaptive weight optimization
procedure described in \citere{Kl94}.
In the multi-channel approach, 
we define a suitable mapping of random numbers into phase-space variables
for each arising propagator structure.
These variables are generated according to distributions that
approximate this specific peaking behaviour of the integrand.
For $\eeffff$ and $\eeffffg$ we identify up to 128 and 928 channels,
respectively, which necessitates an efficient and generic procedure
for the phase-space generation.  

\section{Analytical results}
\label{se:anres}
\subsection{Notation and conventions}
\label{se:not&con}

We consider reactions of the types
\beqar\label{eq:eeffff}
\Pep(p_+,\si_+)+\Pem(p_-,\si_-) &\to& 
f_1(k_1,\si_1)+\bar f_2(k_2,\si_2)+f_3(k_3,\si_3)+\bar f_4(k_4,\si_4),
\qquad\\
\nn
\Pep(p_+,\si_+)+\Pem(p_-,\si_-) &\to& 
f_1(k_1,\si_1)+\bar f_2(k_2,\si_2)+f_3(k_3,\si_3)+\bar f_4(k_4,\si_4)
+\gamma(k_5,\lambda).\\
\label{eq:eeffffg}
\eeqar
The arguments label the momenta $p_\pm$, $k_i$ and helicities
$\si_i=\pm1/2$, $\la=\pm1$ of the corresponding particles. We often
use only the signs to denote the helicities. The fermion masses are
neglected everywhere.

For the Feynman rules we use the conventions of \citere{De93,sm}. In
particular, all fields and momenta are incoming. It is
convenient to use a non-linear gauge-fixing term \cite{nlgauge}
of the form
\beqar\label{eq:nlgauge}
{\cal L}_{\mathrm{fix}} &=& -\left| \partial^\mu W^+_\mu + \ri e
                (A^\mu - \frac{\cw}{\sw} Z^\mu) W^+_\mu
               -\ri \MW \phi^+ \right|^2                        \nn\\*
            & &{}-  \frac{1}{2} (\partial^\mu Z_\mu - \MZ \chi)^2
              - \frac{1}{2} (\partial^\mu A_\mu)^2 \;,
\eeqar
where $\phi^\pm$ and $\chi$ are the would-be Goldstone bosons of the 
$W^\pm$ and $Z$ fields, respectively.
With this choice, the $\phi^\pm W^\mp A$ vertices vanish, and the
bosonic couplings that are relevant for $\eeffffg$ read
\beqar
\setlength{\unitlength}{1pt}
\label{fr:vvv}
\barr{l}
\begin{picture}(90,80)(-50,-38)
\Text(-45,3)[lb]{$V_{\mu},k_{V}$}
\Text(35,27)[rb]{$W^+_{\nu},k_{+}$}
\Text(35,-27)[rt] {$W^-_{\rho},k_{-}$}
\Vertex(0,0){2}
\Photon(0,0)(35,25){2}{3.5}
\Photon(0,0)(35,-25){2}{3.5}
\Photon(0,0)(-45,0){2}{3.5}
\end{picture} 
\earr
&&\barr{l}
=  -\ri e g_{VWW}\left[
g_{\nu\rho}(k_- -k_+)_\mu-2g_{\mu\nu}k_{V,\rho}+2g_{\mu\rho}k_{V,\nu} \right],
\nn\\
\earr\\
\label{fr:vvvv}
\barr{l}
\begin{picture}(90,80)(-50,-36)
\Text(-35,27)[lb]{$A_{\mu}$}
\Text(-35,-27)[lt]{$V_{\nu}$}
\Text(35,27)[rb]{$W^+_{\rho}$}
\Text(35,-27)[rt]{$W^-_{\sigma}$}
\Vertex(0,0){2}
\Photon(0,0)(35,25){2}{3.5}
\Photon(0,0)(35,-25){2}{3.5}
\Photon(0,0)(-35,25){2}{3.5}
\Photon(0,0)(-35,-25){2}{3.5}
\end{picture} 
\earr
&&\barr{l}
= -2\ri e g_{VWW} \, g_{\mu\nu}g_{\rho\sigma},
\earr
\eeqar
with $V=A,Z$, and the coupling factors
\beq
g_{AWW} = 1, \qquad g_{ZWW} = -\frac{\cw}{\sw}.
\eeq
Note that the gauge-boson propagators have the same simple form as in
the 't~Hooft--Feynman gauge, i.e.\ they are proportional to the metric
tensor $g_{\mu\nu}$.
This gauge choice eliminates some diagrams and simplifies others owing
to the simpler structure of the photon--gauge-boson couplings.

The vector-boson--fermion--fermion couplings have the usual form
\beq
\label{fr:Vff}
\barr{l}
\begin{picture}(90,80)(-50,-36)
\Text(-45,5)[lb]{$V_{\mu}$}
\Text(35,27)[rb]{$\bar{f}_{i}$}
\Text(35,-27)[rt]{$f_{j}$}
\Vertex(0,0){2}
\ArrowLine(0,0)(35,25)
\ArrowLine(35,-25)(0,0)
\Photon(0,0)(-45,0){2}{3.5}
\end{picture} 
\earr
\barr{l}
= \disp\ri e \gamma_\mu \sum_\si g^\sigma_{V\bar f_i f_j}\omega_\sigma,
\earr
\eeq
where
$\omega_\pm=(1\pm\gamma_5)/2$.
The corresponding coupling factors read 
\beq
g^\sigma_{A\bar f_i f_i} = -Q_i, \qquad
g^\sigma_{Z\bar f_i f_i} = 
-\frac{\sw}{\cw}Q_i+\frac{I^3_{{\mathrm{w}},i}}{\cw\sw}\delta_{\sigma-}, 
\qquad
g^\sigma_{W\bar f_i f'_i} = \frac{1}{\sqrt{2}\sw}\delta_{\sigma-},
\eeq
where $Q_i$ and $I^3_{{\mathrm{w}},i}=\pm{1/2}$ denote the
relative charge and the weak isospin of the fermion $f_i$,
respectively, and $f'_i$ is the weak-isospin partner of $f_i$.
The colour factor of a fermion $f_i$ is denoted by 
$N^{\mathrm{c}}_{f_i}$, i.e.\
$N^{\mathrm{c}}_{\mathrm{lepton}}=1$ and
$N^{\mathrm{c}}_{\mathrm{quark}}=3$.
 
\subsection[Classification of final states for ${\mathrm e^+ \mathrm e^- \to 4\,f}$]{\boldmath Classification of final states for ${\mathrm e^+ \mathrm e^- \to 4\,f}$}
\label{se:pclass}

The final states for $\eeffff$ have 
already been classified in
\citeres{Ba94,Be94,CERN9601mcgen}. We introduce a classification that
is very close to the one of \citeres{Ba94,CERN9601mcgen}.  It is based
on the production
mechanism, i.e.\ whether the 
reactions proceed via charged-current (CC), or
neutral-current (NC) interactions, or via both interaction types.  The
classification can be performed by considering the quantum numbers of
the final-state fermion pairs.  In the following, $f$ and $F$ denote
different fermions ($f\ne F$) that are neither electrons nor electron
neutrinos
($f,F \ne \Pem,\nu_\Pe$), and their weak-isospin partners are denoted
by $f'$ and $F'$, respectively. We find the following 11 classes of
processes (in parenthesis the corresponding classification of
\citere{CERN9601mcgen} is given):
\renewcommand{\labelenumi}{(\roman{enumi})}
\renewcommand{\labelenumii}{(\alph{enumii})}
\newcommand{\ientry}[2]{\mbox{\rlap{#1}\hspace*{5cm}#2}}
\begin{enumerate}
{\samepage
\item CC reactions:
\begin{enumerate}
\item \ientry{$\Pep\Pem \to f \bar f' F \bar F'$,}%
             {({\em CC11} family),} 
\item \ientry{$\Pep\Pem \to \nu_\Pe \Pep f \bar f'$,}%
             {({\em CC20} family),}
\item \ientry{$\Pep\Pem \to f \bar f' \Pem \bar\nu_\Pe$,}%
             {({\em CC20} family),}
\end{enumerate}
}
\item NC reactions:
\begin{enumerate}
\item \ientry{$\Pep\Pem \to f \bar f F \bar F$,}%
             {({\em NC32} family),}
\item \ientry{$\Pep\Pem \to f \bar f f \bar f$,}%
             {({\em NC4$\cdot$16} family),}
\item \ientry{$\Pep\Pem \to \Pem \Pep f \bar f$,}%
             {({\em NC48} family),}
\item \ientry{$\Pep\Pem \to \Pem \Pep \Pem \Pep$,}%
             {({\em NC4$\cdot$36} family),}
\end{enumerate}
\item Mixed CC/NC reactions:
\begin{enumerate}
\item \ientry{$\Pep\Pem \to f \bar f f' \bar f'$,}%
             {({\em mix43} family),}
\item \ientry{$\Pep\Pem \to \nu_\Pe \bar\nu_\Pe f \bar f$,}%
             {({\em NC21} family),}
\item \ientry{$\Pep\Pem \to \nu_\Pe \bar\nu_\Pe \nu_\Pe \bar\nu_\Pe$,}%
             {({\em NC4$\cdot$9} family),}
\item \ientry{$\Pep\Pem \to \nu_\Pe \bar\nu_\Pe \Pem \Pep$,}%
             {({\em mix56} family).}
\end{enumerate}
\end{enumerate}
The radiation of an additional photon does not change this
classification.

\subsection[Generic diagrams and amplitudes for 
${\mathrm e^+ \mathrm e^- \to 4\,f}$]
{\boldmath Generic diagrams and amplitudes for 
${\mathrm e^+ \mathrm e^- \to 4\,f}$}
\label{se:ee4f}

In order to explain and to illustrate our generic approach we first
list the results for $\eeffff$. All these processes can be composed
from only two generic diagrams, the abelian and non-abelian diagrams
shown in \reffi{ee4fdiags}. All external fermions $f_{a,\dots,f}$ are
assumed to be incoming, and the momenta and helicities are denoted by
$p_{a,\dots,f}$ and $\si_{a,\dots,f}$, respectively.
\bfi
\begin{center}
\setlength{\unitlength}{1pt}
\begin{picture}(420,150)(0,-20)
\Text(0,120)[lb]{a) abelian graph}
\Text(210,120)[lb]{b) non-abelian graph}
\put(20,-8){
\begin{picture}(150,100)(0,0)
\ArrowLine(35,70)( 5, 95)
\ArrowLine( 5, 5)(35, 30)
\ArrowLine(35,30)(35,70)
\Photon(35,30)(90,20){2}{6}
\Photon(35,70)(90,80){-2}{6}
\Vertex(35,70){2.0}
\Vertex(35,30){2.0}
\Vertex(90,80){2.0}
\Vertex(90,20){2.0}
\ArrowLine(90,80)(120, 95)
\ArrowLine(120,65)(90,80)
\ArrowLine(120, 5)( 90,20)
\ArrowLine( 90,20)(120,35)
\put(55,82){$V_1$}
\put(55,10){$V_2$}
\put( 20,50){$f_g$}
\put(-15,110){$\bar f_a(p_a,\si_a)$}
\put(-15,-10){$f_b(p_b,\si_b)$}
\put(125,90){$\bar f_c(p_c,\si_c)$}
\put(125,65){$f_d(p_d,\si_d)$}
\put(125,30){$\bar f_e(p_e,\si_e)$}
\put(125, 5){$f_f(p_f,\si_f)$}
\end{picture}
}
\put(245,-8){
\begin{picture}(150,100)(0,0)
\ArrowLine( 15,50)(-15, 65)
\ArrowLine(-15,35)( 15, 50)
\Photon(15,50)(60,50){2}{5}
\Photon(60,50)(90,20){-2}{5}
\Photon(60,50)(90,80){2}{5}
\Vertex(15,50){2.0}
\Vertex(60,50){2.0}
\Vertex(90,80){2.0}
\Vertex(90,20){2.0}
\ArrowLine(90,80)(120, 95)
\ArrowLine(120,65)(90,80)
\ArrowLine(120, 5)( 90,20)
\ArrowLine( 90,20)(120,35)
\put(30,58){$V$}
\put(62,70){$W$}
\put(62,18){$W$}
\put(-35,75){$\bar f_a(p_a,\si_a)$}
\put(-35,20){$f_b(p_b,\si_a)$}
\put(125,90){$\bar f_c(p_c,\si_c)$}
\put(125,65){$f_d(p_d,\si_a)$}
\put(125,30){$\bar f_e(p_e,\si_e)$}
\put(125, 5){$f_f(p_f,\si_f)$}
\end{picture}
}
\end{picture}
\end{center}
\caption{Generic diagrams for $\Pep\Pem\to 4f$}
\label{ee4fdiags}
\efi
The helicity amplitudes of these diagrams are calculated within the
Weyl--van~der~Waerden (WvdW) formalism following the conventions of 
\citere{wvdw} (see also references therein). 

\subsubsection{Leptonic and semi-leptonic final states}

We first treat purely leptonic and semi-leptonic final states. In this
case, none of the gauge bosons in the generic graphs of \reffi{ee4fdiags}
can be a gluon, and the colour structure trivially leads to a global
factor $N^{\mathrm{c}}_{f_1}N^{\mathrm{c}}_{f_3}$, which is equal to 1
or 3, after summing the squared amplitude over the colour degrees of 
freedom. The results for the generic amplitudes are
\beqar\label{genfun4fNC}
&& \hspace*{-3em}
{\cal M}^{\si_a,\si_b,\si_c,\si_d,\si_e,\si_f}_{V_1 V_2}
(p_a,p_b,p_c,p_d,p_e,p_f) 
\nn\\*
\hspace*{2em}
&=& -4e^4 \de_{\si_a,-\si_b} \de_{\si_c,-\si_d} \de_{\si_e,-\si_f} \, 
g^{\si_b}_{V_1\bar f_a f_g} g^{\si_b}_{V_2\bar f_g f_b}
g^{\si_d}_{V_1\bar f_c f_d} g^{\si_f}_{V_2\bar f_e f_f}
\nn\\ && {} \times
\frac{P_{V_1}(p_c+p_d) P_{V_2}(p_e+p_f)}{(p_b+p_e+p_f)^2} \,
A^{\si_a,\si_c,\si_e}_2(p_a,p_b,p_c,p_d,p_e,p_f),
\hspace*{3em}
\\[1em] 
&& \hspace*{-3em}
{\cal M}^{\si_a,\si_b,\si_c,\si_d,\si_e,\si_f}_{VWW}
(p_a,p_b,p_c,p_d,p_e,p_f) 
\nn\\*
&=& -4e^4 \de_{\si_a,-\si_b} \de_{\si_c,+} \de_{\si_d,-} 
\de_{\si_e,+} \de_{\si_f,-} \, 
(Q_c-Q_d)g_{VWW} g^{\si_b}_{V\bar f_a f_b} 
g^{-}_{W\bar f_c f_d} g^{-}_{W\bar f_e f_f}
\nn\\ && {} \times
P_V(p_a+p_b) P_W(p_c+p_d) P_W(p_e+p_f) \,
A^{\si_a}_3(p_a,p_b,p_c,p_d,p_e,p_f),
\label{genfun4fCC}
\eeqar
where 
the vector-boson propagators are abbreviated by
\beq
P_V(p) = \frac{1}{p^2-\MV^2}, \qquad V=A,Z,W,g, \qquad M_A = M_g = 0.
\eeq
(The case of the gluon is included for later convenience.)
The auxiliary functions $A^{\si_a,\si_c,\si_e}_2$ and $A^{\si_a}_3$ are
expressed in terms of WvdW spinor products,
\beqar
A^{{+}{+}{+}}_2(p_a,p_b,p_c,p_d,p_e,p_f) &=& 
\spac\cspbf(\cspbd\spbe+\cspdf\spef),
\nn\\
A^{{+}{+}{-}}_2(p_a,p_b,p_c,p_d,p_e,p_f) &=& 
A^{{+}{+}{+}}_2(p_a,p_b,p_c,p_d,p_f,p_e),
\nn\\
A^{{+}{-}{+}}_2(p_a,p_b,p_c,p_d,p_e,p_f) &=& 
A^{{+}{+}{+}}_2(p_a,p_b,p_d,p_c,p_e,p_f),
\nn\\
A^{{+}{-}{-}}_2(p_a,p_b,p_c,p_d,p_e,p_f) &=& 
A^{{+}{+}{+}}_2(p_a,p_b,p_d,p_c,p_f,p_e),
\nn\\
A^{{-},\si_c,\si_d}_2(p_a,p_b,p_c,p_d,p_e,p_f) &=& 
\left(A^{{+},-\si_c,-\si_d}_2(p_a,p_b,p_c,p_d,p_e,p_f)\right)^*,
\\[1em]
A^{+}_3(p_a,p_b,p_c,p_d,p_e,p_f) &=& \phantom{{}+{}}
  \cspbd\cspbf\spab\spce + \cspbd\cspdf\spae\spcd
\nn\\
&& {}
+ \cspbf\cspdf\spac\spef,
\nn\\
A^{-}_3(p_a,p_b,p_c,p_d,p_e,p_f) &=&
A^{+}_3(p_b,p_a,p_c,p_d,p_e,p_f).
\eeqar
The spinor products are defined by
\beq
\langle pq\rangle=\epsilon^{AB}p_A q_B
=2\sqrt{p_0 q_0} \,\Biggl[
{\mathrm{e}}^{-\ri\phi_p}\cos\frac{\theta_p}{2}\sin\frac{\theta_q}{2}
-{\mathrm{e}}^{-\ri\phi_q}\cos\frac{\theta_q}{2}\sin\frac{\theta_p}{2}
\Biggr], 
\eeq
where $p_A$, $q_A$ are the associated momentum spinors for the momenta
\beqar
p^\mu&=&p_0(1,\sin\theta_p\cos\phi_p,\sin\theta_p\sin\phi_p,\cos\theta_p),\nl
q^\mu&=&p_0(1,\sin\theta_q\cos\phi_q,\sin\theta_q\sin\phi_q,\cos\theta_q).
\eeqar 

Incoming fermions are turned into outgoing ones by crossing, which
is performed by inverting the corresponding fermion momenta and
helicities.
If the generic functions are called with negative momenta $-p$, $-q$, it
is understood that only the complex conjugate spinor products get the
corresponding sign change. We illustrate this by 
simple examples:
\beq
\begin{array}[b]{rllll}
A(p,q)   &= \langle pq \rangle
&=\phantom{-}A(p,-q)  
&=\phantom{-}A(-p,q)  
&=A(-p,-q), \\
B(p,q)   &= \langle pq \rangle^*
&=-B(p,-q)  
&=-B(-p,q)  
&=B(-p,-q). 
\end{array}
\eeq
We have checked 
the results for the generic diagrams against those of \citere{Be94}
and found 
agreement.

Using the results for the generic diagrams of \reffi{ee4fdiags}, the
helicity amplitudes for all possible processes involving 
six external fermions can be built up. 
It is convenient to construct first 
the amplitudes for the process types 
CC(a) and NC(a) (see \refse{se:pclass}) in terms of the generic functions
\refeq{genfun4fNC} and \refeq{genfun4fCC}, because these amplitudes 
are the basic subamplitudes of 
the other channels. The full amplitude for each
process type can be built up from those subamplitudes
by appropriate substitutions and linear combinations.

We first list the helicity amplitudes for the CC processes:
\beqar
\label{eq:ee4f_cca}
&& \hspace*{-4em}
{\cal M}^{\si_+,\si_-,\si_1,\si_2,\si_3,\si_4}_{\mathrm{CCa}}
(p_+,p_-,k_1,k_2,k_3,k_4) 
\nn\\ &=& 
{\cal M}^{\si_+,\si_-,-\si_1,-\si_2,-\si_3,-\si_4}_{WW}
(p_+,p_-,-k_1,-k_2,-k_3,-k_4)
\nn\\ && {}
+ \sum_{V=\gamma,Z} \Bigl[ \phantom{{}+{}}
{\cal M}^{\si_+,\si_-,-\si_1,-\si_2,-\si_3,-\si_4}_{VWW}
(p_+,p_-,-k_1,-k_2,-k_3,-k_4)
\nn\\ && \hphantom{{} + \sum_{V=\gamma,Z} \Bigl[}
+ {\cal M}^{-\si_1,-\si_2,\si_+,\si_-,-\si_3,-\si_4}_{VW}
(-k_1,-k_2,p_+,p_-,-k_3,-k_4)
\nn\\ && \hphantom{{} + \sum_{V=\gamma,Z} \Bigl[}
+ {\cal M}^{-\si_3,-\si_4,\si_+,\si_-,-\si_1,-\si_2}_{VW}
(-k_3,-k_4,p_+,p_-,-k_1,-k_2)
\nn\\ && \hphantom{{} + \sum_{V=\gamma,Z} \Bigl[}
+ {\cal M}^{-\si_1,-\si_2,-\si_3,-\si_4,\si_+,\si_-}_{WV}
(-k_1,-k_2,-k_3,-k_4,p_+,p_-)
\nn\\ && \hphantom{{} + \sum_{V=\gamma,Z} \Bigl[}
+ {\cal M}^{-\si_3,-\si_4,-\si_1,-\si_2,\si_+,\si_-}_{WV}
(-k_3,-k_4,-k_1,-k_2,p_+,p_-) \Bigr],
\\[1em]
&& \hspace*{-4em}
{\cal M}^{\si_+,\si_-,\si_1,\si_2,\si_3,\si_4}_{\mathrm{CCb}}
(p_+,p_-,k_1,k_2,k_3,k_4) 
\nn\\ &=& \phantom{{}+{}}
{\cal M}^{\si_+,\si_-,\si_1,\si_2,\si_3,\si_4}_{\mathrm{CCa}}
(p_+,p_-,k_1,k_2,k_3,k_4) 
\nn\\ && {}
- {\cal M}^{\si_+,-\si_2,\si_1,-\si_-,\si_3,\si_4}_{\mathrm{CCa}}
(p_+,-k_2,k_1,-p_-,k_3,k_4),
\\[1em]
&& \hspace*{-4em}
{\cal M}^{\si_+,\si_-,\si_1,\si_2,\si_3,\si_4}_{\mathrm{CCc}}
(p_+,p_-,k_1,k_2,k_3,k_4) 
\nn\\ &=& \phantom{{}+{}}
{\cal M}^{\si_+,\si_-,\si_1,\si_2,\si_3,\si_4}_{\mathrm{CCa}}
(p_+,p_-,k_1,k_2,k_3,k_4) 
\nn\\ && {}
- {\cal M}^{-\si_3,\si_-,\si_1,\si_2,-\si_+,\si_4}_{\mathrm{CCa}}
(-k_3,p_-,k_1,k_2,-p_+,k_4).
\eeqar
The ones for the NC processes are given by
\beqar
\label{eq:ee4f_nca}
&& \hspace*{-4em}
{\cal M}^{\si_+,\si_-,\si_1,\si_2,\si_3,\si_4}_{\mathrm{NCa}}
(p_+,p_-,k_1,k_2,k_3,k_4) 
\nn\\ &=& \sum_{V_1,V_2=\gamma,Z} \Bigl[ \phantom{{}+{}}
{\cal M}^{\si_+,\si_-,-\si_1,-\si_2,-\si_3,-\si_4}_{V_1 V_2}
(p_+,p_-,-k_1,-k_2,-k_3,-k_4)
\nn\\ && \hphantom{\sum_{V_1,V_2=\gamma,Z} \Bigl[}
+ {\cal M}^{\si_+,\si_-,-\si_3,-\si_4,-\si_1,-\si_2}_{V_1 V_2}
(p_+,p_-,-k_3,-k_4,-k_1,-k_2)
\nn\\ && \hphantom{\sum_{V_1,V_2=\gamma,Z} \Bigl[}
+ {\cal M}^{-\si_1,-\si_2,\si_+,\si_-,-\si_3,-\si_4}_{V_1 V_2}
(-k_1,-k_2,p_+,p_-,-k_3,-k_4)
\nn\\ && \hphantom{\sum_{V_1,V_2=\gamma,Z} \Bigl[}
+ {\cal M}^{-\si_3,-\si_4,\si_+,\si_-,-\si_1,-\si_2}_{V_1 V_2}
(-k_3,-k_4,p_+,p_-,-k_1,-k_2)
\nn\\ && \hphantom{\sum_{V_1,V_2=\gamma,Z} \Bigl[}
+ {\cal M}^{-\si_1,-\si_2,-\si_3,-\si_4,\si_+,\si_-}_{V_1 V_2}
(-k_1,-k_2,-k_3,-k_4,p_+,p_-)
\nn\\ && \hphantom{\sum_{V_1,V_2=\gamma,Z} \Bigl[}
+ {\cal M}^{-\si_3,-\si_4,-\si_1,-\si_2,\si_+,\si_-}_{V_1 V_2}
(-k_3,-k_4,-k_1,-k_2,p_+,p_-) \Bigl],
\\[1em]
&& \hspace*{-4em}
{\cal M}^{\si_+,\si_-,\si_1,\si_2,\si_3,\si_4}_{\mathrm{NCb}}
(p_+,p_-,k_1,k_2,k_3,k_4) 
\nn\\* &=& \phantom{{}+{}}
{\cal M}^{\si_+,\si_-,\si_1,\si_2,\si_3,\si_4}_{\mathrm{NCa}}
(p_+,p_-,k_1,k_2,k_3,k_4) 
\nn\\ && {}
- {\cal M}^{\si_+,\si_-,\si_3,\si_2,\si_1,\si_4}_{\mathrm{NCa}}
(p_+,p_-,k_3,k_2,k_1,k_4),
\\[1em]
&& \hspace*{-4em}
{\cal M}^{\si_+,\si_-,\si_1,\si_2,\si_3,\si_4}_{\mathrm{NCc}}
(p_+,p_-,k_1,k_2,k_3,k_4) 
\nn\\* &=& \phantom{{}+{}}
{\cal M}^{\si_+,\si_-,\si_1,\si_2,\si_3,\si_4}_{\mathrm{NCa}}
(p_+,p_-,k_1,k_2,k_3,k_4) 
\nn\\ && {}
- {\cal M}^{-\si_1,\si_-,-\si_+,\si_2,\si_3,\si_4}_{\mathrm{NCa}}
(-k_1,p_-,-p_+,k_2,k_3,k_4),
\\[1em]
&& \hspace*{-4em}
{\cal M}^{\si_+,\si_-,\si_1,\si_2,\si_3,\si_4}_{\mathrm{NCd}}
(p_+,p_-,k_1,k_2,k_3,k_4) 
\nn\\* &=& \phantom{{}+{}}
{\cal M}^{\si_+,\si_-,\si_1,\si_2,\si_3,\si_4}_{\mathrm{NCa}}
(p_+,p_-,k_1,k_2,k_3,k_4) 
\nn\\ && {}
- {\cal M}^{\si_+,\si_-,\si_3,\si_2,\si_1,\si_4}_{\mathrm{NCa}}
(p_+,p_-,k_3,k_2,k_1,k_4)
\nn\\ && {}
- {\cal M}^{-\si_1,\si_-,-\si_+,\si_2,\si_3,\si_4}_{\mathrm{NCa}}
(-k_1,p_-,-p_+,k_2,k_3,k_4)
\nn\\ && {}
- {\cal M}^{-\si_3,\si_-,\si_1,\si_2,-\si_+,\si_4}_{\mathrm{NCa}}
(-k_3,p_-,k_1,k_2,-p_+,k_4) 
\nn\\ && {}
+ {\cal M}^{-\si_1,\si_-,\si_3,\si_2,-\si_+,\si_4}_{\mathrm{NCa}}
(-k_1,p_-,k_3,k_2,-p_+,k_4) 
\nn\\ && {}
+ {\cal M}^{-\si_3,\si_-,-\si_+,\si_2,\si_1,\si_4}_{\mathrm{NCa}}
(-k_3,p_-,-p_+,k_2,k_1,k_4).
\eeqar
Finally, the helicity amplitudes for reactions of mixed CC/NC type read
\beqar
&& \hspace*{-4em}
{\cal M}^{\si_+,\si_-,\si_1,\si_2,\si_3,\si_4}_{\mathrm{CC/NCa}}
(p_+,p_-,k_1,k_2,k_3,k_4) 
\nn\\ &=& \phantom{{}+{}}
{\cal M}^{\si_+,\si_-,\si_1,\si_2,\si_3,\si_4}_{\mathrm{NCa}}
(p_+,p_-,k_1,k_2,k_3,k_4) 
\nn\\ && {}
- {\cal M}^{\si_+,\si_-,\si_1,\si_4,\si_3,\si_2}_{\mathrm{CCa}}
(p_+,p_-,k_1,k_4,k_3,k_2),
\\[1em]
&& \hspace*{-4em}
{\cal M}^{\si_+,\si_-,\si_1,\si_2,\si_3,\si_4}_{\mathrm{CC/NCb}}
(p_+,p_-,k_1,k_2,k_3,k_4) 
\nn\\ &=& \phantom{{}+{}}
{\cal M}^{\si_+,\si_-,\si_1,\si_2,\si_3,\si_4}_{\mathrm{NCa}}
(p_+,p_-,k_1,k_2,k_3,k_4) 
\nn\\ && {}
- {\cal M}^{-\si_3,-\si_4,\si_1,-\si_-,-\si_+,\si_2}_{\mathrm{CCa}}
(-k_3,-k_4,k_1,-p_-,-p_+,k_2),
\\[1em]
&& \hspace*{-4em}
{\cal M}^{\si_+,\si_-,\si_1,\si_2,\si_3,\si_4}_{\mathrm{CC/NCc}}
(p_+,p_-,k_1,k_2,k_3,k_4) 
\nn\\ &=& \phantom{{}+{}}
{\cal M}^{\si_+,\si_-,\si_1,\si_2,\si_3,\si_4}_{\mathrm{NCa}}
(p_+,p_-,k_1,k_2,k_3,k_4) 
\nn\\ && {}
- {\cal M}^{\si_+,\si_-,\si_3,\si_2,\si_1,\si_4}_{\mathrm{NCa}}
(p_+,p_-,k_3,k_2,k_1,k_4) 
\nn\\ && {}
- {\cal M}^{-\si_1,-\si_2,-\si_+,\si_4,\si_3,-\si_-}_{\mathrm{CCa}}
(-k_1,-k_2,k_3,-p_-,-p_+,k_4) 
\nn\\ && {}
+ {\cal M}^{-\si_1,-\si_4,-\si_+,\si_2,\si_3,-\si_-}_{\mathrm{CCa}}
(-k_1,-k_4,k_3,-p_-,-p_+,k_2) 
\nn\\ && {}
+ {\cal M}^{-\si_3,-\si_2,-\si_+,\si_4,\si_1,-\si_-}_{\mathrm{CCa}}
(-k_3,-k_2,k_1,-p_-,-p_+,k_4) 
\nn\\ && {}
- {\cal M}^{-\si_3,-\si_4,-\si_+,\si_2,\si_1,-\si_-}_{\mathrm{CCa}}
(-k_3,-k_4,k_1,-p_-,-p_+,k_2),
\\[1em]
&& \hspace*{-4em}
{\cal M}^{\si_+,\si_-,\si_1,\si_2,\si_3,\si_4}_{\mathrm{CC/NCd}}
(p_+,p_-,k_1,k_2,k_3,k_4) 
\nn\\ &=& \phantom{{}+{}}
{\cal M}^{\si_+,\si_-,\si_1,\si_2,\si_3,\si_4}_{\mathrm{NCa}}
(p_+,p_-,k_1,k_2,k_3,k_4) 
\nn\\ && {}
- {\cal M}^{-\si_3,\si_-,\si_1,\si_2,-\si_+,\si_4}_{\mathrm{NCa}}
(-k_3,p_-,k_1,k_2,-p_+,k_4) 
\nn\\ && {}
- {\cal M}^{\si_+,\si_-,\si_1,\si_4,\si_3,\si_2}_{\mathrm{CCa}}
(p_+,p_-,k_1,k_4,k_3,k_2) 
\nn\\ && {}
+ {\cal M}^{\si_+,-\si_4,\si_1,-\si_-,\si_3,\si_2}_{\mathrm{CCa}}
(p_+,-k_4,k_1,-p_-,k_3,k_2) 
\nn\\ && {}
+ {\cal M}^{-\si_3,\si_-,\si_1,\si_4,-\si_+,\si_2}_{\mathrm{CCa}}
(-k_3,p_-,k_1,k_4,-p_+,k_2) 
\nn\\ && {}
- {\cal M}^{-\si_3,-\si_4,\si_1,-\si_-,-\si_+,\si_2}_{\mathrm{CCa}}
(-k_3,-k_4,k_1,-p_-,-p_+,k_2).
\eeqar
The relative signs between contributions of the basic subamplitudes 
${\cal M}_{\mathrm{CCa}}$ and ${\cal M}_{\mathrm{NCa}}$ to the full matrix
elements account for the sign changes resulting from interchanging
external fermion lines.

For the CC reactions, the amplitudes $\M_{\mathrm{CCa}}$ are the
smallest gauge-invariant subset of diagrams \cite{Bo99}. In the
case of NC reactions, the amplitudes $\M_{\mathrm{NCa}}$ are composed of
three separately gauge-invariant subamplitudes consisting of the first
two lines, the two lines in the middle, and the last two lines of
\refeq{eq:ee4f_nca}.

\subsubsection{Hadronic final states}
\label{se:hadfinstat}

\begin{sloppypar}
  Next we inspect purely hadronic final states, i.e.\ the cases where
  all final-state fermions $f_i$ are quarks. This concerns only the
  channels CC(a), NC(a), NC(b), and CC/NC(a) given in
  \refse{se:pclass}. The colour structure of the quarks leads to two
  kinds of modifications. Firstly, the summation of the squared
  amplitudes over the colour degrees of freedom can become
  non-trivial, and secondly, the possibility of virtual-gluon exchange
  between the quarks has to be taken into account.  More precisely,
  there are diagrams of type (a) in \reffi{ee4fdiags} in which one of
  the gauge bosons $V_{1,2}$ is a gluon. The other gauge boson of
  $V_{1,2}$ can only be a photon or Z~boson, since this boson has to
  couple to the incoming $\Pep\Pem$ pair. Consequently, there is an
  impact of gluon-exchange diagrams only for the channels NC(a),
  NC(b), and CC/NC(a), but not for CC(a).  This can be easily seen by
  inspecting the generic diagrams in \reffi{ee4fdiags}: the presence
  of a gluon exchange requires two quark--antiquark pairs $q\bar q$ in
  the final state.
\end{sloppypar}

We first inspect the colour structure of the purely electroweak
diagrams. Since the colour structure of each diagram contributing to the
basic channels CC(a) and NC(a) is the same, the corresponding amplitudes
factorize into a simple colour part and the ``colour-singlet
amplitudes'' ${\cal M}_{\mathrm{CCa}}$ and ${\cal M}_{\mathrm{NCa}}$,
given in \refeq{eq:ee4f_cca} and \refeq{eq:ee4f_nca}, respectively.
The amplitudes for NC(b) and CC/NC(a) are composed from the ones of CC(a)
and NC(a) in a way that is analogous to the singlet case, but now the
colour indices $c_i$ of the quarks $f_i$ have to be taken into account.
Indicating the electroweak amplitudes for fully hadronic final states 
by ``had,ew'', and writing colour indices explicitly, we get
\beqar
\lefteqn{
{\cal M}^{\si_+,\si_-,\si_1,\si_2,\si_3,\si_4}
_{{\mathrm{CCa,had,ew,}}c_1,c_2,c_3,c_4}
(p_+,p_-,k_1,k_2,k_3,k_4) } \hspace*{7em} &&
\nn\\
&=& \phantom{{}+{}}
{\cal M}^{\si_+,\si_-,\si_1,\si_2,\si_3,\si_4}_{\mathrm{CCa}}
(p_+,p_-,k_1,k_2,k_3,k_4) \, \delta_{c_1 c_2} \delta_{c_3 c_4},
\\[1em]
\lefteqn{
{\cal M}^{\si_+,\si_-,\si_1,\si_2,\si_3,\si_4}
_{{\mathrm{NCa,had,ew,}}c_1,c_2,c_3,c_4}
(p_+,p_-,k_1,k_2,k_3,k_4) } \hspace*{7em} &&
\nn\\
&=& \phantom{{}+{}}
{\cal M}^{\si_+,\si_-,\si_1,\si_2,\si_3,\si_4}_{\mathrm{NCa}}
(p_+,p_-,k_1,k_2,k_3,k_4) \, \delta_{c_1 c_2} \delta_{c_3 c_4},
\\[1em]
\lefteqn{
{\cal M}^{\si_+,\si_-,\si_1,\si_2,\si_3,\si_4}
_{{\mathrm{NCb,had,ew,}}c_1,c_2,c_3,c_4}
(p_+,p_-,k_1,k_2,k_3,k_4) } \hspace*{7em} &&
\nn\\* 
&=& \phantom{{}+{}}
{\cal M}^{\si_+,\si_-,\si_1,\si_2,\si_3,\si_4}_{\mathrm{NCa}}
(p_+,p_-,k_1,k_2,k_3,k_4) \, \delta_{c_1 c_2} \delta_{c_3 c_4}
\nn\\ && {}
- {\cal M}^{\si_+,\si_-,\si_3,\si_2,\si_1,\si_4}_{\mathrm{NCa}}
(p_+,p_-,k_3,k_2,k_1,k_4) \, \delta_{c_3 c_2} \delta_{c_1 c_4},
\label{eq:ee4f_ncbhadew}
\\[1em]
\lefteqn{
{\cal M}^{\si_+,\si_-,\si_1,\si_2,\si_3,\si_4}
_{{\mathrm{CC/NCa,had,ew,}}c_1,c_2,c_3,c_4}
(p_+,p_-,k_1,k_2,k_3,k_4) } \hspace*{7em} &&
\nn\\*
&=& \phantom{{}+{}}
{\cal M}^{\si_+,\si_-,\si_1,\si_2,\si_3,\si_4}_{\mathrm{NCa}}
(p_+,p_-,k_1,k_2,k_3,k_4) \, \delta_{c_1 c_2} \delta_{c_3 c_4}
\nn\\ && {}
- {\cal M}^{\si_+,\si_-,\si_1,\si_4,\si_3,\si_2}_{\mathrm{CCa}}
(p_+,p_-,k_1,k_4,k_3,k_2) \, \delta_{c_1 c_4} \delta_{c_3 c_2}.
\label{eq:ee4f_ccncahadew}
\eeqar

In the calculation of the gluon-exchange diagrams we can also make use
of the ``colour-singlet'' result \refeq{genfun4fNC} for the generic diagram 
(a) of \reffi{ee4fdiags}, after splitting off the colour structure
appropriately. Since each of these diagrams
involves exactly one internal gluon,
exchanged by the two quark lines, the corresponding matrix elements can
be deduced in a simple way from the diagrams in which the gluon is
replaced by a photon. The 
gluon-exchange contributions to the channels NC(b)
and CC/NC(a) can again be composed from the ones for NC(a). Making use of
the auxiliary function
\beqar
\lefteqn{
{\cal M}^{\si_+,\si_-,\si_1,\si_2,\si_3,\si_4}_{\mathrm{g}}
(p_+,p_-,k_1,k_2,k_3,k_4) 
} \hspace*{3em} &&
\nn\\* 
& = \disp\frac{g_{\mathrm{s}}^2}{Q_1 Q_3 e^2} \,
\sum_{V=\gamma,Z} \Bigl[ & \phantom{{}+{}}
{\cal M}^{-\si_1,-\si_2,\si_+,\si_-,-\si_3,-\si_4}_{V\gamma}
(-k_1,-k_2,p_+,p_-,-k_3,-k_4)
\nn\\ && {}
+ {\cal M}^{-\si_3,-\si_4,\si_+,\si_-,-\si_1,-\si_2}_{V\gamma}
(-k_3,-k_4,p_+,p_-,-k_1,-k_2)
\nn\\ && {}
+ {\cal M}^{-\si_1,-\si_2,-\si_3,-\si_4,\si_+,\si_-}_{\gamma V}
(-k_1,-k_2,-k_3,-k_4,p_+,p_-)
\nn\\ && {}
+ {\cal M}^{-\si_3,-\si_4,-\si_1,-\si_2,\si_+,\si_-}_{\gamma V}
(-k_3,-k_4,-k_1,-k_2,p_+,p_-) \Bigl],
\hspace*{3em}
\eeqar
where $g_{\mathrm{s}}=\sqrt{4\pi\alpha_{\mathrm{s}}}$ is the strong
gauge coupling, the matrix elements involving gluon exchange
explicitly read
\beqar
\lefteqn{
{\cal M}^{\si_+,\si_-,\si_1,\si_2,\si_3,\si_4}
_{{\mathrm{NCa,had,gluon,}}c_1,c_2,c_3,c_4}
(p_+,p_-,k_1,k_2,k_3,k_4) } \hspace*{7em} &&
\nn\\* 
&=& \phantom{{}+{}}
{\cal M}^{\si_+,\si_-,\si_1,\si_2,\si_3,\si_4}_{\mathrm{g}}
(p_+,p_-,k_1,k_2,k_3,k_4) 
\; \text\frac{1}{4} \, \lambda^a_{c_1 c_2} \lambda^a_{c_3 c_4},
\\[1em]
\lefteqn{
{\cal M}^{\si_+,\si_-,\si_1,\si_2,\si_3,\si_4}
_{{\mathrm{NCb,had,gluon,}}c_1,c_2,c_3,c_4}
(p_+,p_-,k_1,k_2,k_3,k_4) } \hspace*{7em} &&
\nn\\* 
&=& \phantom{{}+{}}
{\cal M}^{\si_+,\si_-,\si_1,\si_2,\si_3,\si_4}_{\mathrm{g}}
(p_+,p_-,k_1,k_2,k_3,k_4) 
\; \text\frac{1}{4} \, \lambda^a_{c_1 c_2} \lambda^a_{c_3 c_4}
\nn\\ && {}
- {\cal M}^{\si_+,\si_-,\si_3,\si_2,\si_1,\si_4}_{\mathrm{g}}
(p_+,p_-,k_3,k_2,k_1,k_4)
\; \text\frac{1}{4} \, \lambda^a_{c_3 c_2} \lambda^a_{c_1 c_4},
\label{eq:ee4f_ncbhadqcd}
\\[1em]
\lefteqn{
{\cal M}^{\si_+,\si_-,\si_1,\si_2,\si_3,\si_4}
_{{\mathrm{CC/NCa,had,gluon,}}c_1,c_2,c_3,c_4}
(p_+,p_-,k_1,k_2,k_3,k_4) } \hspace*{7em} &&
\nn\\*
&=& \phantom{{}+{}}
{\cal M}^{\si_+,\si_-,\si_1,\si_2,\si_3,\si_4}_{\mathrm{g}}
(p_+,p_-,k_1,k_2,k_3,k_4) 
\; \text\frac{1}{4} \, \lambda^a_{c_1 c_2} \lambda^a_{c_3 c_4}.
\eeqar
The colour structure is easily evaluated by making use of the
completeness relation
$\lambda^a_{ij} \lambda^a_{kl} = -\text\frac{2}{3}\delta_{ij}\delta_{kl}
+2\delta_{il}\delta_{jk}$ for the Gell-Mann matrices $\lambda^a_{ij}$.

\begin{sloppypar}
The complete matrix elements for the fully hadronic channels result
from the sum of the 
purely electroweak and the gluon-exchange  contributions,
\beq
{\cal M}^{\si_+,\si_-,\si_1,\si_2,\si_3,\si_4}
_{{\ldots\mathrm{,had,}}c_1,c_2,c_3,c_4}
= {\cal M}^{\si_+,\si_-,\si_1,\si_2,\si_3,\si_4}
_{{\ldots\mathrm{,had,ew,}}c_1,c_2,c_3,c_4}+
{\cal M}^{\si_+,\si_-,\si_1,\si_2,\si_3,\si_4}
_{{\ldots\mathrm{,had,gluon,}}c_1,c_2,c_3,c_4}.
\eeq
The gluon-exchange contributions are separately gauge-invariant.

For clarity, we explicitly write down the colour-summed squared matrix
elements for the fully hadronic channels. Abbreviating 
${\cal M}_{\dots}
^{\si_+,\si_-,\si_a,\si_b,\si_c,\si_d}(p_+,p_-,k_a,k_b,k_c,k_d)$
by ${\cal M}_{\dots}(a,b,c,d)$ we obtain
\end{sloppypar}
\beqar
\lefteqn{
\sum_{\mathrm{colour}}|{\cal M}_{\mathrm{CCa,had}}(1,2,3,4)|^2 
= 9 |{\cal M}_{\mathrm{CCa}}(1,2,3,4)|^2,
} \hspace*{0em} &&
\\[1em]
\lefteqn{
\sum_{\mathrm{colour}}|{\cal M}_{\mathrm{NCa,had}}(1,2,3,4)|^2 
= 9 |{\cal M}_{\mathrm{NCa}}(1,2,3,4)|^2
+ 2 |{\cal M}_{\mathrm{g}}(1,2,3,4)|^2,
} \hspace*{0em} &&
\\[1em]
\lefteqn{
\sum_{\mathrm{colour}}|{\cal M}_{\mathrm{NCb,had}}(1,2,3,4)|^2 
} \hspace*{0em} &&
\nn\\*
&=& 9 |{\cal M}_{\mathrm{NCa}}(1,2,3,4)|^2
+ 9 |{\cal M}_{\mathrm{NCa}}(3,2,1,4)|^2
- 6\Re\big\{ {\cal M}_{\mathrm{NCa}}(1,2,3,4)
{\cal M}^*_{\mathrm{NCa}}(3,2,1,4) \big\}
\nn\\ && {}
+ 2 |{\cal M}_{\mathrm{g}}(1,2,3,4)|^2
+ 2 |{\cal M}_{\mathrm{g}}(3,2,1,4)|^2
+ \text\frac{4}{3}\Re\big\{ {\cal M}_{\mathrm{g}}(1,2,3,4)
{\cal M}^*_{\mathrm{g}}(3,2,1,4) \big\}
\nn\\ && {}
- 8\Re\big\{ {\cal M}_{\mathrm{NCa}}(1,2,3,4)
{\cal M}^*_{\mathrm{g}}(3,2,1,4) \big\}
- 8\Re\big\{ {\cal M}_{\mathrm{NCa}}(3,2,1,4)
{\cal M}^*_{\mathrm{g}}(1,2,3,4) \big\},
\nn\\*
\\[1em]
\lefteqn{
\sum_{\mathrm{colour}}|{\cal M}_{\mathrm{CC/NCa,had}}(1,2,3,4)|^2 
} \hspace*{0em} &&
\nn\\
&=& 9 |{\cal M}_{\mathrm{NCa}}(1,2,3,4)|^2
+ 9 |{\cal M}_{\mathrm{CCa}}(1,4,3,2)|^2
- 6\Re\big\{ {\cal M}_{\mathrm{NCa}}(1,2,3,4)
{\cal M}^*_{\mathrm{CCa}}(1,4,3,2) \big\}
\nn\\ && {}
+ 2 |{\cal M}_{\mathrm{g}}(1,2,3,4)|^2
- 8\Re\big\{ {\cal M}_{\mathrm{CCa}}(1,4,3,2)
{\cal M}^*_{\mathrm{g}}(1,2,3,4) \big\}.
\eeqar 
Owing to the colour structure of the diagrams, a non-zero interference
between purely electroweak and gluon-exchange contributions is only
possible if the four final-state fermions can be combined into one
single closed fermion line in the squared diagram. This implies that
fermion pairs must couple to different resonances in the
electroweak and the gluon-exchange diagrams,
leading to a global suppression of such interference effects in the
phase-space integration (see \refse{se:QCD}).

\subsection[Generic functions and amplitudes for 
${\mathrm e^+ \mathrm e^- \to 4\,f+\ga}$]
{\boldmath Generic functions and amplitudes for 
${\mathrm e^+ \mathrm e^- \to 4\,f+\ga}$}
\label{se:genfunction}

The generic functions for $\eeffffg$ can be constructed in a similar
way. The idea is to combine the contributions of all those graphs to
one generic function that reduce to the same graph after removing the
radiated photon.  These combined contributions to $\eeffffg$ are
classified in the same way as the diagrams for the corresponding
process $\Pep\Pem\to 4f$, i.e.\ the graphs of \reffi{ee4fdiags} also
represent the generic functions for $\eeffffg$. Finally, all
amplitudes for $\eeffffg$ can again be constructed from only two
generic functions.  Note that the number of individual Feynman
diagrams ranges between 14 and 1008 for the various processes.  We
note that the generic functions can in fact be used to construct the
amplitudes for all processes involving exactly six external fermions
and one external photon, such as $\Pem\Pem\to4f\ga$ and $\Pe^\pm \ga\to5f$.

As a virtue of this approach, the so-defined generic functions fulfill
the QED Ward identity for the external photon, i.e.\ replacing the
photon polarization vector by the photon momentum yields zero for each
generic function. This is simply a consequence of electromagnetic
charge conservation. Consequently, in the actual calculation in the
WvdW formalism the gauge spinor of the photon drops out in each
contribution separately.

Assuming the external fermions as incoming and the photon as outgoing, 
the generic functions read
\beqar
&& \hspace*{-3em}
{\cal M}^{\si_a,\si_b,\si_c,\si_d,\si_e,\si_f,\lambda}_{V_1 V_2}
(Q_a,Q_b,Q_c,Q_d,Q_e,Q_f,p_a,p_b,p_c,p_d,p_e,p_f,k) 
\nn\\*
\hspace*{2em}
&=& -4\sqrt{2}e^5 \de_{\si_a,-\si_b} \de_{\si_c,-\si_d} \de_{\si_e,-\si_f} \, 
g^{\si_b}_{V_1\bar f_a f_g} g^{\si_b}_{V_2\bar f_g f_b}
g^{\si_d}_{V_1\bar f_c f_d} g^{\si_f}_{V_2\bar f_e f_f}
\nn\\ && {} \times
A^{\si_a,\si_c,\si_e,\lambda}_2
(Q_a,Q_b,Q_c,Q_d,Q_e,Q_f,p_a,p_b,p_c,p_d,p_e,p_f,k),
\hspace*{3em}
\\[1em] 
&& \hspace*{-3em}
{\cal M}^{\si_a,\si_b,\si_c,\si_d,\si_e,\si_f,\lambda}_{VWW}
(Q_a,Q_b,Q_c,Q_d,Q_e,Q_f,p_a,p_b,p_c,p_d,p_e,p_f,k) 
\nn\\*
&=& -4\sqrt{2}e^5 \de_{\si_a,-\si_b} \de_{\si_c,+} \de_{\si_d,-} 
\de_{\si_e,+} \de_{\si_f,-} \, 
(Q_c-Q_d)g_{VWW} g^{\si_b}_{V\bar f_a f_b} 
g^{-}_{W\bar f_c f_d} g^{-}_{W\bar f_e f_f}
\nn\\ && {} \times
A^{\si_a,\lambda}_3
(Q_a,Q_b,Q_c,Q_d,Q_e,Q_f,p_a,p_b,p_c,p_d,p_e,p_f,k),
\label{eq:Mgenee4fA}
\eeqar
with the auxiliary functions
\beqar
&& \hspace*{-3em}
A^{{+}{+}{+}{+}}_2(Q_a,Q_b,Q_c,Q_d,Q_e,Q_f,p_a,p_b,p_c,p_d,p_e,p_f,k) 
= -\spac \biggl\{
\nn\\
\hspace*{2em} && 
P_{V_1}(p_c+p_d) P_{V_2}(p_e+p_f) 
\nn\\ 
&& \quad
{} \times \biggl[ \frac{\cspbf}{\spak} \biggl(
\frac{Q_c-Q_d}{(p_b+p_e+p_f)^2} \frac{\spac}{\spck}
(\cspbd \spbe+\cspdf \spef)
\nn\\ 
&& \quad
\hphantom{{} \times \biggl[ \frac{\cspbf}{\spak} \biggl(} {}
+\frac{Q_f-Q_e}{(p_a+p_c+p_d)^2} \frac{\spae}{\spek}
(\cspad \spae+\cspcd \spce) \biggr)
\nn\\ 
&& \quad \hphantom{\times \biggl[}
+\frac{Q_b (\cspad\spae+\cspcd\spce) (\cspbf\spab-\cspfk\spak) }
{(p_a+p_c+p_d)^2\spak\spbk} 
\nn\\ 
&& \quad \hphantom{\times \biggl[}
+\frac{(Q_a+Q_c-Q_d) \cspbf\cspcd\spac (\cspbk\spbe-\cspfk\spef) }
{(p_a+p_c+p_d)^2 (p_b+p_e+p_f)^2 \spak} \biggr]
\nn\\ 
&& {}
-\frac{Q_d-(Q_c-Q_d) 2(p_d\cdot k) P_{V_1}(p_c+p_d)}{(p_b+p_e+p_f)^2}
P_{V_1}(p_c+p_d-k) P_{V_2}(p_e+p_f) \cspbf
\nn\\ 
&& {} \hphantom{{}+{}} \;
\times \frac{ \spcd (\cspbd \spbe+\cspdf \spef)
+\spck (\cspbk \spbe-\cspfk \spef) } {\spck\spdk} 
\nn\\ 
&& {}
+\frac{Q_f-(Q_e-Q_f) 2(p_f\cdot k) P_{V_2}(p_e+p_f)}{(p_a+p_c+p_d)^2}
P_{V_1}(p_c+p_d) P_{V_2}(p_e+p_f-k)
\nn\\ 
&& {} \hphantom{{}+{}} \;
\times \frac{ (\cspbf \spef+\cspbk \spek) (\cspad \spae+\cspcd \spce)}
{\spek\spfk} 
\,\biggr\},
\nn\\
&& \hspace*{-3em}
A^{{+}{+}{-}{+}}_2(Q_a,Q_b,Q_c,Q_d,Q_e,Q_f,p_a,p_b,p_c,p_d,p_e,p_f,k)
\nn\\*
&=& A^{{+}{+}{+}{+}}_2(Q_a,Q_b,Q_c,Q_d,-Q_f,-Q_e,p_a,p_b,p_c,p_d,p_f,p_e,k),
\nn\\
&& \hspace*{-3em}
A^{{+}{-}{+}{+}}_2(Q_a,Q_b,Q_c,Q_d,Q_e,Q_f,p_a,p_b,p_c,p_d,p_e,p_f,k)
\nn\\*
&=& A^{{+}{+}{+}{+}}_2(Q_a,Q_b,-Q_d,-Q_c,Q_e,Q_f,p_a,p_b,p_d,p_c,p_e,p_f,k),
\nn\\
&& \hspace*{-3em}
A^{{+}{-}{-}{+}}_2(Q_a,Q_b,Q_c,Q_d,Q_e,Q_f,p_a,p_b,p_c,p_d,p_e,p_f,k)
\nn\\*
&=& A^{{+}{+}{+}{+}}_2(Q_a,Q_b,-Q_d,-Q_c,-Q_f,-Q_e,p_a,p_b,p_d,p_c,p_f,p_e,k),
\nn\\
&& \hspace*{-3em}
A^{{-}{+}{+}{+}}_2(Q_a,Q_b,Q_c,Q_d,Q_e,Q_f,p_a,p_b,p_c,p_d,p_e,p_f,k)
\nn\\*
&=& A^{{+}{+}{+}{+}}_2(Q_b,Q_a,-Q_e,-Q_f,-Q_c,-Q_d,p_b,p_a,p_e,p_f,p_c,p_d,k),
\nn\\
&& \hspace*{-3em}
A^{{-}{+}{-}{+}}_2(Q_a,Q_b,Q_c,Q_d,Q_e,Q_f,p_a,p_b,p_c,p_d,p_e,p_f,k)
\nn\\*
&=& A^{{+}{+}{+}{+}}_2(Q_b,Q_a,Q_f,Q_e,-Q_c,-Q_d,p_b,p_a,p_f,p_e,p_c,p_d,k),
\nn\\
&& \hspace*{-3em}
A^{{-}{-}{+}{+}}_2(Q_a,Q_b,Q_c,Q_d,Q_e,Q_f,p_a,p_b,p_c,p_d,p_e,p_f,k)
\nn\\*
&=& A^{{+}{+}{+}{+}}_2(Q_b,Q_a,-Q_e,-Q_f,Q_d,Q_c,p_b,p_a,p_e,p_f,p_d,p_c,k),
\nn\\
&& \hspace*{-3em}
A^{{-}{-}{-}{+}}_2(Q_a,Q_b,Q_c,Q_d,Q_e,Q_f,p_a,p_b,p_c,p_d,p_e,p_f,k)
\nn\\*
&=& A^{{+}{+}{+}{+}}_2(Q_b,Q_a,Q_f,Q_e,Q_d,Q_c,p_b,p_a,p_f,p_e,p_d,p_c,k),
\nn\\
&& \hspace*{-3em}
A^{\si_a,\si_c,\si_d,{-}}_2
(Q_a,Q_b,Q_c,Q_d,Q_e,Q_f,p_a,p_b,p_c,p_d,p_e,p_f,k)
\\*
&=& \Big( A^{-\si_a,-\si_c,-\si_d,{+}}_2
(Q_a,Q_b,Q_c,Q_d,Q_e,Q_f,p_a,p_b,p_c,p_d,p_e,p_f,k) \Bigr)^*
\Big|_{P_{V_{1,2}}(p)\to P_{V_{1,2}}^*(p)},
\nn
\label{eq:A2}
\eeqar
and
\beqar
&& \hspace*{-3em}
A^{{+}{+}}_3(Q_a,Q_b,Q_c,Q_d,Q_e,Q_f,p_a,p_b,p_c,p_d,p_e,p_f,k) 
\nn\\*
\hspace*{2em} 
&=& P_V(p_a+p_b) P_W(p_c+p_d) P_W(p_e+p_f) 
\frac{(Q_c-Q_d)\spce}{\spck\spek} 
\nn\\
&& \quad
{} \times ( \hphantom{{}+{}} \cspbd\cspbf\spab\spce +\cspbd\cspdf\spae\spcd 
\nn\\
&& \quad \hphantom{{} \times (}
{} + \cspbf\cspdf\spac\spef )
\nn\\
&& {} +P_V(p_a+p_b-k) P_W(p_c+p_d) P_W(p_e+p_f) \frac{Q_b}{\spak\spbk} 
\nn\\
&& \quad
{} \times \{ \hphantom{{}+{}} \cspdf [ \hphantom{{}+{}}
\spae\spcd (\cspbd\spab-\cspdk\spak)
\nn\\
&& \quad \hphantom{{} \times \{ {}+ \cspdf [}
{} +\spac\spef (\cspbf\spab-\cspfk\spak) ]
\nn\\
&& \quad \hphantom{{} \times \{ }
{} + \spce (\cspbd\spab-\cspdk\spak) (\cspbf\spab-\cspfk\spak) \}
\nn\\
&& {} +P_V(p_a+p_b) P_W(p_c+p_d-k) P_W(p_e+p_f) 
\frac{Q_d-(Q_c-Q_d)2(p_d\cdot k) P_W(p_c+p_d)}{\spck\spdk} 
\nn\\
&& \quad
{} \times \{ \hphantom{{}+{}} 
\cspbf [ \hphantom{{}+{}} \spac\spef (\cspdf\spcd-\cspfk\spck)
\nn\\
&& \quad \hphantom{{} \times \{ \hphantom{{}+{}} \cspbf [ }
{} +\spce\spab (\cspbd\spcd+\cspbk\spck) ]
\nn\\
&& \quad \hphantom{{} \times \{ }
{} + \spae (\cspdf\spcd-\cspfk\spck) (\cspbd\spcd+\cspbk\spck) \}
\nn\\
&& {} +P_V(p_a+p_b) P_W(p_c+p_d) P_W(p_e+p_f-k) 
\frac{Q_f+(Q_f-Q_e)2(p_f\cdot k) P_W(p_e+p_f)}{\spek\spfk} 
\nn\\
&& \quad
{} \times \{ \hphantom{{}+{}} 
\cspbd [ \hphantom{{}+{}} \spce\spab (\cspbf\spef+\cspbk\spek)
\nn\\
&& \quad \phantom{ {} \times \{ \hphantom{{}+{}} \cspbd [ }
{} + \spae\spcd (\cspdf\spef+\cspdk\spek) ]
\nn\\
&& \quad \phantom{ {} \times \{ }
{}+\spac (\cspbf\spef+\cspbk\spek) (\cspdf\spef+\cspdk\spek) \},
\nn\\
&& \hspace*{-3em}
A^{{-}{+}}_3(Q_a,Q_b,Q_c,Q_d,Q_e,Q_f,p_a,p_b,p_c,p_d,p_e,p_f,k) 
\nn\\*
&=& A^{{+}{+}}_3(-Q_b,-Q_a,Q_c,Q_d,Q_e,Q_f,p_b,p_a,p_c,p_d,p_e,p_f,k),
\nn\\
&& \hspace*{-3em}
A^{\si_a,{-}}_3(Q_a,Q_b,Q_c,Q_d,Q_e,Q_f,p_a,p_b,p_c,p_d,p_e,p_f,k) 
\\*
&=& \Bigl( 
A^{-\si_a,{+}}_3(Q_a,Q_b,-Q_d,-Q_c,-Q_f,-Q_e,p_a,p_b,p_d,p_c,p_f,p_e,k) 
\Bigr)^* \Big|_{P_{V,W}(p)\to P_{V,W}^*(p)}.
\nn
\label{eq:A3}
\eeqar
The replacements $P_V\to P_V^*$ after the complex conjugation
in the last lines of \refeq{eq:A2} and \refeq{eq:A3} ensure that the
vector-boson propagators remain unaffected.
Note that the vector-boson masses do not enter explicitly in the above
results, but only via $P_V$. 
In gauges such as the 't~Hooft--Feynman or the unitary gauge this feature 
is obtained only after combining different Feynman graphs for $\eeffffg$; 
in the
non-linear gauge \refeq{eq:nlgauge} this is the case diagram by diagram.

The helicity amplitudes for $\eeffffg$ follow from the
generic functions ${\cal M}_{V_1 V_2}$ and ${\cal M}_{VWW}$ of 
\refeq{eq:Mgenee4fA} in exactly the same way as described in
\refse{se:ee4f} 
for $\Pep\Pem\to 4f$.
This holds also for the gluon-exchange matrix elements and for the
colour factors.
Moreover, the classification of gauge-invariant sets of diagrams
for $\eeffff$ immediately yields such sets for $\eeffffg$, if the additional
photon is attached to all graphs of a set in all possible ways.

We have checked analytically that the electromagnetic Ward identity
for the external photon is fulfilled for each generic contribution
separately. In addition, we have numerically compared the amplitudes
for all processes with amplitudes generated by {\sl Madgraph} \cite{St94}
 for zero width of the vector bosons and found complete agreement.
We could not compare our results with {\sl Madgraph} for finite width,
because {\sl Madgraph} uses the unitary gauge for massive vector-boson
propagators and the `t~Hooft--Feynman gauge for the photon propagators,
while we are using the non-linear gauge \refeq{eq:nlgauge}. Therefore,
the matrix elements differ after introduction of finite vector-boson widths.
While the calculation with {\sl Madgraph} is  fully automized,
in our calculation we have full control over the matrix
element and can, in particular, investigate various implementations of
the finite width.

A comparison of our results with those of \citeres{Ae91,Ae91a}, which include
only the matrix elements that involve two resonant \PW~bosons,
immediately reveals the virtues of our generic approach.

\subsection{Implementation of finite gauge-boson widths}
\label{se:finwidth}

We have implemented the finite widths of the $\PW$ and $\PZ$ bosons in
different ways:
\begin{itemize}
\item{\it fixed width} in all propagators: 
$P_V(p) = [p^2-\MV^2+\ri\MV\GV]^{-1}$,
\item{\it running width} in time-like propagators: 
$P_V(p) = [p^2-\MV^2+\ri p^2(\GV/\MV)\theta(p^2)]^{-1}$,
\item{\it complex-mass scheme:} complex gauge-boson masses everywhere,
  \ie $\sqrt{M_V^2-\ri M_V\Ga_V}$ instead of $M_V$ in the propagators
  and in the couplings. This results, in particular, in a constant
  width in all propagators,
\beq\label{Vprop}
  P_V(p) = [p^2-\MV^2+\ri\MV\Ga_V]^{-1},
\eeq
and in a complex weak mixing angle: 
\beq\label{complangle}
\cw^2=1-\sw^2=   \frac{\MW^2-\ri\MW\GW}{\MZ^2-\ri\MZ\GZ}.
\eeq
\end{itemize}
The virtues and drawbacks of the first two schemes
have been discussed in \citere{Be97c}. 
Both violate $\SU(2)$ gauge invariance, the running width
also $\U(1)$ gauge invariance.  
The complex-mass scheme obeys
all Ward identities and thus gives a consistent description of the
finite-width effects in any tree-level calculation.  While the 
complex-mass scheme 
works in general, it is particularly simple for $\eeffffg$ in the
non-linear gauge \refeq{eq:nlgauge}.  In this case, no couplings
involving explicit gauge-boson masses appear,
and it is sufficient to
introduce the finite gauge-boson widths in the propagators [\cf
\refeq{Vprop}] and to introduce the complex weak mixing angle
\refeq{complangle} in the couplings.  We note that a generalization of
this scheme to higher orders requires to introduce complex mass
counterterms in order to compensate for the complex masses in the
propagators \cite{St90}. We did not consider the
fermion-loop scheme \cite{Ba95,Ar95,Be97c,Pa95}, which is also fully
consistent for lowest-order predictions, since it requires the
calculation of fermionic one-loop corrections to $\eeffffg$ 
which is beyond the scope of this work.

\section{The Monte Carlo program}
\label{se:MC}

The cross section for $\Pep\Pem\to 4f(\gamma)$ is given by
\begin{eqnarray}\label{eq:crosssection}
\rd \sigma &=& \frac{(2 \pi)^{4-3 n}}{2 s}
\left[\prod\limits_{i=1}^n \rd^4 k_i \, 
\delta\left(k_i^2\right) \theta(k_i^0)\right]
\delta^{(4)} \left( p_+ +p_- -\sum_{i=1}^n k_i \right)  
\nonumber \\
&& {} 
\times |{\cal M}(p_+,p_-,k_1,\ldots ,k_n)|^2,
\end{eqnarray}
where 
$n=4,5$ is the number of outgoing particles. 
The helicity amplitudes ${\cal M}$ for $\Pep\Pem\to 4f(\gamma)$
have been calculated in \refses{se:ee4f} and \ref{se:genfunction}.
The phase-space integration is performed
with the help of a Monte Carlo technique, since
the Monte Carlo method allows 
us to calculate a variety of observables
simultaneously and to 
easily implement cuts in order to
account for the experimental situation.

The helicity amplitudes in \refeq{eq:crosssection} 
exhibit a complicated peaking behaviour in different
regions of the integration domain. In order to obtain a numerically stable
result and to reduce the Monte Carlo integration error 
we use a multi-channel Monte Carlo method \cite{Be94,Multichannel}, which 
is briefly outlined in the following.   

Before turning to the multi-channel method, we consider the treatment
of a single channel. We choose a suitable set ${\bf \Phi}$ of $3n-4$
phase-space variables to describe a point in phase space, and
determine the corresponding physical region $V$ and the relation
$k_i({\bf \Phi})$ between the phase-space variables ${\bf \Phi}$ and
the momenta $k_1, \ldots ,k_n$.  The phase-space integration of
\refeq{eq:crosssection} reads
\begin{eqnarray}
I_n &=& \int \rd \sigma=
\int_V \rd {\bf \Phi} \, \rho\Big(k_i({\bf \Phi})\Big) \, 
f\Big(k_i({\bf \Phi})\Big),
\\\nonumber
f\Big(k_i({\bf \Phi})\Big)&=& \frac{(2 \pi)^{4-3 n}}{2 s}
\left|{\cal M}\left(p_+,p_-,k_1({\bf \Phi}),\ldots ,
k_n({\bf \Phi})\right)\right|^2,
\end{eqnarray}
where $\rho$ is the phase-space density. 
For the random generation of the events,
we further transform  
the integration variables ${\bf \Phi}$
to $3n-4$ new variables ${\bf r}=(r_i)$
with a hypercube as integration domain:
${\bf \Phi}={\bf h}({\bf r})$ with $0\le r_i \le 1$.
We obtain
\begin{equation}\label{eq:phint}
I_n = \int_V \rd {\bf \Phi} \, \rho\Big(k_i({\bf \Phi})\Big) \, 
f\Big(k_i({\bf \Phi})\Big) 
= \int_0^1 \rd {\bf r} \, 
\frac{f\Big(k_i({\bf h}({\bf r}))\Big)}
{g\Big(k_i({\bf h}({\bf r}))\Big)} \; ,
\end{equation}
where $g$ is the probability density of events generated in phase
space, defined by
\begin{equation} \label{eq:dens}
\frac{1}{g\Big(k_i({\bf \Phi})\Big)} = \rho\Big(k_i({\bf \Phi})\Big)
\left| \frac{\partial{\bf h}({\bf r})}
{\partial{\bf r}} \right|_{{\bf r}={\bf h}^{-1}({\bf \Phi})} \; .
\end{equation} 
If $f$ varies strongly, the efficiency of the Monte Carlo method
can be considerably enhanced by choosing the
mapping of random numbers ${\bf r}$ to ${\bf \Phi}$ in such a way that 
the resulting density $g$ mimics the behaviour of $|f|$. 
For this {\it importance sampling}, the
choice of ${\bf \Phi}$ is guided by the peaking structure of $f$,
which is determined by the propagators in a characteristic Feynman diagram.

We choose the variables ${\bf \Phi}$ in such a way that the invariants
corresponding to the propagators are included.  Accordingly, we
decompose the $n$-particle final state into $2 \to 2$ scattering
processes with subsequent $1 \to 2$ decays.  The variables
${\bf \Phi}$ consist of Lorentz invariants $s_i, t_i$, defined as the
squares of time- and space-like momenta, respectively, and of polar
and azimuthal angles $\theta_i$ and $\phi_i$, defined in appropriate frames.
A detailed description of the parameterization of an $n$-particle
phase space in terms of invariants and angles can be found in
\refapp{ap:phasespace}.
The parameterization of the invariants $s_i$ and $t_i$ in
${\bf \Phi}={\bf h}({\bf r})$ is chosen in such a way that the
propagator structure of the function $f$ is compensated by a similar
behaviour in the density $g$.  More precisely, if $f$ contains
Breit--Wigner resonances or distributions like $s_i^{-\nu}$, which are
relevant for massless propagators, appropriate parameterizations of
$s_i$ are given by:
\begin{itemize}
\item Breit--Wigner resonances:
\beqar\label{eq:maps1}
s_i &=& \MV^2+\MV\GV\tan[y_1+(y_2-y_1) r_i] 
\\ && \mbox{with } \;
y_{1,2}=\arctan\left(\frac{s_{\min,\max}-\MV^2}{\MV\GV}\right); \nn
\eeqar
\item propagators of massless particles:
\begin{eqnarray}\label{eq:maps2}
\nu\ne 1: \qquad s_i &=& \left[s_{\max}^{1-\nu} r_i
+s_{\min}^{1-\nu}(1-r_i)\right]^{1/(1-\nu)} ,
\nn\\[.5em]
\nu=1: \qquad
s_i &=&\exp\left[\ln(s_{\max}) r_i + \ln(s_{\min})(1-r_i)\right].
\end{eqnarray}
\end{itemize}
For the choice of $\nu$ see \refapp{ap:phasespace}.
The remaining variables
in ${\bf \Phi}={\bf h}({\bf r})$, i.e.\ those for which $f$ is
expected not to exhibit a peaking behaviour, are generated as follows:
\begin{eqnarray}\label{eq:nomaps}
s_i &=& s_{\max} r_i+s_{\min} (1-r_i), \qquad
\phi_i = 2 \pi r_i, \qquad \cos\theta_i =2 r_i-1.
\end{eqnarray}
The absolute values of the 
invariants $t_i$ are generated in the same way as $s_i$.
The resulting density $g$ of events in phase space 
is obtained as the product of the corresponding Jacobians,
as given in \refeq{eq:dens}.
In the \refapp{ap:phasespace}, 
we provide explicit examples for the generation of events
with a specific choice of mappings $k_i({\bf \Phi})$ and ${\bf h}({\bf r})$,
and for the calculation of the corresponding density $g$.

The differential cross sections of the processes
$\Pep\Pem\to 4f$ and especially $\eeffffg$ possess very
complex peaking structures so that the peaks in the integrand 
$f({\bf \Phi})$ in \refeq{eq:phint} cannot be described properly 
by only one single density $g({\bf \Phi})$.
The {\it multi-channel approach} \cite{Be94,Multichannel} 
suggests a solution to this problem. For each peaking structure we choose 
a suitable set ${\bf \Phi}_k$, and accordingly a 
mapping of random numbers $r_i$ into ${\bf \Phi}_k$:
${\bf \Phi}_k={\bf h}_k({\bf r})$ with $0 \le r_i \le 1$,
so that the resulting density
$g_k$ describes this particular peaking behaviour of $f$. 
All densities $g_k$ are combined 
into one density $g_{\mathrm {tot}}$
that is expected to smoothes the integrand over the whole phase-space 
integration region. The phase-space integral of \refeq{eq:phint} reads
\begin{eqnarray}
I_n &=& \sum_{k=1}^M \int_V \rd {\bf \Phi}_k \, 
\rho_k\Big(k_i({\bf \Phi}_k)\Big)  \, 
g_k\Big(k_i({\bf \Phi}_k)\Big) \,
\frac{f\Big(k_i({\bf \Phi}_k)\Big)}{g_{\mathrm{tot}}\Big(k_i({\bf \Phi}_k)\Big)}
=\sum_{k=1}^M \int_0^1 \rd {\bf r} \, 
\frac{f\Big(k_i({\bf h}_k({\bf r}))\Big)}
{g_{\mathrm {tot}}\Big(k_i({\bf h}_k({\bf r}))\Big)},
\hspace*{2em}
\eeqar
with
\beqar\label{eq:mdens}
g_{\mathrm {tot}}\Big(k_i({\bf \Phi}_k)\Big)&=&
\sum_{l=1}^M g_l\Big(k_i({\bf \Phi}_k)\Big)
,\qquad
\frac{1}{g_l\Big(k_i({\bf \Phi}_k)\Big)}=\rho_l\Big(k_i({\bf \Phi}_k)\Big) \, 
\left| \frac{\partial{\bf h}_l({\bf r})}
{\partial {\bf r}}\right|_{{\bf r}={\bf h}_k^{-1}({\bf \Phi}_k)}.
\end{eqnarray}
The different mappings ${\bf h}_k({\bf r})$ are called channels,
and $M$ is the number of all channels.

In order to reduce the Monte Carlo error further, we adopt the method
of weight optimization of \citere{Kl94} and 
introduce {\em a-priori weights} $\alpha_k, k=1,\dots, M$ ($\alpha_k \ge 0$
and $\sum_{k=1}^M \alpha_k=1$).
The channel $k$ that is used to generate the event is picked randomly
with probability $\alpha_k$, i.e.\ 
\begin{eqnarray}
I_n &=& \sum_{k=1}^M \alpha_k \int_V \rd {\bf \Phi}_k \, 
\rho_k\Big(k_i({\bf \Phi}_k)\Big) 
g_k\Big(k_i({\bf \Phi}_k)\Big) 
\, \frac{f\Big(k_i({\bf \Phi}_k)\Big)}
{g_{\mathrm{tot}}\Big(k_i({\bf \Phi}_k)\Big)}
\nonumber \\
&=& \int_0^1 \rd r_0 \, \sum_{k=1}^M 
\theta(r_0-\beta_{k-1})\theta(\beta_k-r_0)
\int_V \rd {\bf \Phi}_k \, \rho_k \Big(k_i({\bf \Phi}_k)\Big) 
g_k\Big(k_i({\bf \Phi}_k)\Big) \, 
\frac{f\Big(k_i({\bf \Phi}_k)\Big)}
{g_{\mathrm{tot}}\Big(k_i({\bf \Phi}_k)\Big)} \nn\\
&=&\int_0^1 \rd r_0\,  
\sum_{k=1}^M \theta(r_0-\beta_{k-1})\theta(\beta_k-r_0)
\int_0^1 \rd {\bf r} \,
\frac{f\Big(k_i({\bf h}_k({\bf r}))\Big)}
{g_{\mathrm{tot}}\Big(k_i({\bf h}_k({\bf r}))\Big)},
\end{eqnarray}
where $\beta_0=0, \beta_j=\sum_{k=1}^j \alpha_k, j=1,\ldots ,M-1$, 
$\beta_M=\sum_{k=1}^M \alpha_k=1$, and
\beq
g_{\mathrm{tot}}\Big(k_i({\bf \Phi}_k)\Big)=
\sum_{l=1}^M \alpha_l g_l\Big(k_i({\bf \Phi}_k)\Big),
\eeq
is the total density of the event.  

For the processes $\Pep \Pem \to 4 f$ we have between 6 and 128
different channels, for $\eeffffg$ between 14 and 928 channels.
Each channel smoothes a particular combination of propagators that
results from a characteristic Feynman diagram.  We have written
phase-space generators in a generic way for several classes of
channels determined by the chosen set of invariants $s_i, t_i$.  The
channels within one class differ in the choice of the mappings
\refeqs{eq:maps1}, \refeqf{eq:maps2}, and \refeqf{eq:nomaps} and the
order of the external particles.  We did not include special
channels for interference contributions.

The $\alpha_k$-dependence of the quantity 
\begin{equation}
W({\bf \alpha})= \frac{1}{N} \sum_{j=1}^N [w(r_0^j,{{\bf r}}^{\,j})]^2,
\end{equation}
where $w=f/g_{\mathrm{tot}}$ is the weight assigned to the Monte Carlo point 
$(r_0^j, {\bf r}^{\,j})$ of the $j$th event,
can be exploited to minimize the expected Monte Carlo error 
\begin{equation}
\delta \bar I_n=\sqrt{\frac{W({\bf \alpha})-\bar{I}_n^2}{N}},
\end{equation} 
with the Monte Carlo estimate of $I_n$ 
\begin{equation}
\bar I_n= \frac{1}{N} \sum_{j=1}^N w(r_0^j,{\bf r}^{\,j})
\end{equation} 
by trying to choose an optimal set of a-priori weights.
We perform the search for an optimal set of $\alpha_k$ by using an
{\it adaptive optimization} method, as described in \citere{Kl94}. 
After a certain number of generated events a new set of 
a-priori weights $\alpha_k^{\mathrm{new}}$ is calculated according to
\begin{eqnarray}
\alpha_k^{\mathrm{new}} &\propto& \alpha_k 
\sqrt{\frac{1}{N}\sum_{j=1}^N \, 
\frac{g_k\Big(k_i({\bf h}_k({{\bf r}}^{\,j}))\Big) \, 
[w(r_0^j,{{\bf r}}^{\,j})]^2}
{g_{\mathrm{tot}}\Big(k_i({\bf h}_k({{\bf r}}^{\,j}))\Big)}}, \qquad
\sum_{k=1}^M \alpha_k^{\mathrm{new}}=1.
\end{eqnarray}

Based on the above approach, we have written two independent Monte
Carlo programs. While the general strategy is similar, 
the programs differ in the explicit phase-space generation.

\section{Numerical results}
\label{se:numres}

If not stated otherwise we use the complex-mass scheme and the 
following parameters:
\beq
\begin{array}[b]{rlrl}
\al =& 1/128.89, & \qquad \al_s =& 0.12, \\
\MW =& 80.26\GeV,& \GW =& 2.05\GeV, \\
\MZ =& 91.1884\GeV,& \GZ =& 2.46\GeV.
\end{array}
\eeq
In the complex-mass scheme, the weak mixing angle is defined in
\refeq{complangle}, in all other schemes it 
is fixed by $\cw=\MW/\MZ$, $\sw^2=1-\cw^2$.

The energy in the centre-of-mass (CM) system of the incoming electron
and positron
is denoted by $\sqrt{s}$.
Concerning the phase-space integration, we apply
the canonical cuts of the ADLO/TH detector, 
\beq
\begin{array}[b]{rlrlrl}
\theta (l,\mathrm{beam})> & 10^\circ, & \qquad
\theta( l, l^\prime)> & 5^\circ, & \qquad 
\theta( l, q)> & 5^\circ, \\
\theta (\ga,\mathrm{beam})> & 1^\circ, &
\theta( \ga, l)> & 5^\circ, & 
\theta( \ga, q)> & 5^\circ, \\
E_\ga> & 0.1\GeV, & E_l> & 1\GeV, & E_q> & 3\GeV, \\
m(q,q')> & 5\GeV,
\end{array}
\label{eq:canonicalcuts}
\eeq
where $\theta(i,j)$ specifies the angle between the particles $i$ and
$j$ in the CM system, 
and $l$, $q$, $\ga$, and ``beam'' denote charged leptons,
quarks, photons, and the beam electrons or positrons, respectively.
The invariant mass of a quark pair $qq'$ is denoted by 
$m(q,q')$.
The cuts coincide with those defined in \citere{CERN9601mcgen},
except for the additional angular cut between charged leptons.
The canonical cuts exclude all collinear and infrared singularities 
from phase space for all processes.  

Although our helicity amplitudes and Monte Carlo programs
allow for a treatment of arbitrary polarization configurations, we
consider only unpolarized quantities.

All results are produced with $10^7$ events. 
The calculation of the cross section for
$\Pep\Pem\to\Pep\Pem\mu^+\mu^-$ requires
about 50 minutes on a DEC ALPHA workstation with 500 MHz, 
the calculation of  the cross section for
$\Pep\Pem\to\Pep\Pem\mu^+\mu^-\ga$ takes about 5 hours.
The numbers in 
parentheses in the following tables 
correspond to the statistical errors 
of the results of the Monte Carlo integrations.

\subsection{Comparison with existing results}

In order to compare our results for $\eeffff$ with Tables 6--8 
of \citere{CERN9601table},
we use the corresponding set of phase-space cuts and input parameters,
\ie the canonical cuts defined in \refeq{eq:canonicalcuts},
a CM energy of $\sqrt{s}=190\GeV$, and the parameters 
$\al=\al (2 \MW)=1/128.07$, $\al_s=0.12$, 
$\MW=80.23\GeV$, $\GW=2.0337\GeV$, $\MZ=91.1888\GeV$, and $\GZ=2.4974\GeV$.
The value of $\sw$, 
which enters the couplings, is calculated from  
$\al (2 \MW)/(2 \sw^2)=\GF \MW^2/(\pi \sqrt{2})$
with $\GF=1.16639\times 10^{-5} \GeV^{-2}$.

\begin{table}
\renewcommand{\arraystretch}{1.1}
\newdimen\digitwidth
\setbox0=\hbox{0}
\digitwidth=\wd0
\catcode`!=\active
\def!{\kern\digitwidth}
\newdimen\dotwidth
\setbox0=\hbox{$.$}
\dotwidth=\wd0
\catcode`?=\active
\def?{\kern\dotwidth}
\begin{center}
{\begin{tabular}{|c|r@{}l|r@{}l|r@{}l|}
\hline
$\si/\fba$ & 
\multicolumn{2}{c|}{$\begin{array}{c}
                     \Pep\Pem \to 4 f \\
                     \mbox{running width}
                     \end{array}$} &
\multicolumn{2}{c|}{$\begin{array}{c}
                     \Pep\Pem \to 4 f \\
                     \mbox{constant width}
                     \end{array}$} &
\multicolumn{2}{c|}{$\begin{array}{c}
                     \eeffffg \\
                     \mbox{constant width}
                     \end{array}$} 
\\\hline\hline
$\Pne \Pnebar \Pem \Pep$
&\hspace{1.15cm}$    256.7 $&$( 3)$
&\hspace{1.15cm}$    257.1 $&$( 7)$
&\hspace{0.85cm}$     89.4 $&$( 2)$
\\\hline
$\nu_\mu \mu^+ \Pem \Pnebar$
&$    227.4 $&$( 1)$
&$    227.5 $&$( 1)$
&$     79.1 $&$( 1)$
\\\hline
$\nu_\mu \bar{\nu}_\mu \mu^- \mu^+$
&$    228.7 $&$( 1)$
&$    228.8 $&$( 1)$
&$     81.0 $&$( 2)$
\\\hline
$\nu_\mu \mu^+ \tau^- \bar{\nu}_\tau$
&$   218.55 $&$( 9)$
&$   218.57 $&$( 9)$
&$     76.7 $&$( 1)$
\\\hline
$\Pem \Pep \Pem \Pep$
&$    109.1 $&$( 3)$
&$    109.4 $&$( 3)$
&$     38.8 $&$( 4)$
\\\hline
$\Pem \Pep \mu^- \mu^+$
&$    116.6 $&$( 3)$
&$    116.4 $&$( 3)$
&$     43.4 $&$( 4)$
\\\hline
$\mu^- \mu^+ \mu^- \mu^+$
&$    5.478 $&$( 5)$
&$    5.478 $&$( 5)$
&$     3.37 $&$( 1)$
\\\hline
$\mu^- \mu^+ \tau^- \tau^+$
&$    11.02 $&$( 1)$
&$    11.02 $&$( 1)$
&$     6.78 $&$( 3)$
\\\hline
$\Pem \Pep \nu_\mu \bar{\nu}_\mu $
&$   14.174 $&$( 9)$
&$   14.150 $&$( 9)$
&$     5.36 $&$( 1)$
\\\hline
$\Pne \Pnebar \mu^- \mu^+$
&$    17.78 $&$( 6)$
&$    17.73 $&$( 6)$
&$     6.63 $&$( 2)$
\\\hline
$\nu_\tau \bar{\nu}_\tau\mu^- \mu^+$
&$   10.108 $&$( 8)$
&$   10.103 $&$( 8)$
&$    4.259 $&$( 9)$
\\\hline
$\Pne \Pnebar \Pne \Pnebar$
&$    4.089 $&$( 1)$
&$    4.082 $&$( 1)$
&$   0.7278 $&$( 7)$
\\\hline
$\Pne \Pnebar \nu_\mu \bar{\nu}_\mu$
&$    8.354 $&$( 2)$
&$    8.337 $&$( 2)$
&$    1.512 $&$( 1)$
\\\hline
$\nu_\mu \bar{\nu}_\mu \nu_\mu \bar{\nu}_\mu $
&$    4.069 $&$( 1)$
&$    4.057 $&$( 1)$
&$   0.7434 $&$( 7)$
\\\hline
$\nu_\mu \bar{\nu}_\mu \nu_\tau \bar{\nu}_\tau$
&$    8.241 $&$( 2)$
&$    8.218 $&$( 2)$
&$    1.511 $&$( 1)$
\\\hline
$\Pu\, \Pdbar\, \Pem \Pnebar$
&$    693.5 $&$( 3)$
&$    693.6 $&$( 3)$
&$    220.8 $&$( 4)$
\\\hline
$\Pu\, \Pdbar\, \mu^- \bar{\nu}_\mu$
&$    666.7 $&$( 3)$
&$    666.7 $&$( 3)$
&$    214.5 $&$( 4)$
\\\hline
$\Pem \Pep \Pu\, \Pubar$
&$    86.87 $&$( 9)$
&$    86.82 $&$( 9)$
&$     32.3 $&$( 2)$
\\\hline
$\Pem \Pep \Pd\, \Pdbar$
&$    43.02 $&$( 4)$
&$    42.95 $&$( 4)$
&$    16.17 $&$( 8)$
\\\hline
$\Pu\, \Pubar\, \mu^- \mu^+$
&$    24.69 $&$( 2)$
&$    24.69 $&$( 2)$
&$    12.70 $&$( 4)$
\\\hline
$\Pd\, \Pdbar\, \mu^- \mu^+$
&$    23.73 $&$( 1)$
&$    23.73 $&$( 1)$
&$    10.43 $&$( 2)$
\\\hline
$\Pne \Pnebar \Pu\, \Pubar$
&$    24.00 $&$( 2)$
&$    23.95 $&$( 2)$
&$     6.84 $&$( 1)$
\\\hline
$\Pne \Pnebar \Pd\, \Pdbar$
&$   20.657 $&$( 8)$
&$    20.62 $&$( 1)$
&$    4.319 $&$( 6)$
\\\hline
$\Pu\, \Pubar\, \nu_\mu \bar{\nu}_\mu$
&$   21.080 $&$( 5)$
&$   21.050 $&$( 5)$
&$    6.018 $&$( 9)$
\\\hline
$\Pd\, \Pdbar\, \nu_\mu \bar{\nu}_\mu$
&$   19.863 $&$( 5)!!!$
&$   19.817 $&$( 5)!!!$
&$    4.156 $&$( 5)!!$
\\\hline
$\Pu\, \Pubar\, \Pd\, \Pdbar$
&\multicolumn{2}{c|}{$   2064.1 ( 9),   !2140.8 ( 9)$}
&\multicolumn{2}{c|}{$   2064.3 ( 9),   !!?2141 ( 1)$}
&\multicolumn{2}{c|}{$   !?615 ( 1),     !!?672 ( 1)$}
\\\hline
$\Pu\, \Pdbar\, \Ps\, \Pcbar$
&$   2015.2 $&$( 8)$
&$   2015.3 $&$( 8)$
&$      598 $&$( 1)$
\\\hline
$\Pu\, \Pubar\, \Pu\, \Pubar$
&\multicolumn{2}{c|}{$   25.738 ( 7),    !!71.28 ( 4)$}
&\multicolumn{2}{c|}{$   25.721 ( 7),    !!71.30 ( 4)$}
&\multicolumn{2}{c|}{$    !9.78 ( 2),    !! 42.1 ( 1)$}
\\\hline
$\Pd\, \Pdbar\, \Pd\, \Pdbar$
&\multicolumn{2}{c|}{$   23.494 ( 6),    !!51.35 ( 3)$}
&\multicolumn{2}{c|}{$   23.448 ( 6),    !!51.32 ( 3)$}
&\multicolumn{2}{c|}{$    5.527 ( 7),    !28.68 ( 4)$}
\\\hline
$\Pu\, \Pubar\, \Pc\, \Pcbar$
&\multicolumn{2}{c|}{$    !51.61 ( 1),   !144.72 ( 9)$}
&\multicolumn{2}{c|}{$    !51.57 ( 1),   !144.75 ( 9)$}
&\multicolumn{2}{c|}{$    19.61 ( 4),    !! 86.1 ( 2)$}
\\\hline
$\Pu\, \Pubar\, \Ps\, \Psbar$
&\multicolumn{2}{c|}{$    !49.68 ( 1),   !126.52 ( 8)$}
&\multicolumn{2}{c|}{$    !49.62 ( 1),   !126.52 ( 8)$}
&\multicolumn{2}{c|}{$    15.17 ( 2),    !! 75.1 ( 2)$}
\\\hline
$\Pd\, \Pdbar\, \Ps\, \Psbar$
&\multicolumn{2}{c|}{$    !47.13 ( 1),   !104.79 ( 6)$}
&\multicolumn{2}{c|}{$    !47.02 ( 1),   !104.74 ( 6)$}
&\multicolumn{2}{c|}{$    11.10 ( 2),    !! 59.2 ( 1)$}
\\\hline
\end{tabular}}
\end{center}
\caption[]{Integrated cross sections for all representative  processes
  $\eeffff$ with running widths and constant widths and for the
  corresponding processes $\eeffffg$ with constant widths.
  If two numbers are given, the first  results from
pure electroweak diagrams and the second involves in addition 
gluon-exchange contributions.}
\label{ta:yellowreport}
\end{table}
In \refta{ta:yellowreport}, we list the integrated cross sections for
various processes $\eeffff$ with running widths and
constant widths,
and for the corresponding processes $\eeffffg$ 
with constant widths. For processes involving gluon-exchange
diagrams we give the cross sections resulting from the purely
electroweak diagrams and those including the gluon-exchange
contributions. The latter results include also the interference terms
between purely electroweak and gluon-exchange diagrams.
In \refta{ta:yellowreport} we provide a complete list of processes
for vanishing fermion masses. All processes $\eeffff(\ga)$ not explicitly
listed are equivalent to one of the given processes.

For NC processes $\eeffff$ 
with four neutrinos or four quarks in the final state
we find small deviations of roughly $0.2\%$ between the results with
constant and running widths. Assuming that a running width has been
used in \citere{CERN9601table}, we find very good agreement.

Unfortunately we cannot compare with most of the publications
\cite{vO94,vO96,Fu94,Ca97} for the bremsstrahlung processes
$\eeffffg$. In those papers, either the cuts are not (completely) specified,
or collinear photon emission is not excluded, and the corresponding
fermion-mass effects are taken into account.  Note that the
contributions of collinear photons dominate the results given there.

We have compared our results with the ones given in \citeres{Ae91,Ae91a},
where the total cross sections for $\eeffffg$ have been calculated for
purely leptonic and semi-leptonic final states. 
As done in
\citeres{Ae91,Ae91a} only diagrams involving two resonant \PW~bosons have
been taken into account
for this comparison. Table~\ref{ta:aepplitable} contains our
results corresponding to \refta{ta:aepplitable} of \citere{Ae91}.
Based on \citeres{Ae91,Ae91a},
we have chosen $\sqrt{s}=200\GeV$ and the input parameters
$\al=1/137.03599$, $\MW=80.9\GeV$, $\GW=2.14\GeV$, $\MZ=91.16\GeV$, 
$\GZ=2.46\GeV$, $\sw$ 
obtained from $\al/(2 \sw^2)=\GF \MW^2/(\pi \sqrt{2})$ with
$\GF=1.16637\times 10^{-5} \GeV^{-2}$, and constant gauge-boson
widths.  
The energy of the photon is
required to be larger than $E_{\ga,\min}$,
and the angle between the
photon and any charged fermion must be larger than
$\theta_{\ga,\min}$. A maximal photon energy is required,
$E_\ga<60\GeV$, in order to exclude contributions from the $\PZ$
resonance. Our results 
are consistent with those of \citeres{Ae91,Ae91a} within
the statistical error of 1\% given there. In some cases we find
deviations of 2\%.%
\footnote{Note that the input specified in \citeres{Ae91,Ae91a} is not
  completely clear even if the information of both publications is
  combined.}
\begin{table}
\newdimen\digitwidth
\setbox0=\hbox{0}
\digitwidth=\wd0
\catcode`!=\active
\def!{\kern\digitwidth}
\begin{center}
{\begin{tabular}{|c|c|r@{}l|r@{}l|r@{}l|r@{}l|}
\hline
$\si/\mathrm{fb}$ &
$\theta_{\ga,\min}, E_{\ga,\min}$ &
\multicolumn{2}{c|}{$1\GeV$} &
\multicolumn{2}{c|}{$5\GeV$} & 
\multicolumn{2}{c|}{$10\GeV$} & 
\multicolumn{2}{c|}{$15\GeV$} \\
\hline\hline
&$ !1^\circ  $
&$    53.54 $&$( 8)$
&$    27.57 $&$( 3)$
&$    16.96 $&$( 2)$
&$    11.22 $&$( 2)$
\\\cline{2-10}
leptonic
&$!5^\circ  $
&$    32.65 $&$( 4)$
&$    16.98 $&$( 3)$
&$    10.48 $&$( 2)$
&$     6.94 $&$( 1)$
\\\cline{2-10}
process
&$10^\circ $
&$    23.48 $&$( 3)$
&$    12.30 $&$( 2)$
&$     7.61 $&$( 2)$
&$     5.04 $&$( 1)$
\\\cline{2-10}
&$15^\circ $
&$    18.03 $&$( 2)$
&$     9.51 $&$( 2)$
&$     5.90 $&$( 1)$
&$     3.90 $&$( 1)$
\\\hline\hline
&$!1^\circ  $
&$    141.9 $&$( 2)$
&$    71.90 $&$( 8)$
&$    43.56 $&$( 5)$
&$    28.26 $&$( 4)$
\\\cline{2-10}
semi-leptonic
&$!5^\circ  $
&$     86.8 $&$( 1)$
&$    44.25 $&$( 6)$
&$    26.78 $&$( 4)$
&$    17.40 $&$( 3)$
\\\cline{2-10}
process
&$10^\circ $
&$    62.29 $&$( 7)$
&$    31.92 $&$( 5)$
&$    19.40 $&$( 4)$
&$    12.61 $&$( 3)$
\\\cline{2-10}
&$15^\circ $
&$    47.42 $&$( 5)$
&$    24.50 $&$( 4)$
&$    14.97 $&$( 3)$
&$     9.77 $&$( 2)$
\\\hline
\end{tabular}}
\end{center}
\caption[]{Comparison with Table 2 of \citere{Ae91}:
Cross sections resulting from diagrams involving two resonant W~bosons
for purely 
leptonic and semi-leptonic final states  and several photon separation
cuts}
\label{ta:aepplitable}
\end{table}

\subsection{Comparison of finite-width schemes}
\label{se:width}

{}As discussed in \citeres{Ba95,Ar95,Be97c,Pa95},
particular care has to be
taken when implementing the finite gauge-boson widths.  Differences
between results obtained with running or constant widths can already
be seen in \refta{ta:yellowreport}, where a typical LEP2 energy is
considered.
\begin{table}
\begin{center}
{\begin{tabular}{|c|c|r@{}l|r@{}l|r@{}l|r@{}l|}
\hline
\multicolumn{1}{|c|}{$\si/\fba$} &
$\sqrt{s}$ & 
\multicolumn{2}{c|}{$189\GeV$} & 
\multicolumn{2}{c|}{$500\GeV$} &
\multicolumn{2}{c|}{$2\TeV$} & 
\multicolumn{2}{c|}{$10\TeV$}
\\\hline\hline
& constant width
&$    703.5 $&$( 3)$
&$    237.4 $&$( 1)$
&$    13.99 $&$( 2)$
&$    0.624 $&$( 3)$
\\\cline{2-10}
$\Pep \Pem \to \Pu\, \Pdbar\, \mu^- \bar{\nu}_\mu $
& running width
&$    703.4 $&$( 3)$
&$    238.9 $&$( 1)$
&$    34.39 $&$( 3)$
&$    498.8 $&$( 1)$
\\\cline{2-10}
& complex-mass scheme
&$    703.1 $&$( 3)$
&$    237.3 $&$( 1)$
&$    13.98 $&$( 2)$
&$    0.624 $&$( 3)$
\\\hline\hline
& constant width
&$    224.0 $&$( 4)$
&$     83.4 $&$( 3)$
&$     6.98 $&$( 5)$
&$    0.457 $&$( 6)$
\\\cline{2-10}
$\Pep \Pem \to \Pu\, \Pdbar\, \mu^- \bar{\nu}_\mu \,\ga$
& running width
&$    224.6 $&$( 4)$
&$     84.2 $&$( 3)$
&$     19.2 $&$( 1)$
&$      368 $&$( 6)$
\\\cline{2-10}
& complex-mass scheme
&$    223.9 $&$( 4)$
&$     83.3 $&$( 3)$
&$     6.98 $&$( 5)$
&$    0.460 $&$( 6)$
\\\hline\hline
& constant width
&$    730.2 $&$( 3)$
&$    395.3 $&$( 2)$
&$    211.0 $&$( 2)$
&$    32.38 $&$( 6)$
\\\cline{2-10}
$\Pep \Pem \to \Pu\, \Pdbar\, \Pem \Pnebar$
& running width
&$    729.8 $&$( 3)$
&$    396.9 $&$( 2)$
&$    231.5 $&$( 2)$
&$    530.2 $&$( 6)$
\\\cline{2-10}
& complex-mass scheme
&$    729.8 $&$( 3)$
&$    395.1 $&$( 2)$
&$    210.9 $&$( 2)$
&$    32.37 $&$( 6)$
\\\hline\hline
& constant width
&$    230.0 $&$( 4)$
&$    136.5 $&$( 5)$
&$     84.0 $&$( 7)$
&$     16.8 $&$( 5)$
\\\cline{2-10}
$\Pep \Pem \to \Pu\, \Pdbar\, \Pe^- \Pnebar \,\ga $
& running width
&$    230.6 $&$( 4)$
&$    137.3 $&$( 5)$
&$     95.7 $&$( 7)$
&$      379 $&$( 6)$
\\\cline{2-10}
& complex-mass scheme
&$    229.9 $&$( 4)$
&$    136.4 $&$( 5)$
&$     84.1 $&$( 6)$
&$     16.8 $&$( 5)$
\\\hline
\end{tabular}}
\end{center}
\caption[]{Comparison of different width schemes for several processes 
and energies}
\label{ta:width}
\end{table}
In \refta{ta:width} we compare predictions for integrated cross
sections obtained 
by using a constant width, a running width, or the
complex-mass scheme for several energies. We consider
two semi-leptonic final states for $\eeffff(\ga)$.
The numbers show that the constant width and the complex-mass scheme
yield the same results within the statistical accuracy for
$\eeffff$ and
$\eeffffg$. In contrast, the results with the running width produce
totally wrong results for high energies.  The difference of the
running width with respect to the other implementations of the finite
width is up to $1\%$ already for $500\GeV$. Thus, the running width
should not be used for linear-collider energies.  As already stated
above, our default treatment of the finite width is the complex-mass
scheme in this chapter.

\subsection{Survey of photon-energy spectra}
\label{se:photonspectra}

\begin{figure}
\centerline{
\setlength{\unitlength}{1cm}
\begin{picture}(7.2,7)
\put(0,0){\includegraphics{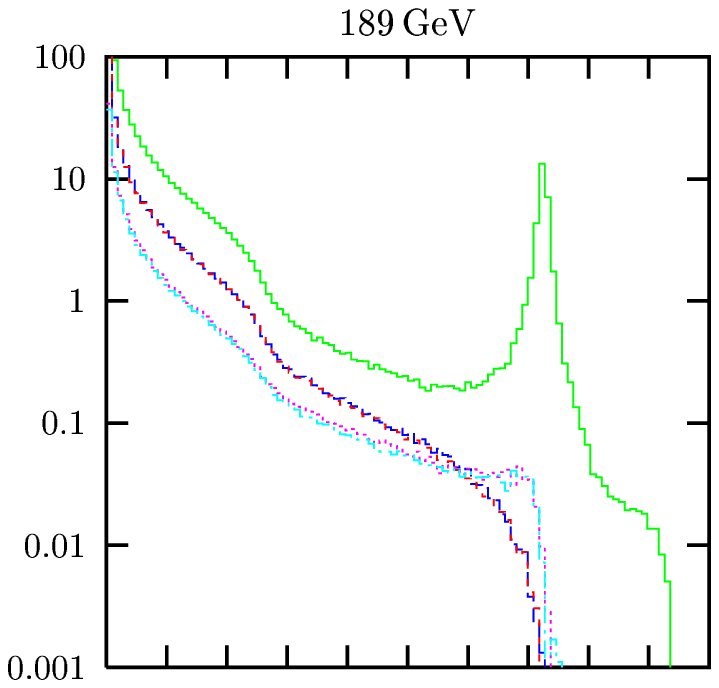}}
\put(-0.3,4.9){\makebox(1,1)[c]{$\frac{\rd \si}{\rd E_\ga}/
                              \frac{\fba}\GeV$}}
\end{picture}
\begin{picture}(7.2,7)
\put(0,0){\includegraphics{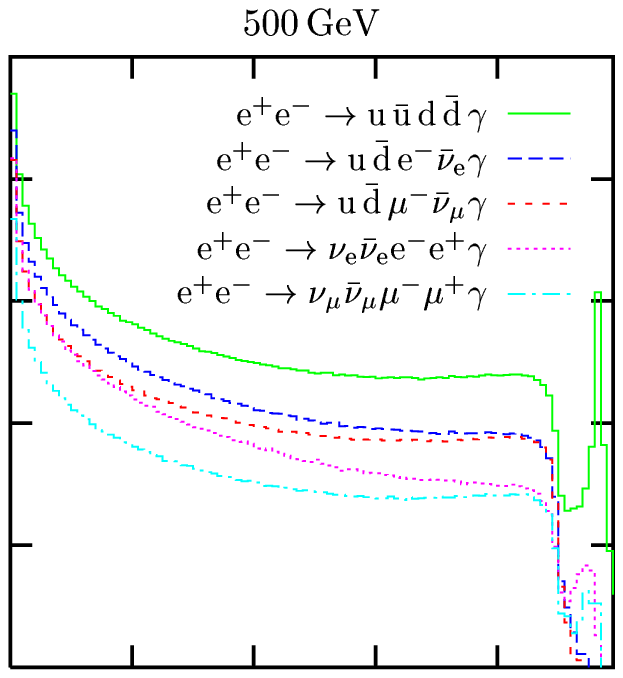}}
\end{picture}
}
\centerline{
\setlength{\unitlength}{1cm}
\begin{picture}(7.2,7)
\put(0,0){\includegraphics{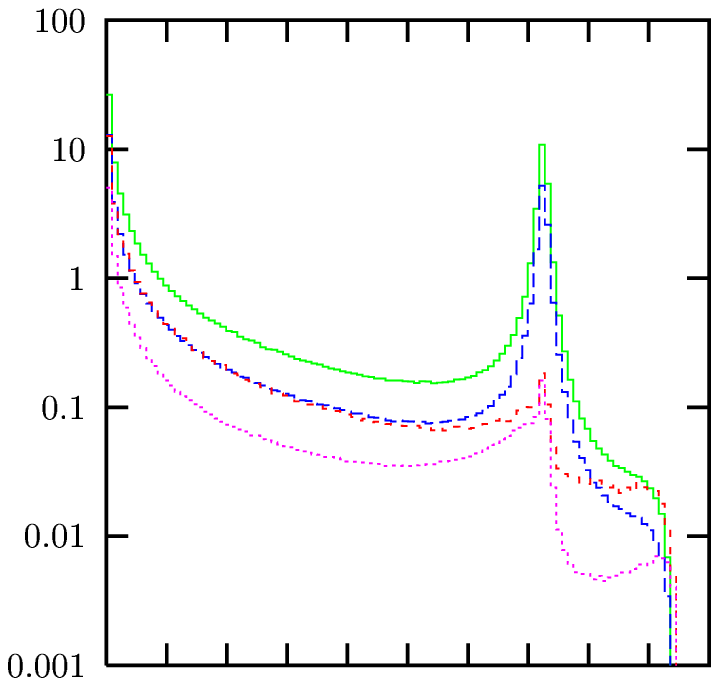}}
\put(-0.3,5.15){\makebox(1,1)[c]{$\frac{\rd \si}{\rd E_\ga}/
                              \frac{\fba}\GeV$}}
\end{picture}
\begin{picture}(7.2,7)
\put(0,0){\includegraphics{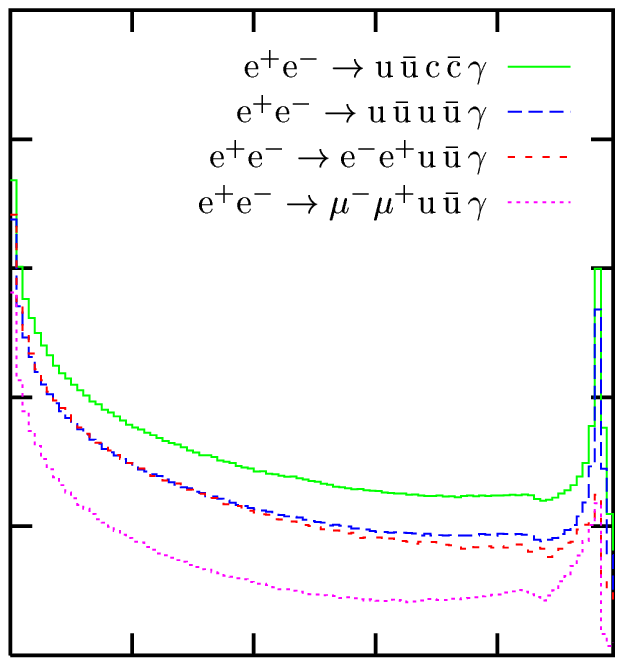}}
\end{picture}
}
\centerline{
\setlength{\unitlength}{1cm}
\begin{picture}(7.2,7)
\put(0,0){\includegraphics{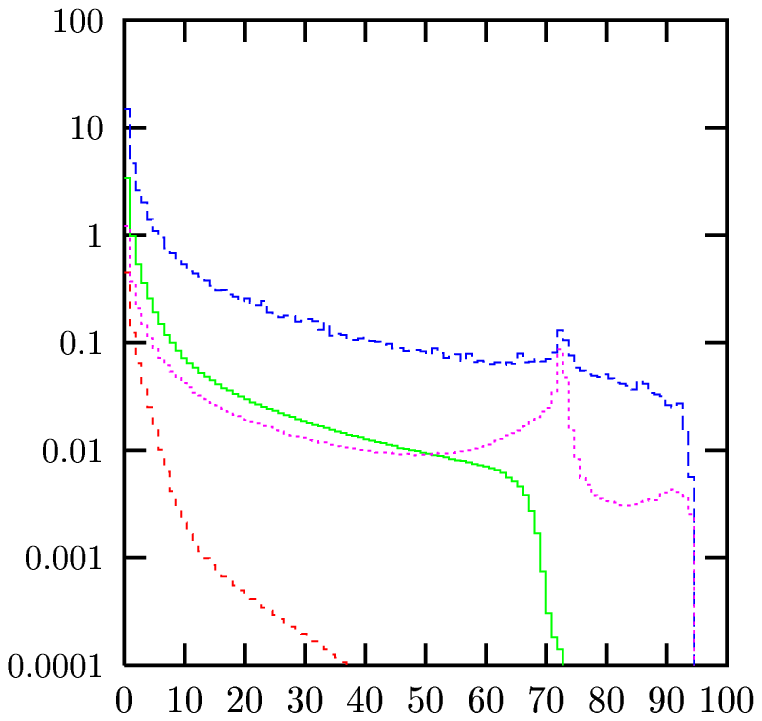}}
\put(-0.3,5.5){\makebox(1,1)[c]{$\frac{\rd \si}{\rd E_\ga}/
                              \frac{\fba}\GeV$}}
\put(4.5,-0.3){\makebox(1,1)[cc]{{$E_\ga/\GeV$}}}
\end{picture}
\begin{picture}(7.2,7)
\put(0,0){\includegraphics{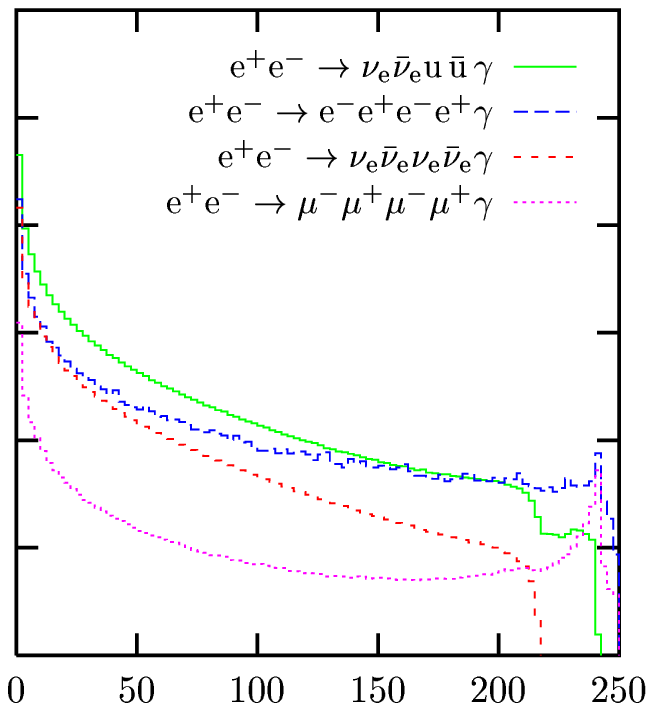}}
\put(3.5,-0.3){\makebox(1,1)[cc]{{$E_\ga/\GeV$}}}
\end{picture}
}
\caption[]{Photon-energy spectra for several processes and 
for  $\sqrt{s}=189\GeV$ and $500\GeV$}
\label{fi:photonspectra}
\end{figure}

\begin{figure}
{
\begin{center}
\begin{picture}(200,120)(-5,0)
\ArrowLine(0,10)(40,50)
\ArrowLine(40,70)(0,110)
\Photon(50,60)(150,60){2}{11}
\Photon(50,63)(120,90){2}{8}
\Photon(50,57)(120,30){2}{8}
\Vertex(120,90){2}
\Vertex(120,30){2}
\ArrowLine(120,90)(150,110)
\ArrowLine(150,70)(120,90)
\ArrowLine(120,30)(150,50)
\ArrowLine(150,10)(120,30)
\GCirc(50,60){20}{0}
\put(95,24){\makebox(1,1)[c]{$V_2$}}
\put(95,96){\makebox(1,1)[c]{$V_1$}}
\put(158,60){\makebox(1,1)[c]{$\gamma$}}
\Text(-10,120)[rt]{a)}
\end{picture}
\begin{picture}(200,120)
\ArrowLine(0,10)(40,50)
\ArrowLine(40,70)(0,110)
\Photon(50,63)(150,100){2}{12}
\Photon(50,57)(110,30){2}{8}
\Vertex(110,30){2}
\ArrowLine(110,30)(140,50)\ArrowLine(140,50)(170,70)
\ArrowLine(150,10)(110,30)
\Vertex(140,50){2}
\Photon(140,50)(170,30){2}{4}
\Vertex(170,30){2}
\ArrowLine(170,30)(200,50)
\ArrowLine(200,10)(170,30)
\GCirc(50,60){20}{0}
\put(85,26){\makebox(1,1)[c]{$Z$}}
\put(153,28){\makebox(1,1)[c]{$V_3$}}
\put(158,99){\makebox(1,1)[c]{$\gamma$}}
\Text(-10,120)[rt]{b)}
\end{picture}
\end{center}
}
\caption[]{Diagrams for important subprocesses, where
  $V_1,V_2=\PW,\PZ,\ga$, and $V_3=\ga,\Pg$ }
\label{fi:resonance}
\end{figure}
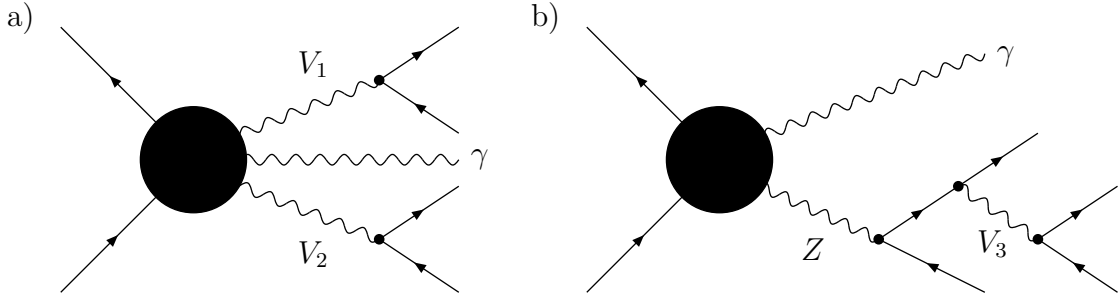

In \reffi{fi:photonspectra} we show the photon-energy spectra of several
processes for the typical LEP2 energy of $189\GeV$ and a possible
linear-collider energy of $500\GeV$.
The upper plots contain CC and CC/NC processes, the plots in the
middle and the lower plots contain NC processes.
Several spectra show threshold or peaking structures. These structures
are caused by diagrams in which the
photon is emitted from the initial state. The two important classes of
diagrams are  shown in \reffi{fi:resonance}.

The first class, shown in \reffi{fi:resonance}a, corresponds
to triple-gauge-boson-production subprocesses which yield dominant
contributions as long as the two virtual gauge bosons $V_1$ and $V_2$
can become simultaneously resonant. If the real photon takes the
energy $E_\ga$, defined
in the CM system, only the energy $\sqrt{s'}$, with
\beq
s' = s-2 \sqrt{s}\, E_\ga,
\eeq
is available for the production of the gauge-boson pair $V_1V_2$.
If at least one of the gauge bosons is massive, 
and if the photon becomes too hard, 
the two gauge bosons cannot be produced
on shell anymore, so that the spectrum falls off for $E_\gamma$ above
the corresponding threshold $E_\gamma^{V_1 V_2}$.
Using the threshold condition for the on-shell production of the
$V_1V_2$ pair,
\beq
\sqrt{s'}>M_{V_1}+M_{V_2},
\eeq
the value of $E_\gamma^{V_1 V_2}$ is determined by 
\beq
E_\gamma^{V_1 V_2} = \frac{s-(M_{V_1}+M_{V_2})^2}{2\sqrt{s}}.
\eeq
The values of the photon energies 
that cause such thresholds
can be found in \refta{ta:resonances}.
\begin{table}
\renewcommand{\arraystretch}{1.1}
\begin{center}
{\begin{tabular}{|c|c|c|c|c|c|c|c|c|}
\hline
$\sqrt{s}$ &
\multicolumn{4}{c|}{$189\GeV$} &
\multicolumn{4}{c|}{$500\GeV$} 
\\ \hline \hline
$V_1 V_2$ & 
$\PW \PW$ & $\PZ \PZ$ & $\ga \PZ$ & $\ga \ga$ &
$\PW \PW$ & $\PZ \PZ$ & $\ga \PZ$ & $\ga \ga$ 
\\ \hline
$E_\ga^{V_1 V_2}/\GeV$ & 
$26.3$ & $6.5$ & $72.5$& $94.5$ &
$224$ & $217$ & $242$ & $250$
\\ \hline
\end{tabular}}
\end{center}
\caption[]{Photon energies $E^{V_1 V_2}_\ga$ corresponding to thresholds}
\label{ta:resonances}
\end{table}
The value $E^{\gamma\gamma}_\ga$ corresponds to the 
upper endpoint of the photon-energy  spectrum, which is given by the beam
energy $\sqrt{s}/2$.
Since $\sqrt{s'}$ is fully determined by $s$ and
$E_\ga$, the contribution of the $V_1 V_2$-production subprocess to the 
$E_\ga$ spectrum qualitatively follows the energy
dependence of the total cross section for $V_1 V_2$ production (\cf
\citere{CERN9601table}, Fig.~1) above the corresponding thresholds.  
The cross sections for $\ga\ga$ and $\ga\PZ$ production strongly
increase with decreasing energy, while the ones for $\PZ\PZ$ and
$\PW\PW$ production are comparably flat.
Thus, the $\ga\ga$ and $\ga\PZ$-production subprocesses introduce contributions
in the photon-energy  spectra with resonance-like structures,
whereas the ones with $\PZ\PZ$ or $\PW\PW$ pairs yield 
edges.

The second class of important diagrams, shown in 
\reffi{fi:resonance}b corresponds to the production of a photon and a
resonant \PZ~boson that decays into four fermions. These diagrams are
important 
if the gauge boson $V_3$ 
is also resonant, \ie a photon or a gluon with small
invariant mass. In this case, the kinematics fixes the energy of the
real photon to
\beq
E_\ga = E_{\ga}^{\ga\PZ}=\frac{s-\MZ^2}{2\sqrt{s}},
\eeq
which corresponds to the $\ga\PZ$ threshold  in
\refta{ta:resonances}. This subprocess gives rise to resonance
structures at $E_{\ga}^{\ga\PZ}$, which are even enhanced by
$\al_{\mathrm{s}}/\al$ in the presence of gluon exchange.

In the photon-energy spectra of \reffi{fi:photonspectra} all these
threshold and resonance effects are visible.  The effect of the $\ga
\PZ$ peak can be nicely seen in different photon-energy spectra, in
particular in those where gluon-exchange diagrams contribute (\cf also
\reffi{fi:QCD}).  The effect of the WW threshold is present in the
upper two plots of \reffi{fi:photonspectra}. In the plot for
$\sqrt{s}=189\GeV$ the threshold for single W production causes the
steep drop of the spectrum for the pure CC processes above $70\GeV$.
Note that the CC cross sections are an order of magnitude larger than
the NC cross sections if the WW channel is open.  The ZZ threshold is
visible in the middle and lower plots for $\sqrt{s}=500\GeV$. The
$\ga$Z threshold (resulting from the graphs of \reffi{fi:resonance}a)
is superimposed on the $\ga \PZ$ peak (resulting from the graphs of
\reffi{fi:resonance}b) and therefore best recognizable in those
channels where the $\ga \PZ$ peak is absent or suppressed, \ie where a
neutrino pair is present in the final state or where at least no
gluon-exchange diagrams contribute.  Processes with four neutrinos in
the final state do not involve photonic diagrams and are therefore
small above the ZZ threshold.  The effects of the
triple-photon-production subprocess appear as a tendency of some
photon-energy spectra to increase near the maximal value of $E_\ga$
for two charged fermion--antifermion pairs in the final state.

\subsection{Triple-gauge-boson-production subprocesses}

\begin{figure}
\centerline{
\setlength{\unitlength}{1.1cm}
\begin{picture}(14.5,6.3)
\put(0.8,0){\includegraphics{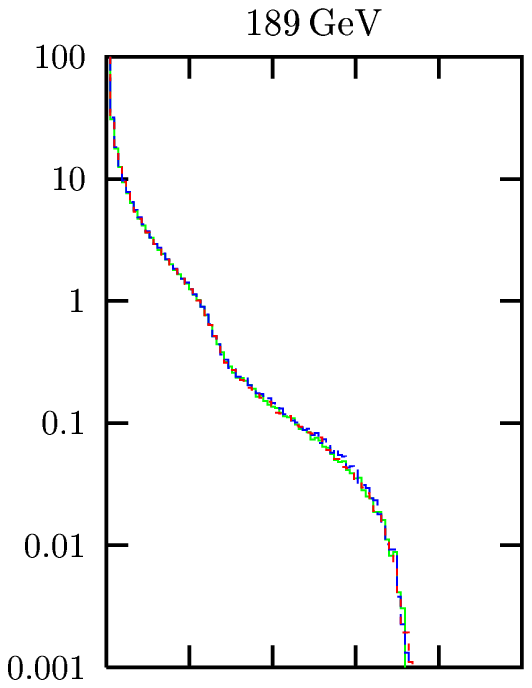}}
\put(5.1,0){\includegraphics{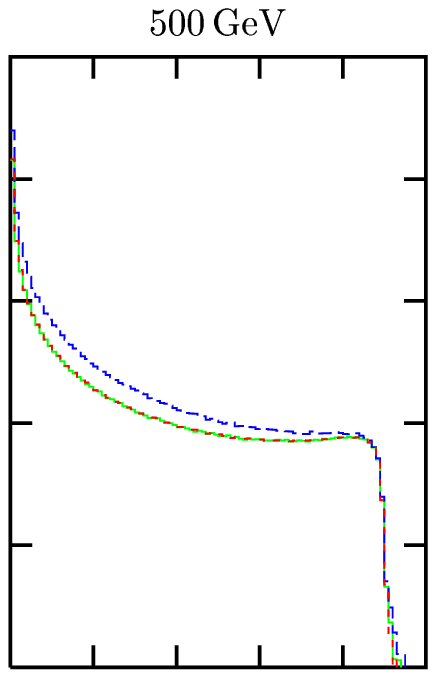}}
\put(9.4,0){\includegraphics{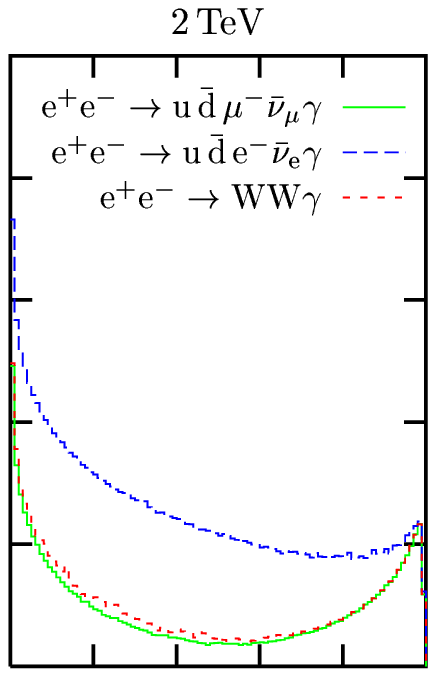}}
\put(0,5.5){\makebox(1,1)[c]{$\frac{\rd \si}{\rd E_\ga}/
                                 \frac{\fba}\GeV$}}
\end{picture}
}
\centerline{
\setlength{\unitlength}{1.1cm}
\begin{picture}(14.5,6.3)
\put(0.8,0){\includegraphics{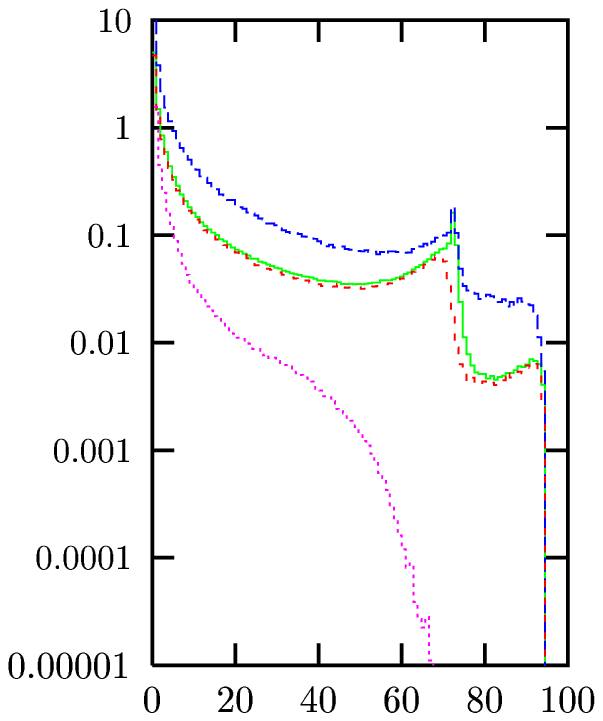}}
\put(5.1,0){\includegraphics{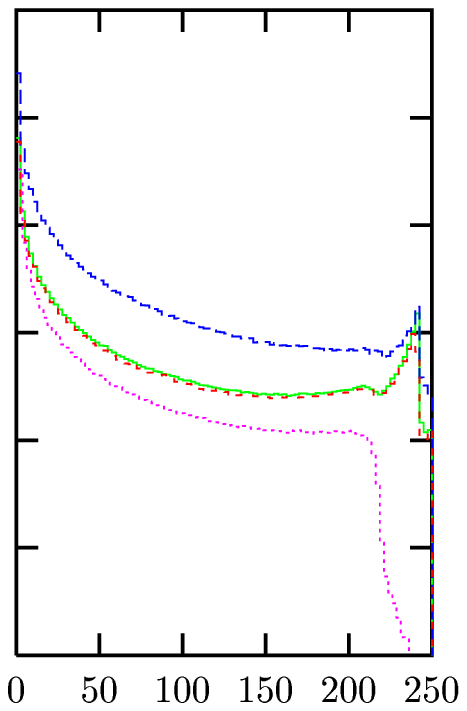}}
\put(9.4,0){\includegraphics{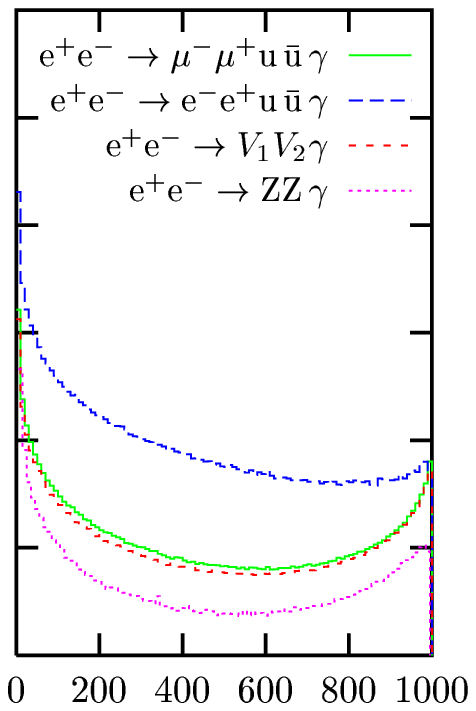}}
\put(0,5.95){\makebox(1,1)[c]{$\frac{\rd \si}{\rd E_\ga}/
                                 \frac{\fba}\GeV$}}
\put(7.5,-0.3){\makebox(1,1)[cc]{{$E_\ga/\GeV$}}}
\end{picture}
}
\caption[]{Photon-energy spectra resulting from the
  triple-gauge-boson-production subprocesses compared to those
  resulting from all diagrams ($V_1V_2$ includes ZZ, $\ga$Z, and
  $\ga\ga$)}
\label{fi:signal}
\end{figure}

In \reffi{fi:signal} we compare predictions that are based on the full
set of diagrams with those that include only the graphs associated
with the triple-gauge-boson-production subprocesses, i.e.\ the graphs
in \reffi{fi:resonance}a.  In addition we consider the contributions
of the $\PZ\PZ\ga$-production subprocess alone.
For CC processes, the photon-energy spectra resulting from the
$\PWp\PWm\ga$-production subprocess are close to those resulting from
all diagrams at LEP2 energies, but large differences are found for
higher energies and $\Pe^\pm$ in the final state. Note that the
spectra are shown on a logarithmic scale.  Even at LEP2 energies the
differences between the predictions for different final states may be
important, as can be seen, for instance, in \refta{ta:yellowreport} by
comparing the cross sections of $\Pep\Pem\to
\Pu\,\Pdbar\,\mu^-\bar\nu_\mu\ga$ and $\Pep\Pem\to
\Pu\,\Pdbar\,\Pem\Pnebar\ga$.  In the case of NC processes, already
for $189\GeV$ the contributions from $\PZ\PZ\ga$, $\PZ\ga\ga$, and
$\ga\ga\ga$ production are not sufficient: in the vicinity of the
$\ga\PZ$ peak sizeable contributions result from the
$\ga\PZ$-production subprocess (\reffi{fi:resonance}b) even for the
$\mu^+\mu^-\Pu\,\Pubar\,\ga$ final state.  For
$\Pep\Pem\to\Pem\Pep\Pu\,\Pubar\,\ga$ other diagrams become dominating
everywhere.  The contribution of $\PZ\PZ\ga$ production is always
small and could only be enhanced by invariant-mass cuts.  Note that
the triple-gauge-boson-production diagrams form a gauge-invariant
subset for NC processes, while this is not the case for CC processes.

\subsection{Relevance of gluon-exchange contributions}
\label{se:QCD}

In the analytical calculation of the matrix elements for
$\eeffff(\ga)$ in \refse{se:anres} we have seen that NC processes
with four quarks in the final state involve, besides purely electroweak,
 also gluon-exchange diagrams. Table \ref{ta:QCD}
illustrates the impact of these diagrams on the integrated cross
sections for a CM energy of $500\GeV$.
\begin{table}
\begin{center}
{\begin{tabular}{|l|r@{}l|r@{}l|r@{}l|r@{}l|}
\hline
\multicolumn{1}{|c|}{$\si/\fba$} & 
\multicolumn{2}{c|}{ew and gluon} & 
\multicolumn{2}{c|}{purely ew} & 
\multicolumn{2}{c|}{gluon} &
\multicolumn{2}{c|}{interference} 
\\\hline\hline
$\Pep \Pem \to \Pu\, \Pubar\, \Pc\, \Pcbar$
&\hspace{0.5cm}$    52.98 $&$( 4)$
&$   21.560 $&$( 6)$
&\hspace{0.2cm}$    31.38 $&$( 3)$
&\hspace{0.35cm}$     0.04 $&$( 5)$
\\\hline
$\Pep \Pem \to \Pu\, \Pubar\, \Pc\, \Pcbar\, \ga$
&$     29.8 $&$( 1)$
&$    10.38 $&$( 4)$
&$     19.6 $&$( 1)$
&$     -0.1 $&$( 1)$
\\\hline
$\Pep \Pem \to \Pu\, \Pubar \,\Pu\, \Pubar$
&$    26.25 $&$( 2)$
&$   10.765 $&$( 3)$
&$    15.34 $&$( 1)$
&$     0.14 $&$( 2)$
\\\hline
$\Pep \Pem \to \Pu\, \Pubar\, \Pu\, \Pubar\, \ga$
&$    14.83 $&$( 7)$
&$     5.16 $&$( 2)$
&$     9.52 $&$( 5)$
&$     0.15 $&$( 9)$
\\\hline
$\Pep \Pem \to \Pd\, \Pdbar\, \Pu\, \Pubar$
&$    901.2 $&$( 6)$
&$    876.4 $&$( 5)$
&$    24.24 $&$( 2)$
&$      0.6 $&$( 8)$
\\\hline
$\Pep \Pem \to \Pd\, \Pdbar\, \Pu\, \Pubar\, \ga$
&$      290 $&$( 1)$
&$      275 $&$( 1)$
&$    14.82 $&$( 8)$
&$        0 $&$( 1)$
\\\hline
\end{tabular}}
\end{center}
\caption[]{Full lowest order cross section (ew and gluon) and
  contributions of purely electroweak diagrams (ew), of
  gluon-exchange diagrams (gluon), and their interference for $500\GeV$}
\label{ta:QCD}
\end{table}
The results for the interference are obtained by subtracting the
purely electroweak and the gluon contribution from the total cross
section.  For pure NC processes the contributions of gluon-exchange
diagrams dominate over the purely electroweak graphs. This can be
understood from the fact that the gluon-exchange diagrams are enhanced
by the strong coupling constant, and, as discussed in
\refse{se:photonspectra}, that the diagrams with gluons replaced by
photons yield a sizeable contribution to the cross section.  For the
mixed CC/NC processes the purely electroweak diagrams dominate the
cross section. Here, the contributions from the
$\PWp\PWm\ga$-production subprocess are large compared to all other
diagrams, even if the latter are enhanced by the strong coupling.  At
$500\GeV$ the gluon-exchange diagrams contribute to the cross section
at the level of several per cent. The interference contributions are
relatively small.  As discussed at the end of \refse{se:hadfinstat},
this is due to the fact that interfering electroweak and
gluon-exchange diagrams involve different resonances. Note that the
interference vanishes for $\Pep \Pem \to
\Pu\,\Pubar\,\Pc\,\Pcbar\,\ga$, and the corresponding numbers in
\refta{ta:QCD} are only due to the Monte Carlo integration error.

\begin{figure}
\centerline{
\setlength{\unitlength}{1cm}
\begin{picture}(7.2,8)
\put(0,0){\includegraphics{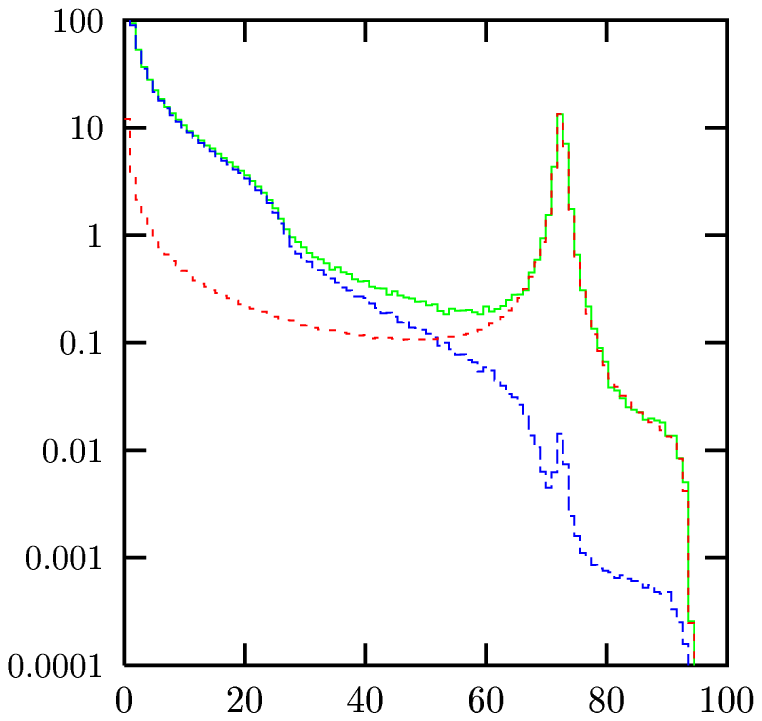}}
\put(-0.3,5.2){\makebox(1,1)[c]{$\frac{\rd \si}{\rd E_\ga}/
                              \frac{\fba}\GeV$}}
\put(4.5,-0.3){\makebox(1,1)[cc]{{$E_\ga/\GeV$}}}
\put(5.2,6.7){\makebox(1,1)[cc]{\small
        {$\Pep \Pem \to \Pu\,\Pubar\,\Pd\,\Pdbar\,\ga$}}}
\end{picture}
\begin{picture}(7.2,8)
\put(0,0){\includegraphics{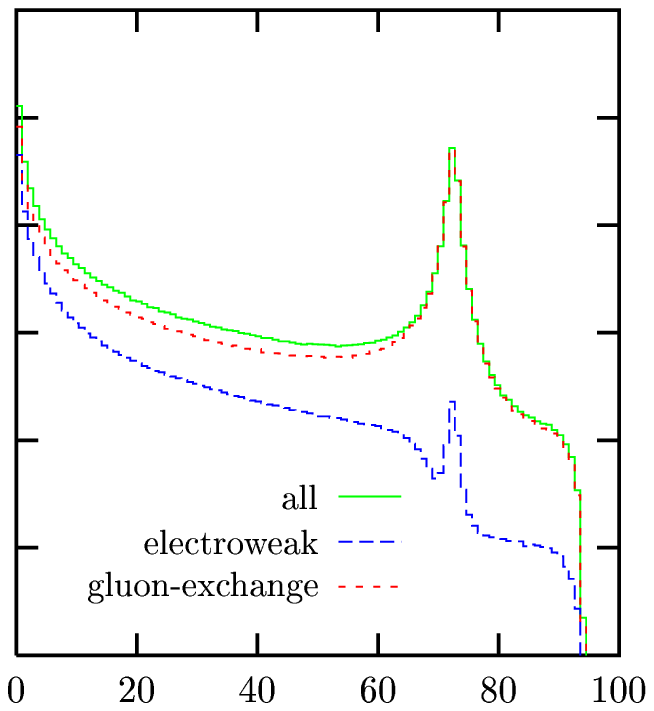}}
\put(3.5,-0.3){\makebox(1,1)[cc]{{$E_\ga/\GeV$}}}
\put(4.5,6.7){\makebox(1,1)[cc]{
        {\small$\Pep \Pem \to \Pu\,\Pubar\,\Pu\,\Pubar\,\ga$}}}
\end{picture}
}
\caption[]{Electroweak and gluon-exchange contributions to the
  photon-energy spectra for $\Pep \Pem \to
  \Pu\,\Pubar\,\Pd\,\Pdbar\,\ga$ and $\Pep \Pem \to
  \Pu\,\Pubar\,\Pu\,\Pubar\,\ga$ at $\sqrt{s}=189\GeV$}
\label{fi:QCD}
\end{figure}

In \reffi{fi:QCD} we show the photon-energy spectra for the processes
$\Pep \Pem \to \Pu\,\Pubar\,\Pd\,\Pdbar\,\ga$ and $\Pep \Pem \to
\Pu\,\Pubar\,\Pu\,\Pubar\,\ga$ together with the separate
contributions from purely electroweak and gluon-exchange diagrams.
The pure electroweak contributions are similar to the ones for 
$\Pep \Pem \to \Pu\,\Pdbar\,\mu^-\bar\nu_\mu \,\ga$ and $\Pep \Pem \to
\Pem \Pep \Pu\, \Pubar \,\ga$ in \reffi{fi:photonspectra}.  For the NC
process $\Pep\Pem \to \Pu\,\Pubar\,\Pu\,\Pubar\,\ga$, the
photon-energy spectrum is dominated by the gluon-exchange
contribution, which shows a strong peak at $72.5\GeV$ owing to the
$\ga\PZ$-production subprocess.    For the CC/NC process $\Pep \Pem \to
\Pu\,\Pubar\,\Pd\,\Pdbar\,\ga$, the electroweak diagrams dominate
below the WW threshold, whereas the gluon-exchange diagrams dominate
at the $\ga\PZ$ peak and above.  The interference between purely
electroweak and gluon-exchange diagrams is generally small.

\chapter{\boldmath Non-factorizable photonic corrections to 
$\mathrm e^+ \mathrm e^- \to \mathrm W^+ \mathrm W^- \to 4 f$}
\label{ch:nfc}

In this chapter we define and explicitly calculate the 
non-factorizable photonic corrections. 
It has been shown that they vanish in inclusive
quantities, \ie if the invariant masses of both $\PW$~bosons are
integrated out \cite{Fa94}. However, for non-inclusive quantities
these corrections do not vanish in general.  The non-factorizable
photonic corrections have already been investigated by two groups.
Melnikov and Yakovlev \cite{Me96} have given the analytical results
only in an implicit form and restrict the numerical evaluation to a
special phase-space configuration. Beenakker, Berends and Chapovsky
have provided both the complete formulae and an adequate numerical
evaluation \cite{Be97a,Be97b}, but do not find agreement with all
results of \citere{Me96}. For this reason, it is worth-while to
present the results of a third independent calculation.
The material of this chapter has been published in \citeres{De97,De98}.

We start out by discussing the definition of the virtual and real photonic
non-factoriz\-able corrections in double-pole approximation in detail.
Since only soft photons are relevant in double-pole approximation, 
the virtual non-factoriz\-able correction is just a factor to the 
lowest-order cross section. For the corresponding real correction, the
situation is similar, but in addition an integration over the photon 
momentum has to be performed. This requires a specification of the
phase-space parameterization,
which includes, in particular, the invariant masses of the \PW~bosons.
Usually these are defined via 
the invariant masses of the respective final-state fermion pairs
and are chosen as independent variables
\cite{Me96,Be97a,Be97b}. Experimentally, however,
the invariant mass of a \PW~boson is identified with the
invariant mass of the associated jet pair that necessarily includes
soft and collinear photons. Therefore,
the influence of the choice for the invariant masses of the \PW~bosons
on the non-factorizable corrections should be investigated 
in order to provide sound predictions for physical situations.

Besides the non-factorizable doubly-resonant corrections, the most
important effect of the instability of the \PW~bosons is the
modification of the Coulomb singularity. Since the off-shell Coulomb
singularity results from a scalar integral that also contributes to
the doubly-resonant non-factorizable corrections, it seems to be
natural to approximate this integral in such a way that both effects
are simultaneously included. This requires going beyond the strict
double-pole approximation.

\section{Definition of the approximation}
\label{sedefapp}

\subsection{Conventions and notations}
\label{se:cono}

We discuss corrections to the process
\beq\label{process}
\Pep(p_+) + \Pem(p_-) \;\to\; \PWp(k_+) + \PWm(k_-) \;\to\; 
f_1(k_1) + \bar f_2(k_2) + f_3(k_3) + \bar f_4(k_4).
\eeq
The relative charges of the 
fermions $f_i$ are represented by $Q_i$ with $i=1,\ldots,4$.
The masses of the external fermions, $m_i^2=k_i^2$ and
$\Me^2=p_\pm^2$, are neglected,
except where this would lead to mass singularities. 
The momenta of the intermediate \PW~bosons are defined by
\beq
k_+ = k_1+k_2, \qquad k_- = k_3+k_4,
\eeq
their complex mass squared and their respective invariant masses are denoted
by
\beq
M^2 = \MW^2-\ri\MW\Gamma_\PW, \qquad
M_\pm = \sqrt{k_\pm^2},
\eeq
respectively, and we introduce the variables
\beq
K_+ = k_+^2-M^2, \qquad K_- = k_-^2-M^2.
\eeq
Furthermore, we define the following kinematical invariants
\beqar
\begin{array}[b]{rlrl}
t=&(p_\pm-k_\pm)^2,\qquad  &u=&(p_\pm-k_\mp)^2,\\
t_{+ i}=&(p_+-k_i)^2,\qquad&u_{- i}=&(p_--k_i)^2,\qquad i=1,2,\\
t_{- i}=&(p_--k_i)^2,\qquad&u_{+ i}=&(p_+-k_i)^2,\qquad i=3,4,
\end{array}
\eeqar
and
\beqar
s &=& (p_+ + p_-)^2 = (k_+ + k_-)^2, \nl
s_{ij}  &=& (k_i + k_j)^2, \nl
s_{ijk} &=& (k_i + k_j + k_k)^2,
\qquad i,j,k=1,2,3,4, 
\eeqar
which obey the relations
\beqar
s &=& k_+^2 + k_-^2 + s_{13} + s_{14} + s_{23} + s_{24},
\qquad s_{12}=k_+^2, \qquad s_{34}=k_-^2, \nl
s_{ijk} &=& s_{ij} + s_{ik} + s_{jk},
\qquad i,j,k=1,2,3,4.
\eeqar

\subsection{Doubly-resonant virtual corrections}
\label{se:vnfpc}

The aim of this chapter is to evaluate the non-factorizable
corrections to the process \refeq{process} in 
double-pole approximation (DPA). The DPA takes
into account only the leading terms in an expansion around the poles
originating from the two resonant \PW~propagators.

In DPA, the lowest-order matrix element for the process
\refeq{process} factorizes into the matrix element for the on-shell
\PW-pair production,
$\M^\eeWW_\born(p_+,p_-,k_+,k_-)$, the
(transverse parts of the) propagators of these bosons, and the matrix
elements for the decays of these on-shell bosons,
$\M^\Wpff_\born(k_+,k_1,k_2)$ and $\M^\Wmff_\born(k_-,k_3,k_4)$:
\beq\label{mborn}
{\cal M}_{\born} = 
\sum_{\la_+,\la_-}\frac{\M^\eeWW_\born \M^\Wpff_\born \M^\Wmff_\born}{K_+ K_-}.
\eeq
The sum runs over the physical polarizations $\la_\pm$ of the 
\PWpm~bosons. 

The higher-order corrections to \refeq{process} can be separated into
factorizable and non-factorizable contributions \cite{Be96,Ae94,wwrev}.  
In the factorizable contributions the production of two $\PW$~bosons and
their subsequent decays are independent. The corresponding Feynman
diagrams can be split into three parts by cutting only the two
\PW-boson lines. The corresponding matrix element factorizes in the
same way as the lowest-order matrix element \refeq{mborn}.

The non-factorizable corrections comprise all those 
contributions in which W-pair
production and/or the subsequent W decays are not independent.
Obviously, this includes all Feynman diagrams in which a particle is
exchanged between the production subprocess and one of the decay
subprocesses or between the decay subprocesses. Examples for such
manifestly non-factorizable corrections are the diagrams (a), (b), and
(c) in \reffi{virtual_final_final_diagrams}.
\bfi
\begin{center}
\begin{picture}(360,230)(0,0)
\Text(0,220)[lb]{(a) type (\mfp)}
\Text(210,220)[lb]{(b) type (\mfp)}
\Text(0,90)[lb]{(c) type (\ffp)}
\Text(210,90)[lb]{(d) type (\mmp)}
\put(20,115){
\begin{picture}(150,100)(0,0)
\ArrowLine(30,50)( 5, 95)
\ArrowLine( 5, 5)(30, 50)
\Photon(30,50)(90,20){2}{6}
\Photon(30,50)(90,80){-2}{6}
\Vertex(60,65){2.0}
\GCirc(30,50){10}{0}
\Vertex(90,80){2.0}
\Vertex(90,20){2.0}
\ArrowLine(90,80)(120, 95)
\ArrowLine(120,65)(90,80)
\ArrowLine(120, 5)( 90,20)
\ArrowLine( 90,20)(105,27.5)
\ArrowLine(105,27.5)(120,35)
\Vertex(105,27.5){2.0}
\Photon(60,65)(105,27.5){-2}{5}
\put(86,50){$\gamma$}
\put(63,78){$W$}
\put(40,65){$W$}
\put(52,18){$W$}
\put(10, 5){$\mathrm{e}^-(p_-)$}
\put(10,90){$\mathrm{e}^+(p_+)$}
\put(125,90){$f_1(k_1)$}
\put(125,65){$\bar f_2(k_2)$}
\put(125,30){$f_3(k_3)$}
\put(125, 5){$\bar f_4(k_4)$}
\end{picture}
}
\put(210,115){
\begin{picture}(120,100)(0,0)
\ArrowLine(30,50)( 5, 95)
\ArrowLine( 5, 5)(30, 50)
\Photon(30,50)(90,80){-2}{6}
\Photon(30,50)(90,20){2}{6}
\Vertex(60,35){2.0}
\GCirc(30,50){10}{0}
\Vertex(90,80){2.0}
\Vertex(90,20){2.0}
\ArrowLine(90,80)(120, 95)
\ArrowLine(120,65)(105,72.5)
\ArrowLine(105,72.5)(90,80)
\Vertex(105,72.5){2.0}
\ArrowLine(120, 5)(90,20)
\ArrowLine(90,20)(120,35)
\Photon(60,35)(105,72.5){2}{5}
\put(86,46){$\gamma$}
\put(63,13){$W$}
\put(40,24){$W$}
\put(55,73){$W$}
\end{picture}
}
\put(20,-15){
\begin{picture}(120,100)(0,0)
\ArrowLine(30,50)( 5, 95)
\ArrowLine( 5, 5)(30, 50)
\Photon(30,50)(90,80){-2}{6}
\Photon(30,50)(90,20){2}{6}
\GCirc(30,50){10}{0}
\Vertex(90,80){2.0}
\Vertex(90,20){2.0}
\ArrowLine(90,80)(120, 95)
\ArrowLine(120,65)(105,72.5)
\ArrowLine(105,72.5)(90,80)
\Vertex(105,72.5){2.0}
\ArrowLine(120, 5)( 90,20)
\ArrowLine( 90,20)(105,27.5)
\ArrowLine(105,27.5)(120,35)
\Vertex(105,27.5){2.0}
\Photon(105,27.5)(105,72.5){2}{4.5}
\put(93,47){$\gamma$}
\put(55,73){$W$}
\put(55,16){$W$}
\end{picture}
}
\put(210,-15){
\begin{picture}(120,100)(0,0)
\ArrowLine(30,50)( 5, 95)
\ArrowLine( 5, 5)(30, 50)
\Photon(30,50)(90,80){-2}{6}
\Photon(30,50)(90,20){2}{6}
\Photon(70,30)(70,70){2}{3.5}
\Vertex(70,30){2.0}
\Vertex(70,70){2.0}
\GCirc(30,50){10}{0}
\Vertex(90,80){2.0}
\Vertex(90,20){2.0}
\ArrowLine(90,80)(120, 95)
\ArrowLine(120,65)(90,80)
\ArrowLine(120, 5)( 90,20)
\ArrowLine( 90,20)(120,35)
\put(76,47){$\gamma$}
\put(45,68){$W$}
\put(45,22){$W$}
\put(72,83){$W$}
\put(72,11){$W$}
\end{picture}
}
\end{picture}
\end{center}
\caption{Examples of non-factorizable photonic corrections
in ${\cal O}(\alpha)$. The shaded blobs stand for all tree-level graphs
contributing to $\Pep\Pem\to\PWp\PWm$.
Whenever Feynman diagrams with intermediate would-be Goldstone bosons
$\phi^\pm$ instead of $W^\pm$ bosons are relevant, the inclusion of such
graphs is implicitly understood.
}
\label{virtual_final_final_diagrams}
\efi
If the additional exchanged particle is massive, the corresponding 
correction has no double pole for on-shell 
\PW~bosons. However, if a photon is
exchanged between the different subprocesses, this leads to a
doubly-resonant contribution originating from the soft-photon region.
This can be directly seen from the usual soft-photon approximation
(SPA), which yields contributions proportional to the (doubly-resonant)
lowest-order contribution.

The doubly-resonant contributions can be extracted on the basis of a
simple power-counting argument.
For instance, the loop integral corresponding to diagram (c) in
\reffi{virtual_final_final_diagrams} is of the following form:
\beqar\label{vintexp} I&=&
\int\!\rd^4q
\frac{N(q,k_i)}{(q^2-\la^2)[(q-k_3)^2-m_3^2]
  [(q-k_-)^2-\MW^2][(q+k_+)^2-\MW^2][(q+k_2)^2-m_2^2]}\nl &=&
\int\!\rd^4q
\frac{N(q,k_i)}{(q^2-\la^2)(q^2-2qk_3)
  (q^2-2qk_- + k_-^2-\MW^2)(q^2+2qk_+ + k_+^2-\MW^2)(q^2+2qk_2)},
\nln
\eeqar
where we have introduced an infinitesimal photon mass $\la$ to regularize the
infrared (IR) singularity. The function $N(q,k_i)$ involves the
numerator of the Feynman integral, i.e.\ a polynomial in the momenta $q$ and
$k_i$, and possible further denominator factors originating from propagators 
(hidden in the blob of the diagrams)
that are regular for $q=0$ and $k_\pm^2=\MW^2$.  
For on-shell \PW~bosons ($k_\pm^2=\MW^2$), 
the integral has a quadratic IR singularity. For off-shell \PW~bosons,
part of the IR singularity is regularized by the
off-shellness, $k_\pm^2-\MW^2\ne 0$, such that the usual logarithmic IR
singularity remains. 
Vice versa, the off-shell result develops a pole
if either \PW~boson becomes on shell, and is thus doubly-resonant. 
Therefore, the quadratic IR singularity in the on-shell limit 
is characteristic 
of the doubly-resonant non-factorizable contributions. All terms that
involve a factor $q$ in the numerator are less 
IR-singular and therefore do not lead to 
doubly-resonant contributions and
can be omitted. Similarly, $q$ can be neglected in all denominator
factors included in $N(q,k_i)$. 
In summary, $q$ can be put to zero in $N(q,k_i)$ in DPA. 
We have checked this for various examples explicitly.
As a consequence, we are left with only scalar integrals,
and the non-factorizable virtual corrections are proportional to the
lowest-order matrix element. We call the resulting approximation
{\it extended soft-photon approximation} (ESPA). It differs from the
usual SPA only by the fact that $q$ is not
neglected in the resonant \PW~propagators. In ESPA,
diagram (c) in \reffi{virtual_final_final_diagrams} gives the following
contribution to the matrix element:
\beqar\label{vintESPA} 
\M^{\bar f_2 f_3} &=& \ri e^2 Q_2Q_3\M_\born
\int\!\frac{\rd^4q}{(2\pi)^4}
\frac{4k_2k_3}{(q^2-\la^2)(q^2-2qk_3)(q^2+2qk_2)}
\nl &&
\phantom{e^2 Q_2Q_3\M_\born\int\!\frac{\rd^4q}{(2\pi)^4}} \times
\frac{(k_-^2-\MW^2)(k_+^2-\MW^2)}{[(q-k_-)^2-\MW^2][(q+k_+)^2-\MW^2]}.
\eeqar

The $q^2$ terms in the last four denominators are not relevant in
the soft-photon limit and were omitted in \citeres{Me96,Be97a,Be97b}. In fact,
using the above power-counting argument it can easily be seen that the
differences of doubly-resonant 
contributions with and without these $q^2$
terms are non-doubly-resonant. We have chosen to keep the $q^2$ terms,
because we want to use the standard techniques for the evaluation of 
virtual scalar integrals \cite{tH79}. In DPA, \ie if we perform the
limit $k_\pm^2\to\MW^2$ after evaluating the integral, we should
obtain the same result.

In order to arrive at physical results, we have to incorporate the finite
width of the \PW~bosons. In DPA this can be done in 
at least two different ways:

As a first possibility, we perform the integrals for zero width and
afterwards put $k_\pm^2=\MW^2$ where this does not give rise to
singularities. In all other places, \ie in the resonant propagators
and in logarithms of the form $\ln(k_\pm^2-\MW^2+\ieps)$,
we replace $k_\pm^2-\MW^2+\ieps$ by $K_\pm=k_\pm^2-\MW^2+\ri\MW\GW$.
Since the width is only relevant in the on-shell limit, it is clear
that the (physical) on-shell width has to be used.

Alternatively, we introduce the width in the \PW~propagators before
integration. This has to be done with caution. If we introduce the
finite width by resumming \PW-self-energy insertions, the width depends on
the invariant mass of the \PW~boson and thus on the integration
momentum. Fortunately, the contribution we are interested in results
only from the soft-photonic region where the virtual \PW~bosons are
almost on shell. Therefore, we can insert the 
on-shell width inside the loop integral. After
performing the integral, we put $k_\pm^2=\MW^2$ and $\GW=0$ where this
does not lead to singularities. In DPA this gives the same results as
the above treatment.

In the following we write $M^2=\MW^2-\ri\MW\GW$ instead of $\MW^2$ in
the loop integrals. It is always understood that
$M^2$ and $k_\pm^2$ are replaced by $\MW^2$ where possible
after evaluation of the integrals. 

If we implement the width into the integrand, it is clear that only
the part of the integration region with $|q_0|\lsim\GW$ contributes 
in DPA. If $|q_0|\gg\GW$, one of the \PW~propagators must be
non-resonant and the contribution becomes negligible.

Once the width is introduced, it becomes evident that the 
relative error of the DPA is of the order of
$\GW/\mathrm{scale}$. Let $E_\CM$ be the center-of-mass (CM) energy and 
$\De E= E_\CM-2\MW$ be the available kinetic
energy of the \PW~bosons. Then, for $\De E\gsim\MW$ the scale is
given by \MW, for $\GW\lsim\De E\lsim\MW$ it is given by $\De E$ and
for $\De E\lsim\GW$ it is given by $\GW$. This shows that the DPA is
only sensible several $\GW$'s above threshold.
This is simply due to the fact that, close to threshold,
the phase space where both \PW~propagators can
become doubly-resonant is very small,
and the singly-resonant diagrams become important.

\subsection[Definition of the non-factorizable doubly-reso\-nant virtual 
corrections]
{Classification and gauge-independent definition of the non-factorizable 
doubly-reso\-nant virtual corrections}

Manifestly non-factorizable corrections arise from photon exchange
between the final states of the two \PW~bosons (\ffp),
between initial and final state (if), and between one of the intermediate
resonant \PW~bosons and the final state of the other \PW~boson (\mfp). 
Examples for these types of corrections are shown in 
\reffi{virtual_final_final_diagrams} (c),
\reffi{virtual_initial_final_diagrams} (a), and
\reffi{virtual_final_final_diagrams} (a,b), respectively.
\bfi
\begin{center}
\begin{picture}(360,230)(0,0)
\Text(0,220)[lb]{(a) type (if)}
\Text(210,220)[lb]{(b) type (im)}
\Text(0,90)[lb]{(c) type (mf)}
\Text(210,90)[lb]{(d) type (mm)}
\put(20,115){
\begin{picture}(120,100)(0,0)
\ArrowLine(27,55)(15, 75)
\Vertex(15,75){2.0}
\ArrowLine(15,75)( 3, 95)
\ArrowLine( 3, 5)(30, 50)
\Photon(30,50)(90,80){-2}{6}
\Photon(30,50)(90,20){2}{6}
\GCirc(30,50){10}{0}
\Vertex(90,80){2.0}
\Vertex(90,20){2.0}
\ArrowLine(90,80)(105,87.5)
\ArrowLine(105,87.5)(120, 95)
\ArrowLine(120,65)(90,80)
\ArrowLine(120, 5)( 90,20)
\ArrowLine( 90,20)(120,35)
\Vertex(105,87.5){2.0}
\PhotonArc(66.25,36.25)(64.25,52.9,142.9){2}{8}
\put(55,90){$\gamma$}
\put(68,55){$W$}
\put(55,16){$W$}
\end{picture}
}
\put(230,115){
\begin{picture}(150,100)(0,0)
\ArrowLine(27,55)(15, 75)
\Vertex(15,75){2.0}
\ArrowLine(15,75)( 3, 95)
\ArrowLine( 3, 5)(30, 50)
\Photon(30,50)(90,20){2}{6}
\Photon(30,50)(90,80){-2}{6}
\Vertex(60,65){2.0}
\GCirc(30,50){10}{0}
\Vertex(90,80){2.0}
\Vertex(90,20){2.0}
\ArrowLine(90,80)(120, 95)
\ArrowLine(120,65)(90,80)
\ArrowLine(120, 5)( 90,20)
\ArrowLine( 90,20)(120,35)
\PhotonArc(32.5,47.5)(32.596,32.47,122.47){2}{4.5}
\put(36,92){$\gamma$}
\put(75,61){$W$}
\put(51,48){$W$}
\put(52,18){$W$}
\end{picture}
}
\put(20,-15){
\begin{picture}(120,100)(0,0)
\ArrowLine(30,50)( 3, 95)
\ArrowLine( 3, 5)(30, 50)
\Photon(30,50)(90,80){-2}{6}
\Photon(30,50)(90,20){2}{6}
\Vertex(70,70){2.0}
\GCirc(30,50){10}{0}
\Vertex(90,80){2.0}
\Vertex(90,20){2.0}
\ArrowLine(90,80)(105,87.5)
\Vertex(105,87.5){2.0}
\ArrowLine(105,87.5)(120, 95)
\ArrowLine(120,65)(90,80)
\ArrowLine(120, 5)(90,20)
\ArrowLine(90,20)(120,35)
\PhotonArc(87.5,78.75)(19.566,26.565,206.565){2}{6}
\put(57,86){$\gamma$}
\put(77,62){$W$}
\put(50,48){$W$}
\put(55,16){$W$}
\end{picture}
}
\put(230,-15){
\begin{picture}(120,100)(0,0)
\ArrowLine(30,50)( 5, 95)
\ArrowLine( 5, 5)(30, 50)
\Photon(30,50)(90,80){-2}{6}
\Photon(30,50)(90,20){2}{6}
\Vertex(75,72.5){2.0}
\Vertex(50,60){2.0}
\GCirc(30,50){10}{0}
\Vertex(90,80){2.0}
\Vertex(90,20){2.0}
\ArrowLine(90,80)(120, 95)
\ArrowLine(120,65)(90,80)
\ArrowLine(120, 5)( 90,20)
\ArrowLine( 90,20)(120,35)
\PhotonArc(62.5,66.25)(13.975,26.565,206.565){-2}{3.5}
\put(55,90){$\gamma$}
\put(44,45){$W$}
\put(62,54){$W$}
\put(82,64){$W$}
\put(55,16){$W$}
\end{picture}
}
\end{picture}
\end{center}
\caption{Further examples of non-factorizable photonic corrections
in ${\cal O}(\alpha)$.}
\label{virtual_initial_final_diagrams}
\efi
In addition, there are diagrams where the photon 
does not couple to uniquely distinguishable subprocesses.
These contributions can be classified into photon-exchange 
contributions between one of the
intermediate resonant \PW~bosons and the final 
state of the same  
\PW~boson (mf), between the intermediate and the initial state (im), 
between the two intermediate \PW~bosons (\mmp), and within a single W-boson 
line, \ie the photonic part of the \PW-self-energy corrections (mm). 
Diagrams contributing to these types of corrections are given in 
\reffi{virtual_initial_final_diagrams} (c),
\reffi{virtual_initial_final_diagrams} (b),
\reffi{virtual_final_final_diagrams} (d), and
\reffi{virtual_initial_final_diagrams} (d), respectively.
Because the photon coupling to the \PW~boson can be
attributed to the decay or the production subprocesses, these diagrams
involve both factorizable and non-factorizable corrections.

In order to define the non-factorizable corrections, we have to specify
how the factorizable contributions are split off. This should be done
in such a way that the non-factorizable corrections become gauge-independent.
In \citeres{Be97a,Be97b} this was reached by exploiting the fact that 
in ESPA the matrix element can be viewed as a product of 
the lowest-order matrix element with two conserved currents. 
Taking all interferences between
the positively and the negatively charged currents 
arising from the outgoing W~bosons and fermions
gives a gauge-independent result.

We have chosen a different definition of the non-factorizable corrections,
which, however, turns out to be equivalent to the one of \citeres{Be97a,Be97b} 
in DPA. Our approach has the advantage of providing a clear procedure how 
to combine factorizable and non-factorizable contributions to the full
${\cal O}(\alpha)$ correction in DPA. Because the complete matrix
element is gauge-independent order by order, the sum of all
doubly-resonant $\Oa$ corrections must be gauge-independent. On the
other hand, the factorizable doubly-resonant corrections can be defined
by the product of gauge-independent on-shell matrix elements for 
W-pair production and \PW~decays and the 
(transverse parts of the) \PW~propagators, 
\beqar\label{mvirt}
{\cal M}_{\mathrm{f}} &=& 
\sum_{\la_+,\la_-} \frac{1}{K_+ K_-}
\Big( \de\M^\eeWW \M^\Wpff_\born \M^\Wmff_\born
\\ && {}
+ \M^\eeWW_\born \de\M^\Wpff \M^\Wmff_\born
+ \M^\eeWW_\born \M^\Wpff_\born \de\M^\Wmff \Big),
\nn
\eeqar
where $\de\M^\eeWW$, $\de\M^\Wpff$, and $\de\M^\Wmff$ denote the
one-loop amplitudes of the respective subprocesses. We can
define the non-factorizable doubly-resonant corrections by subtracting
the factorizable doubly-resonant corrections from the complete
doubly-resonant corrections. This definition allows 
us to calculate the complete doubly-resonant
corrections by simply adding the factorizable corrections, defined via
the on-shell matrix elements, to our results. Our definition
can be applied diagram by diagram. In this
way, all diagrams that are neither manifestly factorizable nor
manifestly non-factorizable can be split. 
Such diagrams receive doubly-resonant contributions from the complete
range of the photon momentum $q$, and not only from the soft-photon
region. This is obviously due to the presence of two explicit
resonant propagators. However, after subtracting the factorizable 
contributions, all 
doubly-resonant terms that are not IR-singular in the on-shell limit 
cancel exactly, i.e.\ only the soft-photon region contributes. 
Consequently, also in this case 
$q$ can be neglected everywhere except for the denominators that become
IR-singular in the on-shell limit.
As an example, we give the
non-factorizable correction originating from diagram (d) of
\reffi{virtual_final_final_diagrams}:%
\footnote{We use the sign $\sim$ to indicate an equality within 
DPA, \ie up to non-doubly-resonant terms.}
\beqar\label{virtual_coulomb_integral}
\M_{\mathrm{nf}}^{\PWp\PWm} & \sim & \ri e^2\M_\born\biggl\{
\int\!\frac{\rd^4q}{(2\pi)^4}\frac{4k_+k_-}
{q^2[(q + k_+)^2-M^2] [(q - k_-)^2-M^2]}\nl
&& \phantom{e^2\M_\born\biggl[}
-\biggl[\int\!\frac{\rd^4q}{(2\pi)^4}\frac{4k_+k_-}
{(q^2-\la^2)(q^2+2qk_+)(q^2-2qk_-)}
\biggr]_{k_\pm^2=\MW^2}\biggr\}.
\hspace{1em}
\eeqar
This example shows that the on-shell subtraction introduces
additional IR singularities. If the IR singularities in the 
non-factorizable real corrections are regularized in the same way, 
they cancel in the sum.
In \refeq{virtual_coulomb_integral} an infinitesimal photon mass
$\lambda$ is used as IR regulator, but we have repeated the same
calculation also by using a finite W-decay width as IR regulator instead
of $\lambda$, leading to the same results in the sum of virtual and
real photonic corrections.

We illustrate our definition of the non-factorizable corrections also for 
the photonic contribution to the \PW-self-energy correction 
[diagram (d) of \reffi{virtual_initial_final_diagrams}]. 
The non-factorizable part of the \PWp~self-energy reads 
\beqar\label{virtual_WSE_integral}
\M_{\mathrm{nf}}^{\PWp\PWp} & \sim & -\ri e^2\M_\born\biggl\{
\int\!\frac{\rd^4q}{(2\pi)^4}\frac{4k_+^2}{q^2[(q + k_+)^2-M^2](k_+^2-M^2)}\nl
&& \phantom{-\ri e^2\M_\born\biggl\{}
- \frac{1}{k_+^2-M^2}
\biggl[\int\!\frac{\rd^4q}{(2\pi)^4}\frac{4k_+^2}{q^2(q^2+2qk_+)}
\biggr]_{k_+^2=M^2} \nl
&& \phantom{-\ri e^2\M_\born\biggl\{}
+\biggl[\int\!\frac{\rd^4q}{(2\pi)^4}\frac{4k_+^2}{(q^2-\la^2)(q^2+2qk_+)^2}
\biggr]_{k_+^2=\MW^2}\biggr\}.
\eeqar
The first integral results from the off-shell self-energy diagram, the second
from the corresponding mass-renormalization term, and the third
integral is the negative of the on-shell limit of the first two integrals.
The integrals in \refeq{virtual_WSE_integral} are UV-divergent and can
be easily evaluated in dimensional regularization. 

The gauge independence of the non-factorizable corrections has been
ensured by construction. The consistent evaluation of gauge theories
requires, besides gauge independence of the physical matrix elements,
the validity of Ward identities. 
It was found in \citere{Ba95,Ku95,Ar95,Be97c} that the violation of 
Ward identities 
can lead to completely wrong predictions. The procedure described above
for extracting the non-factorizable corrections from the full matrix
element does not lead to problems with the Ward identities that rule the 
gauge cancellations inside matrix elements. This is due to the fact that
the non-factorizable corrections are proportional to the Born matrix 
element. Therefore, if Ward identities and gauge cancellations 
are under control in lowest order, the same is true for the non-factorizable 
corrections.

Finally, we show how our definition of the non-factorizable corrections
can be rephrased in terms of products of appropriately defined currents.
By using 
\beq \label{partfrac}
\frac{1}{(q\pm k_\pm)^2 - M^2} = \frac{1}{q^2\pm 2qk_\pm}
\biggl[1-\frac{k_\pm^2-M^2}{(q\pm k_\pm)^2 - M^2}\biggr],
\eeq
and the fact that in DPA
$k_\pm^2$ can be put to $\MW^2$ before integration
in integrals that do not depend on $M^2$,
the contribution \refeq{virtual_coulomb_integral} can be expressed as
\beqar\label{virtual_coulomb_integral2}
\M_{\mathrm{nf}}^{\PWp\PWm}&\sim& \ri e^2\M_\born
\int\!\frac{\rd^4q}{(2\pi)^4}\frac{4k_+k_-}{(q^2-\la^2)(q^2+2qk_+)(q^2-2qk_-)}
\biggl[-\frac{k_+^2-M^2}{(q+k_+)^2-M^2}
\nl&& \phantom{e^2\M_\born} {}
-\frac{k_-^2-M^2}{(q-k_-)^2-M^2}
+\frac{k_+^2-M^2}{(q+k_+)^2-M^2}\frac{k_-^2-M^2}{(q-k_-)^2-M^2}
\biggr].
\hspace{2em}
\eeqar 

The other non-factorizable corrections that involve photons coupled to
\PW~bosons can be rewritten in a similar way. Finally, all non-factorizable 
virtual corrections can be cast into the following form:
\beqar\label{virtual_nonfac}
\M_{\mathrm{nf}}^{\virt} &\sim& \ri\M_\born \int\!\frac{\rd^4q}{(2\pi)^4}
\frac{1}{q^2-\la^2}\left[j^{\eeWW,\mu}_{\virt,+} j^\Wpff_{\virt,\mu}
+j^{\eeWW,\mu}_{\virt,-} j^\Wmff_{\virt,\mu}
\right. \nl &&
\phantom{e^2\M_\born \int\!\frac{\rd^4q}{(2\pi)^4}\frac{1}{q^2-\la^2}\Big[}
\left.{}
+j^{\Wpff,\mu}_\virt j_{\virt,\mu}^\Wmff\right]
\eeqar
with
\beqar
\label{eq:virtcurr}
j_{\virt,\pm,\mu}^\eeWW &=&
e\biggl( \frac{2k_{+\mu}}{q^2+2q k_+} + \frac{2k_{-\mu}}{q^2-2q k_-}
   \pm \frac{2p_{-\mu}}{q^2\pm 2q p_-} 
   \mp \frac{2p_{+\mu}}{q^2\pm 2q p_+}\biggr)\nlc
j_{\virt,\mu}^\Wpff&=&
e\biggl( Q_1 \frac{2k_{1\mu}}{q^2+2q k_1} - Q_2 \frac{2k_{2\mu}}{q^2+2q k_2}
      - \frac{2k_{+\mu}}{q^2+2q k_+}\biggr)\frac{k_+^2-M^2}{(k_++q)^2-M^2} \nlc
j_{\virt,\mu}^\Wmff&=&
-e\biggl( Q_3 \frac{2k_{3\mu}}{q^2-2q k_3} - Q_4 \frac{2k_{4\mu}}{q^2-2q k_4}
      + \frac{2k_{-\mu}}{q^2-2q k_-}\biggr)\frac{k_-^2-M^2}{(k_--q)^2-M^2}.
\hspace{3em}
\eeqar
The last term in \refeq{virtual_nonfac} originates from the Feynman graphs 
shown in \reffi{virtual_final_final_diagrams} and 
those where the final-state fermions are appropriately interchanged 
[interference terms (\ffp), (\mfp), and (\mmp)].
The contributions involving the current $j_{\virt,\mu}^\eeWW$ contain 
the interference terms (if), (mf), (im), (mm), and the remaining
contributions of (\mfp) and (\mmp).
The contribution of the \PWp~self-energy is given, for instance, 
by the product of the two terms involving $k_{+\mu}$ in 
$j^{\eeWW}_{\virt,\mu}$ and $j_{\virt,\mu}^\Wpff$.

In DPA, the $q^2$ terms in the denominators 
of \refeq{eq:virtcurr} can be neglected, and the
currents are conserved. The currents $j_{\virt,\mu}^\Wpff$
and $j_{\virt,\mu}^\Wmff$ are the ones mentioned in \citeres{Be97a,Be97b}.
For the virtual corrections,
this shows that our definition of non-factorizable doubly-resonant corrections
coincides with the one of \citeres{Be97a,Be97b} in DPA.
 
\subsection{Doubly-resonant real corrections}
\label{se:rnfpc}

The photonic virtual corrections discussed above are IR-singular and
have to be combined with the corresponding real corrections in order
to arrive at a sensible physical result.
The real corrections originate from the process
\beqar\label{eeffffa}
\Pep(p_+) + \Pem(p_-) & \;\to\; & \PWp(k'_+) + \PWm(k'_-)\, [{}+ \ga(q)]
\nl
& \;\to\; &
f_1(k'_1) + \bar f_2(k'_2) + f_3(k'_3) + \bar f_4(k'_4)+ \ga(q). 
\eeqar
Note that we have marked the fermion momenta $k'_i$ by primes in order
to distinguish them from the respective momenta without real photon
emission. 
The momenta of the W~bosons are $k'_+=k'_1+k'_2$ and $k'_-=k'_3+k'_4$
if the photon is emitted in the initial state or in the final state of
the other \PW~boson, and $\bar k_+=k'_1+k'_2+q$ or $\bar k_-=k'_3+k'_4+q$
if the photon is emitted in the final state of the respective \PW~boson.

The non-factorizable corrections 
induced by the process \refeq{eeffffa} arise 
from interferences between diagrams where the photon is
emitted from different subprocesses.
A typical non-factorizable contribution is shown in
\reffi{real_non_factorizable_diagram}. 
\bfi
\centerline{
\begin{picture}(240,100)(0,0)
\ArrowLine(30,50)( 5, 95)
\ArrowLine( 5, 5)(30, 50)
\Photon(30,50)(90,80){-2}{6}
\Photon(30,50)(90,20){2}{6}
\GCirc(30,50){10}{0}
\Vertex(90,80){2.0}
\Vertex(90,20){2.0}
\ArrowLine(90,80)(120, 95)
\ArrowLine(120,65)(105,72.5)
\ArrowLine(105,72.5)(90,80)
\ArrowLine(120, 5)( 90,20)
\ArrowLine( 90,20)(120,35)
\Vertex(105,72.5){2.0}
\PhotonArc(120,65)(15,150,270){2}{3}
\put(55,73){$W$}
\put(55,16){$W$}
\put(100,47){$\gamma$}
\DashLine(120,0)(120,100){6}
\PhotonArc(120,35)(15,-30,90){2}{3}
\Vertex(135,27.5){2.0}
\ArrowLine(150,80)(120,95)
\ArrowLine(120,65)(150,80)
\ArrowLine(120, 5)(150,20)
\ArrowLine(150,20)(135,27.5)
\ArrowLine(135,27.5)(120,35)
\Vertex(150,80){2.0}
\Vertex(150,20){2.0}
\Photon(210,50)(150,80){2}{6}
\Photon(210,50)(150,20){-2}{6}
\ArrowLine(210,50)(235,95)
\ArrowLine(235, 5)(210,50)
\GCirc(210,50){10}{0}
\put(177,73){$W$}
\put(177,16){$W$}
\end{picture}
}
\caption{Example of a non-factorizable real correction.}
\label{real_non_factorizable_diagram}
\efi
Including the integration over the
photon phase space, this contribution 
has the following form:
\beqar\label{rintex}
\Ibr&=&
\int\!\frac{\rd^3\bq}{2q_0}
\frac{N(q,k'_i)\de(p_++p_--k'_1-k'_2-k'_3-k'_4-q)}
{[(q+k'_2)^2-m_2^2][(q+k'_+)^2-M^2](k_-^{\prime 2}-M^2)}
\nn\\*
&& {} \phantom{\int\!\frac{\rd^3\bq}{2q_0}} \times
\Biggl\{ \frac{1}{[(q+k'_3)^2-m_3^2](k_+^{\prime 2}-M^2)[(q+k'_-)^2-M^2]}
\Biggr\}^*
\,\biggr|_{q_0=\sqrt{{\bf q}^2+\la^2}}
\nl
&=&\int\!\frac{\rd^3\bq}{2q_0}
\frac{N(q,k'_i)\de(p_++p_--k'_1-k'_2-k'_3-k'_4-q)}
{(2qk'_2)(2qk'_+ + k_+^{\prime 2}-M^2)(k_-^{\prime 2}-M^2)}
\nn\\*
&& {} \phantom{\int\!\frac{\rd^3\bq}{2q_0}} \times
\frac{1}{(2qk'_3)[k_+^{\prime 2}-(M^*)^2][2qk'_- + k_-^{\prime 2}-(M^*)^2]}
\,\biggr|_{q_0=\sqrt{{\bf q}^2+\la^2}},
\eeqar
where we again use a photon mass $\la$ to regularize the IR
singularities, and $N(q,k'_i)$ has the same meaning as above.
Again, the doubly-resonant contributions are characterized by a
quadratic IR singularity for $k_\pm^{\prime 2}=\MW^2$, 
$\Gamma_\PW\to 0$, and only soft-photon emission is relevant in DPA. 
For this reason, the \PW~bosons are nearly on shell, and the on-shell width
is appropriate. As for the virtual corrections, the introduction
of the width before or after phase-space integration leads to the same
results in DPA. As already indicated in \refeq{rintex},
in the following real integrals we use $M^2$ 
with the understanding that it has to be replaced by $\MW^2$ after
integration where possible.

The aim is to integrate over the photon momentum analytically and to 
relate the fermion momenta $k'_i$ to the ones of the process without photon 
emission, $k_i$. Primed and unprimed momenta differ
by terms of the order of the photon momentum: $k_i'=k_i+\O(q_0)$.
In DPA we can neglect $q$ in $N(q,k'_i)$,
leading to the replacement $N(q,k'_i)\to N(0,k_i)$.
Moreover, we can extend the integration region for $q_0$ to infinity,
because large photon momenta yield negligible contributions in DPA. 
After extension of the integration region the integral becomes
Lorentz-invariant. 

While the correction factor to the lowest-order cross section 
is universal in SPA for all
observables, the correction factor is non-universal in ESPA. In order to
define this correction factor in a unique way, one has to specify the
parameterization of phase space, i.e.\ the variables that are kept fixed
when the photon momentum is integrated over. 
This fact has not been addressed in the literature so far. 

Let us consider this problem in more detail.
It can be traced back to the
appearance of the photon momentum $q$ in the $\de$-function for momentum
conservation. In the usual SPA $q$ is neglected in this $\de$-function, 
which is sensible if the exact matrix element is a slowly varying function
of $q$ in the vicinity of $q=0$. However, in the presence of resonant
propagators, in which $q$ cannot be neglected, the simple omission of $q$
in the  momentum-conservation $\de$-function leads to ambiguous results:
putting $q=0$ in the $\de$-function
and identifying $k'_i$ with $k_i$ in \refeq{rintex} yields
\beqar\label{rintex1}
\Ibr &\to&
\int\!\frac{\rd^3\bq}{2q_0}
\frac{N(0,k_i)\de(p_++p_--k_1-k_2-k_3-k_4)}
{(2qk_2)(2qk_+ + k_+^2-M^2)(k_-^2-M^2)}
\nn\\*
&& {} \phantom{\int\!\frac{\rd^3\bq}{2q_0}} \times
\frac{1}{(2qk_3)[k_+^2-(M^*)^2][2qk_- + k_-^2-(M^*)^2]}
\,\biggr|_{q_0=\sqrt{{\bf q}^2+\la^2}}.
\eeqar 
On the other hand, eliminating $k'_+$ in the denominator of
\refeq{rintex} with the help of the $\de$-function, putting $q=0$ in
the $\de$-function, restoring 
$k'_+$ with the modified $\de$-function, and setting $k'_i\to k_i$ results in
\beqar\label{rintex2}
\Ibr &\to&
\int\!\frac{\rd^3\bq}{2q_0}
\frac{N(0,k_i)\de(p_++p_--k_1-k_2-k_3-k_4)}
{(2qk_2)(k_+^2-M^2)(k_-^2-M^2)}
\nn\\*
&& {} \phantom{\int\!\frac{\rd^3\bq}{2q_0}} \times
\frac{1}{(2qk_3)[-2qk_+ + k_+^2-(M^*)^2][2qk_- + k_-^2-(M^*)^2]}
\,\biggr|_{q_0=\sqrt{{\bf q}^2+\la^2}}.\quad
\eeqar
Both expressions differ by a doubly-resonant contribution. The
difference is in general confined to the \PW~propagators and
originates from the fact that not only soft photons but also photons
with energies of the order of $|k_\pm^2-\MW^2|/\MW$ or, after the
inclusion of the finite width, of order $\GW$ contribute in DPA.
Since photons with finite energies contribute, it is evident
that the integral over the photon momentum depends on the choice of
the phase-space variables that are kept fixed.

As a consequence, one has to choose a definite
parameterization of phase space and to exploit the $\de$-function
carefully, in order to define the non-factorizable corrections uniquely.
For instance, if the vector
$k'_+=k'_1+k'_2$ is kept fixed, the alternative \refeq{rintex2} is
excluded. However, because of momentum conservation,
not all external momenta can be kept fixed independently. 

It is, however, possible to keep, for instance, the invariant masses of the
final-state fermion pairs $k_+^{\prime2}=(k'_1+k'_2)^2$ and
$k_-^{\prime2}=(k'_3+k'_4)^2$ fixed  when integrating over the photon momentum.
If we require $(k'_1+k'_2)^2=k_+^{\prime2}=k_+^2=(k_1+k_2)^2$, we
obtain for the denominator of the \PWp~boson
\beqar\label{Wpprop}
(q+k'_+)^2-M^2 &=& 2qk'_++ k_+^{\prime2} - M^2  
= 2qk_++ k_+^{2} - M^2 + \O(q_0^2) \nl
&=& (q+k_+)^2-M^2 + \O(q_0^2),
\eeqar
where $k_i'=k_i+\O(q_0)$ was used.
Based on the power-counting argument given above,
the terms of order $q_0^2$ can be neglected in DPA, and we find
\beq
(q+k'_+)^2-M^2 \sim (q+k_+)^2-M^2 .
\eeq
If we choose to eliminate $k'_+$, as done in the derivation of 
\refeq{rintex2}, we find, on the other hand,
\beqar
(q+k'_+)^2-M^2 &=& (p_++p_--k'_-)^2-M^2 
=(p_++p_--k_-)^2-M^2 + \O(q_0) \nl
&=&  k_+^2 - M^2 + \O(q_0).
\eeqar
The $\O(q_0)$ terms are relevant in DPA [and in fact given by \refeq{Wpprop}]. 
As a consequence, \refeq{rintex2} is not correct if
we choose to fix $k_+^{\prime2}=k_+^2$ when integrating over the
photon momentum. For fixed $k_+^{\prime2}=k_+^2$, \refeq{Wpprop} leads to the
unique result \refeq{rintex1} for the \PWp~propagator in 
DPA, independently of the other phase-space parameters. 
If we choose, on the other hand, to fix $\bar k_+^2=k_+^2$,
which corresponds to a different 
definition of the invariant mass of the \PWp~boson, we obtain
\beq
(q+k'_+)^2-M^2 = \bar k_+^2 - M^2 ,
\eeq
and thus \refeq{rintex2} instead of \refeq{rintex1}. 
Consequently, the different approaches \refeq{rintex1} and
\refeq{rintex2} correspond to different definitions of the invariant
mass of the \PWp~boson which decays into the fermion pair $(f_1,\bar f_2)$. 
In order to
define the DPA for real radiation, one has to specify at least the
definition of the invariant masses of the \PW~bosons that are kept fixed.
In the following we always fix $k_+^2=(k'_1+k'_2)^2=(k_1+k_2)^2$
and $k_-^2=(k'_3+k'_4)^2=(k_3+k_4)^2$,
as it was also implicitly done in \citeres{Me96,Be97a,Be97b}. 
Once the invariant masses of the \PW~bosons are fixed in this way, the
resulting formulae for the non-factorizable corrections hold
independently of the choice of all other phase-space variables.

We stress that the results obtained within this parameterization of 
phase space differ from those in other parameterizations by
doubly-resonant corrections. 
As already indicated in the introduction, in an experimentally more
realistic approach the invariant masses of W~bosons are identified with
invariant masses of jet pairs, which also include part of the
photon radiation. Since this situation can only be described with 
Monte Carlo programs, 
our results (as well as those of \citeres{Me96,Be97a,Be97b}) should be 
regarded as an estimate of the non-factorizable corrections. 

\subsection[Definition of the non-factorizable doubly-resonant real 
corrections]{Classification and gauge-independent definition of the 
non-factorizable doubly-resonant real corrections}
\label{se:classificationofrealdiagrams}

The doubly-resonant real corrections can be classified in exactly the same 
way as the virtual corrections. For each virtual diagram there is exactly 
a real contribution, which we denote in the same way, e.g.\ \ffp\
refers to all interferences where the photon is emitted by two fermions
corresponding to the two different W~bosons. 

As in the case of the virtual corrections, one has to define the
non-factorizable real corrections in a gauge-independent way. 
To this end, we proceed analogously and 
define the non-factorizable real corrections as the difference of the
complete real corrections and the factorizable real corrections in DPA. 
The factorizable corrections are defined by the products of the matrix
elements for on-shell W-pair production and decay with additional photon
emission. There are three contributions to the factorizable real corrections, 
one where the photon is radiated during the production process and two
where it is emitted during one of the \PW-boson decays. The corresponding 
matrix elements read 
\beqar\label{realfac}
{\cal M}_{\real,1} &=& 
\sum_{\la_+,\la_-}\frac{\M^{\eeWW\ga}_\born\M^\Wpff_\born \M^\Wmff_\born}
{(k_+^2-M^2)(k_-^2-M^2)}\nlc
{\cal M}_{\real,2} &=& 
\sum_{\la_+,\la_-}\frac{\M^\eeWW_\born \M^{\Wpff\ga}_\born\M^\Wmff_\born}
{[(k_++q)^2-M^2](k_-^2-M^2)}\nlc
{\cal M}_{\real,3} &=& 
\sum_{\la_+,\la_-}\frac{\M^\eeWW_\born \M^\Wpff_\born \M^{\Wmff\ga}_\born}
{(k_+^2-M^2)[(k_-+q)^2-M^2]},
\eeqar
in analogy to \refeq{mborn}.
Note that the matrix elements $\M_{\real,2}$ and $\M_{\real,3}$
involve an explicit $q$-dependent propagator.
The factorizable corrections are given by
the squares of these three matrix elements, and 
include by definition no interferences between them.

As an example for the extraction of non-factorizable corrections from real 
diagrams that involve both factorizable and non-factorizable corrections,
we consider the contribution of the diagram in \reffi{real_mixed_diagram}.
\bfi
\centerline{
\begin{picture}(240,100)(0,0)
\ArrowLine(30,50)( 5, 95)
\ArrowLine( 5, 5)(30, 50)
\Photon(30,50)(90,80){-2}{6}
\Photon(30,50)(90,20){2}{6}
\GCirc(30,50){10}{0}
\Vertex(90,80){2.0}
\Vertex(90,20){2.0}
\ArrowLine(90,80)(120, 95)
\ArrowLine(120,65)(90,80)
\ArrowLine(120, 5)( 90,20)
\ArrowLine( 90,20)(120,35)
\Vertex(65,67.5){2.0}
\put(55,73){$W$}
\put(55,16){$W$}
\DashLine(120,0)(120,100){6}
\Photon(65,67.5)(175,32.5){-2}{10}
\put(100,45){$\gamma$}
\Vertex(175,32.5){2.0}
\ArrowLine(150,80)(120,95)
\ArrowLine(120,65)(150,80)
\ArrowLine(120, 5)(150,20)
\ArrowLine(150,20)(120,35)
\Vertex(150,80){2.0}
\Vertex(150,20){2.0}
\Photon(210,50)(150,80){2}{6}
\Photon(210,50)(150,20){-2}{6}
\ArrowLine(210,50)(235,95)
\ArrowLine(235, 5)(210,50)
\GCirc(210,50){10}{0}
\put(177,73){$W$}
\put(177,16){$W$}
\end{picture}
}
\caption{Real bremsstrahlung diagram containing non-factorizable and 
factorizable contributions.}
\label{real_mixed_diagram}
\efi
After subtraction of the factorizable contribution, which 
originates from $|\M_{\real,1}|^2$, it gives rise to
the following correction factor to the 
square $|\M_\born|^2$ of the lowest-order matrix element \refeq{mborn}:
\beqar\label{rWWnfcorr}
\de_{\real}^{\PWp\PWm} &=&e^2 \int\!\frac{\rd^3\bq}{(2\pi)^3 2q_0}\, 
2\Re\biggl\{
\frac{4k_+k_-}{[(k_++q)^2-M^2][(k_-+q)^2-(M^*)^2]}
\nn\\ && \qquad {}
-\biggl[\frac{4 k_+k_-}{2q k_+ 2q k_-}\biggr]_{k_\pm^2=\MW^2}\biggr\}
\bigg|_{q_0=\sqrt{{\bf q}^2+\la^2}}.
\eeqar
Note that the form of the correction factor is
only correct for fixed $(k'_1+k'_2)^2$ and $(k'_3+k'_4)^2$.
For other conventions the off-shell contribution changes,
whereas the on-shell contribution stays the same.
Using the relations \refeq{partfrac} for $q^2=0$,
in DPA we can rewrite \refeq{rWWnfcorr} as
\beqar
\de_{\real}^{\PWp\PWm}
&\sim&-e^2\int\!\frac{\rd^3\bq}{(2\pi)^3 2q_0}\,
2\Re\biggl\{\frac{k_+k_-}{(q k_+)(q k_-)} \biggl[ 
           \frac{k_+^2-M^2}{(k_++q)^2-M^2}
         + \frac{k_-^2-(M^*)^2}{(k_-+q)^2-(M^*)^2}  
\nn\\ && \qquad {}
-\frac{k_+^2-M^2}{(k_++q)^2-M^2} \frac{k_-^2-(M^*)^2}{(k_-+q)^2-(M^*)^2}
\biggr]\biggr\}
\biggr|_{q_0=\sqrt{{\bf q}^2+\la^2}}. 
\eeqar

In the same way, all other contributions  that originate
from a photon coupled to a \PW~boson can be rewritten such that the
complete real non-factorizable corrections can finally be 
expressed as the following correction factor to the lowest-order cross section,
\beqar\label{rnfcorr}
\de_{\real,\nf} &\sim& -\int\!\frac{\rd^3\bq}{(2\pi)^3 2q_0}\,
2\Re\left[j^{\eeWW,\mu}_{\real} (j^\Wpff_{\real,\mu})^*
+j^{\eeWW,\mu}_{\real}(j^\Wmff_{\real,\mu})^*\right. \nl
&& \phantom{-\int\!\frac{\rd^3\bq}{(2\pi)^3 2q_0}\,2\Re\Big[}
\left.\left.{}+j^{\Wpff,\mu}_{\real}(j^\Wmff_{\real,\mu})^*\right]
\right|_{q_0=\sqrt{{\bf q}^2+\la^2}}
\eeqar
with the currents
\beqar
\label{realcurrents}
j_{\real,\mu}^\eeWW &=&
e\biggl( \frac{k_{+\mu}}{q k_+} - \frac{k_{-\mu}}{q k_-}
      + \frac{p_{-\mu}}{q p_-} - \frac{p_{+\mu}}{q p_+}\biggr)\nlc
j_{\real,\mu}^\Wpff&=&
e\biggl( Q_1 \frac{k_{1\mu}}{q k_1} - Q_2 \frac{k_{2\mu}}{q k_2}
      - \frac{k_{+\mu}}{q k_+}\biggr)\frac{k_+^2-M^2}{(k_++q)^2-M^2} ,\nl
j_{\real,\mu}^\Wmff&=&
e\biggl( Q_3 \frac{k_{3\mu}}{q k_3} - Q_4 \frac{k_{4\mu}}{q k_4}
      + \frac{k_{-\mu}}{q k_-}\biggr)\frac{k_-^2-M^2}{(k_-+q)^2-M^2}.
\eeqar  

The factor \refeq{rnfcorr} for the non-factorizable correction
can be viewed as the interference contributions in the square 
of the matrix element ($\veps$ denotes the polarization vector of the photon) 
\beq
{\cal M}_{\real} =  {\cal M}_{\born}\,
\veps^\mu \left(j_{\real,\mu}^\eeWW+j_{\real,\mu}^\Wpff
+j_{\real,\mu}^\Wmff\right),
\eeq
which is just the sum of the three matrix elements $\M_{\real,i}$,
$i=1,2,3$, in ESPA (including radiation from the external fermions
and the internal \PW~bosons). The respective squares
of these three contributions correspond to the factorizable corrections.  
Note that $\M_\born^\eeWW\*\veps^\mu j_{\real,\mu}^\eeWW$ is the
soft-photon matrix element for on-shell \PW-pair production.
Similarly, $\M_\born^\Wpff\*\veps^\mu j_{\real,\mu}^\Wpff$ and
$\M_\born^\Wmff\*\veps^\mu j_{\real,\mu}^\Wmff$ 
correspond to the soft-photon matrix elements for the
decays of the on-shell \PW~bosons, apart from the extra factors
$(k_\pm^2-M^2)/[(k_\pm+q)^2-M^2]$. These factors result from the
definition of the lowest-order matrix element in terms of $k_\pm$ and
from the fact that we 
impose $k_\pm^{\prime2}=k_\pm^2$ and not $\bar k_\pm^2=k_\pm^2$.

Obviously, the currents \refeq{realcurrents} are conserved, and the
corresponding Ward identity for the U(1)$_{\mathrm{em}}$ symmetry of
the emitted photon is fulfilled.

\section{Calculation of scalar integrals}
\label{secalc}

In this section we set our conventions for the scalar integrals.
We describe the reduction of the virtual and real 5-point functions to
4-point functions and indicate how the scalar integrals were evaluated.
More details and the explicit results for the scalar integrals can be 
found in the appendix.

\subsection{Reduction of 5-point functions}

\subsubsection{Reduction of virtual 5-point functions}

The virtual 2-, 3-, 4-, and 5-point functions are defined as
\beqar \label{BCDE0int}
B_0(p_1,m_0,m_1) &=&
\frac{1}{\ri\pi^{2}}\int\!\rd^{4}q\,\frac{1}{N_0 N_1},
\nl
C_0(p_1,p_2,m_0,m_1,m_2) &=&
\frac{1}{\ri\pi^{2}}\int\!\rd^{4}q\,\frac{1}{N_0 N_1 N_2},
\nl
D_{\{0,\mu\}}(p_1,p_2,p_3,m_0,m_1,m_2,m_3) &=&
\frac{1}{\ri\pi^{2}}\int\!\rd^{4}q\,\frac{\{1,q_\mu\}}{N_0 N_1 N_2 N_3},
\nl
E_{\{0,\mu\}}(p_1,p_2,p_3,p_4,m_{0},m_1,m_2,m_3,m_4) &=&
\frac{1}{\ri\pi^{2}}\int\!\rd^{4}q\,\frac{\{1,q_\mu\}}{N_0 N_1 N_2 N_3 N_4},
\eeqar
with the denominator factors
\beq \label{D0Di}
N_{0}= q^{2}-m_{0}^{2}+\ri\epsilon, \qquad
N_{i}= (q+p_{i})^{2}-m_{i}^{2}+\ri\epsilon, \qquad i=1,\ldots,4 ,
\eeq
where $\ri\epsilon$ $(\epsilon>0)$ 
denotes an infinitesimal imaginary part. 

The reduction of the virtual 5-point function to 4-point functions is
based on the fact that in four dimensions the integration momentum
depends linearly on the four external momenta $p_i$ \cite{De93,Me65}.
This gives rise to the identity
\beq\label{detid}
0=\left\vert
\barr{cccc}
2q^{2}  & 2qp_{1}    & \ldots & \;2qp_{4} \\
2p_{1}q & 2p_{1}^{2} & \ldots & \;2p_{1}p_{4} \\
\vdots    & \vdots     & \ddots     &\;\vdots     \\
2p_{4}q & 2p_{4}p_{1} & \ldots &\; 2p_{4}^{2}
\earr
\right\vert =
\left\vert\barr{cccc}
2N_{0}+Y_{00}  & 2qp_{1}    & \ldots & \;2qp_{4} \\
N_{1}-N_{0}+Y_{10}-Y_{00} & 2p_{1}^{2} & \ldots & \;2p_{1}p_{4} \\
\vdots    & \vdots     & \ddots     &\;\vdots     \\
N_{4}-N_{0}+Y_{40}-Y_{00} & 2p_{4}p_{1} & \ldots &\; 2p_{4}^{2}
\earr\right\vert
\eeq
with
\beq\label{defY}
Y_{00} = 2 m_0^2, \quad Y_{i0} = Y_{0i} = m_0^2 + m_i^2-p_i^2, \quad
Y_{ij} = m_i^2 + m_j^2 - (p_i-p_j)^2, \quad i,j=1,2,3,4.
\eeq
Dividing this equation by $N_{0}N_{1}\cdots N_{4}$ and integrating over
$\rd^{4}q$ yields
\beq\label{E0red0}
0 = \frac{1}{\ri\pi^{2}}\int\! \rd^{4}q\, \frac{1}{N_{0}N_{1}\cdots N_{4}}
\left\vert
\barr{cccc}
2N_{0}+Y_{00}  & 2qp_{1}    & \ldots & \;2qp_{4} \\
N_{1}-N_{0}+Y_{10}-Y_{00} & 2p_{1}^{2} & \ldots & \;2p_{1}p_{4} \\
\vdots    & \vdots     & \ddots     &\;\vdots     \\
N_{4}-N_{0}+Y_{40}-Y_{00} & 2p_{4}p_{1} & \ldots &\; 2p_{4}^{2}
\earr
\right\vert.
\eeq
Expanding the determinant along the first column, we obtain
\beqar \label{E0red1}
0&=&\Bigl[2D_{0}(0)+Y_{00}E_0\Bigr]
\left\vert\barr{ccc}
 2p_{1}p_{1}    & \ldots & \;2p_{1}p_{4} \\
 \vdots     & \ddots     &\;\vdots     \\
 2p_{4}p_{1} & \ldots &\; 2p_{4}p_{4}
\earr\right\vert \nl
&&+\disp\sum_{k=1}^{4}(-1)^{k}\Bigl\{D_{\mu}(k) 
-\Bigl[D_{\mu}(0)+p_{4\mu}D_{0}(0)\Bigr]
+p_{4\mu}D_{0}(0)+(Y_{k0}-Y_{00})E_{\mu}\Bigr\}\nl
&&\qquad\qquad\times \left\vert\barr{ccc}
 2p_1^\mu    & \ldots & \;2p_4^\mu \\
 2p_1p_1 & \ldots &\; 2p_1p_4\\
 \vdots     & \ddots     &\;\vdots     \\
 2p_{k-1}p_1 & \ldots &\; 2p_{k-1}p_4\\
 2p_{k+1}p_1 & \ldots &\; 2p_{k+1}p_4\\
 \vdots     & \ddots     &\;\vdots     \\
 2p_4p_1 & \ldots &\; 2p_4p_4
\earr\right\vert,
\eeqar
where $D_0(k)$ denotes the 4-point function that is obtained from the
5-point function $E_0$ by omitting the $k$th propagator $N_k^{-1}$.
The terms involving $p_{4\mu}D_{0}(0)$ have been added for later convenience. 
 
All integrals in \refeq{E0red1} are UV-finite and
Lorentz-covariant. Therefore, the vector integrals possess the
following decompositions
\beqar\label{decomp}
E_\mu &=& \sum_{i=1}^{4} E_i p_{i\mu}, \nl 
D_\mu(k) &=& \sum_{i=1\atop  i\ne k }^4 D_i(k) p_{i\mu}, \qquad
k=1,2,3,4 \nlc
D_\mu(0)+p_{4\mu}D_0(0) &=& \sum_{i=1}^{3} D_i(0) (p_{i}-p_4)_\mu .
\eeqar
The last decomposition becomes obvious after performing a shift
$q\to q-p_4$ in the integral. From \refeq{decomp} it follows
immediately that the terms 
in \refeq{E0red1} that involve $D_\mu(k)$ drop out 
when multiplied with the determinants, because the resulting determinants
vanish. Similarly, $D_\mu(0)+p_{4\mu}D_0(0)$
vanishes after summation over $k$.  Finally, the term
$p_{4\mu}D_0(0)$ contributes only for $k=4$, where it
can be combined with the first term in (\ref{E0red1}).  Rewriting the
resulting equation as a determinant and reinserting the explicit form
of the tensor integrals leads to
\beq \label{eq46}
0 = \frac{1}{\ri\pi^{2}}\int\!\rd^{4}q\,
\frac{1}{N_{0}N_{1}\cdots N_{4}}
\left\vert
\barr{cccc}
N_{0}+Y_{00}  & 2qp_{1}    & \ldots & \;2qp_{4} \\
Y_{10}-Y_{00} & 2p_{1}p_1 & \ldots & \;2p_{1}p_{4} \\
\vdots    & \vdots     & \ddots     &\;\vdots     \\
Y_{40}-Y_{00} & 2p_{4}p_{1} & \ldots &\; 2p_{4}p_4
\earr
\right\vert.
\eeq
Using
\beqar
2p_{i}p_{j}&=& Y_{ij}-Y_{i0}-Y_{0j}+Y_{00},\nl
2qp_{j}&=& N_{j}-N_{0}+Y_{0j}-Y_{00} ,
\eeqar
adding the first column to each of the other columns, and then
enlarging the determinant by one column and one row, this can be written as
\beq\label{E0red2}
0 = \left\vert\barr{cccc}
1 & Y_{00}    & \ldots & \;Y_{04} \\
0 & \quad D_{0}(0)+Y_{00}E_{0} \quad & \ldots & \quad D_{0}(4)+Y_{04}E_{0} \\
0 & Y_{10}-Y_{00} & \ldots &\; Y_{14}-Y_{04}  \\
\vdots    & \vdots     & \ddots     &\;\vdots     \\
0 & Y_{40}-Y_{00} & \ldots &\; Y_{44}-Y_{04}
\earr\right\vert.
\eeq
Equation \refeq{E0red2} is equivalent to
\beq \label{E0red}
0 = \left\vert \barr{cccccc}
-E_{0} &\:D_{0}(0)&\:D_{0}(1)&\:D_{0}(2)&\:D_{0}(3)&\:D_{0}(4)\\
  1   &  Y_{00}   &  Y_{01}   &  Y_{02}   &  Y_{03}   &  Y_{04}   \\
  1   &  Y_{10}   &  Y_{11}   &  Y_{12}   &  Y_{13}   &  Y_{14}   \\
  1   &  Y_{20}   &  Y_{21}   &  Y_{22}   &  Y_{23}   &  Y_{24}   \\
  1   &  Y_{30}   &  Y_{31}   &  Y_{32}   &  Y_{33}   &  Y_{34}   \\
  1   &  Y_{40}   &  Y_{41}   &  Y_{42}   &  Y_{43}   &  Y_{44}
\earr \right\vert,
\eeq
which expresses the scalar 5-point function $E_{0}$ in terms of
five scalar 4-point functions
\beq\label{E0redf}
 E_0 = \frac{1}{\det(Y)} \, \sum_{i=0}^4 \,\det(Y_i)\,D_0(i),
\eeq
where $Y=(Y_{ij})$,
and $Y_i$ is obtained from $Y$ by replacing all
entries in the $i$th column with~1. 

In the special case of an infrared singular 5-point function we have 
\beq
Y_{00} = 2\la^2 \to 0, \qquad Y_{01} = m_1^2-p_1^2 =0, 
\qquad Y_{04} =m_4^2-p_4^2 =0,
\eeq
and the determinants fulfill the relations
\beqar\label{reldetY}
0 &=& \det(Y) - Y_{02}\det(Y_2) - Y_{03}\det(Y_3), \\
0 &=& (Y_{24}-Y_{02})\det(Y) + Y_{02}Y_{14}\det(Y_1)
        + (Y_{02}Y_{34}-Y_{03}Y_{24})\det(Y_3)
       + Y_{02}Y_{44}\det(Y_4), \nl
0 &=& (Y_{31}-Y_{03})\det(Y) + Y_{03}Y_{11}\det(Y_1)
        + (Y_{03}Y_{21}-Y_{02}Y_{31})\det(Y_2)
       + Y_{03}Y_{41}\det(Y_4), \nn
\eeqar
which allow the simplification of \refeq{E0redf}.

\subsubsection{Reduction of real 5-point functions}
 
The real 3-, 4-, and 5-point functions are defined as
\beqar \label{realCDE0}
\Cbr_0(p_1,p_2,\la,m_1,m_2) &=&
\left.\frac{2}{\pi}\int_{q_0<\De E}\!\frac{\rd^{3}{\bf q}}{2q_0}\,
\frac{1}{N'_1 N'_2}\right|_{q_0=\sqrt{{\bf q}^2+\la^2}},
\nl
\Dbr_0(p_1,p_2,p_3,\la,m_{1},m_2,m_3) &=&
\left.\frac{2}{\pi}\int\!\frac{\rd^{3}{\bf q}}{2q_0}\,
\frac{1}{N'_1 N'_2 N'_3}\right|_{q_0=\sqrt{{\bf q}^2+\la^2}},
\nl
\Ebr_0(p_1,p_2,p_3,p_4,\la,m_1,m_2,m_3,m_4) &=&
\left.\frac{2}{\pi}\int\!\frac{\rd^{3}{\bf q}}{2q_0}\,
\frac{1}{N'_1 N'_2 N'_3 N'_4}\right|_{q_0=\sqrt{{\bf q}^2+\la^2}},
\eeqar
with 
\beq \label{D0N'i}
N'_{i}= 2qp_i+p_{i}^{2}-m_{i}^{2}, \qquad i=1,\ldots,4.
\eeq
The shift of the integration boundary of $q_0$ to infinity leads
to an artificial UV divergence in the 3-point function $\Cbr_0$,
which is regularized by an energy cutoff \mbox{$\De E\to\infty$}. In
the following we only need differences of 3-point functions 
that are independent of $\De E$ and Lorentz-invariant.

Because of the appearance of UV-singular integrals in intermediate steps,
the reasoning of the previous section cannot directly be applied to $\Ebr_0$.
Therefore, we rewrite the real 5-point function as an
integral over a closed anticlockwise contour $\C$ in the $q_0$ plane and 
introduce a Lorentz-invariant UV regulator $\La$:
\beq \label{auxint}
\Ebr_0(p_1,p_2,p_3,p_4,\la,m_1,m_2,m_3,m_4)= \lim_{\La\to\infty}
\frac{1}{\ri\pi^{2}}\int_{\C}\rd^{4}q\,\frac{1}{N'_0N'_{1}\cdots N'_{4}}
\frac{-\La^2}{q^2-\La^2}
\eeq
with 
\beq
N'_0 = q^2-\la^2.
\eeq
The contour $\C$ is chosen such that it includes the poles at 
$q_0=\sqrt{\bq^2+\la^2}$ and $q_0=\sqrt{\bq^2+\La^2}$, but none else
(see \reffi{contourC})%
\footnote{\label{fn:powercount}
It is straightforward to check that the naive power counting 
for the UV behaviour in \refeq{auxint} is valid for time-like momenta $p_i$; 
light-like $p_i$ can be treated as a limiting case.
The basic idea for the proof is to deform the contour $\C$ to the vertical
line $\Re\{q_0\}=|\bq|-\epsilon$ with a small $\epsilon>0$, which is
allowed for sufficiently large $|\bq|$, more precisely when 
all particle poles appear left from the line $\Re\{q_0\}=|\bq|-\epsilon$.}.
\bfi
\centerline{
\begin{picture}(400,200)(0,0)
\LongArrow(20,100)(380,100)
\LongArrow(200,0)(200,200)
\put(375,105){$\scriptstyle \Re\{q_0\}$}
\put(210,190){$\scriptstyle \Im\{q_0\}$}
\Vertex( 50,100){2}
\Vertex(150,100){2}
\Vertex(250,100){2}
\Vertex(350,100){2}
\put( 25,80){$\scriptstyle -\sqrt{\bq^2+\La^2}$}
\put(125,80){$\scriptstyle -\sqrt{\bq^2+\la^2}$}
\put(248,80){$\scriptstyle  \sqrt{\bq^2+\la^2}$}
\put(330,80){$\scriptstyle  \sqrt{\bq^2+\La^2}$}
\CArc(250,100)(10,180,0)
\CArc(350,100)(10,180,0)
\ArrowArc(300,100)(60,0,180)
\ArrowArc(300,100)(-40,180,0)
\put(345,150){$\scriptstyle \C$}
\DashLine(245,20)(245,180){4}
\put(225,10){$\scriptstyle \Re\{q_0\}=|\bq|-\epsilon$}
\GCirc(295, 95){2}{0}
\GCirc(220,130){2}{0}
\GCirc(180,100){2}{0}
\GCirc(140, 50){2}{0}
\put(285,105){$\scriptstyle q_0(p_1)$}
\put(210,140){$\scriptstyle q_0(p_4)$}
\put(170,110){$\scriptstyle q_0(p_3)$}
\put(130, 40){$\scriptstyle q_0(p_2)$}
\end{picture} 
}
\caption{Illustration of the contour $\C$ of \refeq{auxint} in the 
complex $q_0$ plane. The open circles indicate the ``particle poles'' 
located at $q_0(p_i)=(2\bq{\bf p}_i-p_i^2+m_i^2)/(2p_{i0})$.}
\label{contourC}
\efi

The integral \refeq{auxint} can be reduced similarly to the virtual
5-point function.  
Owing to the different propagators $N'_i$, \refeq{detid} leads to
\beq\label{rE0red0}
0 = \lim_{\La\to\infty}\frac{1}{\ri\pi^2}\int_{\C}\rd^{4}q\,
\frac{1}{N'_{0}N'_{1}\cdots N'_{4}}\frac{-\La^2}{q^2-\La^2}
\left\vert
\barr{cccc}
2N'_{0}  & 2qp_{1}    & \ldots & \;2qp_{4} \\
N'_{1}+Y_{10} & 2p_{1}^{2} & \ldots & \;2p_{1}p_{4} \\
\vdots    & \vdots     & \ddots     &\;\vdots     \\
N'_{4}+Y_{40} & 2p_{4}p_{1} & \ldots &\; 2p_{4}^{2}
\earr
\right\vert
\eeq
instead of \refeq{E0red0}, with $Y_{ij}$ from \refeq{defY},
and $\lambda^2$ can be set to zero in all $Y_{ij}$, in particular we
have $Y_{00} = 2\la^2 \to 0$. After expanding the determinant along the 
first column and using the Lorentz
decompositions of the integrals where $N'_{k}$ is cancelled, we see
that these terms vanish, and we are left with
\beqar\label{rE0red2}
0&=&\lim_{\La\to\infty}\frac{1}{\ri\pi^2}\int_{\C}\rd^{4}q\,
\frac{1}{N'_{0}N'_{1}\cdots N'_{4}}\frac{-\La^2}{q^2-\La^2}
\left\vert
\barr{cccc}
2N'_{0}  & 2qp_{1}    & \ldots & \;2qp_{4} \\
Y_{10} & 2p_{1}^{2} & \ldots & \;2p_{1}p_{4} \\
\vdots    & \vdots     & \ddots     &\;\vdots     \\
Y_{40} & 2p_{4}p_{1} & \ldots &\; 2p_{4}^{2}
\earr
\right\vert.
\eeqar
Using
\beqar
2p_{i}p_{j}&=& Y_{ij}-Y_{i0}-Y_{0j},\nl
2qp_{j}&=& N'_{j}+Y_{0j},
\eeqar
adding the first column to the other columns and extending the
determinant leads to
\beq
0=\lim_{\La\to\infty}\frac{1}{\ri\pi^2}\int_{\C}\rd^{4}q\,
\frac{1}{N'_{0}N'_{1}\cdots N'_{4}}\frac{-\La^2}{q^2-\La^2}
\left\vert
\barr{ccccc}
1  & 0  &  Y_{01}   & \ldots & \; Y_{04} \\      
0 & 2N'_{0} & N'_{1}+2N'_0+Y_{01}    & \ldots & \;N'_4+2N'_0+Y_{04} \\
0 & Y_{10} & Y_{11}-Y_{01} & \ldots & \; Y_{14}-Y_{04} \\
\vdots & \vdots    & \vdots     & \ddots     &\;\vdots     \\
0 & Y_{40} &  Y_{41}-Y_{01}& \ldots &\;  Y_{44}-Y_{04}
\earr
\right\vert.\nl
\eeq
Subtracting the first row from the second, adding the first row to the
other rows, and exchanging the first two rows, we arrive at
\beq\label{almostE0brred}
0=\lim_{\La\to\infty}\frac{1}{\ri\pi^2}\int_{\C}\rd^{4}q\,
\frac{1}{N'_{0}N'_{1}\cdots N'_{4}}\frac{-\La^2}{q^2-\La^2}
\left\vert
\barr{ccccc}
-1 & 2N'_{0} & N'_{1}+2N'_0    & \ldots & \;N'_4+2N'_0 \\
1  & 0  &  Y_{01}   & \ldots & \; Y_{04} \\      
1 & Y_{10} & Y_{11} & \ldots & \; Y_{14} \\
\vdots & \vdots    & \vdots     & \ddots     &\;\vdots     \\
1 & Y_{40} &  Y_{41}& \ldots &\;  Y_{44}
\earr
\right\vert.
\eeq
Now we perform the contour integral
using power counting for $\Lambda\to\infty$ (see
footnote~\ref{fn:powercount}).
In the contribution of the pole at
$q_0=\sqrt{\bq^2+\la^2}$ we have $N'_0=0$ in the numerator;
the limit $\Lambda\to\infty$ can be trivially taken and reproduces usual
bremsstrahlung integrals, as defined in \refeq{realCDE0}.
In the contribution of the pole at $q_0=\sqrt{\bq^2+\La^2}$ the term containing 
$N'_0$ in the numerator survives and will be calculated below, 
but all other terms vanish after taking the limit $\La\to\infty$. 

Thus, we find
\beq\label{E0brred}
0 \;=\;
\left|\begin{array}{cccccc}
\;-\Ebr_0\; &\; \tilde\Dbr_0(0)\; 
&\; \Dbr_0(1) +\tilde\Dbr_0(0)\; &\; \Dbr_0(2) +\tilde\Dbr_0(0)\;
&\; \Dbr_0(3) +\tilde\Dbr_0(0)\; &\; \Dbr_0(4) +\tilde\Dbr_0(0)\; \\
1 & Y_{00} & Y_{01} & Y_{02} & Y_{03} & Y_{04} \\
1 & Y_{10} & Y_{11} & Y_{12} & Y_{13} & Y_{14} \\
1 & Y_{20} & Y_{21} & Y_{22} & Y_{23} & Y_{24} \\
1 & Y_{30} & Y_{31} & Y_{32} & Y_{33} & Y_{34} \\
1 & Y_{40} & Y_{41} & Y_{42} & Y_{43} & Y_{44}
\end{array}\right|,
\eeq
or
\beq\label{Erred}
\Ebr_0 = \frac{1}{\det(Y)} \,
\biggl\{\det(Y_0)\,\tilde\Dbr_0(0)
+\sum_{i=1}^4 \,\det(Y_i)\,\left[\Dbr_0(i)+\tilde\Dbr_0(0)\right]\biggr\}.
\eeq
Here 
\beq
\tilde\Dbr_0(0) = \lim_{\La\to\infty}\frac{2}{\ri\pi^2}\int_\C\rd^{4}q\,
\frac{1}{N'_{1}\cdots N'_{4}}\frac{-\La^2}{q^2-\La^2},
\eeq
and the 4-point bremsstrahlung integrals $\Dbr_0(i)$, $i = 1,2,3,4$,
result from $\Ebr_0$ by omitting the $i$th denominator $N'_i$.
The result \refeq{E0brred} differs from \refeq{E0red2} only by the 
extra $\tilde\Dbr_0(0)$'s added to the $\Dbr_0(i)$'s.

The integral $\tilde\Dbr_0(0)$ stems from the terms involving $N'_0$
in the numerator in \refeq{almostE0brred} and can be expressed as follows:
\beqar
\tilde\Dbr_0(0) 
 &=&\frac{2}{\ri\pi^2}\int_{\C'}\rd^{4}q'\,
\frac{1}{2q'p_1\cdots 2q'p_4}\frac{-1}{q'^2-1}, 
\eeqar
where the contour $\C'$ surrounds $q'_0=\sqrt{{\bf q'}^2+1}$.
Performing the contour integral over $\rd q'_0$ yields
\beqar
\tilde\Dbr_0(0) &=& \frac{4}{\pi}
\int\frac{\rd^{3}{\bf q'}}{2q'_0}\,
\frac{1}{2q' p_1\cdots 2q' p_4}
\biggr|_{q'_0=\sqrt{{\bf q'}^2+1}}.
\eeqar
Now the vector $q'$ is time-like. Since also the vectors $p_i$ are 
time-like (or at least light-like), the scalar products 
$q'p_i$ cannot become zero. After redefining the momenta,
\beq\label{eq:momredef}
p_i = \si_i \tilde p_i,  \qquad \si_i = \pm 1, \qquad \tilde p_{i0}> 0,
\eeq
and extracting the signs $\si_i$, 
this integral can be evaluated by a Feynman-parameter representation and 
momentum integration in polar coordinates resulting in:
\beqar
\tilde\Dbr_0(0) &=& 
 - \si_1\si_2\si_3\si_4\int_0^\infty\rd x_1\,\rd x_2\,\rd x_3\,\rd x_4\, 
\de\biggl(1-\sum_{i=1}^4 x_i\biggr)
\left[\biggl(\sum_{i=1}^4 x_i \tilde p_i\biggr)^2\right]^{-2}.
\eeqar
This is just the Feynman-parameter representation of a virtual 4-point
function such that we finally obtain
\beq
\tilde\Dbr_0(0) =- \si_1\si_2\si_3\si_4 D_0\left(\tilde p_2-\tilde p_1,\tilde
  p_3-\tilde p_1,\tilde p_4-\tilde p_1,
\sqrt{p_1^2},\sqrt{p_2^2},\sqrt{p_3^2},\sqrt{p_4^2}\right).
\eeq 

\subsubsection{\boldmath{Explicit reduction of the virtual 5-point function 
for the photon exchange between $\bar f_2$ and $f_3$}}

For the photon exchange between $\bar f_2$ and $f_3$ 
the following scalar integrals are relevant:
\beqar\label{virt_ints}
E_0 &=& E_0(-k_3,-k_-,k_+,k_2,\la,m_3,M,M,m_2),
\nl
D_0(0) &=& D_0(-k_4,k_+ +k_3,k_2+k_3,0,M,M,0),
\nl
D_0(1) &=& D_0(-k_-,k_+,k_2,0,M,M,m_2),
\nl
D_0(2) &=& D_0(-k_3,k_+,k_2,\la,m_3,M,m_2),
\nl
D_0(3) &=& D_0(-k_3,-k_-,k_2,\la,m_3,M,m_2)
\nlc
D_0(4) &=& D_0(-k_3,-k_-,k_+,0,m_3,M,M).
\eeqar
Since we neglect the external fermion masses, the last two relations
\refeq{reldetY} simplify to
\beqar\label{reldetYipole}
0 &=& (s_{23}+s_{24})\det(Y) + K_-s_{23}\det(Y_1)
        - \left[ K_+(s_{23}+s_{24})+K_-M_+^2 \right]\det(Y_3), \nl
0 &=& (s_{13}+s_{23})\det(Y) + K_+s_{23}\det(Y_4)
        - \left[ K_-(s_{13}+s_{23})+K_+M_-^2 \right]\det(Y_2).
\eeqar
These relations allow 
us to eliminate $\det(Y_1)$ and $\det(Y_4)$ from \refeq{E0redf}, resulting in:
\begin{eqnarray}\label{vE5redpole}
\nonumber
\lefteqn{E_0(-k_3,-k_-,k_+,k_2,\lambda,m_3,M,M,m_2)=
\frac{\det(Y_0)}
{\det(Y)} D_0(0)}\quad\\
\nonumber
&&{}+ \frac{\det(Y_3)}{\det(Y)K_-s_{23}}
\Big\{[K_+(s_{23}+s_{24})+K_-M_+^2]D_0(1)+K_-s_{23}D_0(3)\Big\}\\
\nonumber
&&{}+ \frac{\det(Y_2)}{\det(Y)K_+s_{23}}
\Big\{[K_-(s_{13}+s_{23})+K_+M_-^2]D_0(4)+K_+s_{23}D_0(2)\Big\}\\
&&
-\frac{s_{13}+s_{23}}{K_+s_{23}} D_0(4)
-\frac{s_{23}+s_{24}}{K_-s_{23}} D_0(1).
\end{eqnarray}
The matrix $Y$ reads
\beqar\label{Yvpole}
Y &=& \left(\begin{array}{@{\ }c@{\ \ }c@{\ \ }c@{\ \ }c@{\ \ }c@{\ }}
{}0      & 0       & -K_-     & -K_+       & 0 \\
{}*      & 0       &  M^2     & \;(-K_+-s_{13}-s_{23})\; & -s_{23} \\
{}*      & *       & 2M^2     & \; 2M^2-s\;& (-K_- -s_{23}-s_{24}) \\
{}*      & *       & *        & 2M^2       & M^2 \\
{}*      & *       & *        & *          & 0 
\end{array}\right).
\eeqar
Neglecting terms that do not contribute to the correction factor in
DPA, the corresponding determinants are given by
\beqar\label{detYvpole}
\det(Y) &\sim& 2s_{23} \left[
K_+K_-s_{14}s_{23} - (K_+\MW^2+K_-s_{13})(K_-\MW^2+K_+s_{24}) \right],
\nl
\det(Y_0) &\sim& -\kappa_\PW^2,
\nl
\det(Y_1)
&\sim& K_+\left[\MW^4(s_{23}-s_{24})
                +(s_{23}+s_{24})(s_{13}s_{24}-s_{14}s_{23})\right]
\nn\\ && {}
+ K_-\MW^2(-\MW^4+2s_{13}s_{23}+s_{13}s_{24}+s_{14}s_{23}),
\nl
\det(Y_2)
&\sim& -s_{23}\left[ K_+(\MW^4+s_{13}s_{24}-s_{14}s_{23})
                +2K_-\MW^2s_{13} \right],
\nl
\det(Y_3) &=& \det(Y_2)\big|_{K_+\leftrightarrow K_-,
                s_{13}\leftrightarrow s_{24}},
\nl
\det(Y_4) &=& \det(Y_1)\big|_{K_+\leftrightarrow K_-,
                s_{13}\leftrightarrow s_{24}},
\eeqar
where the 
shorthand $\kappa_\PW$ is defined in \refapp{prelim}.

\subsubsection{\boldmath{Explicit reduction of the real 5-point function for 
the photon exchange between $\bar f_2$ and $f_3$}}

In this case the integrals appearing in \refeq{Erred} read
\beqar\label{real_ints}
\Ebr_0 &=& \Ebr_0(k_3,k_-,k_+,k_2,\la,m_3,M^*,M,m_2),
\nl
\tilde\Dbr_0(0) &=&  
-D_0(k_4,k_+-k_3,k_2-k_3,0,M_-,M_+,0),
\nl
\Dbr_0(1) &=& \Dbr_0(k_-,k_+,k_2,0,M^*,M,m_2),
\nl
\Dbr_0(2) &=& \Dbr_0(k_3,k_+,k_2,\la,m_3,M,m_2),
\nl
\Dbr_0(3) &=& \Dbr_0(k_3,k_-,k_2,\la,m_3,M^*,m_2)
\nlc
\Dbr_0(4) &=& \Dbr_0(k_3,k_-,k_+,0,m_3,M^*,M).
\eeqar
In analogy to the virtual  5-point function, we can express the real
5-point function with the help of the relations (we denote the matrix $Y$ for 
the real 5-point function with a prime, in order to distinguish it from the 
one for the virtual 5-point function):
\beqar
0 &=&
(s_{23}+s_{24})\det(\Ybr) + K_-^*s_{23}\det(\Ybr_1)
        - \left[ K_+(s_{23}+s_{24})-K_-^*M_+^2 \right]\det(\Ybr_3), \nl
0 &=&
(s_{13}+s_{23})\det(\Ybr) + K_+s_{23}\det(\Ybr_4)
        - \left[ K_-^*(s_{13}+s_{23})-K_+M_-^2 \right]\det(\Ybr_2),\quad
\eeqar
by
\begin{eqnarray}\label{rE5redpole}
\lefteqn{\Ebr_0(k_3,k_-,k_+,k_2,\la,m_3,M^*,M,m_2) =
\frac{\det(\Ybr_0)}{\det(\Ybr)} \tilde\Dbr_0(0)}\nl
\nonumber
&&{}+ \frac{\det(\Ybr_3)}{\det(\Ybr)K_-^*s_{23}}
\biggl\{\Bigl[K_+(s_{23}+s_{24})-K_-^*M_+^2\Bigr]
\Bigl[\Dbr_0(1)+\tilde\Dbr_0(0)\Bigr]
+K_-^*s_{23}\Bigl[\Dbr_0(3)+\tilde\Dbr_0(0)\Bigr]\biggr\}\\
\nonumber
&&
{}+ \frac{\det(\Ybr_2)}{\det(\Ybr)K_+s_{23}}
\biggl\{\Bigl[K_-^*(s_{13}+s_{23})-K_+M_-^2\Bigr]
\Bigl[\Dbr_0(4)+\tilde\Dbr_0(0)\Bigr]
+K_+s_{23}\Bigl[\Dbr_0(2)+\tilde\Dbr_0(0)\Bigr]\biggr\}\\
&&{}
-\frac{s_{13}+s_{23}}{K_+s_{23}} \Bigl[\Dbr_0(4)+\tilde\Dbr_0(0)\Bigr]
-\frac{s_{23}+s_{24}}{K_-^*s_{23}} \Bigl[\Dbr_0(1)+\tilde\Dbr_0(0)\Bigr].
\end{eqnarray}
For the matrix $\Ybr$ we find
\beqar
\Ybr &=& \left(\begin{array}{ccccc}
\ 0 & 0 & -K_-^* & -K_+                   & 0 \\
\ * & 0 & (M^*)^2  & (-K_++s_{13}+s_{23}) & s_{23} \\
\ * & * & 2(M^*)^2 & \;(-2K_+-2K_-^*+s-(M^*)^2-M^2)\; 
                                          & (-K_-^*+s_{23}+s_{24}) \\
\ * & * &   *    & 2M^2                   & M^2 \\
\ * \ & \ * \ & \ * \ & \ * \ & 0
\end{array}\right).
\nln
\eeqar
Replacing $-K_-^*$  by $K_-$ and $(M^*)^2$ by $M^2$ and multiplying the 
second and third columns and rows by $-1$, this becomes equal to  
\refeq{Yvpole} in DPA.

In DPA, $\tilde\Dbr_0(0)$ can be neglected in the terms
$\Dbr_0(i)+\tilde\Dbr_0(0)$ in \refeq{Erred} and \refeq{rE5redpole},
and the reduction of the real 5-point function becomes 
algebraically identical to the reduction of the virtual 5-point function,
apart from the differences in signs of some momenta.
As a consequence, the results for the virtual corrections can be translated
to the real case if we substitute $K_-\to -K_-^*$ in all 
algebraic factors such as the determinants and $E_0\to\Ebr_0$,
$D_0(0)\to \tilde\Dbr_0(0)$, 
$D_0(1)\to -\Dbr_0(1)$, $D_0(2)\to -\Dbr_0(2)$, $D_0(3)\to
\Dbr_0(3)$ and $D_0(4)\to\Dbr_0(4)$.
In particular, the determinants are related by
\beqar\label{eq:detrels}
\det(\Ybr) &\sim& +\det(Y)\Big|_{K_-\to -K_-^*} ,\qquad
\det(\Ybr_0) \sim +\det(Y_0) \sim -\kappa_\PW^2                \nlc
\det(\Ybr_1) &\sim& -\det(Y_1)\Big|_{K_-\to -K_-^*},\qquad 
\det(\Ybr_2) \sim -\det(Y_2)\Big|_{K_-\to -K_-^*} \nlc
\det(\Ybr_3) &\sim& +\det(Y_3)\Big|_{K_-\to -K_-^*}, \qquad
\det(\Ybr_4) \sim +\det(Y_4)\Big|_{K_-\to -K_-^*}. 
\eeqar

\subsection{Calculation of 3- and 4-point functions}

The scalar loop integrals have been evaluated following the methods 
of \citere{tH79}. Our explicit results are listed in \refapp{appvirtual}.
For vanishing \PW-boson width they agree with the general results
of \citere{tH79}. For finite \PW-boson width the virtual 4-point functions 
are in agreement with those of \citeres{Be97a,Be97b}
in DPA. This shows explicitly that the $q^2$ terms in the
W-boson and fermion propagators are irrelevant in DPA. 

An evaluation of the bremsstrahlung integrals, which follows
closely the techniques for calculating loop integrals, is sketched in
\refapp{appbrcal}. 
The final results in DPA are listed in \refapp{appreal}, and the 4-point
functions agree with those of \citeres{Be97a,Be97b}.
We have analytically reproduced all exact results for the 
occurring bremsstrahlung 3- and 4-point integrals 
by independent methods.
In addition, we have evaluated the IR-finite integrals $\Dbr_0(1)$,
$\Dbr_0(4)$, and $\Ebr_0-\Dbr_0(3)/K_+$ by a direct 
Monte Carlo integration over the photon momentum,
yielding perfect agreement with our exact
analytical results for these integrals. Note that this, in particular,
checks the reduction of the bremsstrahlung 5-point function
described in the previous section.

\begin{sloppypar}
Our results for the 3-point functions cannot directly be compared
with those of \citeres{Be97a,Be97b}, 
because different approaches have been used.
While our results are IR-singular owing to the subtracted on-shell
integrals, the results of \citeres{Be97a,Be97b} are 
artificially UV-singular owing to the neglect of $q^2$ in the 
W~propagators. However, when adding the real and virtual 3-point functions 
the two results agree. This confirms that
our definition of the non-factorizable corrections is equivalent to the
one of \citeres{Be97a,Be97b}
in DPA. Thus, it turns out that in DPA the subtraction of the on-shell
contribution is effectively equivalent to the neglect of the $q^2$ terms
in all but the photon propagators.
\end{sloppypar}

\section{Analytic results for the non-factorizable corrections}
\label{seanres}

\subsection{General properties  of non-factorizable corrections}

In \citere{Me96} it was shown from the integral representation that the
non-factorizable corrections associated with 
photon exchange between initial and final state vanish in DPA. This was
confirmed in \citeres{Be97a,Be97b}. Via explicit evaluation of all integrals
we have checked that the cancellation between virtual and real
integrals takes place 
diagram by diagram once the factorizable contributions are subtracted.
In this way all interference terms (if), (mf), (im), and (mm) drop
out. Examples for the virtual Feynman diagrams contributing to these
types of corrections are shown in \reffi{virtual_initial_final_diagrams}.

The only non-vanishing non-factorizable corrections are due to the 
contributions (\ffp), (\mfp), and (\mmp). The corresponding virtual
diagrams are shown in \reffi{virtual_final_final_diagrams},
apart from permutations of the final-state fermions. Two of the 
corresponding real diagrams are pictured in
\reffis{real_non_factorizable_diagram} and \ref{real_mixed_diagram}.
Since these corrections depend only on $s$-channel invariants, 
the non-factorizable corrections are independent of the
production  angle of the \PW~bosons,
as was also pointed out in \citeres{Be97a,Be97b}.

\subsection{Generic form of the correction factor}
\label{gfcf}

The non-factorizable corrections $\rd\si_\nf$ to the fully differential 
lowest-order cross-section $\rd\si_\born$ resulting from the matrix element
\refeq{mborn} take the form of a correction factor to the
lowest-order cross-section:
\beq
\rd \si_\nf = \de_\nf \, \rd\si_\born.
\eeq
Upon splitting the contributions that result from photons coupled to
the \PW~bosons according to
$1=Q_\PWp=Q_1-Q_2$ and $1=-Q_\PWm=Q_4-Q_3$ into contributions
associated with definite final-state fermions, the
complete correction factor to the lowest-order cross-section can be written as
\beq\label{nfcorrfac}
\delta_\nf = \sum_{a=1,2} \, \sum_{b=3,4} \, (-1)^{a+b+1} \, Q_a Q_b \,
\frac{\alpha}{\pi} \, \Re\{\Delta(k_+,k_a;k_-,k_b)\}.
\eeq
In the following, only $\Delta=\Delta(k_+,k_2;k_-,k_3)$ is given; 
the other terms follow by obvious substitutions. As discussed above, $\De$
gets contributions from {\it intermediate--intermediate} 
($\Delta_{\mmp}$), {\it intermediate--final}
($\Delta_{\mfp}$), and {\it final--final}
($\Delta_{\ffp}$) interactions:
\beq\label{eq:Delta}
\Delta = \Delta_{\mmp}+\Delta_{\mfp}+\Delta_{\ffp}.
\eeq
The quantity ($\Delta_{\mmp}$ is independent of the final-state
fermions. The individual contributions read
\beqar
\label{eq:dmm}
\Delta_{\mmp} &\sim& (2\MW^2-s)\biggl\{
C_0(k_+,-k_-,0,M,M)
- \Bigl[C_0(k_+,-k_-,\la,\MW,\MW)\Bigr]_{k_\pm^2=\MW^2} 
\nn\\ && {}
\phantom{(2\MW^2-s)\biggl[}
- \Cbr_0(k_+,k_-,0,M,M^*) 
+ \Bigl[\Cbr_0(k_+,k_-,\la,\MW,\MW)\Bigr]_{k_\pm^2=\MW^2} \biggr\},
\hspace{2em}
\\[.5em]
\Delta_{\mfp} &\sim& 
- (s_{23}+s_{24})\Bigl[ K_+D_0(1)-K_+\Dbr_0(1) \Bigr]
- (s_{13}+s_{23})\Bigl[ K_-D_0(4)-K_-^*\Dbr_0(4) \Bigr],
\\[.5em]
\Delta_{\ffp} &\sim& 
-K_+s_{23}\Bigl[ K_-E_0 -K_-^*\Ebr_0\Bigr],
\eeqar
with the arguments of the 5- and 4-point functions as defined in
\refeq{virt_ints} and \refeq{real_ints}. 

The sum $\Delta_{\mfp}+\Delta_{\ffp}$ can be
simplified by inserting the decompositions of the 5-point functions
\refeq{vE5redpole} and \refeq{rE5redpole}. In DPA this leads to
\beqar\label{denf1}
\Delta_{\mfp}+\Delta_{\ffp} &\sim&
-\frac{K_+K_-s_{23}\det(Y_0)}{\det(Y)}D_0(0)
+\frac{K_+K_-^*s_{23}\det(\Ybr_0)}{\det(\Ybr)}\tilde\Dbr_0(0)
\nn\\ && {}
-\frac{K_+\det(Y_3)}{\det(Y)}
\left\{[K_+(s_{23}+s_{24})+K_-\MW^2]D_0(1) + K_-s_{23}D_0(3)\right\} 
\nn\\ && {}
-\frac{K_-\det(Y_2)}{\det(Y)}
\left\{[K_-(s_{13}+s_{23})+K_+\MW^2]D_0(4) + K_+s_{23}D_0(2)\right\}
\nn\\ && {}
+ \frac{K_+\det(\Ybr_3)}{\det(\Ybr)}
\left\{ [K_+(s_{23}+s_{24})-K_-^*\MW^2]\Dbr_0(1) + K_-^*s_{23}\Dbr_0(3) 
\right\}
\nn\\ && {}
+ \frac{K_-^*\det(\Ybr_2)}{\det(\Ybr)}
\left\{ [K_-^*(s_{13}+s_{23})-K_+\MW^2]\Dbr_0(4) + K_+s_{23}\Dbr_0(2) 
\right\}.\quad
\eeqar
Note that $\Delta_{\mfp}$ is exactly cancelled by the contributions 
of the last two terms in \refeq{vE5redpole} and \refeq{rE5redpole}. 

Inserting the expressions for the scalar integrals from \refapp{appscalint}
and using the first of the relations \refeq{reldetY},
various terms, notably all IR divergences and mass-singular logarithms, 
cancel between the real and virtual corrections, and in DPA we are left with
\begin{eqnarray}
\label{eq:dmmDPA}
\Delta_{\mmp} &\sim& 
\frac{2\MW^2-s}{s\beta_\PW}\biggl[
-\cLi\biggl(\frac{K_-}{K_+},x_\PW\biggr)
+\cLi\biggl(\frac{K_-}{K_+},x_\PW^{-1}\biggr)
+\cLi\biggl(-\frac{K_-^*}{K_+},x_\PW\biggr)
\nn\\ && {}
-\cLi\biggl(-\frac{K_-^*}{K_+},x_\PW^{-1}\biggr)
+ 2\pi\ri \ln\biggl(\frac{K_+ +K_-^*x_\PW}{\ri\MW^2}\biggr)
\biggr]
\; + \; \mbox{imaginary parts,}
\hspace{2em}
\label{eq:DmmDPA}
\\[1em]
\Delta_{\mfp}+\Delta_{\ffp} &\sim &
-\frac{K_+K_-s_{23}\det(Y_0)}{\det(Y)}D_0(0)
- \frac{K_+ \det(Y_3)}{\det( Y)} F_3
- \frac{K_- \det(Y_2)}{\det( Y)} F_2
\nn\\ && {}
+\frac{K_+K_-^*s_{23}\det(\Ybr_0)}{\det(\Ybr)}\tilde\Dbr_0(0)
+ \frac{\Kp \det(\Ybr_3)}{\det( \Ybr)} \F_3
+ \frac{\Km \det(\Ybr_2)}{\det( \Ybr)} \F_2
\nn\\ && {}
+ \; \mbox{imaginary parts,}
\label{eq:DmfDPA}
\end{eqnarray}
with $D_0(0)$ and $\tilde\Dbr_0(0)$ 
given in \refeq{D00} and \refeq{Dbr00}, respectively, and
\beqar\label{Fi}
F_3&=&
-2\cLi\biggl(\frac{K_+}{K_-},-\frac{s_{23}+s_{24}}{M_W^2}-\ri\epsilon\biggr)
+\sum\limits_{\tau=\pm 1}\biggl[
\cLi\biggl(\frac{K_+}{K_-},\xW^\tau\biggr)-
\cLi\biggl(-\frac{M_W^2}{s_{23}+s_{24}}+\ri\epsilon,\xW^\tau\biggr)\biggr]
\nn\\ && {}
-\Li\biggl(-\frac{s_{24}}{s_{23}}\biggr)
+2\ln\biggl(-\frac{s_{23}}{M_W^2}-\ri \epsilon\biggr)
  \ln\biggl(-\frac{K_-}{M_W^2}\biggr)
- \ln^2\biggl(-\frac{s_{23}+s_{24}}{M_W^2}-\ri \epsilon\biggr),
\nl[1em]
\F_3&=&
F_3\Big|_{K_- \to -K_-^*}
+ 2\pi\ri \biggl[
2\ln\biggl(1-\frac{K_+}{K_-^*}\frac{s_{23}+s_{24}}{\MW^2}\biggr)
-\ln\biggl(1+\frac{K_+}{K_-^* x_\PW}\biggr)
-\ln\biggl(1+\frac{x_\PW \MW^2}{s_{23}+s_{24}}\biggr)
\nn\\ && {}
\qquad\qquad\qquad
+\ln\biggl(\frac{\ri K_-^*}{s_{23}+s_{24}}\biggr)
\biggr],
\nl[1em]
F_2&=& F_3\Big|_{K_+\leftrightarrow K_-,s_{24}\leftrightarrow s_{13}}, 
\nl[1em]
\F_2&=& F_2\Big|_{K_- \to -K_-^*}
+ 2\pi\ri \biggl\{
\ln\biggl(1+\frac{K_-^*x_\PW}{K_+}\biggr) 
-\ln\biggl(1+\frac{x_\PW \MW^2}{s_{13}+s_{23}}\biggr)
+\ln\biggl[\frac{K_+s_{23}}{\ri\MW^2(s_{13}+s_{23})}\biggr]
\biggr\}. 
\nn\\
\end{eqnarray}
The variables $\beta_\PW$, $x_\PW$ and the dilogarithms $\Li$,
$\cLi$ are defined in \refapp{prelim}.

The above results contain logarithms and dilogarithms
the arguments of which depend on $K_\pm$. It is interesting to see whether 
enhanced logarithms of the form $\ln (K_\pm/\MW^2)$ appear
in the limits $K_\pm\to0$. It turns out that such logarithms are absent 
from the non-factorizable corrections,
irrespective of the ratio in which the two limits $K_\pm\to0$ are realized.

Moreover, the non-factorizable corrections vanish in the high-energy limit.
This feature of the non-factorizable corrections has been checked 
by analytical and numerical calculations.

\subsection{Non-factorizable virtual corrections}
\label{se:vnfc}

In \refch{ch:radcorr} the factorizable 
and non-factorizable radiative corrections to four-fermion production
are investigated. 
In order to overcome the problem of overlapping resonances for the DPA
of the real corrections discussed in \refse{se:overlappingresonances}, 
the complete real process, involving resonant and non-resonant diagrams, 
is evaluated by Monte Carlo integration. 
On the other hand, only the doubly-resonant part of the 
virtual corrections is taken into account.
Therefore, the results of 
the non-factorizable virtual corrections are required separately.

As in \refse{gfcf}, the non-factorizable virtual corrections 
factorize to the lowest-order cross section: 
\begin{eqnarray}
\rd \si_\nf^\virt = \de_\nf^\virt \, \rd\si_\born
\end{eqnarray}
with
\beq\label{nfcorrfacvirt}
\delta_\nf^\virt = \sum_{a=1,2} \, \sum_{b=3,4} \, (-1)^{a+b+1} \, Q_a Q_b \,
\frac{\alpha}{\pi} \, \Re\{\Delta^\virt(k_+,k_a;k_-,k_b)\}.
\eeq
The explicit results
of the term $\Delta^\virt=\Delta^\virt (k_+,k_2;k_-,k_3)$ 
are listed in the following:
\beq\label{eq:Deltavirt}
\Delta^\virt = \Delta_{\mm}^\virt+\Delta_{\mf}^\virt+
\Delta_{\im}^\virt+\Delta_{\mmp}^\virt+\Delta_{\ifp}^\virt+
\Delta_{\mfp}^\virt+\Delta_{\ffp}^\virt.
\eeq
with the individual contributions of the diagrams
shown in \reffi{virtual_final_final_diagrams}
and \reffi{virtual_initial_final_diagrams}
\beqar
\Delta_{\mm}^\virt &\sim& \frac{2 \MW^2}{k_+^2-M^2} \left\{
B_0(k_+,0,M)
- \Bigl[B_0\left(k_+,0,M\right)\Bigr]_{k_+^2=M^2} \right\}
\nn\\ && {}
- 2 \MW^2 \Bigl[B_0^\prime \left(k_+,\la,\MW\right)\Bigr]_{k_+^2=\MW^2} 
+ (k_+\leftrightarrow k_-),
\hspace{2em}
\\[.5em]
\Delta_{\mf}^\virt &\sim& 
\MW^2 \left\{C_0(k_2,k_+,0,m_2,M)
     - \Bigl[C_0(k_2,k_+,\la,m_2,\MW)\Bigr]_{k_+^2=\MW^2}\right\}
\nn\\ && {}
+ (k_2 \leftrightarrow k_3, k_+ \leftrightarrow k_-) ,
\hspace{2em}
\\[.5em]
\Delta_{\im}^\virt &\sim&  
  2k_+p_+ \left\{C_0(p_+,k_+,0,\Me,M)
- \Bigl[C_0(p_+,k_+,\la,\Me,\MW)\Bigr]_{k_+^2=\MW^2}\right\}
\nn\\ && {}
+ (p_+ \leftrightarrow p_-, k_+ \leftrightarrow k_-)
- (k_+ \leftrightarrow k_-)
- (p_+ \leftrightarrow p_-) ,
\hspace{2em}
\\[.5em]
\Delta_{\mmp}^\virt &\sim& - 2 k_+ k_- \left\{
  C_0(k_+,-k_-,0,M,M)
- \Bigl[C_0(k_+,-k_-,\la,\MW,\MW)\Bigr]_{k_\pm^2=\MW^2}
\right\}, \qquad\\[.5em]
\Delta_{\ifp}^\virt &\sim& 
  2p_+k_2 K_+ D_0(p_+,k_+,k_2,\la,\Me,M,m_2)
\nn\\ && {}
- (p_+ \leftrightarrow p_-)
- (k_2 \leftrightarrow k_3, k_+ \leftrightarrow k_-)
+ (k_2 \leftrightarrow k_3, k_+ \leftrightarrow k_-, p_+ \leftrightarrow p_-),
\hspace{2em}
\\[.5em]
\Delta_{\mfp}^\virt &\sim& 
- 2 k_2 k_- K_+ D_0(1) - 2 k_3 k_+ K_- D_0(4),\\[.5em]
\Delta_{\ffp}^\virt &\sim& 
- 2 k_2 k_3 K_+ K_- E_0.
\eeqar
The other terms, $\Delta^\virt(k_+,k_a;k_-,k_b)$, can be obtained by 
obvious substitutions.

Contributions of the diagrams (\ifp ), (\im ), (\mf ), and (\mm ) 
of \reffi{virtual_initial_final_diagrams}
cancel in the sum of virtual and real corrections in DPA
but remain for the virtual corrections.
After combining several diagrams the result becomes relatively simple:
\begin{eqnarray}
\Delta_{\mm}^\virt &\sim& 
- 2 \ln\biggl(\frac{K_+}{\MW^2}\biggr)
- 2 \ln\biggl(\frac{K_-}{\MW^2}\biggr)
+ 4 \ln\biggl(\frac{\la}{\MW}\biggr)
+4,
\\[.5em]
\Delta_{\mmp}^\virt &\sim& 
\frac{2\MW^2-s}{s\beta_\PW}\biggl[
-\cLi\biggl(\frac{K_-}{K_+},x_\PW\biggr)
+\cLi\biggl(\frac{K_-}{K_+},x_\PW^{-1}\biggr)
+\Li(1-x_\PW^2)+\pi^2 +\ln^2(-x_\PW)
\nn\\ && {}
+2\ln\biggl(\frac{K_+}{\lambda\MW}\biggr)\ln(x_\PW) -2\pi\ri\ln(1-x_\PW^2)
\biggr]
\; + \; \mbox{imaginary parts},
\end{eqnarray}
and
\begin{eqnarray}
\lefteqn{
  \Delta_{\mf}^\virt + \Delta_{\im}^\virt + \Delta_{\ifp}^\virt
+ \Delta_{\mfp}^\virt + \Delta_{\ffp}^\virt \sim
  \ln\biggl(\frac{\la^2}{\MW^2}\biggr) 
  \ln\biggl(\frac{-s_{23}-\ri \epsilon}{\MW^2}\biggr) 
}
\nn\\ && {}
-\frac{K_+K_-s_{23}\det(Y_0)}{\det(Y)}D_0(0)
- \frac{K_+ \det(Y_3)}{\det( Y)} F_3
- \frac{K_- \det(Y_2)}{\det( Y)} F_2
\nn\\ && {}
+2\ln\biggl(\frac{-K_+}{\la \MW}\biggr) 
  \ln\biggl(\frac{u_{-2}(\MW^2-t)}{t_{+2}(\MW^2-u)}\biggr) 
+2\ln\biggl(\frac{-K_-}{\la \MW}\biggr) 
  \ln\biggl(\frac{u_{+3}(\MW^2-t)}{t_{-3}(\MW^2-u)}\biggr) 
\nn\\ && {}
+ \Li\biggl(1-\frac{t-\MW^2}{t_{+2}}\biggr)
+ \Li\biggl(1-\frac{t-\MW^2}{t_{-3}}\biggr)
\nn\\ && {}
- \Li\biggl(1-\frac{u-\MW^2}{u_{+3}}\biggr)
- \Li\biggl(1-\frac{u-\MW^2}{u_{-2}}\biggr).
\end{eqnarray}
where $D_0(0)$ can be found in \refeq{D00} and the variables 
$\beta_\PW$, $x_\PW$ and the dilogarithms $\Li$,
$\cLi$ are defined in \refapp{prelim}.

All mass singularities vanish in the non-factorizable virtual 
correction factor $\delta_{\mathrm {nf}}^{\virt}$. 
Infrared singularities remain
which have to be cancelled by the corresponding real corrections.

\subsection{Inclusion of the exact off-shell Coulomb singularity}
\label{se43}

For non-relativistic \PW~bosons, \ie for 
a small on-shell velocity $\be_\PW$, the long range of
the Coulomb interaction leads to a large radiative correction, known as
the Coulomb singularity. For on-shell \PW~bosons,
this correction behaves like $1/\be_\PW$
near threshold, but including the instability of the 
\PW~bosons the long-range interaction is effectively truncated,
and the $1/\beta_\PW$ singularity is regularized. Therefore, for realistic
predictions in the threshold region, the on-shell Coulomb singularity
should be replaced by the corresponding off-shell correction.
The precise form of this off-shell Coulomb singularity \cite{Coul} 
reveals that corrections of some per cent occur even a few \PW-decay 
widths above threshold. As explained above, the DPA becomes valid only 
several widths above threshold. Nevertheless, there exists an overlapping 
region in which the inclusion of the Coulomb singularity within the 
double-pole approach is reasonable.

The Feynman graph relevant for the Coulomb singularity is diagram (d) 
of \reffi{virtual_final_final_diagrams}. The
non-factorizable corrections contain just the difference between 
the off-shell and the on-shell contributions of this diagram. Therefore,
the difference between off-shell and on-shell Coulomb singularity 
is in principle included in $\Delta_{\mathrm{mm}'}$, as defined in
\refeq{eq:dmm}. The genuine form of
$\Delta_{\mathrm{mm}'}$ in DPA, which is given by \refeq{eq:DmmDPA},
does not contain 
the full effect of the Coulomb singularity, because 
in both $C_0$ functions of \refeq{eq:dmm} the on-shell limit
$K_\pm\to 0$ was taken under the assumption of a finite $\be_\PW$.
In order to include the correct difference 
between off-shell and on-shell Coulomb singularity
in $\Delta_{\mathrm{mm}'}$, the on-shell limit of the $C_0$ functions of 
\refeq{eq:dmm} has to be taken for arbitrary $\be_\PW$. 
The full off-shell Coulomb singularity can be included by adding 
\beq\label{demmcoul}
\frac{2\MW^2-s}{s} \biggl[
\frac{2\pi\ri}{\betap} \ln\biggl(
\frac{\betaM+\Delta_M-\betap}{\betaM+\Delta_M+\betap}\biggr) 
-\frac{2\pi\ri}{\betaW} \ln\biggl(
\frac{K_+ + K_- + s\betaW\Delta_M}{2\betaW^2s}\biggr) \biggr]
\eeq
to the genuine DPA result \refeq{eq:dmmDPA} for $\Delta_{\mathrm{mm}'}$.
The quantities $\betaM$, $\betap$, and $\Delta_M$ are defined in
\refeq{abbrev}.  
After combination with the factorizable doubly-resonant corrections,
all doubly-resonant corrections and the correct off-shell Coulomb singularity
are included. The on-shell Coulomb
singularity contained in the factorizable corrections is compensated
by a corresponding contribution in the non-factorizable ones. 
Note, however, that this subtracted on-shell Coulomb singularity appears 
as an artefact if the non-factorizable corrections are discussed separately.
 
\subsection{Non-local cancellations}

In \citere{Fa94} it was pointed out that the non-factorizable
photonic corrections completely cancel in DPA if the phase-space 
integration over both invariant masses of the W~bosons is performed.
This cancellation is due to a symmetry of the non-factorizable corrections.

The lowest-order cross section in DPA
is symmetric with respect to the ``reflections'' 
\beqar
\label{reflect}
K_+=(k_+^2-\MW^2)+\ri\MW\Gamma_\PW 
&\;\leftrightarrow\;& -(k_+^2-\MW^2)+\ri\MW\Gamma_\PW=-K_+^*, \nn\\
K_-=(k_-^2-\MW^2)+\ri\MW\Gamma_\PW 
&\;\leftrightarrow\;& -(k_-^2-\MW^2)+\ri\MW\Gamma_\PW=-K_-^*. 
\eeqar
Therefore, $\De$ can be 
symmetrized in $K_+ \to -K_+^*$ or $K_- \to -K_-^*$ if the respective 
invariant mass is integrated out. For instance, if $k_-^2$ is integrated
out, $\De$ can be replaced by
\beqar
\label{kmaverage}
\lefteqn{
\frac{1}{2}\left(\Delta+\Delta\Big|_{K_- \leftrightarrow -K_-^*}\right) } && 
\nn\\*
& \;\sim\; &
\ri\pi \biggl\{ \;
\frac{s-2\MW^2}{\beta_\PW s}
\ln\biggl(\frac{K_- x_\PW+K_+^*}{K_- x_\PW-K_+}\biggr)
\nn\\ && {} \qquad
+ K_+s_{23}\kappa_\PW 
\biggl[ \frac{K_-}{\det(Y)}-\frac{K_-^*}{\det(\Ybr)} \biggr]
\biggl[ \ln\biggl(-x_\PW\frac{s_{23}}{\MW^2}\biggr)
+\ln\biggl(1+\frac{s_{13}}{s_{23}}(1-z)\biggr)
\nn\\ && {} \qquad\qquad
+\ln\biggl(1+\frac{s_{24}}{s_{23}}(1-z)\biggr) 
-\ln\biggl(1+\frac{s_{13}}{\MW^2}z x_\PW\biggr)
-\ln\biggl(1+\frac{s_{24}}{\MW^2}z x_\PW\biggr) \biggr]
\nn\\ && {} \qquad
-\biggl[\frac{K_- \det(Y_2)}{\det(Y)}
+\frac{K_-^* \det(\Ybr_2)}{\det(\Ybr)}\biggr] 
\biggl[ \ln\biggl(x_\PW+\frac{s_{13}+s_{23}}{\MW^2}\biggr)
-\ln\biggl(-x_\PW\frac{s_{23}}{\MW^2}\biggr) \biggr]
\nn\\ && {} \qquad
-\frac{K_+\det(Y_3)}{\det(Y)}\biggl[
\ln\biggl(x_\PW+\frac{s_{23}+s_{24}}{\MW^2}\biggr)
-2\ln\biggl(1+\frac{K_+}{K_-}\frac{s_{23}+s_{24}}{\MW^2}\biggr) \biggr]
\nn\\ && {} \qquad
-\frac{K_+ \det(\Ybr_3)}{\det(\Ybr)}\biggl[
\ln\biggl(x_\PW+\frac{s_{23}+s_{24}}{\MW^2}\biggr)
-2\ln\biggl(1-\frac{K_+}{K_-^*}\frac{s_{23}+s_{24}}{\MW^2}\biggr) \biggr]
\nn\\ && {} \qquad
+\frac{2s_{23}\MW^2(K_+^2 s_{24}-K_-^2 s_{13})}{\det(Y)}
\ln\biggl(1-\frac{K_+}{K_- x_\PW}\biggr)
\nn\\ && {} \qquad
+\frac{2s_{23}\MW^2(K_+^2 s_{24}-K_-^{*2} s_{13})}{\det(\Ybr)}
\ln\biggl(1+\frac{K_+}{K_-^* x_\PW}\biggr)
\; \biggr\}
\nn\\ && {} 
+ \; \mbox{imaginary parts},
\eeqar
with $z$ defined in \refeq{eq:xtilde}.
Note that this expression is considerably simpler than the full result 
for $\De$. In particular, all dilogarithms have dropped out. 

Symmetrizing \refeq{kmaverage} also in $K_+ \leftrightarrow -K_+^*$
leads to further simplifications if the relations \refeq{eq:detrels} for
the determinants are used. Under the assumptions that
$(s_{13}+s_{23})>-\MW^2 x_\PW$, $(s_{23}+s_{24})>-\MW^2 x_\PW$, 
and that $\kappa_\PW$ is imaginary, the real part of the result vanishes.
These assumptions are fulfilled on resonance, $k_\pm^2=\MW^2$; 
off resonance, there are regions in phase space 
where the assumptions are
violated. The volume of those regions of phase space is suppressed 
by factors $|k_\pm^2-\MW^2|/\MW^2$ and thus negligible in DPA. Therefore 
we can use the above assumptions and find that $\De$ vanishes in DPA
after averaging over the four points in the
$(k_+^2,k_-^2)$ plane that are related by the reflections \refeq{reflect}:
\beq\label{eq:Dcanc}
\De + \De\Big|_{K_+\to-K_+^*} + \De\Big|_{K_-\to-K_-^*} 
+ \De\Big|_{K_\pm\to-K_\pm^*} \sim 0.
\eeq
This explicitly confirms the results of \citere{Fa94}.
In particular, the non-factorizable corrections vanish in the limit 
$|k_\pm^2-\MW^2|\ll\GW\MW$, \ie for on-shell \PW~bosons.

The above considerations lead to the following simplified recipe for
the non-factorizable corrections to single invariant-mass distributions,
\ie  as long as at least one of the invariant masses $k_\pm^2$ is 
integrated out: the full factor $\De$ can be replaced according to
\beq
\label{recipe}
\De \;\to\; 
\frac{1}{2}\left(\Delta+\Delta\Big|_{K_- \leftrightarrow -K_-^*}\right) 
+ \frac{1}{2}\left(\Delta+\Delta\Big|_{K_+ \leftrightarrow -K_+^*}\right),
\eeq
where the first term on the r.h.s.\ is given in \refeq{kmaverage}, and
the second follows from the first by interchanging $K_+ \leftrightarrow
K_-$ and $s_{13} \leftrightarrow s_{24}$. Note that no double counting
is introduced, since the first (second) term does not contribute if
$k_+^2$ ($k_-^2$) is integrated out. 
In order to introduce the exact Coulomb singularity, one simply has to add the 
additional term of \refeq{demmcoul} 
to \refeq{kmaverage} and \refeq{recipe}.

\subsection{Non-factorizable corrections to related processes}

Since all non-factorizable corrections involving the initial $\Pep\Pem$ 
state cancel, the above results for the correction factor also apply 
to other \PW-pair production processes, such as 
$\gamma\gamma\to\PW\PW\to 4\,$fermions and $q\bar q\to\PW\PW\to 4\,$fermions.

The presented analytical results can also be carried over to
\PZ-pair-mediated four-fermion production, $\Pep\Pem\to\PZ\PZ\to
4\,$fermions.  In this case, $\Delta_{\ffp}$ yields the complete
non-factorizable correction, where all quantities such as $\MW$ and
$\Ga_\PW$ defined for the W~bosons are to be replaced by the ones for
the Z~bosons.  The fact that $Q_1=Q_2$ and $Q_3=Q_4$ has two important
consequences. Firstly, it implies the cancellation of mass
singularities contained in $\Delta_{\ffp}$ when all contributions are
summed as in \refeq{nfcorrfac}.  Secondly, it leads to the
antisymmetry of $\delta_\nf$ under each of the interchanges
$k_1\leftrightarrow k_2$ and $k_3\leftrightarrow k_4$. 
It is interesting to note that \refeq{nfcorrfac} with \refeq{eq:Delta}
are directly applicable, since 
$\De_{\mfp}$ and $\De_{\mmp}$ cancel in the sum of \refeq{nfcorrfac}. 
Therefore, a practical way to calculate the non-factorizable corrections
to $\Pep\Pem\to\PZ\PZ\to4\,$fermions consists in taking 
$\Delta_{\ffp}+\De_{\mfp}$ from \refeq{eq:DmfDPA} and \refeq{Fi}, and
setting $\De_{\mmp}$ to zero.

\section{Numerical results}
\label{senumres}

We used the parameters
\beq\label{params}
\alpha ^{-1}=137.0359895, \quad \MZ=91.187\GeV, \quad
\MW=80.22\GeV, \quad \GW = 2.08\GeV,
\eeq
which coincide with those of \citere{Be97a},
for the numerical evaluation.

In order to exclude errors, we have written two
independent programs for the correction factor \refeq{nfcorrfac} and
compared all building blocks numerically.
These subroutines are implemented in the Monte Carlo generator
{\tt EXCALIBUR} \cite{Be94}%
\footnote{Since the Monte Carlo program of \refch{ch:treelevel} was 
programmed after the results of the non-factorizable corrections 
were published, the Monte Carlo program EXCALIBUR was used for the
results of this chapter.}
as a correction factor to the three
(doubly-resonant) W-pair-production signal diagrams.
In all numerical results below, only these signal diagrams are included,
and no phase-space cuts have been applied.

If not stated otherwise the results for the figures 
have been obtained from 50 million
phase-space points using the histogram 
routine of {\tt EXCALIBUR} with 40 bins for each figure.
For each entry in the tables,
10 million phase-space points were generated. 

\subsection{Comparison with existing results}

The non-factorizable photonic corrections have already been evaluated
by two groups. As was noted in the introduction, the
authors of \citeres{Be97a,Be97b} have not confirmed%
\footnote{While the result of \citeres{Be97a,Be97b} (as our result) for
the complete non-factorizable correction
is free of mass-singular logarithms, the result of \citere{Me96} 
contains logarithms of ratios of fermion masses. However, the authors of 
\citere{Me96} have informed us \cite{myprivcom} that the results of 
\citere{Me96} and \citeres{Be97a,Be97b} agree for equal fermion pairs in the 
final state.} 
the results of \citere{Me96}. Therefore, we first compare our 
findings with the results of these two groups.

\newcommand{\bk}{{\bf k}}
Melnikov and Yakovlev \cite{Me96} give the relative non-factorizable
corrections only to the completely differential cross section for the
process $\Pep\Pem\to\PW\PW\to\Pne\Pep\Pem\Pane$ as a function of the
invariant mass $M_+$ of the $\Pne\Pep$ system for all other
phase-space parameters fixed. All momenta of the final-state fermions
lie in a plane, and the momenta $\bk_2$ and $\bk_3$ of the final-state
positron and electron point into opposite directions. The angle
between the $\PWm$~boson and the positron is fixed to
$\theta_{\PWm\Pep}=150^\circ$, and the CM energy is chosen
as $\sqrt{s}=180\GeV$. The invariant mass of the $\Pem\Pane$ system
takes the values $M_-=78$ and $82\GeV$. The other parameters are
$\alpha=1/137$, $\MW=80\GeV$, and $\GW=2.0\GeV$.
In \reffi{fimy} we show our results for the non-factorizable photonic
corrections for this phase-space configuration.
\bfi
\centerline{
\setlength{\unitlength}{1cm}
\begin{picture}(7.5,7.8)
\put(0,0){\includegraphics{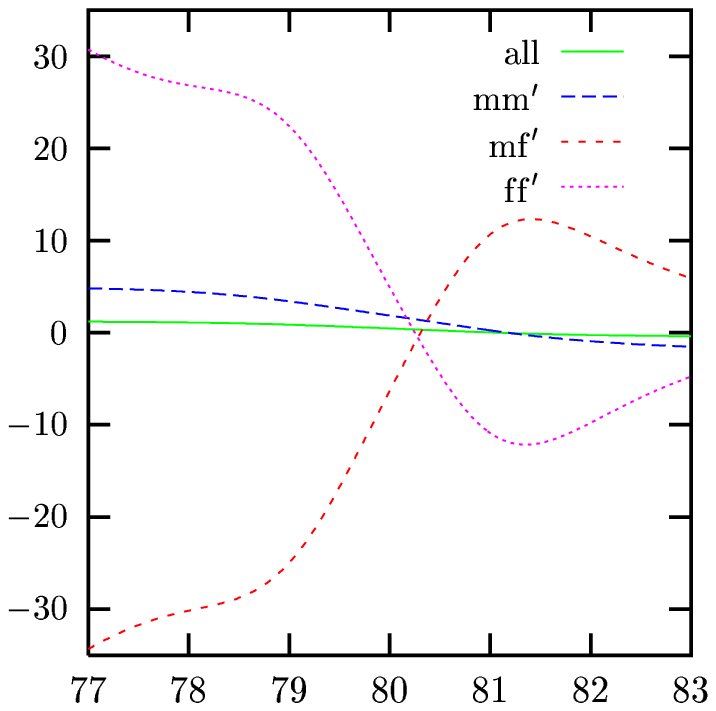}}
\put(0,5.0){\makebox(1,1)[c]{$\de_\nf/\%$}}
\put(4.5,-0.3){\makebox(1,1)[cc]{{$M_+/{\rm GeV}$}}}
\put(4.6,2){$M_-=78\GeV$}
\end{picture}
\begin{picture}(7.5,7.8)
\put(0,0){\includegraphics{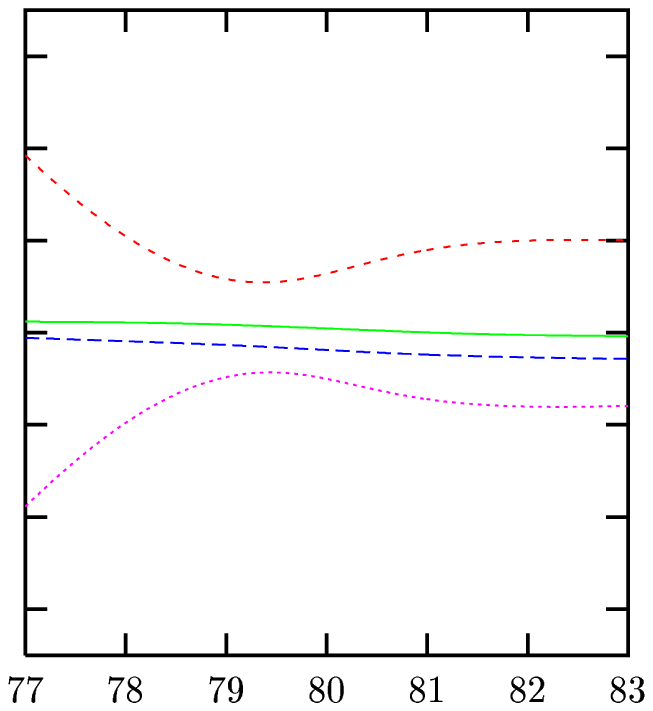}}
\put(3.5,-0.3){\makebox(1,1)[cc]{{$M_+/{\rm GeV}$}}}
\put(4,2){$M_-=82\GeV$}
\end{picture}
}
\caption{Relative non-factorizable correction factor to the
  differential cross section for the phase-space configuration
  given in the text.
  The curves labelled \mmp, \mfp, and \ffp~correspond
  to the curves A, B, and C of \protect\citere{Me96}, respectively.}
\label{fimy}
\efi
The intermediate--intermediate (\mmp)
corrections agree reasonably with those of \citere{Me96}, 
but the other curves differ qualitatively and
quantitatively%
\footnote{\samepage{Although not stated in \citere{Me96}, mass-singular parts
have been dropped there in the numerical evaluation \cite{myprivcom}
rendering a thorough comparison of the (\mfp) and (\ffp) parts
impossible. Comparing the sum of all three contributions, i.e.\ the
complete non-factorizable correction factor, our result differs from 
the sum read off from \citere{Me96}.}}.
We mention that \reffi{fimy} has been reproduced \cite{bbcprivcom} by the
authors of \citeres{Be97a,Be97b} within the expected level of accuracy.
While the individual contributions shown in
\reffi{fimy} are at the level of 10\% owing to mass singularities, the
sum, which is free of mass singularities, is below 1.2\%.

Beenakker et al.\ \cite{Be97a} have evaluated the relative non-factorizable
corrections to the distributions 
$\rd\sigma/\rd M_+\rd M_-$, $\rd\sigma/\rd M_+$, and
$\rd\sigma/\rd M_{\mathrm{av}}$, where $M_{\mathrm{av}}=(M_-+M_+)/2$.
Our corresponding results for the set of parameters given in \refeq{params}
are shown in \reffi{fi3bbc} for the single
invariant-mass distributions and in 
\refta{tab1bbc} for the double invariant-mass distribution. 
\bfi
\centerline{
\setlength{\unitlength}{1cm}
\begin{picture}(10,7.8)
\put(0,0){\includegraphics{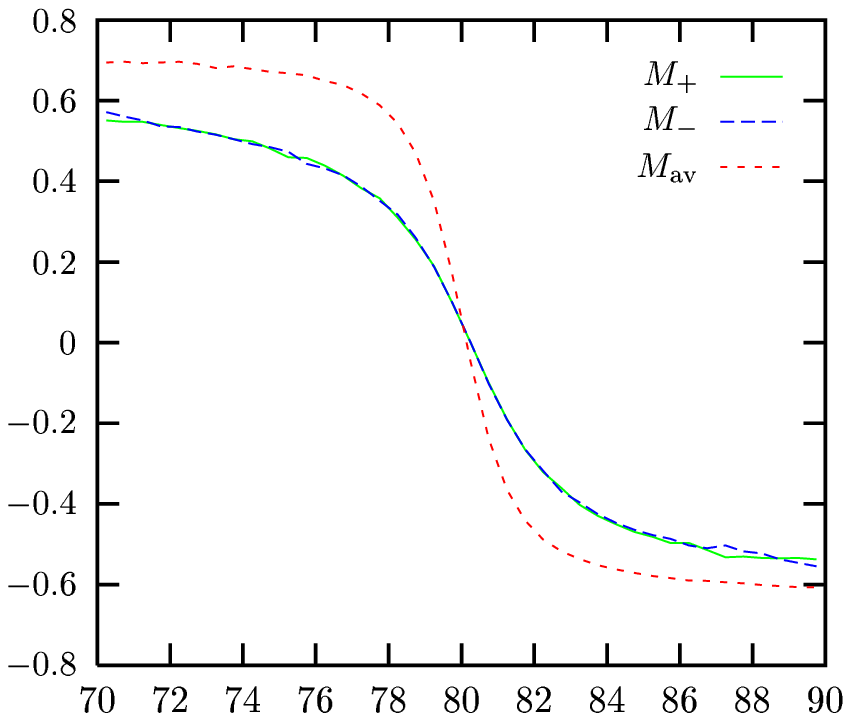}}
\put(-0.5,5.0){\makebox(1,1)[c]{$\de_\nf/\%$}}
\put(4.7,-0.3){\makebox(1,1)[cc]{{$M/{\rm GeV}$}}}
\end{picture}
}
\caption[]{Relative non-factorizable corrections %$\delta_{\nf}(M)$ 
  to the single invariant-mass distributions
  $\protect\rd\sigma/\protect\rd M_\pm$ and
  $\protect\rd\sigma/\protect\rd M_{\protect\mathrm{av}}$ for the
    CM energy $\sqrt{s}=184\protect\GeV$.}
\label{fi3bbc}
\efi
\newcommand{\mspc}{\phantom{-}}
\begin{table}
\renewcommand{\arraycolsep}{2ex}
\[
  \begin{array}{||c||c|c|c|c|c|c|c||}    \hline \hline
    \raisebox{-3mm}{$\Delta_+$}
    & \multicolumn{7}{|c||}{\raisebox{-1mm}{$\Delta_-$}}\\
    \cline{2-8}
      & -1    & -1/2  & -1/4  &  0    & 1/4   & 1/2   & 1     \\
    \hline  \hline
 -1  &\mspc0.81 &\mspc0.64 &\mspc0.52 &\mspc0.38 &\mspc0.22 &\mspc0.07 & -0.16\\
-1/2 &\mspc0.64 &\mspc0.52 &\mspc0.40 &\mspc0.24 &\mspc0.08 &    -0.07 & -0.25\\
-1/4 &\mspc0.52 &\mspc0.40 &\mspc0.28 &\mspc0.13 &    -0.02 &    -0.15 & -0.31\\
  0  &\mspc0.38 &\mspc0.24 &\mspc0.13 &\mspc0.00 &    -0.13 &    -0.24 & -0.37\\
 1/2 &\mspc0.22 &\mspc0.08 &    -0.02 &    -0.13 &    -0.24 &    -0.32 & -0.43\\
 1/4 &\mspc0.07 &    -0.07 &    -0.15 &    -0.24 &    -0.32 &    -0.39 & -0.48\\
  1  &    -0.16 &    -0.25 &    -0.31 &    -0.37 &    -0.43 &    -0.48 & -0.54\\
    \hline \hline
  \end{array}
\]
\caption[]{Relative non-factorizable corrections 
           in per cent to the double invariant-mass distribution
           $\rd\sigma/\rd M_+\rd M_-$ for the CM energy 
           $\sqrt{s}=184\GeV$ and various values of $M_{\pm}$
           specified in terms of their
           distance from $\MW$ in units of $\GW$,
           \ie $\Delta_{\pm} = (M_{\pm}-\MW)/\GW$.}
\label{tab1bbc}
\end{table}%
The deviation between the distributions $\rd\sigma/\rd M_+$ and
$\rd\sigma/\rd M_-$ in \reffi{fi3bbc}, which should be identical,
gives an indication on the Monte Carlo error of our calculation.  The
single and double invariant-mass distributions agree very well with
those of \citere{Be97a}. 
The worst agreement is found for small invariant masses and amounts to
0.03\%. In fact, the agreement is better than expected,
in view of the fact that our results differ from those of
\citere{Be97a} by non-doubly resonant corrections. 
In the numerical evaluations of \citere{Be97a} the phase space and the
Born matrix element are taken entirely on shell \cite{bbcprivcom}.
Moreover, the scalar integrals are parameterized by scalar invariants
different from ours, leading to differences of the order
of $|k_\pm^2-\MW^2|/\MW^2$.

In \citere{Be97a}, additionally, the decay-angular distribution 
$\rd\sigma/\rd M_-\rd M_+\rd\cos\theta_{\PWp\Pep}$ has been considered,
where $\theta_{\PWp\Pep}$ is the decay angle between $\bk_+$ and $\bk_2$ 
in the laboratory system. 
Our results for this angular distribution are shown in \reffi{fi2bbc}.
\bfi
\centerline{
\setlength{\unitlength}{1cm}
\begin{picture}(10,7.8)
\put(0,0){\includegraphics{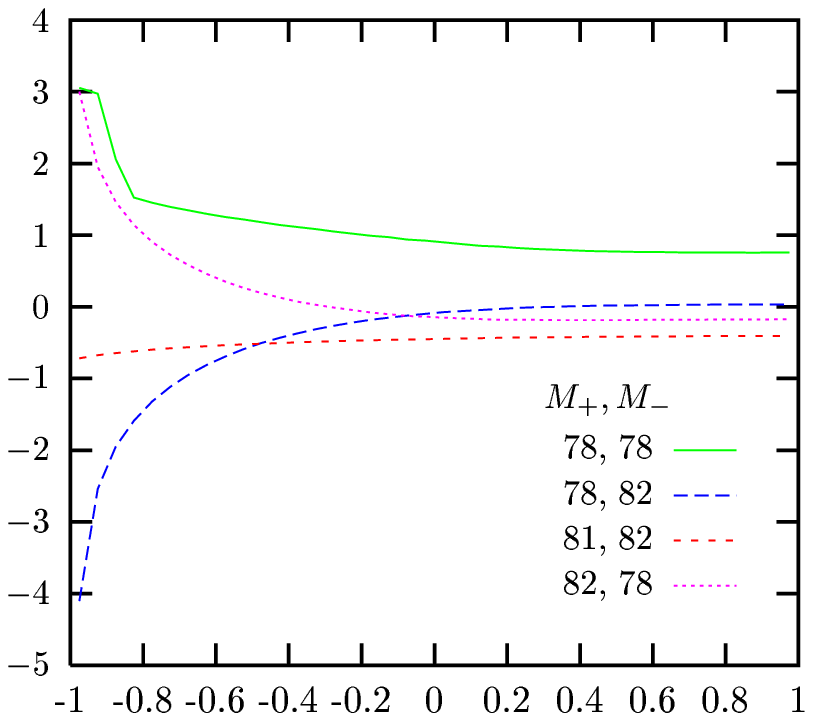}}
\put(-0.5,5.0){\makebox(1,1)[c]{$\de_\nf/\%$}}
\put(4.7,-0.3){\makebox(1,1)[cc]{{$\cos\theta_{\PWp\Pep}$}}}
\end{picture}
}
\caption[]{Relative non-factorizable corrections to the decay-angular
           distribution $\rd\sigma/\rd M_-\rd M_+\rd\cos\theta_{\PWp\Pep}$
           for fixed values of the invariant masses $M_{\pm}$ 
           and the CM energy $\sqrt{s}=184\protect\GeV$.}
\label{fi2bbc}
\efi
The cross section is small for $\cos\theta_{\PWp\Pep}\sim-1$, where
the corrections are largest. 
Unfortunately the corresponding figure in \citere{Be97a} is not correct 
\cite{bbcprivcom}. The authors of \citere{Be97a} have provided a 
corrected figure, which agrees reasonably well with \reffi{fi2bbc}, but 
does not show the kinks in the curve for $M_\pm=78\GeV$. 
The kinks are due to a logarithmic Landau singularity in the 
4-point functions.
If one employs the on-shell parameterization of phase space, 
as in \citere{Be97a}, the Landau
singularities appear at the boundary of phase space. Although no kinks
appear in the physical phase space in this case, the Landau
singularities still give rise to large corrections for
$\cos\theta_{\PWp\Pep}\sim-1$. 
Since the kinks appear in a region where the cross section is
small, they are not relevant for phenomenology.
The issue of the kinks is further discussed in \refse{se:ambig}.

\subsection{Numerical results for leptonic final state}

In \reffi{energyplot} we show the non-factorizable corrections to the
single invariant-mass distribution $\rd\sigma/\rd M_+$ for various
CM energies.  While the corrections reach up to 1.3\% for $\sqrt{s}=172\GeV$,
they decrease with increasing energy and are less than 0.04\%
for $\sqrt{s}=300\GeV$. 
\bfi
\centerline{
\setlength{\unitlength}{1cm}
\begin{picture}(10,7.8)
\put(0,0){\includegraphics{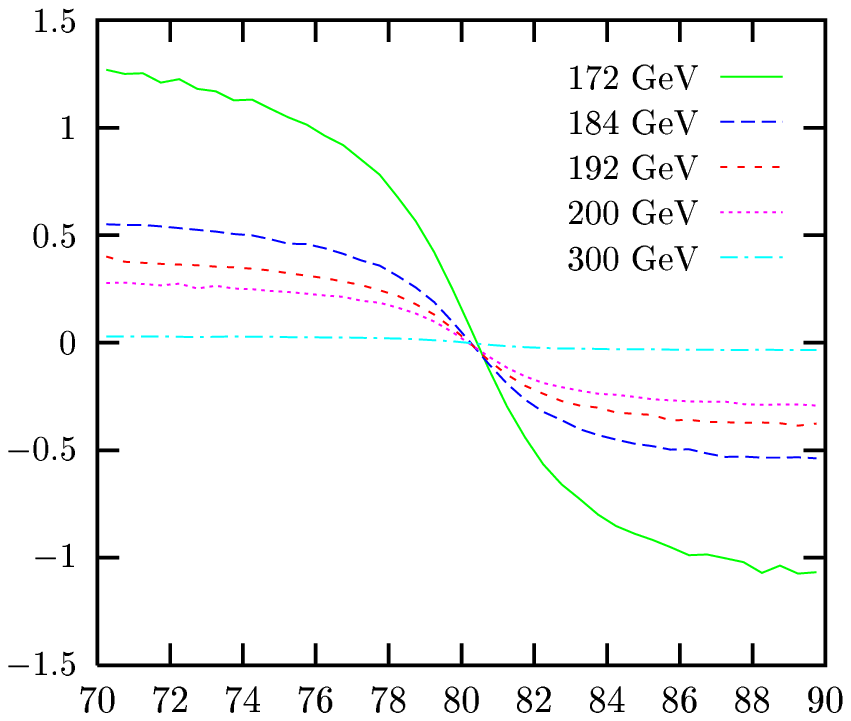}}
\put(-0.5,5.0){\makebox(1,1)[c]{$\de_\nf/\%$}}
\put(4.7,-0.3){\makebox(1,1)[cc]{{$M_+/{\rm GeV}$}}}
\end{picture}
}
\caption[]{Relative non-factorizable corrections  
  to the single invariant-mass distribution 
  $\protect\rd\sigma/\protect\rd M_+$ for
  various CM energies.}
\label{energyplot}
\efi
Note that the shape of the corrections is exactly what is naively
expected. If a photon is emitted in the final state, the invariant mass 
of the fermion pair is smaller than the invariant mass of the resonant
\PW~boson, which is given by the invariant mass of the fermion pair plus 
photon. Since we calculate the corrections as a function of the invariant
masses of the fermion pairs, the cross section tends to increase
for small invariant masses and decrease for large invariant masses.

The non-factorizable corrections distort the invariant-mass
distribution and thus lead to a shift in the \PW-boson mass determined
from the direct reconstruction of the decay products with respect to the
actual \PW-boson mass. This shift can be estimated by the
displacement of the maximum of the single-invariant-mass distribution
caused by the corrections shown in \reffi{energyplot}. 
To this end, we determine
the slope of the corrections for $M_+=\MW$, multiply this linearized
correction to a simple Breit-Wigner factor, and determine the shift 
$\Delta M_+$ of the maximum. The smallness of the correction allows us
to evaluate $\Delta M_+$ in linear approximation, leading to the simple
formula
\beq
\Delta M_+ = 
\biggl(\frac{\rd\delta_{\nf}}{\rd M_+}\biggr)\bigg|_{M_+=\MW} 
\frac{\GW^2}{8}.  
\eeq
Extracting the slope from our numerical results we obtain the mass shifts 
shown in \refta{ta:shifts}.

In \reffis{phiplot} and \ref{thetaepem} we show the 
effect of the non-factorizable corrections on various angular distributions. 
\bfi
\centerline{
\setlength{\unitlength}{1cm}
\begin{picture}(10,7.8)
\put(0,0){\includegraphics{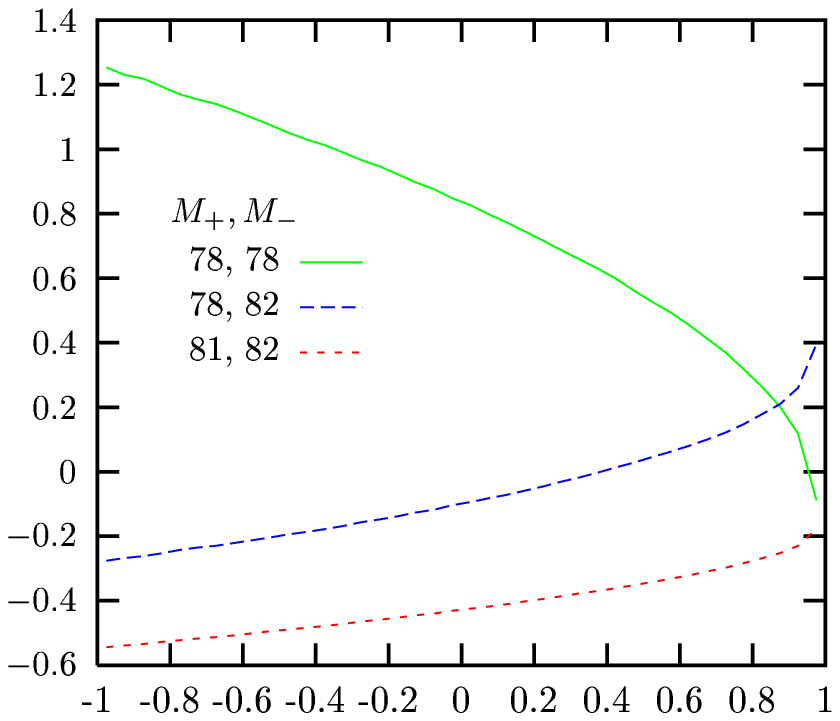}}
\put(-0.5,5.0){\makebox(1,1)[c]{$\de_\nf/\%$}}
\put(4.7,-0.3){\makebox(1,1)[cc]{{$\cos\phi$}}}
\end{picture}
}
\caption[]{Relative non-factorizable corrections to the angular
             distribution $\protect\rd\sigma/\protect\rd M_-\rd M_+\rd\cos\phi$
             for fixed values of the invariant masses $M_{\pm}$ 
             and the CM energy $\sqrt{s}=184\protect\GeV$.}
\label{phiplot}
\efi
\bfi
\centerline{
\setlength{\unitlength}{1cm}
\begin{picture}(10,7.8)
\put(0,0){\includegraphics{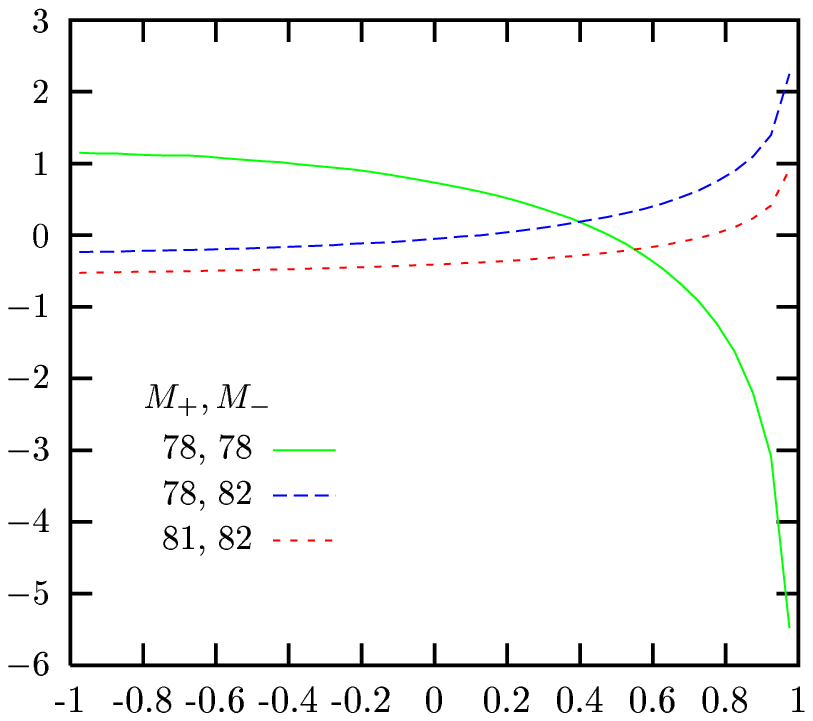}}
\put(-0.5,5.0){\makebox(1,1)[c]{$\de_\nf/\%$}}
\put(4.7,-0.3){\makebox(1,1)[cc]{{$\cos\theta_{\Pep\Pem}$}}}
\end{picture}
}
\caption[]{Relative non-factorizable corrections to the angular distribution 
             $\protect\rd\sigma/\protect\rd M_-\rd M_+\rd\cos\theta_{\Pep\Pem}$
             for fixed values of the invariant masses $M_{\pm}$ 
             and a CM energy $\sqrt{s}=184\protect\GeV$.}
\label{thetaepem}
\efi
\begin{table}
$$
\renewcommand{\arraycolsep}{2ex}
\begin{array}{|c|r|r|r|r|r|}
\hline
\sqrt{s}/\mathrm{GeV}    &   172  &    184   &  192  &  200  &   300 \\
\hline
\Delta M_+/\mathrm{MeV}  &  -2.0  &   -1.1   & -0.8  & -0.6  &  -0.09
\\
\hline
\end{array}
$$
\caption{Shift of the maximum of the single invariant-mass
  distributions $\rd\si/\rd M_+$ induced by the non-factorizable
  corrections at various CM energies.}
\label{ta:shifts}
\end{table}
Since the non-factorizable corrections are independent of the production angle
of the \PW~bosons, it suffices to consider distributions
involving the angles of the final-state fermions. 
We define all angles in the laboratory system,
which is the CM system of the production process. 
The distribution over the angle $\phi$ between the two planes spanned
by the momenta of the two fermion pairs in which the \PW~bosons decay, \ie
\beq
\cos\phi = \frac{(\bk_1\times \bk_2)(\bk_3\times \bk_4)}
{|\bk_1\times \bk_2||\bk_3\times \bk_4|},
\eeq
is presented in \reffi{phiplot}. 
The corrections are of the order of 1\% or less. Like
the $\phi$ distribution, the distribution over the angle between positron 
and electron $\theta_{\Pep\Pem}$ (\reffi{thetaepem}) is symmetric under the
interchange of $M_+$ and $M_-$. 
As for the $\theta_{\Pep\PWp}$ distribution (\reffi{fi2bbc}), 
the corrections reach several per
cent in the region where the cross section is small.

The distribution in the electron energy $E_\Pem$ is considered in \reffi{Eem}.
\bfi
\centerline{
\setlength{\unitlength}{1cm}
\begin{picture}(10,7.8)
\put(0,0){\includegraphics{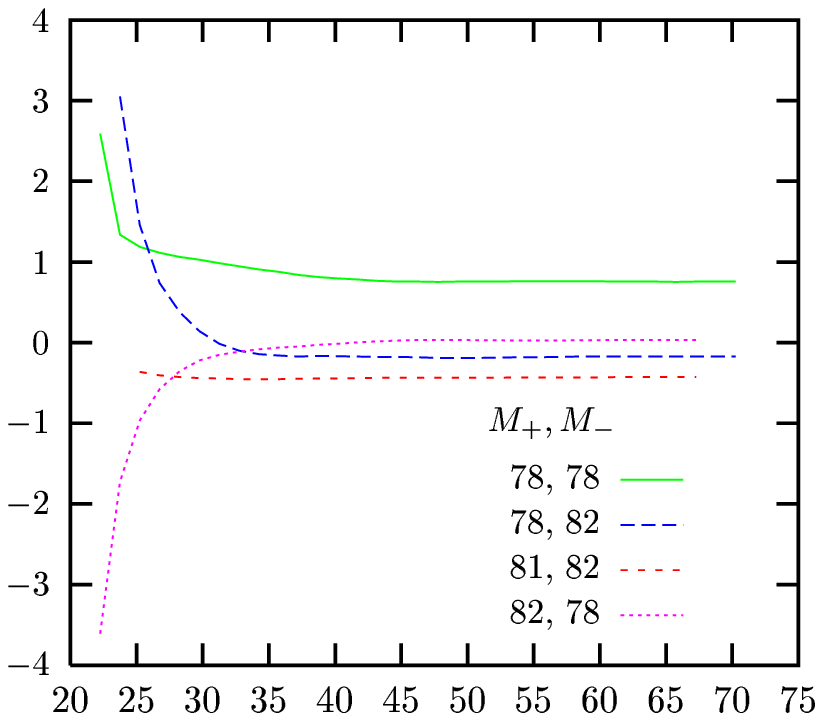}}
\put(-0.5,5.0){\makebox(1,1)[c]{$\de_\nf/\%$}}
\put(4.7,-0.3){\makebox(1,1)[cc]{{$E_\Pem/\GeV$}}}
\end{picture}
}
\caption[]{Relative non-factorizable corrections to the electron-energy
distribution 
$\protect\rd\sigma/\protect\rd M_-\rd M_+\rd E_\Pem$
for fixed values of the invariant masses $M_{\pm}$ 
and a CM energy $\sqrt{s}=184\protect\GeV$.}
\label{Eem}
\efi
The corrections are typically of the order of 1\%. Again the 
corrections become large where the cross section is small.

In \refse{se43} we have introduced a correction term that includes the
full off-shell Cou\-lomb singularity. The results for the
non-factorizable corrections with this improvement
are compared with those of the pure DPA in \reffi{bbcfig3coul}.
\bfi
\centerline{
\setlength{\unitlength}{1cm}
\begin{picture}(10,7.8)
\put(0,0){\includegraphics{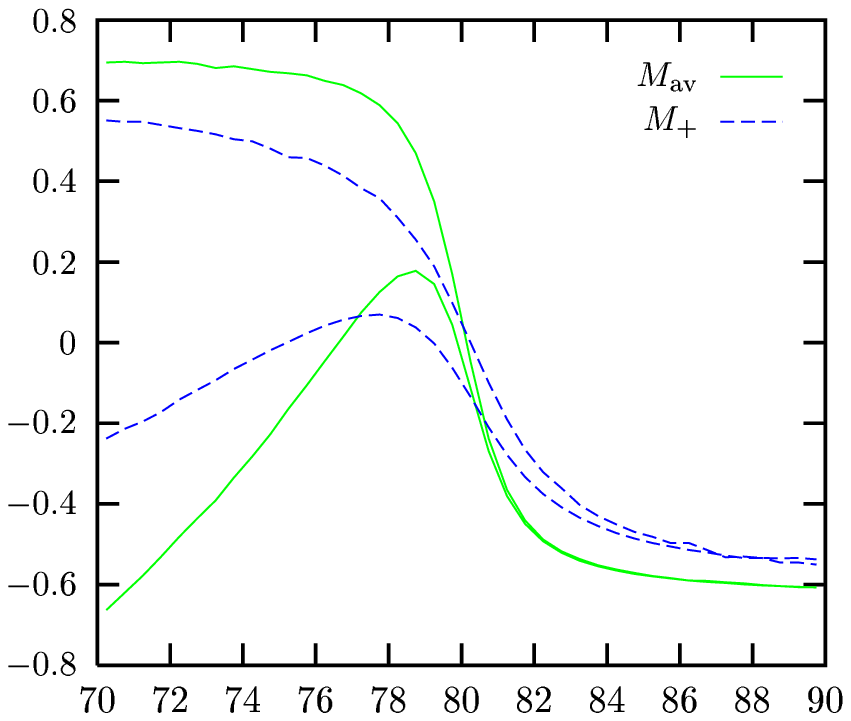}}
\put(-0.5,5.0){\makebox(1,1)[c]{$\de_\nf/\%$}}
\put(4.7,-0.3){\makebox(1,1)[cc]{{$M/{\rm GeV}$}}}
\end{picture}
}
\caption[]{Relative non-factorizable corrections 
  to the single invariant-mass distributions
  $\protect\rd\sigma/\protect\rd M_+$ and
  $\protect\rd\sigma/\protect\rd M_{\protect\mathrm{av}}$ for a CM
    energy $\sqrt{s}=184\protect\GeV$ with (lower curves) and without 
    (upper curves) improved Coulomb-singularity treatment.}
\label{bbcfig3coul}
\efi
\begin{table}
\renewcommand{\arraycolsep}{2ex}
\[
  \begin{array}{||c||c|c|c|c|c|c|c||}    \hline \hline
    \raisebox{-3mm}{$\Delta_+$}
    & \multicolumn{7}{|c||}{\raisebox{-1mm}{$\Delta_-$}}\\
    \cline{2-8}
         & -1    & -1/2  & -1/4  &  0    & 1/4   & 1/2   & 1     \\
    \hline  \hline
  -1   & \mspc0.39 &   \mspc 0.31 &   \mspc 0.23 &   \mspc 0.15 &   \mspc 0.05 &   -0.04 &   -0.20\\
 -1/2  &\mspc0.31 &\mspc0.28 &\mspc0.20 &\mspc0.10 &   -0.02 & -0.13 & -0.27\\
 -1/4  &\mspc0.23 &\mspc0.20 &\mspc0.13 &\mspc0.03 &   -0.09 & -0.19 & -0.33\\
   0   &\mspc0.15 &\mspc0.10 &\mspc0.03 &   -0.08  &   -0.18 & -0.27 & -0.38\\
  1/2  &\mspc0.05 &   -0.02  &   -0.09  &   -0.18  &   -0.27 & -0.35 & -0.44\\
  1/4  &   -0.04  &   -0.13  &   -0.19  &   -0.27  &   -0.35 & -0.41 & -0.49\\
   1   &   -0.20  &   -0.27  &   -0.33  &   -0.38  &   -0.44 & -0.49 & -0.54\\
    \hline \hline
  \end{array}
\]
\caption[]{Same as in \refta{tab1bbc} but with improved Coulomb-singularity 
treatment.}
\label{tab1bbcwc}
\end{table}%
For $\sqrt{s}=184\GeV$
the additional terms shift the non-factorizable corrections by up to 1.4\%
for $\protect\rd\sigma/\protect\rd M_{\protect\mathrm{av}}$ and
by up to 0.8\% for $\protect\rd\sigma/\protect\rd M_+$
for small invariant masses, whereas for large invariant masses
there is practically no effect.
The difference originates essentially from the differences between $1/\betap$ 
and $1/\betaW$ in \refeq{demmcoul}. For large invariant masses, 
the explicit logarithms in \refeq{demmcoul} are small, \ie
the Coulomb singularity correction is minuscule, and
this difference practically makes no effect. For small invariant
masses, the logarithms are approximately $\ri\pi$ and the
difference causes the effect seen in \reffi{bbcfig3coul}. 
In \refta{tab1bbcwc} we show the non-factorizable corrections to the
double invariant-mass distribution, as in \refta{tab1bbc}, but now with
the improved Coulomb-singularity treatment.
We find a difference of up to half a per cent for small invariant masses 
but no effect for large ones.
We mention that the difference between the entries in \reftas{tab1bbcwc} 
and \ref{tab1bbc} is directly given by the contribution \refeq{demmcoul}
to $\Delta_{\mathrm{mm}'}$, without any influence of the phase-space
integration.

\subsection{Discussion of intrinsic ambiguities}
\label{se:ambig}

In the results presented so far, all scalar integrals were
parameterized by $s$, $s_{23}$, $s_{13}$, $s_{24}$, and $k_\pm^2$
(parameterization 1). In DPA, however, the parameters of the
scalar integrals are only fixed up to terms of order $k_\pm^2-\MW^2$. 
We can for example parameterize the scalar integrals in terms of $s$,
$s_{23}$, $s_{123}$, $s_{234}$, and $k_\pm^2$ (parameterization 2) instead. 
As a third parameterization, we fix all scalar invariants except for
$k_\pm^2$ by their on-shell values, corresponding exactly to the
approach of \citere{Be97a}. The results of these three parameterizations 
differ by non-doubly-resonant corrections. 

The difference between parameterizations 1 and 2 is illustrated
in \reffi{vglparamM1} for the single invariant-mass distribution.
\bfi
\centerline{
\setlength{\unitlength}{1cm}
\begin{picture}(10,7.8)
\put(0,0){\includegraphics{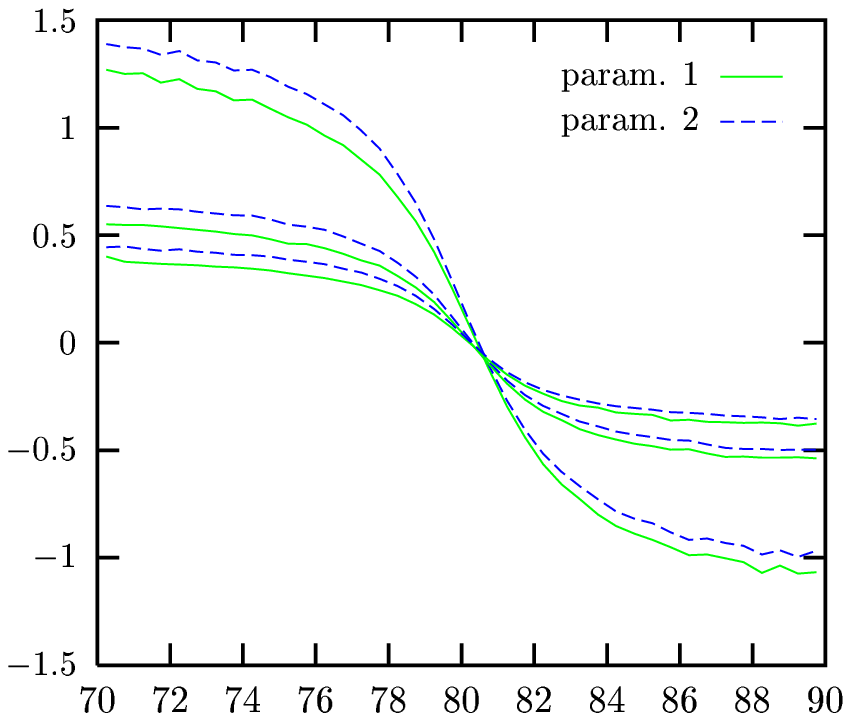}}
\put(-0.5,5.0){\makebox(1,1)[c]{$\de_\nf/\%$}}
\put(4.7,-0.3){\makebox(1,1)[cc]{{$M_+/\protect\mathrm{GeV}$}}}
\put(6.9,4.0){$\scriptstyle\sqrt{s}=192\GeV$}
\put(6.9,2.8){$\scriptstyle\phantom{\sqrt{s}=}184\GeV$}
\put(6.9,1.7){$\scriptstyle\phantom{\sqrt{s}=}172\GeV$}
\end{picture}
}
\caption{Relative non-factorizable corrections  
  to the single invariant-mass distribution
  $\protect\rd\sigma/\protect\rd M_+$ for 
  the CM energies 
  $172$, $184$, and $192\GeV$ and two different parameterizations.}
\label{vglparamM1}
\efi
\bfi
\centerline{
\setlength{\unitlength}{1cm}
\begin{picture}(14.5,7.8)
\put(0.5,0){\includegraphics{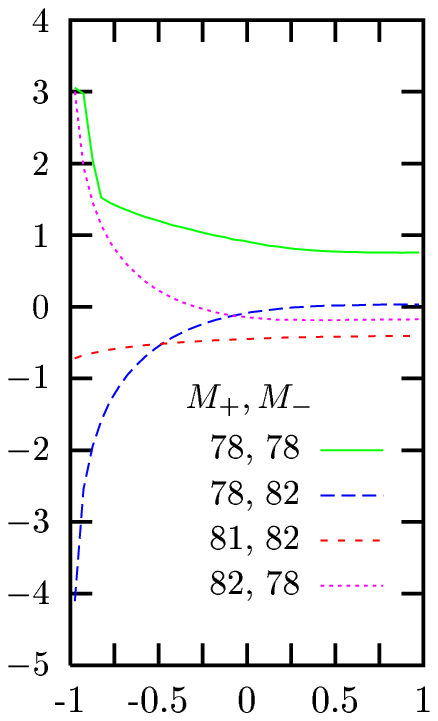}}
\put(4.8,0){\includegraphics{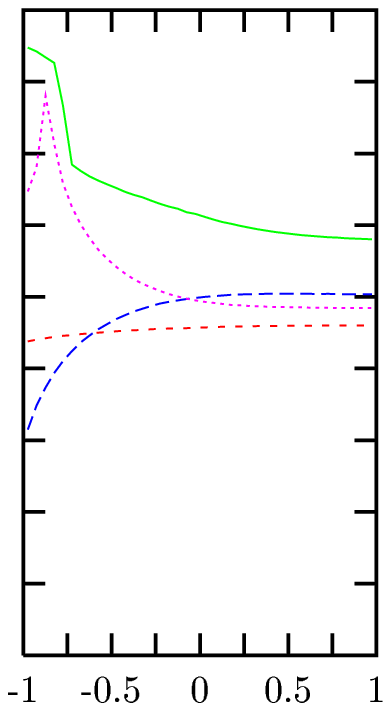}}
\put(9.1,0){\includegraphics{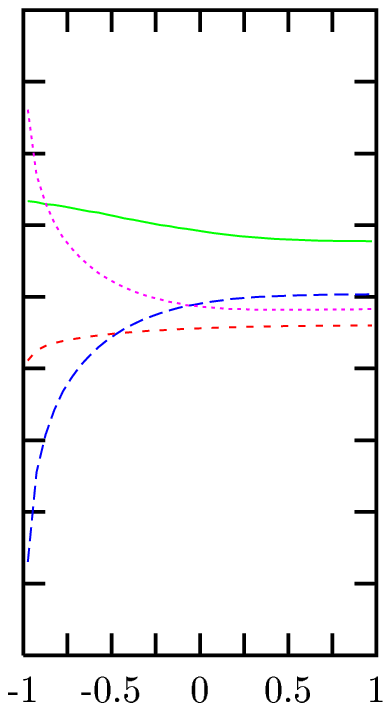}}
\put(-0.2,5.0){\makebox(1,1)[c]{$\de_\nf/\%$}}
\put(7.0,-0.3){\makebox(1,1)[cc]{{$\cos\theta_{\PWp\Pep}$}}}
\put( 2.6,7.0){param.~1}
\put( 6.9,7.0){param.~2}
\put(11.2,7.0){param.~3}
\end{picture}
}
\caption[]{Relative non-factorizable corrections
             to the decay-angular
             distribution $\rd\sigma/\rd M_-\rd M_+\rd\cos\theta_{\PWp\Pep}$
             for fixed values of the invariant masses $M_{\pm}$ 
             and a CM energy $\sqrt{s}=184\protect\GeV$
             using three different parameterizations, as specified in
             the text.}
\label{vglparamtheta1}
\efi
The difference amounts to $\sim 0.1\%$.
Note that for an invariant mass $M_+ = 70\GeV$ we have 
$\al|\MW^2-k_+^2|/\MW^2\sim 0.002$ and would thus expect absolute
changes in the non-factorizable corrections at this level.  

For the non-factorizable corrections to the angular distributions,
uncertainties of the same order are to be expected. The only exceptions
are the distributions over the decay angles $\theta_{\PWp\Pep}$ and
$\theta_{\PWm\Pem}$. Let us explain this for $\theta_{\PWp\Pep}$ in more
detail: the non-factorizable correction contains the term
$2\pi\ri\ln[1+x_\PW\MW^2/(s_{13}+s_{23})]$, which can be evaluated by taking
$(s_{13}+s_{23})$ directly or $(s_{123}-\MW^2)$ as input. This
parameterization ambiguity can lead to larger uncertainties, because the
above logarithm can become singular, and the location of this Landau
singularity is shifted by the ambiguity. Since there is a one-to-one
correspondence between $s_{123}$ and $\theta_{\PWp\Pep}$ for fixed $s$
and $k_+^2$, this logarithmic singularity is washed out if the angular 
integration over $\theta_{\PWp\Pep}$ is performed, but appears as a kink 
structure in the angular distribution over this angle.
Fig.~\ref{vglparamtheta1} shows the non-factorizable corrections to
this angular distribution for the three parameterizations.
For $M_+ = 78\GeV$ we still have $\alpha|k_+^2-\MW^2|/\MW^2\sim 0.0004$. 
Apart from the regions where the Landau singularities appear, this is
indeed of the order of the differences between the three parameterizations.
When considering the Landau singularities, one should realize that the
parameterization ambiguity of the locations of the singularities is not
suppressed by a factor $\alpha$, \ie different parameterizations shift
the locations at the level of $|k_+^2-\MW^2|/\MW^2\sim 0.05$.
However, the impact of the corresponding kinks on observables is again
suppressed with $\alpha|k_+^2-\MW^2|/\MW^2$ if the angles are integrated over, 
since the singularities can only appear near the boundary of phase space 
and disappear from phase space exactly on resonance. 
Since the cross section is small where the Landau singularity appears,
the effect is phenomenologically irrelevant.
\subsection{Comparison between leptonic, semi-leptonic, 
and hadronic final state}

The non-factorizable corrections to the invariant-mass distributions 
are different for different final states and in general also for the
intermediate \PWp~ and \PWm~bosons.% 
\footnote{% 
In \citere{Be97a,Be97b} and in the preprint version of \citere{De97}
it has been argued that
the relative non-factorizable corrections to pure invariant-mass
distributions are identical for all final states in
$\Pep\Pem\to\PW\PW\to 4\,$fermions and vanish for Z-pair-mediated
four-fermion production. This was deduced from the assumption that
(up to charge factors)
the non-factorizable corrections become symmetric under the separate
interchanges $k_1\leftrightarrow k_2$ and $k_3\leftrightarrow k_4$
after integration over all decay angles. 
Although the function $\Delta$ for the relative correction has this
property,
this assumption is
not correct, because the differential lowest-order cross section is not
symmetric under these interchanges.}
The invariant-mass distributions to the intermediate $\PWpm$~bosons 
coincide only if the complete process is CP-symmetric. In this context,
CP symmetry does not distinguish between the different fermion generations,
since we work in double-pole approximation and neglect fermion masses;
in other words, the argument also applies to final states like 
$\nu_\Pe\Pep\mu^-\bar\nu_\mu$ and $\Pu\,\Pdbar\,\Ps\,\Pcbar$, which are not
CP-symmetric in the strict sense.
Thus, we end up with equal distributions for the $\PWpm$~bosons in the purely
leptonic and purely hadronic channels, respectively, but not in the 
semi-leptonic case. 

\bfi
\centerline{
\setlength{\unitlength}{1cm}
\begin{picture}(16,7.8)
\put(0,0){\includegraphics{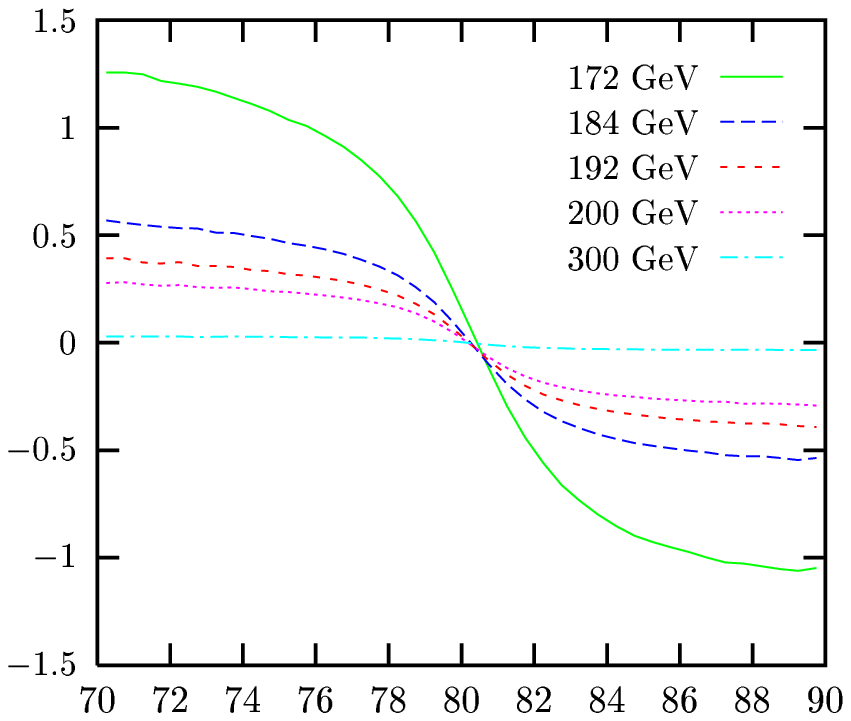}}
\put(0,0){\includegraphics{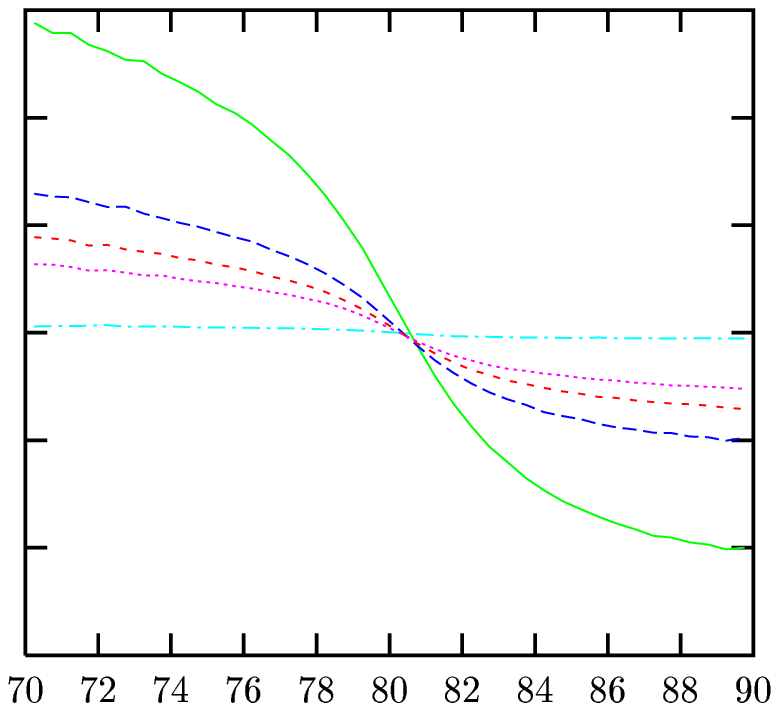}}
\put(0.0,5.2){\makebox(1,1)[c]{$\de_\nf/\%$}}
\put(4.7,-0.3){\makebox(1,1)[cc]{{$M_\pm/{\rm GeV}$}}}
\put(11.7,-0.3){\makebox(1,1)[cc]{{$M_\pm/{\rm GeV}$}}}
\put( 2.8,7.3){$\Pep\Pem\to\PW\PW\to\nu_\ell \ell^+\ell^{\prime-}\nu_{\ell'}$}
\put(10.0,7.3){$\Pep\Pem\to\PW\PW\to u\bar d d'\bar u'$}
\end{picture}
}
\vspace*{3mm}
\centerline{
\setlength{\unitlength}{1cm}
\begin{picture}(16,7.8)
\put(0,0){\includegraphics{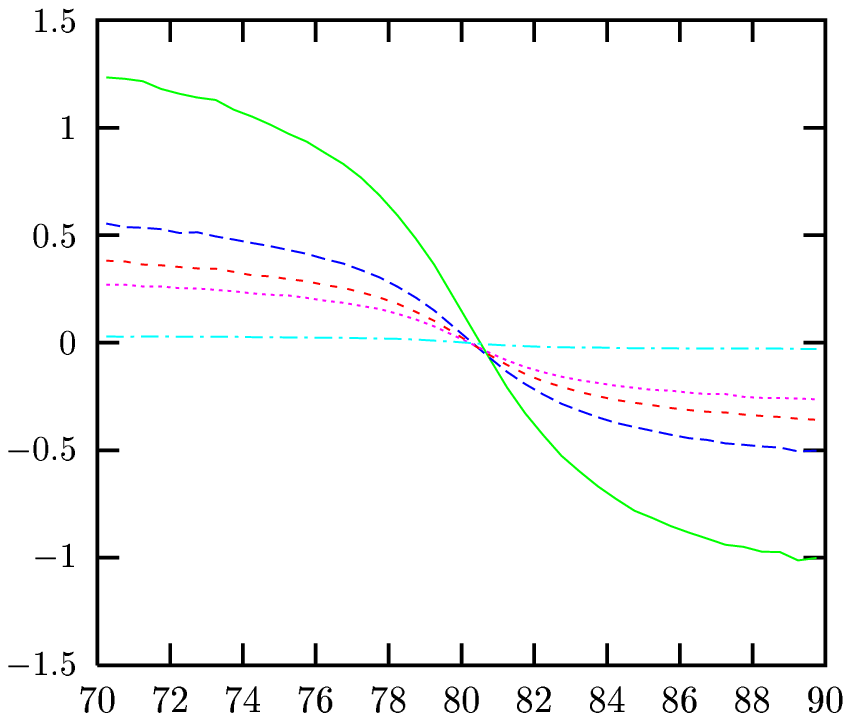}}
\put(0,0){\includegraphics{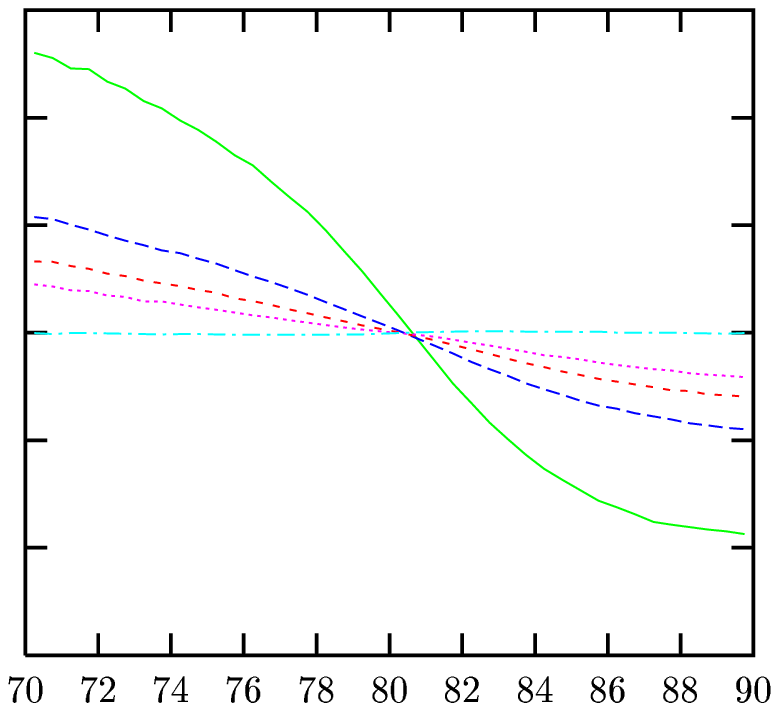}}
\put(0.0,5.2){\makebox(1,1)[c]{$\de_\nf/\%$}}
\put(4.7,-0.3){\makebox(1,1)[cc]{{$M_+/{\rm GeV}$}}}
\put(11.7,-0.3){\makebox(1,1)[cc]{{$M_-/{\rm GeV}$}}}
\put( 2.8,7.3){$\Pep\Pem\to\PW\PW\to\nu_\ell \ell^+ d\bar u$}
\put(10.0,7.3){$\Pep\Pem\to\PW\PW\to\nu_\ell \ell^+ d\bar u$}
\end{picture}
}
\caption[]{Relative non-factorizable corrections to the single-% 
invariant-mass distributions $\protect\rd\sigma/\protect\rd M_\pm$ 
for $\Pep\Pem\to\PW\PW\to 4\,$fermions with different final states
for various centre-of-mass energies.}
\label{Winvmass}
\efi
Fig.~\ref{Winvmass} shows the non-factorizable corrections to 
the single-invariant-mass distributions for leptonic,
hadronic, and semi-leptonic final states at various centre-of-mass
energies.
We observe the same qualitative features for all final states;
the corrections are positive below resonance and negative above. 
Quantitatively the differences between the corrections to the different
final states are small; we note that the slopes of the
corrections on resonance, which are responsible for the shift in the
maximum of the distribution, are maximal for the leptonic final state.
Therefore, we conclude that the W-boson mass determination by
invariant-mass reconstruction at LEP2 is not significantly influenced by
non-factorizable corrections.

The authors of \citere{Be97a,Be97b} have also calculated \cite{bbcprivcom} the 
non-factorizable
corrections to the single-invariant-mass distributions shown in
\reffi{Winvmass} for $\sqrt{s}=172\GeV$ and $184\GeV$. They find good
agreement with our results for positive invariant masses. However,
their corrections are antisymmetric and therefore differ from our
results for negative invariant masses. The differences are of the
order of non-doubly-resonant corrections and due to
different parameterizations of the corrections. 

In the previous sections, we investigated the influence of the 
non-factorizing corrections on various angular and energy distributions 
with fixed invariant masses for the final-state fermion pairs.
We have repeated
this analysis for hadronic and semi-leptonic final states and found
corrections of the same order of magnitude, viz.\ of typically 1\% 
at LEP2 energies.

\subsection{Numerical results for Z-pair production}

For the production channels via a resonant Z-boson pair,
$\Pep\Pem\to\PZ\PZ\to 4\,$fermions, we have $f_1=f_2$ and $f_3=f_4$.
Owing to Bose symmetry the lowest-order cross section $\rd\si_\born$ 
is invariant under the set of interchanges
$(k_1,k_2)\leftrightarrow(k_3,k_4)$. 
This symmetry, which is respected by the non-factorizable corrections,
implies that the single-invariant-mass distributions to each of the 
final-state fermion pairs of the two Z-boson decays are equal.
CP invariance leads to the additional symmetry with respect to
$(p_+,k_1,k_2)\leftrightarrow(p_-,k_4,k_3)$; after integration over the
Z-pair production angle this substitution reduces to
$(k_1,k_2)\leftrightarrow(k_4,k_3)$.
In view of non-factorizable corrections it is also interesting to
inspect the behaviour of $\rd\si_\born$ under the replacements 
$k_1\leftrightarrow k_2$ and $k_3\leftrightarrow k_4$ separately, since
terms in $\rd\si_\born$ that are symmetric in at least one of these
substitutions do not contribute to $\rd\si_\nf$ 
if all decay angles are integrated over.
This is a direct consequence of the antisymmetry of $\de_\nf$ in  
each of the substitutions 
$k_1\leftrightarrow k_2$ and $k_3\leftrightarrow k_4$, which follows
from \refeq{nfcorrfac} and $Q_1=Q_2$, $Q_3=Q_4$.
 
In order to study the behaviour of $\rd\si_\born$ under the
replacements $k_1\leftrightarrow k_2$ and $k_3\leftrightarrow k_4$, it
is convenient to consider the helicity amplitudes for the two signal
diagrams for $\Pep\Pem\to\PZ\PZ\to 4\,$fermions, which contain two
resonant Z-boson propagators. These amplitudes are proportional to the
right- and left-handed couplings $g_i^\pm=v_i\mp a_i$ of each fermion
$f_i=f_1,f_3$ to the Z~boson.  As can be seen from the explicit form
of the amplitudes, the substitution $k_1\leftrightarrow k_2$
transforms the helicity amplitudes to those with reversed helicities
of the fermions $f_1$ and $\bar f_2=\bar f_1$ apart from changing the
couplings $g_1^\pm$ into $g_1^\mp$.  Therefore, the differential
lowest-order cross section, i.e.\ the squared helicity amplitudes
summed over all final-state polarizations, can be split into two
parts: one is symmetric in $k_1\leftrightarrow k_2$ and proportional
to $[(g_1^+)^2+(g_1^-)^2]/2= v_1^2+a_1^2$, the other is anti-symmetric
and proportional to $[(g_1^-)^2-(g_1^+)^2]/2=2v_1 a_1$.  The analogous
reasoning applies to the substitution $k_3\leftrightarrow k_4$. After
performing the angular integrations, we finally find that the
lowest-order cross section is proportional to
$(v_1^2+a_1^2)(v_3^2+a_3^2)$, and the non-factorizable correction
proportional to $4Q_1 v_1 a_1 Q_3 v_3 a_3$, where the charge factors
$Q_i$ stem from the correction factor $\de_\nf$. Comparing pure
invariant-mass distributions for different final states, the ratios of
the non-factorizable corrections should be of the same order of
magnitude as the ratios
of the corresponding coupling factors,
\beq
F = \left|\frac{4Q_1 v_1 a_1 Q_3 v_3 a_3}{(v_1^2+a_1^2)(v_3^2+a_3^2)}\right|.
\label{eq:F}
\eeq
The factors $F$ take the following values:
\beq
\arraycolsep 8pt
\begin{array}{|c||c|c|c|c|c|c|}
\hline
f_1 f_3 & \ell\ell & \ell u & \ell d & uu   & ud   & dd 
\\ \hline
F &       0.04     & 0.09   & 0.06   & 0.21 & 0.14 & 0.10
\\ \hline
\end{array}
\label{eq:Ftab}
\eeq
where $\ell$, $u$, $d$ generically refer to leptons, up-type quarks and
down-type quarks, respectively. The reason for the
smallness of the factors $F$ is different for leptons and quarks: for
leptons the suppression is due to the small coupling $v_i$ 
to the vector current, for quarks the factor $F$ is reduced by the 
relative charges $Q_i$. 

\bfi
\centerline{
\setlength{\unitlength}{1cm}
\begin{picture}(9,7.5)
\put(0,0){\includegraphics{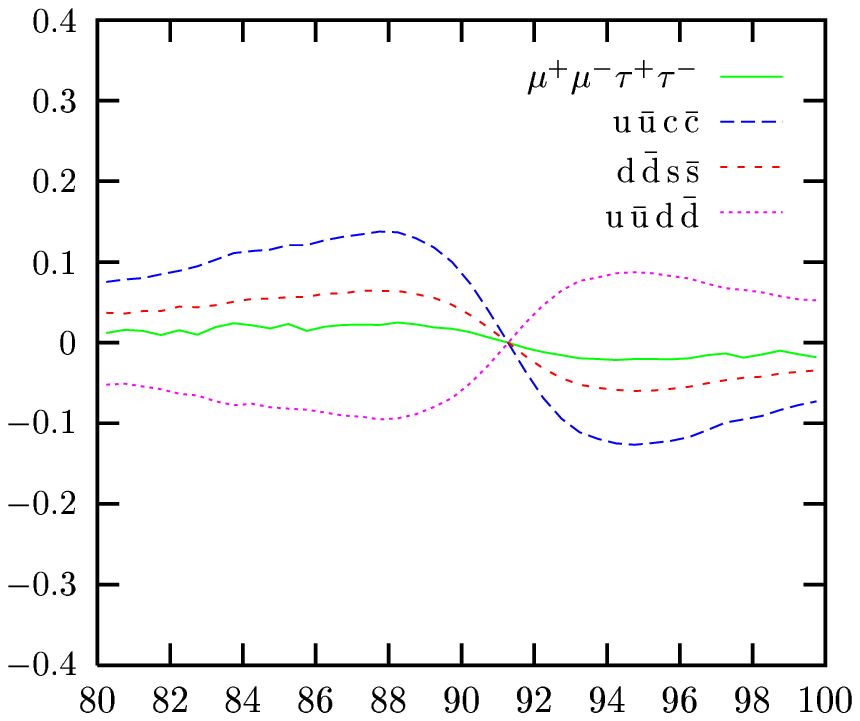}}
\put(0.0,5.2){\makebox(1,1)[c]{$\de_\nf/\%$}}
\put(4.7,-0.3){\makebox(1,1)[cc]{{$M_{1,2}/{\rm GeV}$}}}
\end{picture}
}
\caption[]{Relative non-factorizable corrections to the single-% 
invariant-mass distributions $\protect\rd\sigma/\protect\rd M_{1,2}$ 
for $\Pep\Pem\to\PZ\PZ\to 4\,$fermions with different final states
for $\sqrt{s}=192\GeV$.}
\label{Zinvmass}
\efi
Fig.~\ref{Zinvmass} shows the non-factorizable corrections to the
single-invariant-mass distributions $\rd\sigma/\rd M_{1,2}$, where
$M_{1,2}$ denote the invariant masses of the first and second
fermion--anti-fermion pairs, respectively. The ratios of the different curves 
are indeed of the order of magnitude of the ratios of the factors $F$ given in
\refeq{eq:Ftab}. 
For equal signs of $Q_1$ and $Q_3$ the shape of the corrections is
similar to the shape of the corrections to $\Pep\Pem\to\PW\PW\to
4\,$fermions, for opposite signs of $Q_1$ and $Q_3$ the 
shape is reversed.
The corrections by themselves are very small and
phenomenologically unimportant. The smallness of these corrections can
be qualitatively understood by comparing the factors $F$ of
$\refeq{eq:F}$ for the ratios of the couplings with the corresponding one
for the W-pair-mediated processes. For $\Pep\Pem\to\PW\PW\to 4\,$leptons
we simply have $F=1$, because in the LEP2 energy range the purely
left-handed $t$-channel diagram dominates the cross section, and no
systematic compensations are induced by symmetries. Therefore, the
factors in $\refeq{eq:Ftab}$ should directly give an estimate for the
suppression of $\de_\nf$ for $\Pep\Pem\to\PZ\PZ\to 4\,$fermions with
respect to four-lepton production via a W-boson pair. Comparing the
corrections for energies with the same distance from the respective
on-shell pair-production thresholds, i.e.\ the curve for
$\sqrt{s}=184\GeV$ in the W-boson case (\reffi{Winvmass}) with the
curves for $\sqrt{s}=192\GeV$ in the Z-boson case (\reffi{Zinvmass}), we
find reasonable agreement with our expectation. 
The authors of \citere{Be97a,Be97b} have reproduced the corrections shown in
\reffi{Zinvmass} with good agreement \cite{bbcprivcom}.

\bfi
\centerline{
\setlength{\unitlength}{1cm}
\begin{picture}(16,7.5)
\put(0,0){\includegraphics{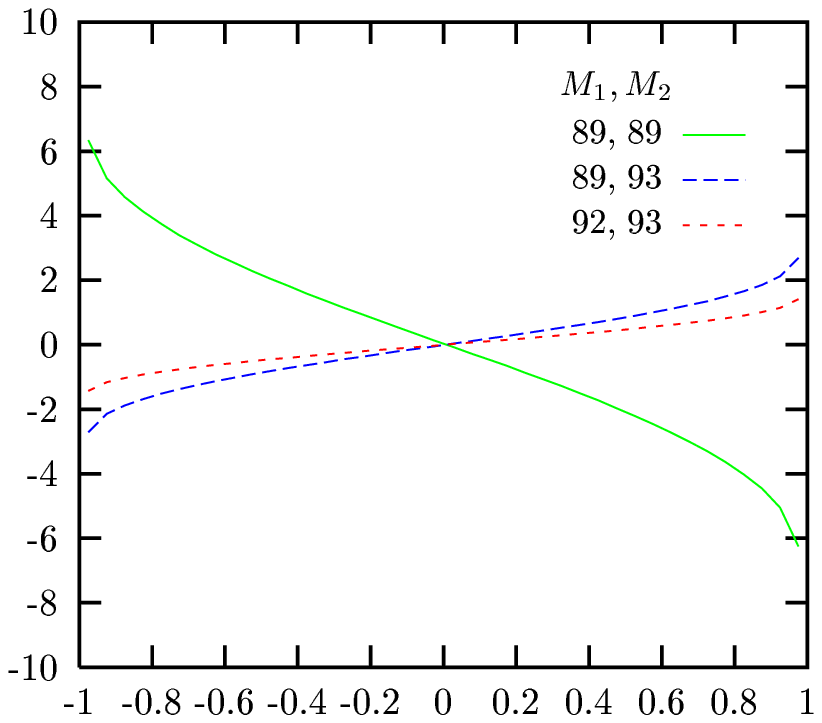}}
\put(0,0){\includegraphics{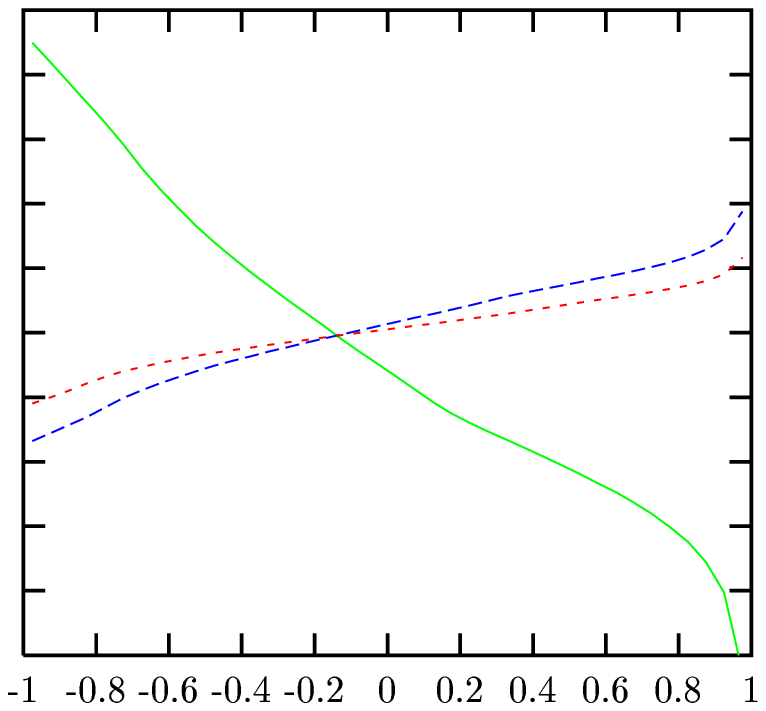}}
\put(0.0,5.2){\makebox(1,1)[c]{$\de_\nf/\%$}}
\put(4.7,-0.3){\makebox(1,1)[cc]{{$\cos\phi$}}}
\put(11.7,-0.3){\makebox(1,1)[cc]{{$\cos\theta_{\mu^+\tau^-}$}}}
\end{picture}
}
\caption[]{Relative non-factorizable corrections to the angular distributions 
$\protect\rd\sigma/\protect\rd M_1\protect\rd M_2\protect\rd\cos\phi$ and
$\protect\rd\sigma/\protect\rd M_1\protect\rd M_2\protect
\rd\cos\theta_{\mu^+\tau^-}$ 
in $\Pep\Pem\to\PZ\PZ\to\mu^-\mu^+\tau^-\tau^+$ 
for fixed values of the invariant masses $M_{1,2}$ and $\sqrt{s}=192\GeV$.}
\label{Zangles}
\efi
Finally, we inspect the impact of non-factorizable corrections to some
angular distributions in Z-pair-mediated four-fermion production for fixed 
values of the invariant masses $M_{1,2}$.
Since the presence of the suppression factor $F$ relies
on the assumption that the
phase-space integration is symmetric under 
$k_1\leftrightarrow k_2$ and $k_3\leftrightarrow k_4$, this suppression
in general does not apply to angular distributions. However, 
partial suppressions occur, e.g., if the integration is still symmetric 
under one of these substitutions and, in particular, for quarks in the 
final state because of their smaller charges.
Two examples for angular distributions without any suppression 
are illustrated in \reffi{Zangles} for the purely leptonic final state 
$\mu^-\mu^+\tau^-\tau^+$. The angle $\phi$ is defined by the two planes 
spanned by the momenta of the two fermion pairs in which the Z~bosons decay,
\renewcommand{\bk}{{\bf k}}
\beq
\cos\phi = \frac{(\bk_1\times \bk_2)(\bk_3\times \bk_4)}
{|\bk_1\times \bk_2||\bk_3\times \bk_4|},
\label{eq:phi}
\eeq
and $\theta_{\mu^+\tau^-}$ denotes the angle between the momenta of the
$\mu^+$ and the $\tau^-$, respectively. The shapes of the curves in
\reffi{Zangles}, specifically the curves for the distribution in
$\cos\phi$, nicely reflect the approximate anti-symmetric behaviour in the
angular dependence, which leads to the suppression in the invariant-mass
distributions. The size of the corrections turns out to be at the level 
of a few per cent, i.e.\ they are not necessarily negligible in precision 
predictions. Note, however, that the cross section for \PZ-pair
production is only one tenth of the \PW-pair production cross section.

\chapter{\boldmath Radiative corrections to 
$\mathrm e^+ \mathrm e^- \to \mathrm W^+ \mathrm W^- \to 4 f$}
\label{ch:radcorr}

As discussed in the introduction, the cross section for 
$\Pep \Pem \to \PW^+ \PW^- \to 4 f$ should be known with an accuracy of $1\%$ 
or better in order to cope with the experimental precision at LEP2.
This requires the calculation of the $\Oa$ corrections to the processes
$\Pep \Pem \to \PW^+ \PW^- \to 4 f$.

The leading radiative corrections, like the running of the electromagnetic 
coupling constant, universal corrections associated with the $\rho$ parameter, 
the Coulomb singularity, and the resummed leading-logarithmic corrections 
from initial-state radiation (ISR), are already implemented in several event 
generators (see \citere{CERN9601mcgen} and references therein).
The neglected non-leading corrections can be estimated from the on-shell 
\PW-pair production \cite{Be96,Bo92}, where the full one-loop calculation
differs from the one including only these leading effects by about  
$1$-$2\%$ at LEP2 energies and $10$-$20\%$ in the $\TeV$ range.
Hence, the leading radiative corrections are in general not 
sufficient to match the experimental accuracy of LEP2.

The calculation of the full $\Oa$~corrections is extremely complicated 
since the number of diagrams is of the order of $10^3$-$10^4$ including
one-loop integrals with up to six propagators \cite{Vi98}.
The numerical computation of these corrections is highly non-trivial because 
the tensor reduction to scalar integrals and the calculation of the one-loop 
integrals with five and more propagators cause numerical instabilities. 
Furthermore, the numerical integration via Monte Carlo techniques
is rather slow for such complex calculations.

Approximations for radiative corrections beyond the leading level have been
calculated by two groups.
A first calculation of the complete doubly-resonant $\Oa$~radiative 
corrections to the four-fermion processes 
$\mathrm e^+ \mathrm e^- \to \mathrm W^+ \mathrm W^- \to 4 f$ 
has been discussed in \citere{Be99}. 
There a semi-analytic approach has been used with different matrix elements 
for different phase-space regions and for different observables. 
Moreover, the four-fermion phase space has been factorized into the phase 
space of the on-shell $\PW$-pair production and of the on-shell $\PW$ decays 
and the integrations of the two invariant masses of the intermediate 
$\PW$ bosons.
For the numerical discussion, only results for leptonic processes have been 
included.
The authors of \citere{Be99} have found a large shift of the Breit--Wigner 
line shape due to logarithms of the form $\ln (m_f^2/s)$ resulting from 
final-state radiation (FSR). 
These logarithms are due to the absence of collinear photons 
in the definition of the invariant mass of the $\PW$ boson.
In realistic observables, collinear photons cannot be resolved from 
charged fermions, except for muons, and have to be included in the 
reconstructed invariant mass of the corresponding $\PW$ boson.

In \citere{Ja98}, a four-fermion event generator has been presented, 
including all $\Oa$ radiative corrections to the on-shell 
$\PW$-pair production, $\Pep \Pem \to \PW^+ \PW^-$, 
with exponentiation of the universal corrections from photon radiation 
off initial-state $\Pe^\pm$ and off the $\PW^\pm$ bosons.
Leading-logarithmic corrections from FSR have been
included in this Monte Carlo generator in \citere{Ja99}.
However, non-leading $\Oa$ corrections to the $\PW$ decays 
and the non-factorizable corrections are missing.
The results of \citere{Be99} have been qualitatively 
confirmed by the calculations of \citere{Ja99}.

Hence, a Monte Carlo generator including the complete 
doubly-resonant $\Oa$ corrections 
is needed in order to match the accuracy of LEP2 
and to take into account realistic experimental situations.

\section{Strategy of the calculation}

In this chapter, we consider the virtual corrections to the processes
\beq
\Pep(p_+) + \Pem(p_-) \;\to\; \PWp(k_+) + \PWm(k_-) \;\to\; 
f_1(k_1) + \bar f_2(k_2) + f_3(k_3) + \bar f_4(k_4)
\eeq
and the complete bremsstrahlung processes
\beq
\Pep(p_+) + \Pem(p_-) \;\to\; 
f_1(k_1) + \bar f_2(k_2) + f_3(k_3) + \bar f_4(k_4) + \ga(q),
\eeq
where the relative charges of the 
fermions $f_i$ are represented by $Q_i$ with $i=1,\ldots,4$.
The masses of the external fermions are neglected,
except where this would lead to mass singularities.
For the virtual corrections, the momenta of the intermediate 
\PW~bosons read
\beq
k_+ = k_1+k_2, \qquad k_- = k_3+k_4.
\eeq
Furthermore, the square of the complex $\PW$-boson mass is defined by
$M^2 = \MW^2-\ri\MW\Gamma_\PW$, and the center-of-mass (CM) energy 
is $\sqrt{s}$.

\subsection{Doubly-resonant virtual corrections}

The diagrams of four-fermion production can be 
classified into the doubly-resonant, singly-resonant, and non-resonant 
diagrams according to the number of resonant $\PW$-boson propagators.
Hence, the amplitude of the virtual corrections can be written in the
following way after implementation of the finite $\PW$-boson width:
\begin{eqnarray}
\label{eq:matrixelementee4f}
\M_{\virt}&=&
\underbrace{\frac{R_{+-}(k_+^2,k_-^2,\theta)}{(k_+^2-M^2)(k_-^2-M^2)}}_{
\mbox{doubly-resonant}}
+\underbrace{\frac{R_+(k_+^2,k_-^2,\theta)}{k_+^2-M^2}
+\frac{R_-(k_+^2,k_-^2,\theta)}{k_-^2-M^2}}_{\mbox{single-resonant}}
+\underbrace{N(k_+^2,k_-^2,\theta)}_{\mbox{non-resonant}},\qquad
\end{eqnarray}
where the variable $\theta$ symbolizes all phase-space variables, 
except for the invariant masses $k_\pm^2$ of the $\PW$ bosons.

Since the matrix elements of non-doubly-resonant diagrams are 
suppressed by a factor $\al \GW \ln(\dots)/(\pi \MW)\approx 0.1 \%$, 
a reasonable approach is to include only the doubly-resonant 
diagrams into the calculation.
However, the naive inclusion of the doubly-resonant diagrams
yields gauge-dependent results and the reliability of this
approximation is unclear.

In order to separate the doubly-resonant corrections in 
a gauge-invariant way, the pole scheme has been proposed in 
\citere{polescheme,Ae94},
where the matrix element is expanded about the complex $\PW$-boson masses.
Since the complete matrix element is gauge-invariant, the single terms 
of the pole expansion are also gauge-invariant.
If only the leading term is kept, the expansion is known as 
the double-pole approximation (DPA):
\begin{eqnarray}
\label{eq:DPAvirt}
\M_{\virt}^{\mathrm{DPA}}&=&
\frac{R_{+-}(M^2,M^2,\theta)}{(k_+^2-M^2)(k_-^2-M^2)}.
\end{eqnarray}

Note that the residue
is taken at the complex pole resulting in complex kinematical invariants.  
In order to avoid the calculation of one-loop integrals for 
complex invariants, the $\PW$ width is neglected in the numerator of 
\refeq{eq:DPAvirt}, where the neglected terms are suppressed 
by a factor $\GW/\MW$ and thus negligible in DPA.

For the application of the DPA on the virtual corrections,
a set of eight independent phase-space variables 
including $k_\pm^2$ has to be chosen, which
determine the momenta of the final-state fermions uniquely.
For several choices of the parameterization of the phase space,
the DPA differs only in non-doubly-resonant contributions.
Note that in general events can be outside of the  
physical phase-space boundaries for on-shell $\PW$ bosons
(see \refse{se:ambig}).
A relatively simple choice that generates only events within 
the physical phase-space domain is given in \refapp{ap:onshell}.

The DPA is not valid near threshold since the phase-space region 
where both $\PW$ bosons are resonant is suppressed by the kinematical factor 
$\lambda^{\frac{1}{2}}(s,k_+^2,k_-^2)$ (see \refeq{eq:decay}).
Therefore, the singly-resonant corrections are as important as 
the doubly-resonant corrections at threshold.
On the other hand, for some processes the non-doubly-resonant 
contributions can be enhanced by nearly on-shell virtual photons.
This enhancement can be avoided by introducing appropriate phase-space 
cuts in the calculation.

In DPA the virtual corrections can be classified into factorizable 
and non-factorizable ones \cite{Be96,Ae94,wwrev}.
The square of the matrix element of the virtual corrections reads in DPA
\begin{eqnarray}
|{\cal M}_{\mathrm {virt}}^{\mathrm{DPA}}|^2&=&|{\cal M}_{\mathrm f}|^2
+|{\cal M}_{\mathrm {Born}}|^2 \delta_{\mathrm {nf}}^{\mathrm {virt}}.
\end{eqnarray}
where ${\cal M}_{\mathrm f}$, ${\cal M}_{\mathrm {Born}}$, and  
$\delta_{\mathrm {nf}}^{\mathrm {virt}}$ are the matrix elements 
of the factorizable corrections, 
the Born matrix element defined in \refeq{mborn}, and the 
non-factorizable correction factor, respectively.

\subsubsection{Factorizable virtual corrections}

\begin{figure}
\centerline{
\begin{picture}(200,105)(0,0)
\ArrowLine(30,50)( 5, 95)
\ArrowLine( 5, 5)(30, 50)
\Photon(30,50)(150,80){2}{11}
\Photon(30,50)(150,20){2}{11}
\ArrowLine(150,80)(190, 95)
\ArrowLine(190,65)(150,80)
\ArrowLine(190, 5)(150,20)
\ArrowLine(150,20)(190,35)
\GCirc(30,50){10}{0}
\GCirc(90,65){10}{1}
\GCirc(90,35){10}{1}
\GCirc(150,80){10}{0}
\GCirc(150,20){10}{0}
\DashLine( 70,0)( 70,100){2}
\DashLine(110,0)(110,100){2}
\put(50,26){$\PW$}
\put(50,68){$\PW$}
\put(115,13){$\PW$}
\put(115,82){$\PW$}
\put(-12, 0){$\Pem$}
\put(-12,95){$\Pep$}
\put(195, 1){$\bar f_4$}
\put(195,34){$f_3$}
\put(195,60){$\bar f_2$}
\put(195,95){$f_1$}
\put(-25,-15){\footnotesize On-shell production}
\put(120,-15){\footnotesize On-shell decays}
\end{picture}
} 
\vspace*{1em}
\caption{Factorizable corrections to $\eeffff$}
\label{fi:factorizable}
\efi

The factorizable virtual corrections are defined
by the product of the on-shell matrix elements of the
W-pair production and the \PW~decays and the (transverse parts of the) 
\PW~propagators (see \reffi{fi:factorizable}): 
\beqar
\label{eq:mvirt}
{\cal M}_{\mathrm{f}} &=& 
\sum_{\la_+,\la_-} \frac{1}{(k_+^2-M^2)(k_-^2-M^2)}
\Big( \de\M^\eeWW \M^\Wpff_\born \M^\Wmff_\born
\\ && {}
+ \M^\eeWW_\born \de\M^\Wpff \M^\Wmff_\born
+ \M^\eeWW_\born \M^\Wpff_\born \de\M^\Wmff \Big),
\nn
\eeqar
where $\de\M^\eeWW$, $\de\M^\Wpff$, and $\de\M^\Wmff$ denote the
one-loop amplitudes of the respective subprocesses.
The sum runs over the physical polarizations $\la_\pm$ of the \PWpm~bosons. 

Since the on-shell $\PW$-pair production and the on-shell $\PW$-decays 
are gauge-invariant processes, the factorizable corrections are also
gauge-invariant.
The results of the on-shell $\PW$-pair production and the 
on-shell $\PW$-boson decay are explicitly given in \citere{rcwprod}
and \citere{rcwdecay}, respectively, and can be implemented in the 
Monte Carlo program,
where we have to take care of the spin correlations of the 
$\PW$ bosons in the production and decay subprocesses. 
The virtual corrections are build up by scalar form factors,
which include all one-loop integrals, and standard matrix elements,
which depend on the polarization vectors and spinors of the 
external particles.
In DPA the form factors depend exclusively on the $\PW$-production angle
and can be evaluated very fast by an expansion in Legendre polynomials
\cite{De99}.

\subsubsection{Non-factorizable virtual corrections}

The non-factorizable virtual corrections are defined as the difference 
between the virtual corrections to the complete four-fermion process in DPA
and the factorizable virtual corrections.
A representative set of diagrams contributing to the 
non-factorizable corrections are shown in
\reffis{virtual_final_final_diagrams} and 
\ref{virtual_initial_final_diagrams}.
The matrix element of the non-factorizable virtual corrections 
$\M_{\mathrm{nf}}^{\mathrm {virt}}$ factorize to the Born matrix element in DPA
(see \refch{ch:nfc} for more details): 
\begin{equation}
\M_{\mathrm{nf}}^{\mathrm {virt}}=
\M_{\born} \delta_{\mathrm {nf}}^{\mathrm {virt}},
\end{equation}
where the correction factor $\delta_{\mathrm {nf}}^{\mathrm {virt}}$ 
is relatively simple and explicitly given in \refse{se:vnfc}.
Since the factorizable and the complete virtual corrections are 
gauge-invariant, the non-factorizable corrections are also gauge-invariant.

Note that the non-factorizable corrections involve logarithms 
of the form $\ln(k_\pm^2-M^2)$, which become singular in the limit 
$k_\pm^2\to M^2$.
Therefore, these corrections are calculated for off-shell phase space, 
\ie $k_\pm^2\ne M^2$, while the off-shellness 
$(k_\pm^2-M^2)$ is neglected whenever possible.

\subsection[On the definition of the reconstructed $\mathrm W$-boson mass]
{\boldmath On the definition of the reconstructed $\mathrm W$-boson mass}
\label{se:wmass}

In \refch{ch:nfc}, the non-factorizable virtual and real corrections 
have been calculated in DPA, where the 
integration over the photon momentum of the real corrections is 
performed analytically.
As for the virtual corrections, the non-factorizable real corrections 
depend on the choice of the independent phase-space parameters in DPA.
These parameters are fixed while the integration over the 
photon momentum is performed. 
To apply the DPA on the non-factorizing real corrections, 
two of these parameters are identified with 
the invariant masses of the $\PW$ bosons.
Owing to the presence of the bremsstrahlung photon the definition
of these invariant masses is not unique.
For different definitions 
the result for the real non-factorizable corrections differs by
doubly-resonant contributions, as discussed in detail in \refse{se:rnfpc}.  
The reason is that the resonant $\PW$-boson 
propagator $1/[(k_++q)^2-M^2]$ is constant, if the invariant mass 
is defined by $(k_++q)^2$, but depends on the photon momentum 
$q$ for the invariant mass $k_+^2$.
Therefore, the calculable observables 
are restricted to the actual parameterization of the phase space
in a semi-analytic calculation.
For instance, only invariant-mass distributions, where the $\PW$-boson
masses are defined by $k_\pm^2$, can be calculated with the results
of \refch{ch:nfc}.

A DPA including all $\Oa$ corrections for four-fermion production 
has been worked out in \citere{Be99}.
As in \refch{ch:nfc}, a semi-analytic approach has been used in 
\citere{Be99}, where, for the invariant mass distributions, 
the $\PW$-boson masses have been defined by 
invariant masses of the fermion--antifermion pairs, \ie $M_\pm^2=k^2_\pm$,
resulting in large corrections from collinear radiation of 
bremsstrahlung photons off final-state fermions. 
These originate from logarithms of the form $\ln(m_f^2/s)$
and yielding large distortions of the peak position of the 
Breit--Wigner line shape at the CM energy $184 \GeV$
of $-77 \MeV$, $-38 \MeV$, and $-20 \MeV$ for $\Pep \nu_\Pe$,
$\mu^+ \nu_\mu$, and $\tau^+ \nu_\tau$, respectively.
In practice, the invariant masses of $\PW$ bosons have to be reconstructed 
in a more realistic way.

In realistic experimental situations, the final-state quarks are 
observed as jets.
Photons radiated collinear to these jets cannot be 
resolved, and the photon momentum is included in the jet momentum.
On the other hand, the momentum of the neutrino can only be calculated
from the missing momentum of the final-state particles.
In this way the neutrino momentum includes also the momenta 
due to emission of photons collinear to the beam \cite{Ba98}.
Note that the reconstruction of the neutrino momentum is possible 
for semi-leptonic final states, but not for purely leptonic final states, 
where two neutrinos are involved. 
Furthermore, it is very difficult to separate collinear photons 
from electrons.
However, photons can be resolved from muons even when they are collinear.
Therefore, only a few observables, like the invariant-mass distribution 
$\rd \si/\rd M_-$ with $M_-^2=k_-^2$ of the process 
$\Pep \Pem \to \Pu \, \Pdbar \, \mu^- \bar\nu_\mu$, 
are directly sensitive to the distortion due to large corrections from FSR.
In any case, a Monte Carlo generator is required in order to take into 
account realistic experimental situations.

\subsection{Overlapping resonances in the bremsstrahlung process}
\label{se:overlappingresonances}

\begin{figure}
{
\begin{center}
\begin{picture}(190,105)(0,0)
\ArrowLine(27,55)(3, 95)
\ArrowLine( 3, 5)(30, 50)
\Photon(30,50)(90,80){-2}{6}
\Photon(30,50)(90,20){2}{6}
\GCirc(30,50){10}{0}
\Vertex(90,80){2.0}
\Vertex(90,20){2.0}
\ArrowLine(90,80)(120,95)
\ArrowLine(120,65)(90,80)
\ArrowLine(120, 5)( 90,20)
\ArrowLine( 90,20)(120,35)
\Vertex(60,65){2}
\Photon(60,65)(120,50){2}{6}
\put(35,65){$\PW $}
\put(70,80){$\PW $}
\put(55,16){$\PW $}
\put(125,47.5){$\ga$}
\put(125,92){$f_1$}
\put(125,62){$\bar{f}_2$}
\put(125,30){$f_3$}
\put(125,0){$\bar{f}_4$}
\put(-15,92){$\Pep$}
\put(-15,0){$\Pem$}
\end{picture}
\end{center}
}
\caption[]{Diagram with overlapping resonances}
\label{fi:overlres}
\end{figure}

The definition of a suitable approximation 
for the real corrections is made difficult by 
overlapping resonances of the $\PW$-bosons. 
In the bremsstrahlung process, the $\PW$-boson propagators differ 
depending on whether the bremsstrahlung photon is emitted from
initial-state particles or from the decay products of the $\PW$ boson.
Both types of propagators are present in \reffi{fi:overlres}
where the poles of the propagators are located at 
$k^2_-=M^2$, $k^2_+=M^2$, and $(k_++q)^2=M^2$. 
Therefore, the expansion of the cross section about 
$k_+^2=M^2$ and $k_-^2=M^2$, like in the case of the virtual corrections,
is not suitable. 

The matrix element of the bremsstrahlung process can be written in the 
following way (see \refse{se:classificationofrealdiagrams}):
\begin{eqnarray}
\label{eq:decompreal}
\nn
{\cal M}_{\real}&=&
\frac{{\cal R}(k_+,k_-,q)}{(k^2_+-M^2)(k^2_--M^2)}+
\frac{{\cal R}_+\left(k_+,k_-,q\right)}
{[(k_++q)^2-M^2](k^2_--M^2)}\\
&&+\frac{{\cal R}_-\left(k_+,k_-,q\right)}
{(k^2_+-M^2)[(k_-+q)^2-M^2]}+{\cal N}(k_+,k_-,q),
\end{eqnarray}
where ${\cal N}$ includes the matrix elements of all 
non-doubly-resonant diagrams.
The factorizable corrections of the cross section are the squares of 
the single terms corresponding to ${\cal R}$, ${\cal R}_+$, and 
${\cal R}_-$, while the non-factorizable corrections are the 
interferences between these terms.

In the hard photon region, $E_\ga \gg \GW$, the two
resonances of the $\PW^+$ bosons located at $k_+^2=M^2$ and 
$(k_++q)^2=M^2$ are well separated in phase space.
Only the factorizable corrections contribute in this region in DPA,
and the non-factorizable corrections vanish in DPA.
For soft photons, $E_\ga \ll \GW$,
the resonances almost coincide, and the photon momentum can be neglected
in the resonant propagators.
However, for semi-soft photons with an energy  $E_\ga=\O(\GW)$, 
the resonances overlap.
Thus the definition of an appropriate approximation is rather 
complicated and, hence, the reliability becomes unclear.

\subsection{Inclusion of the real corrections}

In order to avoid the problem of overlapping resonances, 
the bremsstrahlung process is taken into account exactly.

The bremsstrahlung processes are already studied 
in \refch{ch:treelevel} including all diagrams.
For the numerical discussion, the processes 
$\Pep \Pem \to \nu_\mu \mu^+ \tau^- \bar{\nu}_\tau$,
$\Pep \Pem \to \Pu \, \Pdbar \mu^- \bar{\nu}_\mu$, and
$\Pep \Pem \to \Pu \, \Pdbar \, \Ps \, \Pcbar $
are considered.
These are the processes with the smallest number
of diagrams within the leptonic, semi-leptonic, and hadronic process classes.
Note that the radiative corrections for any
leptonic process are equivalent to the radiative corrections
for $\Pep \Pem \to \nu_\mu \mu^+ \tau^- \bar{\nu}_\tau$ in DPA,
in the absence of logarithms $\ln(m_f^2/s)$ from FSR.
The same is valid for $\Pep \Pem \to \Pu \, \Pdbar \, \mu^- \bar{\nu}_\mu$ and
$\Pep \Pem \to \Pu \, \Pdbar \, \Ps \, \Pcbar$
concerning semi-leptonic and hadronic processes, respectively.

\section{Subtraction method}
\label{se:subtraction}

When calculating radiative corrections for four-fermion production, 
one has to take care of the singularities of
the real and virtual radiative corrections. 
According to the Bloch--Nordsieck theorem \cite{Bl37} 
the infrared singularities cancel between the virtual and real part. 
Furthermore, collinear singularities, \ie mass singularities, 
appear in the radiative corrections and 
are regularized by the fermion masses which are neglected otherwise.
These singularities originate from collinear photon emission from light 
external fermions and show up as large logarithms $\ln(m_f^2/s)$.
For inclusive enough observables, these collinear singularities 
cancel between the real and virtual radiative corrections owing to the 
Kinoshita--Lee--Nauenberg theorem \cite{Ki62}, except for 
the mass singularities remaining from initial-state radiation (ISR) or
resulting from renormalization.
Two methods are exist for the treatment of the 
singularities in a Monte Carlo generator: 
{\it phase-space slicing} and the {\it subtraction method}. 

Phase-space slicing splits the integration domain by a small separation cut 
into a part which includes all singularities and a finite part. 
For infrared singularities the splitting is usually done by 
a cut on the photon energy, for 
collinear singularities by a separation angle between the photon
and the particle which emits the photon.
The integral of the singular part over the photon momentum is analytically 
performed with an appropriate regularization, and 
terms of order of the cut parameter are neglected.
In this way the singularities are extracted and can be added 
to the virtual part of the radiative corrections.
Both the singular and the finite part of the radiative cross section 
depend logarithmically on the cut parameter.
This dependence must cancel in the complete result. 
The logarithms of the separation cut are determined in the singular part 
by analytical integration.
On the other hand, they must be compensated by numerical calculation of 
the corresponding logarithms included in the finite part of the cross section.
Therefore, the cut has to be small enough that terms of order of
the cut parameter can be neglected in the singular part, but large enough 
to obtain numerically stable results.

In contrast to phase-space slicing, the subtraction method 
requires no cut parameter.
We use the subtraction method 
since we expect a better convergence behaviour. 
This method has been mostly applied for NLO predictions in 
massless QCD \cite{Ca96,subtraction}%
\footnote{The subtraction method for massive particles can be found 
in \citeres{Di99,Ma92}.}
where the singularities are usually regularized dimensionally.
In Electroweak Standard Model processes it is more convenient to 
introduce an infinitesimal photon mass as a regularization parameter 
for infrared singularities and to use small fermion masses for
the regularization of the collinear singularities.
In order to apply the subtraction method to four-fermion production,
this method is formulated for mass regularization.

In the following, only singularities for photon emission off
nearly massless external fermions are considered.
Other singularities like collinear singularities of 
diagrams where a virtual photon decays into two external fermions
have to be excluded by appropriate cuts. 
The subtraction method described in the following is based on
the {\it dipole formalism} of S.~Catani and M.H.~Seymour \cite{Ca96}.
The results of this section have been worked out independently by 
S.~Dittmaier \cite{Di99}.
Comparing both results we find full consistence.

In the dipole formalism, various mappings are constructed from 
the five-particle phase space of the bremsstrahlung process $\eeffffg$ 
into the four-particle phase space of the non-radiative process $\eeffff$.
These mappings are required to obtain a process-independent
formulation of the subtraction method. 
Each mapping corresponds to a certain singular behaviour of the 
differential cross section.
According to these mappings, the  
five-particle phase space $\Phi^{(5)}$ splits into a 
four-particle phase space $\Phi^{(4)}$ and a remaining 
one-particle phase space, which contains the singularities:
\begin{eqnarray}
\label{eq:generalsplitting}
\int \rd \Phi^{(5)} &=& 
\int_0^1 \rd x_1 \rd x_2 \int \rd \Phi^{(4)} (x_1 p_1, x_2 p_2)
\int \rd \Phi_{ik}
\end{eqnarray}  
with $i,k=1,\dots6$.
In the following, particles $1$, $2$ are initial-state fermions 
and $3,\dots ,6$ final-state fermions.
For the discussion of the subtraction method, 
all particle momenta are denoted by $p_i$ and the photon momentum by $q$.
 
The arguments of $\rd \Phi^{(4)}(x_1 p_1, x_2 p_2)$ indicate
that the final-state momenta are calculated for incoming momenta 
$x_1 p_1$ and $x_2 p_2$.
The one-particle phase space is decomposed 
into integrations over $\Phi_{ik}$ and over the momentum 
fractions of the incoming momenta $x_1$ and $x_2$.
In this way, all different splittings of the five-particle phase space 
fit in this formula.
The variables $x_1$ and $x_2$ are partly fixed by 
$\delta$-distributions included in definition of $\Phi_{ik}$. 

For each mapping a subtraction term $\V_{ik}$ is constructed in such a way 
that it matches the singular behaviour of the cross section of 
the bremsstrahlung process in a certain phase-space region
and that it can be analytically integrated over $\Phi_{ik}$.
The subtraction term is subtracted from the real corrections and 
added to the virtual corrections after analytic integration over $\Phi_{ik}$:
\begin{eqnarray}
\label{eq:subtraction}
\nonumber
\lefteqn{\int \rd \Phi^{(5)} 
\left|\M^{(5)}_{\mathrm{Born}} \right|^2\O \left(\Phi^{(5)}\right)+
\int \rd \Phi^{(4)} 
2 \Re \left\{\M^{(4)}_{\mathrm{virt}}\M^{(4)*}_{\mathrm{Born}}\right\} 
\O \left(\Phi^{(4)}\right)}\\
\nonumber
&=&
\int \rd \Phi^{(5)}
\Bigg[\left|\M^{(5)}_{\mathrm{Born}}\right|^2 \O \left(\Phi^{(5)}\right)
-\sum\limits_{i,k=1 \atop i\ne k}^6 \V_{ik} 
\,\O \left(\Phi^{(4)}\left(\Phi^{(5)}\right)\right)\Bigg]\\\
\nn
&& {}
+\int_0^1 \rd x_1 \rd x_2 \int \rd \Phi^{(4)} (x_1 p_1, x_2 p_2) 
\, \O \left(\Phi^{(4)}\right)\\
&& {} \quad \times
\Bigg\{2 \Re \left(\M^{(4)}_{\mathrm{virt}}\M^{(4)*}_{\mathrm{Born}}\right) 
\delta (1-x_1) \delta (1-x_2)
+\sum\limits_{i,k=1 \atop i\ne k}^6 \left[\int \rd \Phi_{ik} \, \V_{ik}\right] 
\Bigg\},
\end{eqnarray}
where $\M^{(4)}_{\mathrm{Born}}$, $\M^{(5)}_{\mathrm{Born}}$, and
$\M^{(4)}_{\mathrm{virt}}$ are the matrix elements of the tree-level process,
the bremsstrahlung process, and the virtual corrections, respectively.
The experimental situation, \eg cuts, are included in the 
definition of the observable $\O$.
Note that the observable $\O $ depends on the momenta of the 
four-particle or five-particle phase spaces.
All integrations are performed numerically, 
except for the integration over $\Phi_{ik}$ which is performed analytically.

In this way, a part of the real corrections 
including the singularities is transferred to the virtual corrections. 
The difference between the cross section of the bremsstrahlung process
and the subtraction terms includes no singularities and 
hence can be integrated over the whole phase space.
Note that it is not allowed to exclude the soft-photon region from the 
integration domain where the cross section of the bremsstrahlung process
becomes infrared singular,
since the infrared singularities have to cancel between the real 
and virtual corrections.
This is in accordance with the fact, that it is not possible to 
separate experimentally photons
with infinitesimal small energies from charged particles.

The situation for collinear singularities is different since 
collinear singularities
show up as large logarithms of small but finite fermion masses.
If fermion masses are used as regularization parameters,
the phase-space region where the photon becomes collinear 
to an external fermion has to be included in the integration domain. 

\subsection{Behaviour of the cross section for collinear photons}

As a first step, the collinear limit of the cross section for ISR is 
considered.
It is convenient to calculate the collinear limit of the cross section
in an axial gauge, where the photon polarization sum runs only over 
transverse polarizations and hence interference contributions do not
involve collinear singularities.
A photon with momentum $q$ is emitted from an unpolarized 
initial-state fermion with momentum $p_a$, mass $m_a$, and 
relative charge $Q_a$.
The relevant part of the cross section multiplied by the polarization sum 
of the photon reads
\begin{eqnarray}
\label{eq:calcsplitt}
\nn
\lefteqn{\left[\frac{\ri}{\ps_a-\qs-m_a} (- \ri e Q_a)\gamma_\mu (\ps_a+m_a) 
\ri e Q_a\gamma_\nu \frac{-\ri}{\ps_a-\qs-m_a}\right]
\left[-g^{\mu \nu}+\frac{q^\mu n^\nu+q^\nu n^\mu}{q n}\right]}
\hspace*{1cm}\\
\nn
&=&
e^2 Q_a^2\left[
\frac{(1-x)\ps_a-\qs}{p_aq}\left(\frac{x}{1-x}-\frac{m_a^2}{p_a q}\right)
+\frac{m_a}{p_aq}\left(\frac{2}{1-x}-\frac{m_a^2}{p_a q}\right)
+\frac{\ns}{n q}\right]\\
&&
+\frac{e^2 Q_a^2}{p_aq} 
\left(\frac{1}{x}P(x)-\frac{m_a^2}{p_aq}\right) x\ps_a
\end{eqnarray}
with $n^2=0$, $q n\ne0$, $x=1-(n q)/(n p_a)$, and 
the splitting function is given by
\begin{eqnarray}
P(x)&=&\frac{1+x^2}{1-x}.
\end{eqnarray}
The arbitrary vector $n$ is introduced to define the polarization of the 
photon. 

The factor $(1-x)\ps_a-\qs$ vanishes in the collinear limit. 
Only the last term on the right-hand side proportional to $x \ps_a$ 
is singular in the collinear limit $q\to (1-x) p_a$. 
Hence, the collinear limit for photon radiation off unpolarized 
initial-state particles yields
\begin{eqnarray}
\left|\M^{(5)}_{\mathrm{collinear}}(p_a,q)\right|^2&=&
\frac{e^2 Q_a^2}{p_a q}
\left[\frac{1}{x}P(x)-\frac{m_a^2}{p_aq}\right]
\left|\M^{(4)}_{\mathrm{Born}}(x p_a)\right|^2.
\end{eqnarray}
The behaviour of the cross section for polarized particles 
in the collinear limit can be found in \citere{Kl87}:
\begin{eqnarray}
\nn
\left|\M^{(5)}_{\mathrm{collinear}}(p_a,\sigma_a,q)\right|^2&=&
\frac{e^2 Q_a^2}{p_a q}\Bigg\{
\left[\frac{1}{x}P(x)-\frac{1+x^2}{x}\frac{m_a^2}{2 p_aq}\right]
\left|\M^{(4)}_{\mathrm{Born}}(x p_a,\sigma_a)\right|^2\\
&& {}
\phantom{\frac{e^2 Q_a^2}{p_a q}\Bigg\{}
+\frac{(1-x)^2}{x}\frac{m_a^2}{2 p_aq}
\left|\M^{(4)}_{\mathrm{Born}}(x p_a,-\sigma_a)\right|^2\Bigg\}
\end{eqnarray}
with the helicity $\sigma_a$ of the initial-state fermion $a$.
In the following all results are calculated 
for helicity eigenstates of the external fermions.

For massive fermions the collinear limit
involves a helicity-flip term proportional to the square of the fermion mass,
called finite-mass term. 
Note that if the cross section of the bremsstrahlung process is calculated
for vanishing fermion masses the finite-mass terms are missing 
and have to be added to the cross section of the bremsstrahlung process
or skipped in the corresponding subtraction terms 
in \refeq{eq:subtraction}.

In order to be able to combine the collinear and soft limit
in \refse{se:subtractionterm}, we use charge conservation
\begin{eqnarray}
\label{eq:cc}
\kappa_i Q_i&=&-\sum\limits_{k\atop i\ne k} \kappa_{k} Q_{k}
\end{eqnarray}
to obtain 
\begin{eqnarray}
\label{eq:collineari}
\nn
\left|\M^{(5)}_{\mathrm{collinear}}(p_a,\sigma_a,q)\right|^2&=&
-\sum\limits_{k\atop k \ne a} 
\frac{\kappa_a \kappa_k e^2 Q_a Q_k}{p_a q}
\Bigg\{\left[\frac{1}{x}P(x)-\frac{1+x^2}{x}\frac{m_a^2}{2 p_aq}\right]
\left|\M^{(4)}_{\mathrm{Born}}(x p_a,\sigma_a)\right|^2\\
&& {}
\phantom{-\sum\limits_{k\atop k \ne a} 
\frac{\kappa_a \kappa_k e^2 Q_a Q_k}{p_a q}\Bigg\{}
+\frac{(1-x)^2}{x}\frac{m_a^2}{2 p_aq}
\left|\M^{(4)}_{\mathrm{Born}}(x p_a,-\sigma_a)\right|^2\Bigg\}.\qquad\quad
\end{eqnarray}
The sign $\kappa_i$ refers to charge flow of fermion $i$
into or out of the diagram, respectively, which is illustrated 
in \reffi{fi:flow}. 

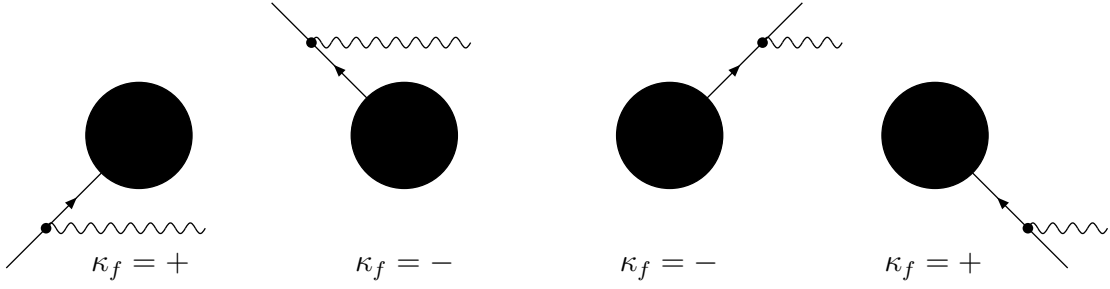
\begin{figure}
{
\begin{center}
\begin{picture}(300,100)
\ArrowLine(-50,0)(0,50)
\GCirc(0,50){20}{0}
\ArrowLine(100,50)(50,100)
\GCirc(100,50){20}{0}
\ArrowLine(200,50)(250,100)
\GCirc(200,50){20}{0}
\ArrowLine(350,0)(300,50)
\GCirc(300,50){20}{0}
\Photon(-35,15)(25,15){2}{8}
\Vertex(-35,15){2}
\Photon(65,85)(125,85){2}{8}
\Vertex(65,85){2}
\Photon(235,85)(265,85){2}{4}
\Vertex(235,85){2}
\Photon(335,15)(365,15){2}{4}
\Vertex(335,15){2}
\put(0,0){\makebox(1,1)[c]{$\kappa_f=+$}}
\put(100,0){\makebox(1,1)[c]{$\kappa_f=-$}}
\put(200,0){\makebox(1,1)[c]{$\kappa_f=-$}}
\put(300,0){\makebox(1,1)[c]{$\kappa_f=+$}}
\end{picture}
\end{center}
}
\caption[]{Value of $\kappa_f$ for the different photon radiation off fermions}
\label{fi:flow}
\end{figure}

For FSR, $q$ has to be replaced 
by $-q$:
\begin{eqnarray} 
\label{eq:collinearf}
\nn
\left|\M^{(5)}_{\mathrm{collinear}}(p_i,\sigma_i,q)\right|^2&=&
-\sum\limits_{k\atop k \ne i} 
\frac{\kappa_i \kappa_k e^2 Q_i Q_k}{p_i q}
\Bigg\{\left[P(z)-\frac{1+z^2}{z}\frac{m_i^2}{2 p_i q}\right]
\left|\M^{(4)}_{\mathrm{Born}}\left(\frac{p_i}{z},\sigma_i \right)\right|^2\\
&& {}
\phantom{-\sum\limits_{k\atop k \ne i} 
\frac{\kappa_i \kappa_k e^2 Q_i Q_k}{p_i q}\Bigg\{}
 +\frac{(1-z)^2}{z}\frac{m_i^2}{2 p_i q} 
\left|\M^{(4)}_{\mathrm{Born}}\left(\frac{p_i}{z},-\sigma_i \right)\right|^2
\Bigg\}\qquad
\end{eqnarray}
with $z=(n p_i)/(n p_i+n q)$, 
where the particle $i$ is a final-state particle.

\subsection{Soft-photon approximation of the cross section}

The infrared limit is given by the well known soft-photon approximation
\begin{eqnarray}
\left|\M^{(5)}_{\mathrm {soft}}\right|^2&=&
-\sum\limits_{i,k}\frac{\kappa_i \kappa_k e^2 Q_i Q_k p_i p_k}
{(p_i q)(p_k q)}\left|\M^{(4)}_{\mathrm{Born}}\right|^2.
\end{eqnarray}
After applying partial fractioning
\begin{eqnarray}
\frac{p_i p_k}{(p_i q)(p_k q)}&=&
\frac{p_i p_k}{[(p_i+p_k)q] (p_k q)}+\frac{p_i p_k}{[(p_i+p_k)q] (p_i q)}
\end{eqnarray}
and distributing the finite-mass term with $i=k$ 
with the help of charge conservation \refeq{eq:cc} to the
terms with $i\ne k$, the soft limit reads
\begin{eqnarray}
\label{eq:soft} 
\left|\M^{(5)}_{\mathrm {soft}}\right|^2&=&
-\sum\limits_{i,k\atop i\ne k}
\frac{\kappa_i \kappa_k e^2 Q_i Q_k}{p_i q} \left[
\frac{2 p_i p_k}{(p_i+p_k)q}-\frac{m_i^2}{p_i q}\right]
\left|\M^{(4)}_{\mathrm{Born}}\right|^2.
\end{eqnarray}
In this way, the soft-photon approximation \refeq{eq:soft} is written 
in a similar way as the collinear limits \refeq{eq:collineari} 
and \refeq{eq:collinearf}.

\subsection{Construction of the subtraction terms}
\label{se:subtractionterm}

\begin{figure}
{
\begin{center}
\begin{picture}(400,250)
\Line(100,200)(150,250)
\Line(100,200)(150,150)
\Photon(125,225)(150,225){2}{4}
\Vertex(125,225){2}
\GCirc(100,200){20}{0}
\put(100,150){\makebox(1,1)[c]{$\V_{ik}$}}
\put(160,250){\makebox(1,1)[l]{$i$}}
\put(160,150){\makebox(1,1)[l]{$k$}}
\Line(300,200)(350,250)
\Line(300,200)(250,150)
\Photon(325,225)(350,225){2}{4}
\Vertex(325,225){2}
\GCirc(300,200){20}{0}
\put(300,150){\makebox(1,1)[c]{$\V_{ia}$}}
\put(360,250){\makebox(1,1)[l]{$i$}}
\put(240,150){\makebox(1,1)[r]{$a$}}
\Line(100,50)(50,100)
\Line(100,50)(150,0)
\Photon(65,85)(150,85){2}{10}
\Vertex(65,85){2}
\GCirc(100,50){20}{0}
\put(100,0){\makebox(1,1)[c]{$\V_{ak}$}}
\put(40,100){\makebox(1,1)[r]{$a$}}
\put(160,0){\makebox(1,1)[l]{$k$}}
\Line(300,50)(250,100)
\Line(300,50)(250,0)
\Photon(265,85)(350,85){2}{10}
\Vertex(265,85){2}
\GCirc(300,50){20}{0}
\put(300,0){\makebox(1,1)[c]{$\V_{ab}$}}
\put(240,100){\makebox(1,1)[r]{$a$}}
\put(240,0){\makebox(1,1)[r]{$b$}}
\end{picture}
\end{center}
}
\caption[]{Diagrams for different combinations of initial-state
or final-state emitter and spectator}
\label{fi:subterm}
\end{figure}
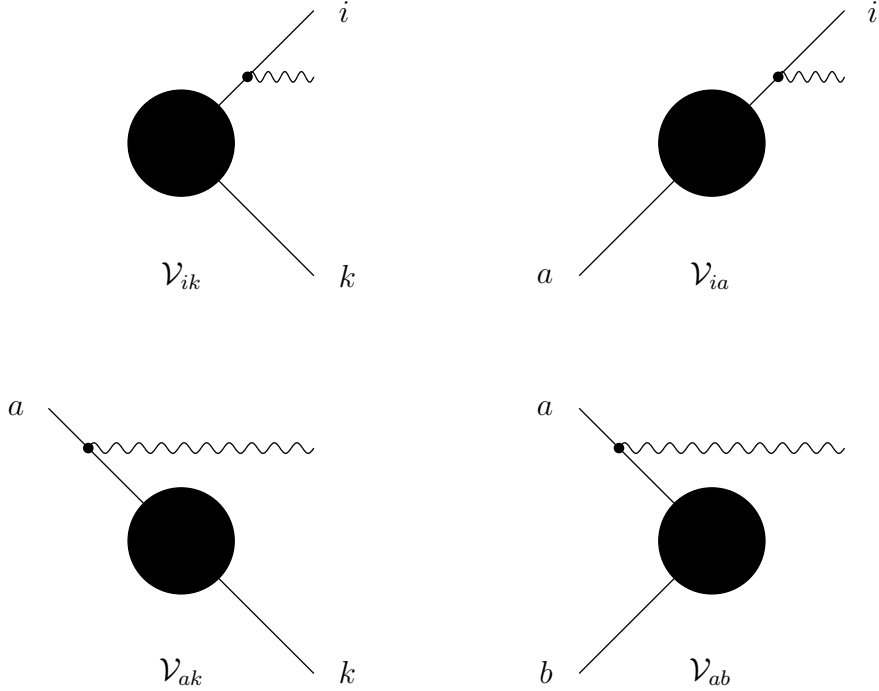

The subtraction terms have to be constructed in such a way that they 
reproduce both the soft and collinear limit of the cross section.
The soft-photon limit \refeq{eq:soft} depends on two momenta $p_i$ and $p_k$.
Particle $i$ is called emitter and $k$ spectator,
since the soft-photon approximation is singular 
if the photon becomes collinear to the emitter, but not if the 
photon is collinear to the spectator.
Since the mass of the spectator regularizes no singularities,
it is neglected in the following calculations.
The diagrams corresponding to the four possible combinations 
of initial-state or final-state emitter and spectator are shown 
in \reffi{fi:subterm}.

Although the following results are formulated for four-fermion production,
they can be applied to all processes where the external particles are 
light charged fermions and any other massless or massive neutral particles. 
Note that it is some freedom in the definition of the subtraction term 
and the separation of the one-particle phase space.
In the following, all initial-state particles are denoted by the indices
$a$ or $b$ with $a\ne b$ and $a,b=1,2$.

\subsubsection{Final-state emitter with final-state spectator}

Firstly, the case is considered where the emitter and spectator are
final-state particles.
A suitable subtraction term is defined by%
\footnote{Note that the actual definition of the subtraction term 
of \citere{Di99} 
differs by a factor $1-\yik$. 
}
\begin{eqnarray}
\label{eq:ff}
\nonumber
\V_{ik}&=&-\frac{\kappa_i \kappa_k e^2 Q_i Q_k}{p_i q}
\Bigg\{\left[\frac{2}{1-\zik (1-\yik)}-(1+\zik)-\frac{1+\zik^2}{\zik} 
\frac{m_i^2}{2 p_i q}\right]
\left|\M^{(4)}_{\mathrm{Born}}(\tp_i,\sigma_i,\tp_k)\right|^2\\
&& {}
\phantom{-\frac{\kappa_i \kappa_k e^2 Q_i Q_k}{p_i q}\Bigg\{}
+\frac{(1-\zik)^2}{\zik} 
\frac{m_i^2}{2 p_i q}
\left|\M^{(4)}_{\mathrm{Born}}(\tp_i,-\sigma_i,\tp_k)\right|^2\Bigg\}
\end{eqnarray}
with the momenta of the four-particle phase space
\begin{eqnarray}
\tp_i&=&p_i+q-\frac{\yik}{1-\yik} p_k,\qquad
\tp_k=\frac{1}{1-\yik} p_k
\end{eqnarray}
and the variables
\begin{eqnarray}
\zik&=&\frac{p_i p_k}{p_i p_k+p_k q},\qquad
\yik=\frac{p_i q}{p_i p_k+p_i q+p_k q},
\end{eqnarray}
where the indices $i$ and $k$
mark the emitter and the spectator, respectively. 

The subtraction term obeys the soft limit \refeq{eq:soft} for
$\zik\to 1$ and $\yik\to 0$, 
which can be verified with the help of 
\begin{eqnarray}
\frac{1}{1-\zik (1-\yik)}&=&\frac{p_i q+p_k q+p_i p_k}{(p_i+p_k) q}.
\end{eqnarray}
Furthermore, 
if the momentum $n$ after \refeq{eq:collinearf} 
is identified with the momentum of the spectator $p_k$,
the subtraction term reproduces the cross section
in the collinear limit, \ie $\yik \to 0$, 
$\zik\to z$, and $\tp_i\to p_i/z$, \refeq{eq:collinearf}.

The subtraction term is proportional to 
the Born matrix element $\M^{(4)}_{\mathrm{Born}}$ which
depends on the momenta $\tp_i$ and $\tp_k$ 
of the four-particle phase space.
These momenta are fixed for the analytical integration over
$\Phi_{ik}$. 
The momenta $\tp_i$ and $\tp_k$ are chosen in such a way that they 
fulfil momentum conservation
\begin{eqnarray}
0&=&K+p_i+p_k+q=K+\tp_i+\tp_k
\end{eqnarray}
and the on-shell conditions of the four-particle phase space
\begin{eqnarray}
\tm_i^2&=&m_i^2+\la^2\to 0,\qquad \tm_k^2=0.
\end{eqnarray}
The momentum $K$ is the sum of the momenta of the external particles, 
except for the emitter, the spectator, and the photon. 
The infinitesimal photon mass is denoted by $\la$.

The next step is to relate the momenta of the five-particle phase space 
to the momenta of the four-particle phase space and 
the variables $\zik$ and $\yik$ of the 
one-particle phase space. Therefore, we apply the mappings
\begin{eqnarray}
&p_i\mapsto \yik,\zik,k_\bot&: 
p_i=\left(\frac{\yik}{1-\yik}\frac{1}{1-\zik}
-\frac{\zik}{1-\zik}\frac{\la^2}{p_k q}\right) p_k
+\frac{\zik}{1-\zik} q+k_\bot,\\
&p_k\mapsto \tp_k&:  p_k=(1-\yik) \tp_k,\\
\label{eq:mapff}
&q\mapsto \tp_i&:
q=(1-\zik) \tp_i-\yik \zik \tp_k
+\frac{\zik}{1-\zik}\frac{\la^2}{\tp_i \tp_k}\tp_k
-(1-\zik) k_\bot\qquad
\end{eqnarray}
with $k_\bot p_k=k_\bot q=0$
and find the splitting of the five-particle phase space 
into the four-particle phase space
determined by the momenta $\tp_i$ and $\tp_k$ and the remaining 
one-particle phase space:
\begin{eqnarray}
\label{eq:splittingff}
\nn
\lefteqn{\int \frac{\rd^4 p_i}{(2 \pi)^3} 
\frac{\rd^4 p_k}{(2 \pi)^3} 
\frac{\rd^4 q}{(2 \pi)^3} \delta(p_i^2-m_i^2)\theta(p_{i,0}) 
\delta(p_k^2)\theta(p_{k,0}) \delta(q^2-\la^2)\theta(q_0) 
\delta^{(4)}(K-p_i-p_k-q)}\\
\nn
&=&\int_0^1 \rd x_1 \rd x_2 \int 
\frac{\rd^4 \tp_i}{(2 \pi)^3} 
\frac{\rd^4 \tp_k}{(2 \pi)^3}
\delta(\tp_i^2-m_i^2-\la^2) \theta(\tp_{i,0}) \delta(\tp_k^2) 
\theta(\tp_{k,0}) \delta^{(4)}(K-\tp_i-\tp_k)
\int \rd \Phi_{ik}\\
\end{eqnarray}
with the integral over the separated one-particle phase space
\begin{eqnarray}
\label{eq:oneff}
\nn
\int \rd \Phi_{ik}&=&
\frac{\tp_i \tp_k}{8 \pi^2} \delta(1-x_1) \delta(1-x_2)
\int_0^{2 \pi} \frac{\rd \phi_{k_\bot}}{2 \pi}
\int\limits_0^1 \rd \zik 
\int\limits_0^1 \rd \yik (1-\yik)\\
&& {} \times
\theta \left(\yik-\frac{\la^2 \zik^2+m_i^2 (1-\zik)^2}
{\zik (1-\zik) 2 \tp_i \tp_k}\right).
\end{eqnarray}
The additional integrations over $x_1$ and $x_2$ are introduced 
in order to match the definitions of \refeq{eq:generalsplitting}.
The masses $\la$ and $m_i$ in the $\delta$-distribution of
\refeq{eq:splittingff} can be neglected since they do not regularize 
singularities.
The step function $\theta$ in \refeq{eq:oneff} results from the
requirement $k_\bot^2<0$ of \refeq{eq:mapff} and regularizes all 
singularities.
It can be verified in the rest frame of $p_i+p_k$
that $k_\bot$ is a space-like vector.
The variable $\phi_{k_\bot}$ is the solid angle of 
$k_\bot$ with respect to $p_k$ in the rest frame of $p_i+p_k$.
Since the subtraction term does not depend on $\phi_{k_\bot}$
the integration over $\phi_{k_\bot}$ yields $2 \pi$. 
Note that the Born cross section included in the subtraction term $\V_{ik}$
depends exclusively on the momenta of the four-particle phase space,
which are fixed for the integration over $\zik$ and $\yik$.

Finally, the subtraction term is integrated over the 
one-particle phase space yielding
\begin{eqnarray}
\label{eq:intsubff}
\nn
\lefteqn{\int \rd \Phi_{ik} \V_{ik} = 
- \frac{\kappa_i \kappa_k e^2 Q_i Q_k}{8 \pi^2} 
\delta(1-x_1) \delta(1-x_2)}\\
\nn
&& {} \times
\left\{\left[ 
\L(\la^2,m_i^2,2 \tp_i \tp_k)+\frac{5}{2}-\frac{2}{3}\pi^2\right]
\left|\M^{(4)}_{\mathrm{Born}}(\tp_i,\sigma_i,\tp_k)\right|^2
+\frac{1}{2}\left|\M^{(4)}_{\mathrm{Born}}(\tp_i,-\sigma_i,\tp_k)\right|^2
\right\}\\
&& {}
+\O(\la)+\O(m_i)
\end{eqnarray}
with
\begin{eqnarray} 
\label{eq:defL}
\L(\la^2,m_i^2,2 \tp_i \tp_k)&=&
\frac{1}{2}\ln^2 \left(\frac{\la^2}{2 \tp_i \tp_k}\right)
-\frac{1}{2}\ln^2 \left(\frac{\la^2}{m_i^2}\right)
+\frac{3}{2}\ln\left(\frac{m_i^2}{2 \tp_i \tp_k}\right)
+\ln\left(\frac{\la^2}{m_i^2}\right),\qquad
\end{eqnarray}
where the photon mass is neglected with respect to the 
fermion masses, \ie $\la \ll m_i$.

\subsubsection{Final-state emitter with initial-state spectator}

Next, the subtraction term for final-state emitter and 
initial-state spectator is considered. 
The subtraction term with the correct collinear and soft limit
is defined by 
\begin{eqnarray}
\nn
\V_{ia}&=&-\frac{\kappa_i \kappa_a e^2 Q_i Q_a}{p_i q}
\Bigg\{
\left[\frac{2}{2-\zia-\xia}-1-\zia-\frac{1+\zia^2}{\zia}
\frac{m_i^2}{2 p_i q}\right]
\frac{1}{\xia}\left|\M^{(4)}_{\mathrm{Born}}(\tp_i,\sigma_i,\tp_a)\right|^2
\\
&& {}
\phantom{-\frac{\kappa_i \kappa_a e^2 Q_i Q_a}{p_i q}\Bigg\{}
+\frac{(1-\zia)^2}{\zia}\frac{m_i^2}{2 p_i q}
\frac{1}{\xia}\left|\M^{(4)}_{\mathrm{Born}}
(\tp_i,-\sigma_i,\tp_a)\right|^2\Bigg\}
\end{eqnarray}
with the momenta of the four-particle phase space
\begin{eqnarray}
\label{eq:def1}
\tp_i&=&p_i+q-(1-\xia) p_a,\qquad \tp_a= \xia p_a,
\end{eqnarray}
and the variables
\begin{eqnarray}
\label{eq:def2}
\zia&=&\frac{p_i p_a}{p_i p_a+p_a q},\qquad
\xia=\frac{p_i p_a+p_a q-p_i q}{p_i p_a+p_a q}.
\end{eqnarray}
Similarly to the previous section, the momenta $\tp_i$ and $\tp_a$ 
fulfil momentum conservation and 
the on-shell requirement for the external particles 
of the four-particle phase space.

In order to separate the integral over the one-particle phase space 
from the integral over the five-particle phase space, the 
following mappings are used:
\begin{eqnarray}
\label{eq:mapfi1}
&p_i\mapsto \xia,\zia,k_\bot&: 
p_i=\left(\frac{1-\xia}{1-\zia}
-\frac{\zia}{1-\zia}\frac{\la^2}{p_a q}\right) p_a
+\frac{\zia}{1-\zia} q+k_\bot, \\
\label{eq:mapfi2}
&q\mapsto \tp_i&:
q=(1-\zia) \tp_i-(1-\xia) \zia p_a
+\frac{\zia}{1-\zia}\frac{\la^2}{\tp_i p_a} p_a
-(1-\zia) k_\bot\qquad\quad
\end{eqnarray}
with $k_\bot p_a=k_\bot q=0$.
They result in the splitting of the phase space
\begin{eqnarray}
\label{eq:splittingfi}
\nn
\lefteqn{\int \frac{\rd^4 p_i}{(2 \pi)^3} 
\frac{\rd^4 q}{(2 \pi)^3} \delta(p_i^2-m_i^2) \theta(p_{i,0}) 
\delta(q^2-\la^2) \theta(q_0) \delta^{(4)}(K+p_a-p_i-q)}\\
&=&\int_0^1 \rd x_a \rd x_b \int \frac{\rd^4 \tp_i}{(2 \pi)^3}
\delta(\tp_i^2-m_i^2-\la^2) \theta(\tp_{i,0}) \delta^{(4)}(K+\tp_a-\tp_i)
\int \rd \Phi_{ia}
\end{eqnarray}
with
\begin{eqnarray}
\label{eq:onefi}
\nn
\int \rd \Phi_{ia}&=&
\frac{\tp_i p_a}{8 \pi^2} \delta(\xia-x_a) \delta(1-x_b)
\int_0^{2 \pi} \frac{\rd \phi_{k_\bot}}{2 \pi}
\int\limits_0^1 \rd \zia\\
&& {} \times
\theta \left((1-\xia)-\frac{\la^2 \zia^2+m_i^2 (1-\zia)^2}
{\zia (1-\zia) 2 \tp_i p_a}\right),
\end{eqnarray}
where $\phi_{k_\bot}$ is the solid angle with respect to $p_a$ in the 
rest frame of $p_a+p_i$.
Since the spectator is an initial-state particle, the four-particle 
phase space becomes $x_a$-dependent and the one-particle 
phase space is included in the $x_a$-integration. 
As in the previous case the $\theta$-function originates from 
$k_\bot^2<0$ of the mapping \refeq{eq:mapfi2} and regularizes all 
singularities.
Furthermore, the masses $m_i$ and $\la$ can be omitted in the
$\delta$-distribution of \refeq{eq:splittingfi}.

Since the subtraction term does not depend of $\phi_{k_\bot}$, 
the integration over $\phi_{k_\bot}$ yields $2 \pi$. 
Moreover, the Born cross section included in the subtraction term
does not depend on $\zia$, but is a function of $\xia$.
In order to avoid the analytic integration over the
Born cross section, the subtraction term is integrated for fixed
Born cross section at $\xia=1$.
The $\xia$-dependence of the Born cross section is taken into account
with the help of the $+$-distribution, which is defined as usual by
\begin{eqnarray}
\int_0^1 \rd x \, g(x) [f(x)]_+ &=& \int_0^1 \rd x [g(x)-g(1)] f(x),
\end{eqnarray}
where $g(x)$ is an arbitrary test function.

The integration of subtraction term yields
\begin{eqnarray}
\label{eq:intsubia}
\nn
\lefteqn{\int \rd \Phi_{ia} \V_{ia}=
- \frac{\kappa_i \kappa_a e^2 Q_i Q_a}{8 \pi^2}
\delta(1-x_b)}
\\\nn
&& {} \times\Bigg\{
\left\{\left[
\L(\la^2,m_i^2,2 \tp_i \tp_a)
+1-\frac{\pi^2}{2}\right]\delta(1-x_a)
+\left[\frac{2}{1-x_a}\ln\left(\frac{2-x_a}{1-x_a}\right)-
\frac{3}{2}\frac{1}{1-x_a}\right]_+\right\}\\\nn
&&{}
\phantom{\times \Bigg\{}
\times
\frac{1}{x_a}\left|\M^{(4)}_{\mathrm{Born}}(\tp_i,\sigma_i,\tp_a)\right|^2
+\frac{1}{2}\delta(1-x_a)
\left|\M^{(4)}_{\mathrm{Born}}(\tp_i,-\sigma_i,\tp_a)\right|^2\Bigg\}\\
&&{}+\O(\la)+\O(m_i)
\end{eqnarray}
with $\la \ll m_i$ and the function $\L$ defined in \refeq{eq:defL}. 
The mass and infrared singularities of \refeq{eq:intsubia} 
have a similar form as in \refeq{eq:intsubff}.
Note that the +-distribution acts on the matrix element,
the observable $\O$ (see \refeq{eq:subtraction}), and the momenta $\tp_i$ and 
$\tp_a$.
Here and in the following case, 
the momenta of the four-particle phase-space are implicitly 
understood to be functions of $x_a$ with $x_a=\xia$.

\subsubsection{Initial-state emitter with final-state spectator}

Borrowing the previous definitions of the momenta of the four-particle 
phase space \refeq{eq:def1} and of the variables $\xia$ and $\zia$
\refeq{eq:def2}, an appropriate subtraction term is defined by
\begin{eqnarray}
\nonumber
\V_{ai}&=&
-\frac{\kappa_a \kappa_i e^2 Q_a Q_i}{p_a q} 
\Bigg\{\left[
\frac{2}{2-\xia-\zia}-1-\xia-(1+\xia^2) \frac{m_a^2}{2 p_a q}\right]
\frac{1}{\xia}\left|\M^{(4)}_{\mathrm{Born}}(\tp_a,\sigma_a,\tp_i)\right|^2\\
&& {}
\phantom{-\frac{\kappa_a \kappa_i e^2 Q_a Q_i}{p_a q} \Bigg\{}
+(1-\xia)^2 \frac{m_a^2}{2 p_a q}
\frac{1}{\xia}
\left|\M^{(4)}_{\mathrm{Born}}(\tp_a,-\sigma_a,\tp_i)\right|^2\Bigg\}.
\end{eqnarray}

As in the previous case, the mappings
\begin{eqnarray}
&q\mapsto \xia,\zia,k_\bot&: 
q=\frac{1-\xia}{\zia} p_a
+\left(\frac{1-\zia}{\zia}-\frac{1-\xia}{\zia}\frac{m_a^2}{p_a p_i}\right) 
p_i+k_\bot,\\
\nonumber
&p_{i}\mapsto \tp_{i}&:
p_i=\left\{\zia \tp_i-(1-\xia) (1-\zia) p_a-\zia k_\bot\right\}\\
&& {} \hspace{3em}
\times\left\{\frac{\zia [\tp_i p_a+(1-\xia)m_a^2]}
{\zia \tp_i p_a-(1-\zia)(1-\xia)m_a^2}\right\}
\end{eqnarray}
with $k_\bot p_a=k_\bot p_i=0$
result in the splitting of the phase-space integral
\begin{eqnarray}
\label{eq:splittingif}
\nn
\lefteqn{\int \frac{\rd^4 p_i}{(2 \pi)^3} 
\frac{\rd^4 q}{(2 \pi)^3} \delta(p_i^2) \theta(p_{i,0}) 
\delta(q^2-\la^2) \theta(q_0) \delta^{(4)}(K+p_a-p_i-q)}\\
&=&\int_0^1 \rd x_a \rd x_b \int 
\frac{\rd^4 \tp_i}{(2 \pi)^3}
\delta(\tp_i^2-\la^2+(1-x_a)^2 m_a^2) \theta(\tp_{i,0})
\delta^{(4)}(K+\tp_a-\tp_i)
\int \rd \Phi_{ai}\qquad
\end{eqnarray}
with
\begin{eqnarray} 
\label{eq:oneif}
\nonumber
\int \rd \Phi_{ai}&=&\int_0^{2 \pi} \frac{\rd \phi_{k_\bot}}{2 \pi}
\int\limits_0^1 \rd \xia 
\int\limits_0^1 \rd \zia \delta(x_a-\xia)\delta(1-x_b)
\\\nn
&& {} \times
\frac{\tp_i p_a+(1-\xia) m_a^2}{8 \pi^2}
\left\{\frac{\zia [\tp_i p_a+(1-\xia)m_a^2]}
{\zia \tp_i p_a-(1-\zia) (1-\xia)m_a^2]}
\right\}\\
&& {} \times
\theta \left(1-\frac{m_i^2 \zia^2+m_a^2 (1-\xia)^2 [(1-\zia)^2+\zia^2]}
{(1-\zia) \zia (1-\xia) 2 \tp_i p_a}\right).
\end{eqnarray}
The variable $\phi_{k_\bot}$ is the solid angle between 
$p_a$ and $k_\bot$ in the rest frame of $p_a+p_i$ and 
the masses $\la$ and $m_a$ can be omitted in \refeq{eq:splittingif}.
The $\theta$-function results from $k_\bot^2<0$.

The integration of the subtraction term over $\zia$ and $\xia$ 
for fixed $p_a$ and $\tp_i$ provides
\begin{eqnarray}\nn
\label{eq:intsubai}
\lefteqn{\int \rd \Phi_{ai} \V_{ai} =
- \frac{\kappa_a \kappa_i e^2 Q_a Q_i}{8 \pi^2} \delta(1-x_b)}\\\nn 
&& {} 
\times \Bigg\{\Bigg\{\left[
\L(\la^2,m_a^2,2 \tp_a \tp_i)
-\frac{1}{4}-\frac{\pi^2}{6}\right]\delta (1-x_a)
+\left[\frac{2}{1-x_a}\ln\left(\frac{1-x_a}{2-x_a}\right)\right]_+\\\nn
&& {} 
\phantom{+\Bigg\{\Bigg\{}
-\left[P(x_a)\left(\ln(1-x_a)+1\right)\right]_+
-\left[P(x_a)\right]_+\ln\left(\frac{m_a^2}{2 p_a \tp_i}\right)\Bigg\}
\frac{1}{x_a} \left|\M^{(4)}_{\mathrm{Born}}(\tp_a,\sigma_i,\tp_i)\right|^2
\\\nn && {}
\phantom{+\Bigg(}
+\left\{\frac{1}{2}\delta(1-x_a)+[1-x_a]_+\right\}\frac{1}{x_a}
\left|\M^{(4)}_{\mathrm{Born}}(\tp_a,-\sigma_a,\tp_i)\right|^2\Bigg\}\\
&&{}
+\O(\la)+\O(m_a)
\end{eqnarray}
with $\la \ll m_a$ and the function $\L$ defined in \refeq{eq:defL}.
The $x_a$-independent singularities are 
similar to the previous integrated subtraction term.
Extra $x_a$-dependent mass singularities appear 
in \refeq{eq:intsubai} and are proportional to the
splitting function $P(x_a)$.

While in \citere{Di99} the momentum $\tp_a$ has been fixed for
the $\xia$-integration, here we fix the momentum $p_a$.
Although the actual form of the subtraction terms are different,
the results for integrated observables are the same.
One can reproduce the result of \citere{Di99} with
\begin{eqnarray}
\nn
\lefteqn{\int_0^1 \rd x_a 
\left[P(x_a)\right]_+\ln\left(\frac{m_a^2}{2 p_a \tp_i}\right) g(x_a)}\\
&=&\int_0^1 \rd x_a \left\{
\left[P(x_a)\right]_+\ln\left(\frac{m_a^2}{2 \tp_a \tp_i}\right)
+\left(\frac{5}{4}-\frac{\pi^2}{3}\right) \delta(1-x_a)\right\} g(x_a).
\end{eqnarray}

\subsubsection{Initial-state emitter with initial-state spectator}

Finally, the subtraction terms for initial-state
emitter and spectator are considered.
The subtraction term reads
\begin{eqnarray}
\nn
\lefteqn{\V_{ab}=
-\frac{\kappa_a \kappa_b e^2 Q_a Q_b}{p_a q}
\Bigg\{
\left[\frac{2}{1-\xab}-1-\xab-(1+\xab^2) \frac{m_a^2}{2 p_a q}\right]
\frac{1}{\xab}
\left|\M^{(4)}_{\mathrm{Born}}(\tp_a,\sigma_a,\tp_b)\right|^2}
&& \hspace{15cm}
\\
\lefteqn{\phantom{\V_{ab}=-\frac{\kappa_a \kappa_b e^2 Q_a Q_b}{p_a q}\Bigg\{}
+(1-\xab)^2 \frac{m_a^2}{2 p_a q}
\frac{1}{\xab}
\left|\M^{(4)}_{\mathrm{Born}}(\tp_a,-\sigma_a,\tp_b)\right|^2 \Bigg\}}
\end{eqnarray}
with the momenta of the four-particle phase space
\begin{eqnarray}
\tp_a&=&\xab p_a,\qquad \tp_b=p_b
\end{eqnarray}
and the variables
\begin{eqnarray}
\vab&=&\frac{p_a q}{p_a p_b},\qquad
\xab=\frac{p_a p_b-p_a q-p_b q}{p_a p_b}.
\end{eqnarray}
Since the momenta of the four-particle phase space 
have to fulfil momentum conservation,
the momenta $p_i$ of the final-state particles are also modified:
\begin{eqnarray}
\label{eq:ptilde}
\tp_i^\mu&=&p_i^\mu-\frac{(K+\tK) p_i}{(K+\tK) K}(K+\tK)^\mu
+\frac{2 K p_i}{K^2}\tK^\mu
\end{eqnarray}
with following auxiliary momenta
\begin{eqnarray}
K&=&p_a+p_b-q=\sum_{i=3}^6 p_i, \qquad 
\tK=\tp_a+\tp_b=\sum_{i=3}^6 \tp_i.
\end{eqnarray} 
Equation \refeq{eq:ptilde} is a proper Lorentz transformation of the 
momenta $p_i$ into $\tp_i$ in the massless limit.%
\footnote{The mapping of the final-state particles coincides with 
\citere{Ca96} for vanishing fermion masses.}

The mapping from the five-particle phase space into the four-particle 
phase space reads
\begin{eqnarray}
&q\mapsto \xab,\vab,k_\bot&: 
q=\left[\vab-(1-\xab-\vab)\frac{m_a^2}{p_a p_b}\right] p_b+(1-\xab-\vab) p_a
+k_\bot \qquad
\end{eqnarray}
with $k_\bot p_a=k_\bot p_b=0$.

Hence, the splitting of the phase-space integral takes the form 
\begin{eqnarray}
\label{eq:splittingii}
\nn
\lefteqn{\prod\limits_i \int 
\frac{\rd^4 p_i}{(2 \pi)^3} 
\frac{\rd^4 q}{(2 \pi)^3} \delta(p_i^2-m_i^2) 
\theta(p_{i,0})
\delta(q^2-\la^2) \theta(q_0) \delta^{(4)}(p_a+p_b-\mbox{$\sum_i$} p_i-q)}\\
&=&\int_0^1 \rd x_a \rd x_b \prod\limits_i 
\int \frac{\rd^4 \tp_i}{(2 \pi)^3}
\delta(\tp_i^2-\tilde{m}_i^2) \theta(\tp_{i,0}) 
\delta^{(4)}(\tp_a+\tp_b-\mbox{$\sum_i$} \tp_i)
\int \rd \Phi_{ab}
\end{eqnarray}
with
\begin{eqnarray}
\nn
\int \rd \Phi_{ab} &=&\frac{p_a p_b}{8 \pi^2}
\int_0^{2 \pi} \frac{\rd \phi_{k_\bot}}{2 \pi}
\int\limits_{-\frac{m_a^2+\la^2}{2 p_a p_b}}^1 \rd \xab
\int\limits_0^{1-\xab} \rd \vab \delta(x_a - \xab) \delta(1-x_b)\\
&& {}
\times \theta \left(\vab-\frac{(1-\xab-\vab)^2 m_a^2+\la^2}
{(1-\xab-\vab) 2 p_a p_b}\right).
\end{eqnarray}
The variable $\phi_{k_\bot}$ is the solid angle between 
$p_a$ and $k_\bot$ in the rest frame of $p_a+p_b$.
The masses $\tilde{m}_i$ of the final-state particles of the four-particle 
phase space, 
\begin{eqnarray}
\nn
\tilde{m}_i^2&=&m_i^2-[(1-\xab^2) m_a^2+\la^2]\\
&& {}
\times\left[
\left(\frac{K p_i+\tK p_i}{K^2+K\tK}\right)^2+
4\left(\frac{K p_i}{K^2}\right)^2-
4\frac{(K p_i+\tK p_i)K p_i}{(K^2+K\tK)K^2}\right],
\end{eqnarray}
can be replaced in \refeq{eq:splittingii} by the masses $m_i$ of the 
final-state particles of the five-particle phase space 
after neglecting $m_a$ and $\la$.
The $\theta$-function is due to the requirement $k_\bot^2<0$. 

Finally, the integrated subtraction term is given:
\begin{eqnarray}
\nn\label{eq:intsubii}
\lefteqn{\int \rd \Phi_{ab} \V_{ab}=
- \frac{\kappa_a \kappa_b e^2 Q_a Q_b}{8 \pi^2}\delta(1-x_b)}\\\nn
&&{}\times \Bigg\{\left\{
\left[\L(\la^2,m_a^2,2 \tp_a \tp_b)
+\frac{3}{2}-\frac{\pi^2}{3}\right]\delta(1-x_a)
+\left[P(x_a)\right]_+
\left[\ln\left(\frac{2 p_a p_b}{m_a^2}\right)-1\right]
\right\}\\\nn
&&{}
\phantom{\times \Bigg\{}
\times\frac{1}{x_a}
\left|\M^{(4)}_{\mathrm{Born}}(\tp_a,\sigma_a,\tp_b)\right|^2
+\left\{\frac{1}{2}\delta(1-x_a)+[1-x_a]_+\right\}
\frac{1}{x_a}
\left|\M^{(4)}_{\mathrm{Born}}(\tp_a,-\sigma_a,\tp_b)\right|^2\Bigg\}\\
&& {}
+\O(\la)+\O(m_a)
\end{eqnarray}
with $\la \ll m_a$ and the function $\L$ defined in \refeq{eq:defL}.
The momentum $\tp_a$ is implicitly understood to be a function of $x_a$ 
with $x_a=\xab$.

\subsection{Remarks to four-fermion production}

Since the virtual corrections are calculated in DPA, 
the integrated subtraction terms
that are added to the virtual radiative corrections have to be treated
in the right way to achieve the correct cancellations of the
singularities.
However, it is not possible to apply the DPA to the whole integrated 
subtraction terms
because the $x_a$-dependent parts are evaluated for reduced
CM energies $\sqrt{x_a s}$, which are below the $\PW$-pair production 
threshold for small $x_a$, where the DPA is not possible.
Therefore, the integrated subtraction terms have to be divided into two parts,
one including all $x_a$-dependent terms and a second which is 
evaluated in DPA including all infrared and mass singularities,
except for the $x_a$-dependent mass singularities resulting from ISR. 
Here, the following decomposition is used:
\begin{eqnarray}
\label{eq:DPApart1}
\int \Phi_{ik}\V_{ik}&\sim&V_{ik}^{\mathrm{DPA}}
+V_{ik}^{\mbox{\tiny off-shell}},\\
\nn
V_{ik}^{\mathrm{DPA}}&=&- \frac{\kappa_i \kappa_k e^2 Q_i Q_k}{8 \pi^2}
\delta(1-x_1)\delta(1-x_2)
\L(\la^2,m_i^2,2 \tp^{\mathrm{on}}_i \tp^{\mathrm{on}}_k)\\
\label{eq:DPApart2}
&& {} \times
\left|\M^{(4)}_{\mathrm{Born}}(\tp^{\mathrm{on}}_i,\sigma_i,
\tp^{\mathrm{on}}_k)\right|^2
\end{eqnarray}
with $i,k=1,\dots,6$, and the on-shell momenta defined in \refeq{eq:onshell}. 
The function $\L$ is defined in \refeq{eq:defL}.
Different choices of the decomposition \refeq{eq:DPApart1} 
of the subtraction terms differ only in non-doubly-resonant contributions.

The singularities of \refeq{eq:DPApart2} cancel with the 
singularities of the virtual corrections,
while $x_a$-dependent mass singularities remain in 
$V_{ik}^{\mbox{\tiny off-shell}}$ from ISR.
Moreover, in the integral over the five-particle phase space
of \refeq{eq:subtraction}, the cuts applied to the bremsstrahlung process and 
to the individual subtraction terms are different,
because the observable $\O$ depends on the five-particle phase space 
for the bremsstrahlung process, but on the four-particle phase space 
in the case of the subtraction terms.
Hence, one should take care of the correct implementation of cuts 
in the phase-space generators, if the 
subtraction terms are involved in the calculation.

\section{Numerical results}
\label{se:radcorrresults}

For the numerical discussion, the radiative corrections are evaluated as 
described in the previous sections, \ie
the virtual corrections are calculated in DPA,
and the complete bremsstrahlung process $\eeffffg$ is included 
for the real corrections.
Although polarized cross sections can be calculated in our approach, 
only unpolarized cross sections are considered in this thesis.

The fixed-width scheme and following input parameters are used 
for the numerical discussion: 
\beq
\label{eq:input1}
\begin{array}[b]{rlrlrl}
\al =& 1/137.0359895, & \qquad
\GF =& 1.16639 \times 10^{-5} \GeV^{-2}, &\qquad 
\MH =& 300\GeV,  \\
\MW =& 80.26 \GeV, & \qquad
\GW =& 2.08174 \GeV, & \qquad
\MZ =& 91.1884 \GeV, \\
\GZ =& 2.4971 \GeV,
\end{array}
\eeq
except for \refse{se:testsub} where the input is given explicitly.
If not stated otherwise, the weak mixing angle is defined by  
$\cw=\MW/\MZ$, $\sw^2=1-\cw^2$, and the following fermion masses are used:
\beq
\begin{array}[b]{rlrlrl}
\label{eq:input2}
\Me =& 0.51099906 \MeV, & \qquad
\Mmy=& 105.658389 \MeV, & \qquad
\Mta=& 1.7771 \GeV, \\
\Mu =& 47 \MeV, & \qquad
\Mc =& 1.55 \GeV, & \qquad
\Mt =& 165.26 \GeV, \\
\Md =& 47 \MeV, & \qquad
\Ms =& 150 \MeV, & \qquad
\Mb =& 4.7 \GeV,
\end{array}
\eeq
where the light quark masses are adjusted in such a way 
that the experimentally measured hadronic 
vacuum polarization is reproduced. 

The input parameters coincide with those of \citere{Be99}, except for the 
additional $\PZ$-boson width.
The finite $\PZ$ width is required, since
the momenta of the bremsstrahlung process are generated
for off-shell $\PW$ bosons.
For vanishing $\PZ$-boson width, the matrix element of 
the bremsstrahlung process becomes singular 
if the real photon has the energy $E_\ga=\sqrt{s}-\MZ$.
This singularity is due to diagrams where the incoming electron and 
positron annihilate into a virtual $\PZ$ boson after radiation of a
bremsstrahlung photon.
The virtual corrections are evaluated for vanishing $\PZ$-boson width. 

If not stated otherwise, we use the $\GF$ parameterization, 
where the Fermi constant $\GF $ and the fine structure constant 
$\al$ are related by
\begin{eqnarray}
\label{eq:gmuscheme}
\GF &=&\frac{\al \pi}{\sqrt{2} \MW^2 \sw^2}\frac{1}
{1-\Delta r}.
\end{eqnarray}
The symbol $\Delta r$ denotes the radiative corrections 
to the muon decay.
The cross section in $\GF$ parameterization can be obtained from the 
results in the on-shell renormalization scheme in the following way:
\begin{eqnarray}
\label{eq:gmucrosssection}
\nn
\rd \si^{\GF}&=& \rd \si_{\born}^{\GF} (1+\delta^{\GF}) 
\end{eqnarray}
with
\begin{eqnarray}
\rd \si_{\born}^{\GF}&=&\frac{\rd \si^{\al}_{\born}}{(1-\Delta r)^4},\qquad
\delta^{\GF} =\frac{\rd \si^{\al}_{\virt}+\rd \si^{\al}_{\real}}
{\rd \si^{\al}_{\born}}-4 \Delta r^{\mbox{\tiny 1-loop}}.
\end{eqnarray}
The quantities $\si^\al_{\born}$, $\si^\al_{\virt}$, and 
$\si^\al_{\real}$ denote 
the tree-level cross section, the virtual, and the real corrections,
respectively, calculated in the one-shell renormalization scheme 
with the free parameters $\al(0)$, $\MW$, $\MZ$, $\MH$, and $m_f$.
The value of $\Delta r$ can be calculated from \refeq{eq:gmuscheme} and 
the one-loop corrections to the muon decay yield
$\Delta r^{\mbox{\tiny 1-loop}}=0.0373994$ for the set of 
input parameters \refeq{eq:input1} and \refeq{eq:input2}.

\subsection{Total cross section and angular distributions}

In order to compare our results with the results of \citere{Be99}, 
the doubly-resonant electroweak $\Oa$ radiative 
corrections are calculated for the leptonic process 
$\Pep \Pem \to \nu_\mu \mu^+ \tau^- \bar\nu_\tau$ and 
the LEP2 energy $184 \GeV$. 
In \citere{Be99}, the DPA is applied to both the matrix element 
and the four-fermion 
phase space, except for the Breit--Wigner propagators. 
The integrations of the Breit--Wigner propagators over the 
invariant masses $k_\pm^2$ are extended to the full range 
$(-\infty, +\infty)$, resulting in   
\begin{equation}
\int_{-\infty}^{\infty} \rd k_\pm^2
\frac{1}{k_\pm^2-\MW^2+\ri \MW \GW}= \frac{\pi}{\MW \GW}.
\end{equation}

In contrast to \citere{Be99},
the radiative corrections in our approach are evaluated 
as described in the previous sections with the 
exact bremsstrahlung process, and the virtual corrections are taken
in DPA.
Whereas in \citere{Be99}, the momenta are 
generated for on-shell $\PW$ bosons, we use the exact off-shell phase space
for the calculation of the radiative corrections.
The DPA-Born cross section is evaluated in the same way as done 
in \citere{Be99}, in particular with on-shell phase space, 
in order to have a common normalization of the relative corrections.
Moreover, the real photon is recombined with a charged 
final-state fermion if their invariant mass is smaller than all other
invariant masses $m(\ga,f)$, where $\ga$ denotes the photon and $f$ 
an initial- or final-state charged fermion. 

\subsubsection{CM-energy dependence of the total cross section}
\label{se:CMenergy}

\begin{figure}
\centerline{
\setlength{\unitlength}{1cm}
\begin{picture}(7.5,8.5)
\put(-1,0){\includegraphics{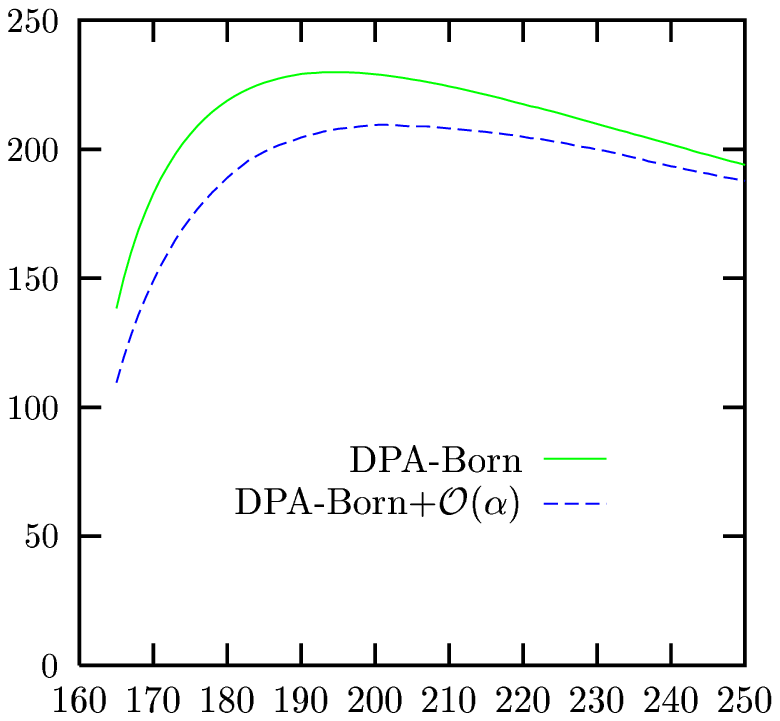}}
\put(0,7.7){\makebox(1,1)[l]{$\si/\fba$}}
\put(4,-0.3){\makebox(1,1)[cc]{{$\sqrt{s}/\GeV$}}}
\end{picture}
\setlength{\unitlength}{1cm}
\begin{picture}(7.5,8.5)
\put(-1,0){\includegraphics{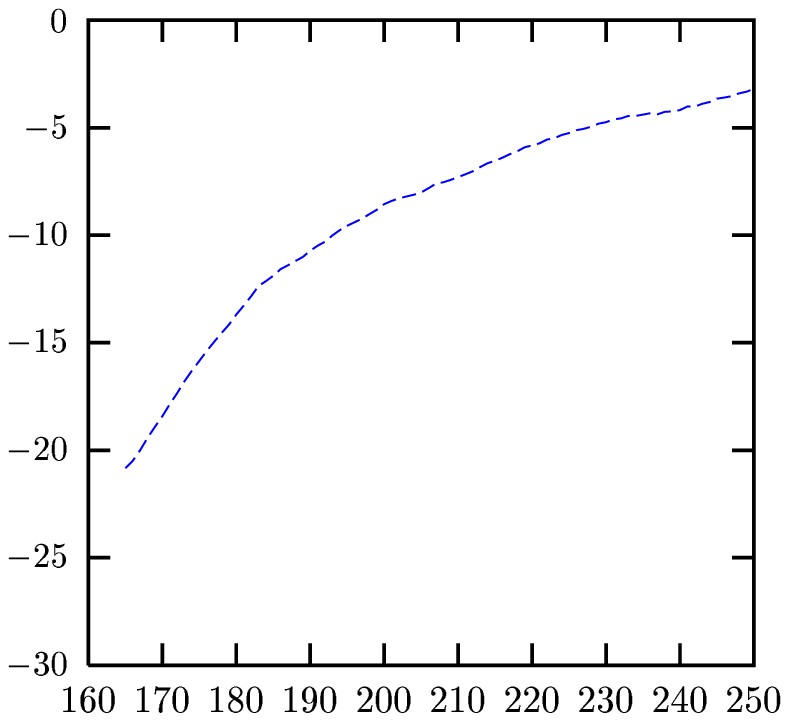}}
\put(0,7.7){\makebox(1,1)[l]{$\delta/\%$}}
\put(4,-0.3){\makebox(1,1)[cc]{{$\sqrt{s}/\GeV$}}}
\end{picture}
}
\caption[]{CM-energy dependence of the total cross section and the 
relative corrections for the process 
$\Pep \Pem \to \nu_\mu \mu^+ \tau^- \bar\nu_\tau$}
\label{fi:BBC1}
\end{figure}

In \reffi{fi:BBC1} the total cross section 
and the corresponding relative correction factor $\delta$ are given
as a function of the CM energy. 
The radiative corrections are large and negative, especially close to the
$\PW$-pair threshold.
This effect is due to real-photon ISR, which effectively 
reduces the available energy of the $\PW$-pair production subprocess,
combined with the fact, that near the $\PW$-pair threshold 
the cross section is rapidly decreasing with decreasing energy.
The large corrections result from diagrams of the bremsstrahlung process 
where the real photon is emitted from an initial-state 
electron or positron.

The DPA-Born cross section agrees very well with Fig.~8 of \citere{Be99}.
The curve for the cross section including $\Oa$ corrections
agrees for high energies, but differs for low energies.
For instance, the cross section at $165 \GeV$ is 
about $7\%$ larger than the cross section taken from Fig.~8 
of \citere{Be99}.
Above $190 \GeV$ the relative corrections agree very well
with the results of Fig.~9 in \citere{Be99}.
Note that the dominating corrections resulting from ISR
are evaluated in \citere{Be99} in DPA with on-shell phase space, 
\ie $k_\pm^2=\MW^2$, but calculated in our approach with the full
off-shell kinematics.
This could account for the deviation to \citere{Be99}.

\subsubsection{Production-angle distribution}

\begin{figure}
\centerline{
\setlength{\unitlength}{1cm}
\begin{picture}(7.5,8.5)
\put(-1,0){\includegraphics{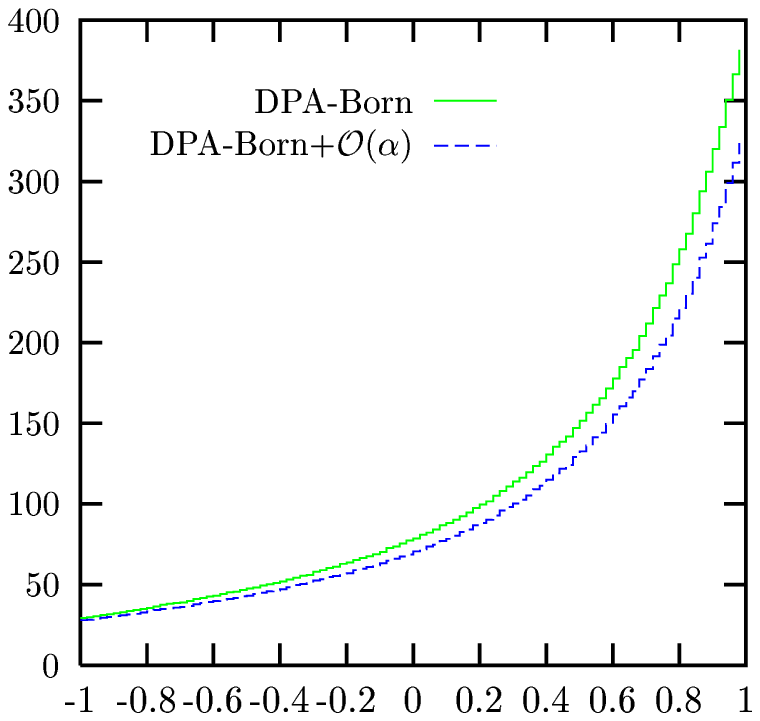}}
\put(0,7.7){\makebox(1,1)[l]
{$\frac{\rd\si}{\rd\cos \theta_{\Pep \PW^+}}/\fba$}}
\put(4,-0.3){\makebox(1,1)[cc]{{$\cos \theta_{\Pep \PW^+}$}}}
\end{picture}
\setlength{\unitlength}{1cm}
\begin{picture}(7.5,8.5)
\put(-1,0){\includegraphics{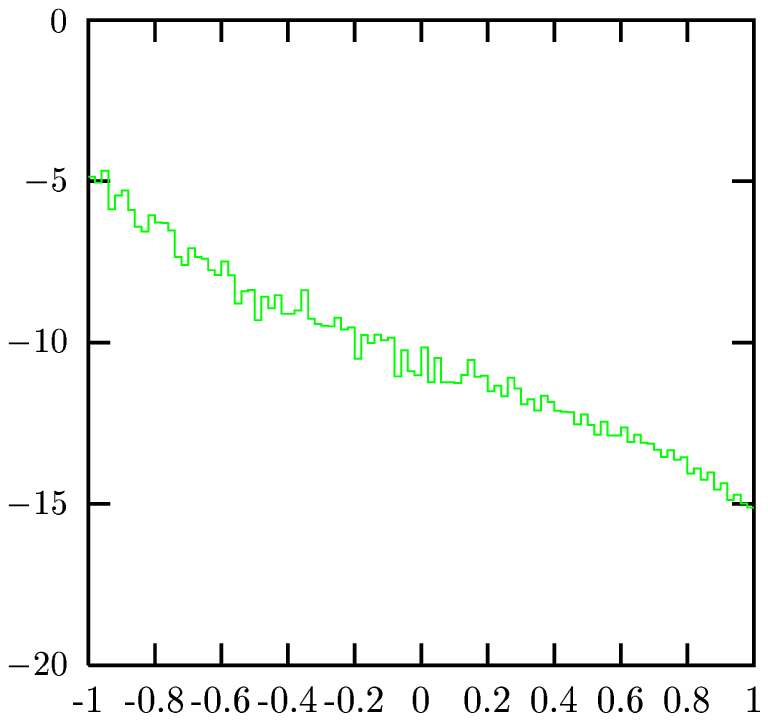}}
\put(0,7.7){\makebox(1,1)[l]{$\delta/\%$}}
\put(4,-0.3){\makebox(1,1)[cc]{{$\cos \theta_{\Pep \PW^+}$}}}
\end{picture}
}
\caption[]{Production-angle distribution for the 
CM energy $\sqrt{s}=184 \GeV$ and the process 
$\Pep \Pem \to \nu_\mu \mu^+ \tau^- \bar\nu_\tau$}
\label{fi:BBC2}
\end{figure}

The production-angle distribution is shown in \reffi{fi:BBC2}.
This distribution is, in particular, important in order to 
get more strict bound on the non-standard triple-gauge-boson couplings.
The radiative corrections are negative, and increase in size with
decreasing production angles.
The origin of the distortion of the distribution 
can be traced back to hard initial-state photonic corrections.
Hard-photon emission boosts the CM system of the $\PW$ bosons, causing a 
migration of events from regions with large cross section 
in the CM system (\eg forward direction) to regions with small 
cross section in the laboratory system (\eg backward direction).

Our result based on the DPA-Born cross section agrees very well 
with Fig.~10 of \citere{Be99}.
The relative corrections are about one per cent larger than the 
results of \citere{Be99} for small scattering angles and agree for large 
scattering angles . 
Note that the total cross section of \citere{Be99} is about one per cent 
smaller than our result at $\sqrt{s}=184 \GeV$.

\subsubsection{Decay-angle distribution}

\begin{figure}
\centerline{
\setlength{\unitlength}{1cm}
\begin{picture}(7.5,8.5)
\put(-1,0){\includegraphics{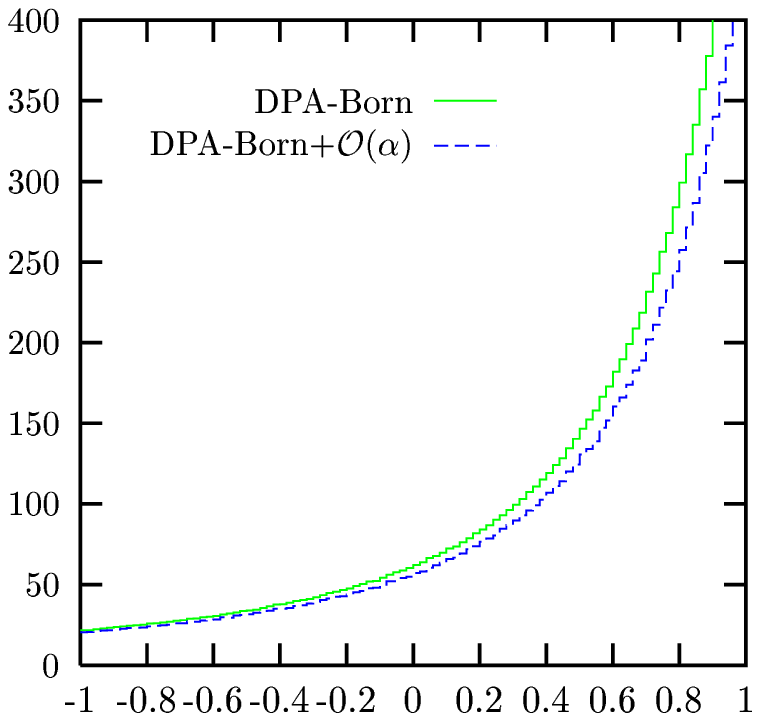}}
\put(0,7.7){\makebox(1,1)[l]
{$\frac{\rd \si}{\rd \cos \theta_{\PW^+ \mu^+}}/\fba$}}
\put(4,-0.3){\makebox(1,1)[cc]{{$\cos \theta_{\PW^+ \mu^+}$}}}
\end{picture}
\setlength{\unitlength}{1cm}
\begin{picture}(7.5,8.5)
\put(-1,0){\includegraphics{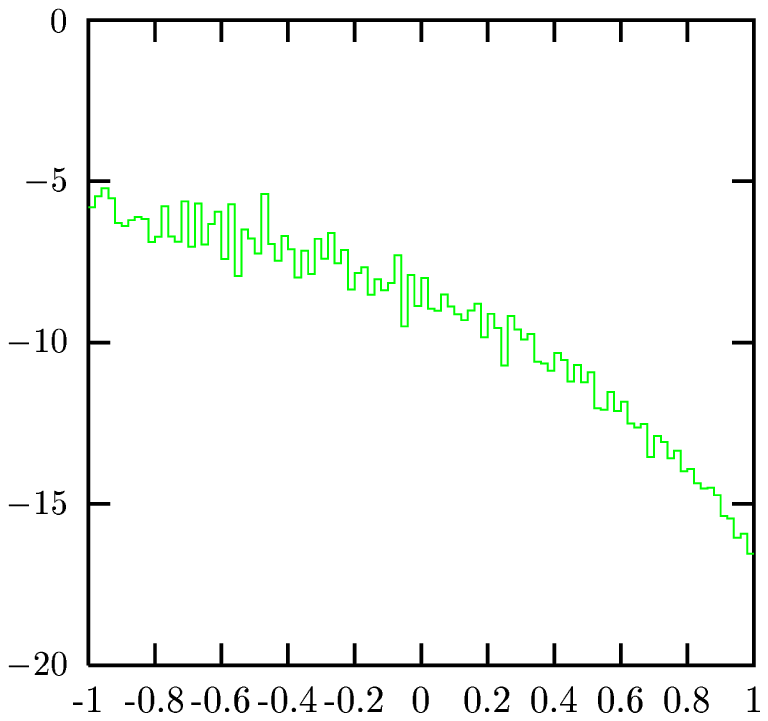}}
\put(0,7.7){\makebox(1,1)[l]{$\delta/\%$}}
\put(4,-0.3){\makebox(1,1)[cc]{{$\cos \theta_{\PW^+ \mu^+}$}}}
\end{picture}
}
\caption[]{Decay-angle distribution of the $\mu^+$ with respect 
to the $\PW^+$ in the laboratory frame for the CM 
energy $\sqrt{s}=184 \GeV$ and the process 
$\Pep \Pem \to \nu_\mu \mu^+ \tau^- \bar\nu_\tau$}
\label{fi:BBC3}
\end{figure}

The decay-angle distribution of the final-state $\mu^+$ with respect 
to the $\PW^+$ boson in the laboratory frame can be found in \reffi{fi:BBC3}.
As in the case of the  production-angle distribution, 
the large corrections are mainly due to hard-photon boost effects.
The radiative corrections are negative and large for small decay angles.

The differential cross sections of \reffi{fi:BBC3} agree very 
well with Fig.~16 of \citere{Be99}.
Deviations of about one per cent are visible from the relative
radiative corrections shown in Fig.~17 of \citere{Be99}.

\subsection{Invariant-mass distribution}

For the definition of realistic observables the momentum of collinear 
photons have to be recombined with the momentum of the nearest fermion, 
except for a muon in the final state, as discussed in \refse{se:wmass}.
Otherwise logarithms of the form $\ln (m_f^2/s)$ remain from FSR in the
invariant-mass distributions.
These logarithms are not calculable in our approach, which is 
explained in the following: 
we use the subtraction method, where all mass singularities, 
which appear for inclusive observables, are transferred 
from the real part to the virtual part of the radiative corrections 
with the help of subtraction terms and
the fermion masses are neglected everywhere, 
except for the mass singularities.
For observables, which are not inclusive in the collinear region,
like invariant-mass distributions without recombination cuts, 
the phase-space integration over the difference of the bremsstrahlung 
cross section and the corresponding 
subtraction terms in \refeq{eq:subtraction} becomes divergent
for vanishing fermion masses.
Therefore, a comparison with the invariant-mass distribution 
of \citere{Be99} is not possible, since the calculation is
performed without recombination cuts.

The following photon recombination procedure is used.
All photons within a cone of $5^\circ$ around the beams are discarded.
If the event is within the recombination cuts, the photon is recombined 
with the nearest final-state charged fermion, more precisely, with the 
fermion that has the smallest invariant mass with the photon. 
Finally, all events are discarded if a charged final-state fermion 
is within a cone of  $10^\circ$ around the beams.
The recombination cuts read:
\begin{description}
\item[recomb a:] $m(f, \ga )<5 \GeV$,
\item[recomb b:] $m(f, \ga )<25 \GeV$,
\end{description}
where $m(f,\ga)$ denotes the invariant mass of 
a final-state charged fermion and the photon.

For the invariant-mass distributions, the Born cross section is 
calculated with the complete matrix element and off-shell kinematics.
The radiative corrections are evaluated as in the previous section.

The invariant-mass distributions are shown in \reffis{fi:invariantmass184m}, 
\ref{fi:invariantmass500m}, \ref{fi:invariantmass184p}, 
and \ref{fi:invariantmass500p} for two recombination cuts and 
two CM energies:
the LEP2 energy $184 \GeV$ and a possible linear collider energy $500 \GeV$.
The results include the leptonic process 
$\Pep \Pem \to \nu_\mu \mu^+ \tau^- \bar{\nu}_\tau$,
the semi-leptonic process 
$\Pep \Pem \to \Pu \, \Pdbar \, \mu^- \bar{\nu}_\mu$, 
and the hadronic process $\Pep \Pem \to \Pu \, \Pdbar \, \Ps \, \Pcbar$.
Note that the radiative corrections of all leptonic processes 
are equivalent in DPA if no mass singularities remain from FSR.
The same is true for the semi-leptonic and hadronic process classes.

As discussed in \refse{se:CMenergy}, the large negative radiative 
corrections for $184 \GeV$ originate from hard-photon ISR, 
which reduces the available CM energy for the $\PW$-pair production 
subprocess, combined with the fact that
the total cross section of the $\PW$-pair production subprocess
is steeply decreasing for decreasing CM energy near threshold.
Since the linear-collider energy $500 \GeV$ is far away from threshold
and the cross section of the $\PW$-pair production is slowly decreasing 
for increasing energies for $\sqrt{s}\lsim 500 \GeV$, 
the relative corrections are small and positive.

\begin{figure}
\centerline{
\setlength{\unitlength}{1cm}
\begin{picture}(7.5,8.5)
\put(-1,0){\includegraphics{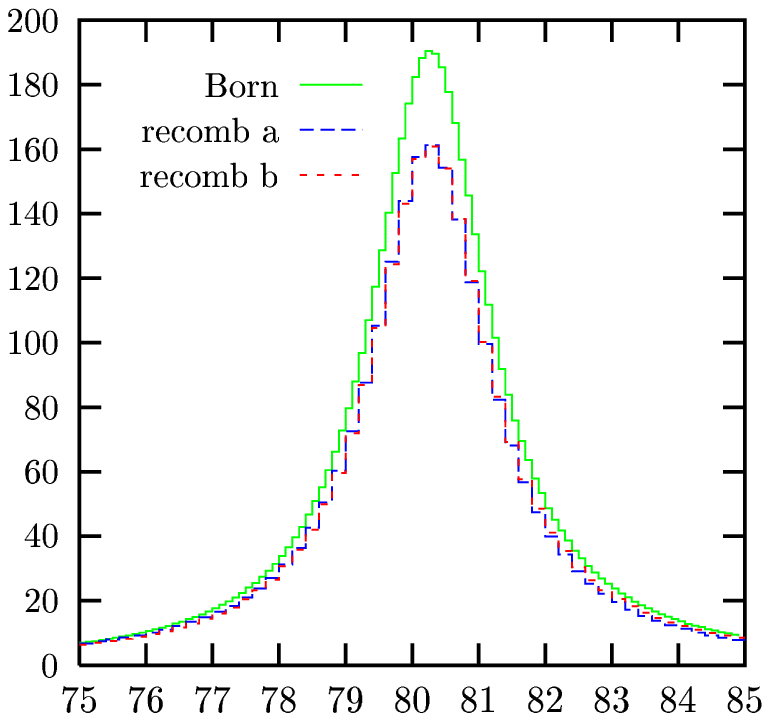}}
\put(0,7.7){\makebox(1,1)[l]{$\frac{\rd \si}{\rd M_{-}}/\frac{\fba}{\GeV}$}}
\put(4,-0.3){\makebox(1,1)[cc]{{$M_{-}/\GeV$}}}
\end{picture}
\setlength{\unitlength}{1cm}
\begin{picture}(7.5,8.5)
\put(-1,0){\includegraphics{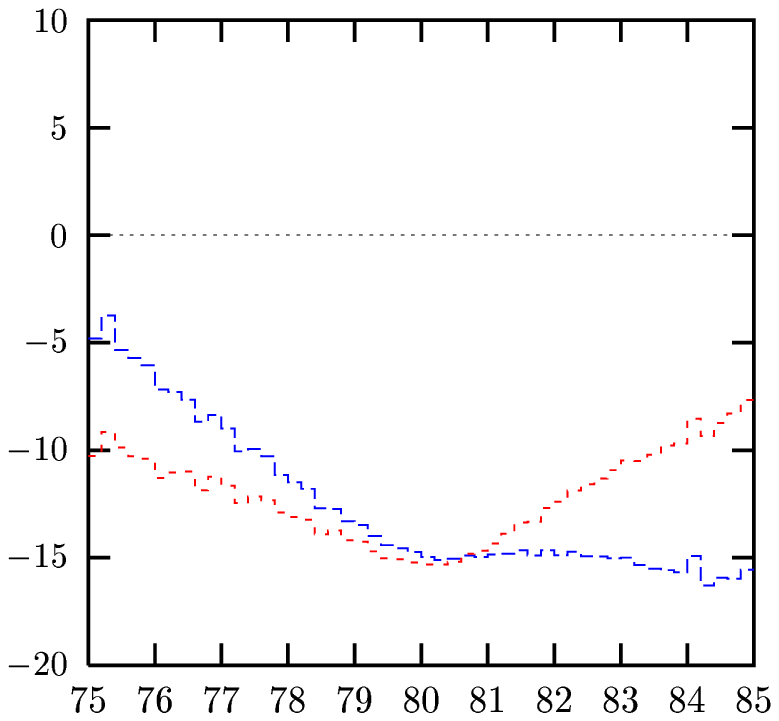}}
\put(2.4,6.6){\makebox(1,1)[l]
{$\Pep \Pem \to \Pu \, \Pdbar \, \mu^- \bar{\nu}_\mu$}}
\put(0,7.7){\makebox(1,1)[l]{$\delta/\%$}}
\put(4,-0.3){\makebox(1,1)[cc]{{$M_{-}/\GeV$}}}
\end{picture}
}
\caption[]{Invariant-mass distributions for different photon-recombination 
cuts and energy $\sqrt{s}=184 \GeV$}
\label{fi:invariantmass184m}
\end{figure}

\begin{figure}
\centerline{
\setlength{\unitlength}{1cm}
\begin{picture}(7.5,8.5)
\put(-1,0){\includegraphics{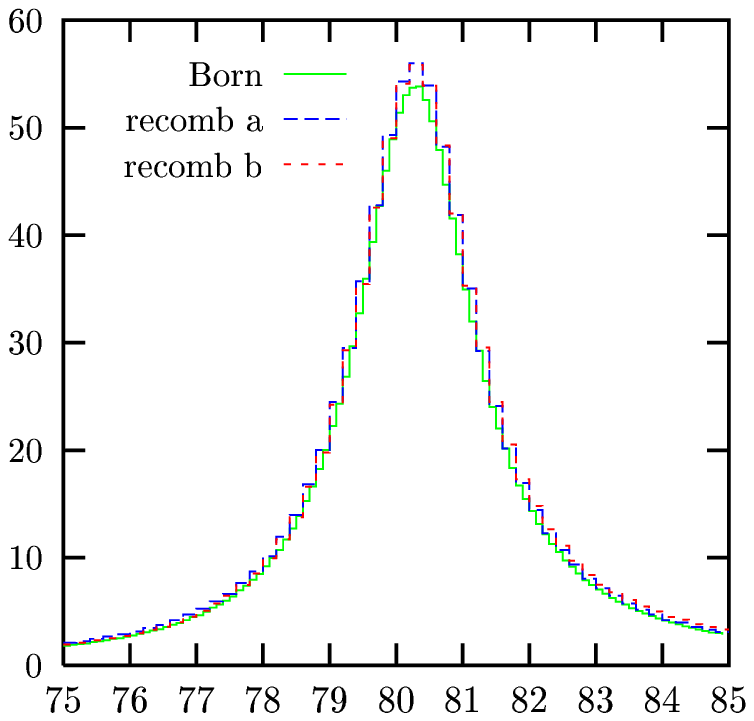}}
\put(0,7.7){\makebox(1,1)[l]{$\frac{\rd \si}{\rd M_{-}}/\frac{\fba}{\GeV}$}}
\put(4,-0.3){\makebox(1,1)[cc]{{$M_{-}/\GeV$}}}
\end{picture}
\setlength{\unitlength}{1cm}
\begin{picture}(7.5,8.5)
\put(-1,0){\includegraphics{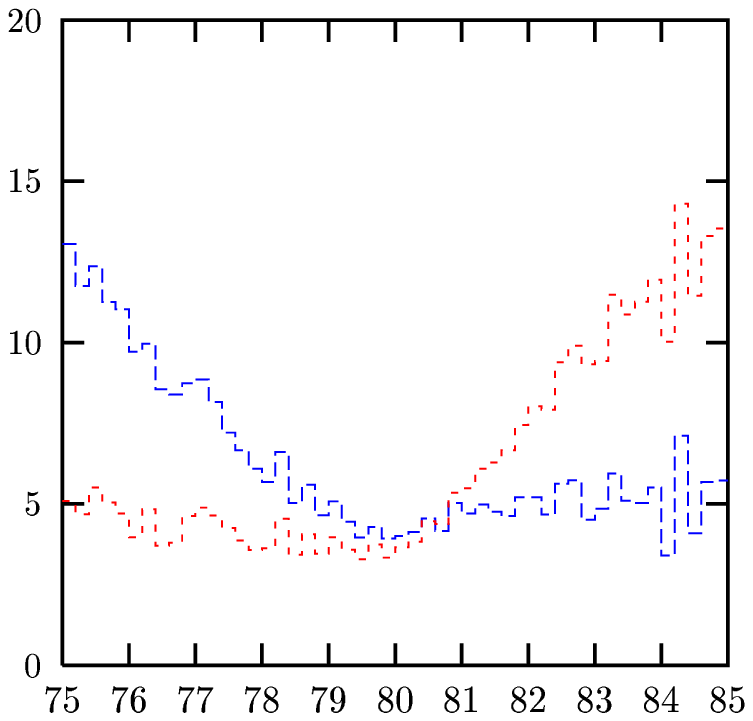}}
\put(2.4,6.6){\makebox(1,1)[l]
{$\Pep \Pem \to \Pu \, \Pdbar \, \mu^- \bar{\nu}_\mu$}}
\put(0,7.7){\makebox(1,1)[l]{$\delta/\%$}}
\put(4,-0.3){\makebox(1,1)[cc]{{$M_{-}/\GeV$}}}
\end{picture}
}
\caption[]{Invariant-mass distributions for different photon-recombination 
cuts and energy $\sqrt{s}=500 \GeV$}
\label{fi:invariantmass500m}
\end{figure}

Apart from the large reduction of the cross section for $184 \GeV$
due to ISR, a distortion of the Breit--Wigner line shape is caused by FSR, 
which can be explained as follows:
if the event is outside the recombination cuts,
the invariant $\PW$-boson masses are defined by the momenta of the 
corresponding fermion--antifermion pairs:
\begin{eqnarray}
M_+^2&=&k_+^2=(k_1+k_2)^2, \qquad M_-^2=k_-^2=(k_3+k_4)^2,
\end{eqnarray}
where $k_1,\dots ,k_4$ are the momenta of the final-state fermions.
For FSR, one of the two resonant $\PW$ bosons decays 
not only into a fermion--antifermion pair, but also in a 
bremsstrahlung photon.
Thus, the corresponding $\PW$-boson propagator, 
\begin{equation}
\label{eq:Pprop}
P\left((k_\pm+q)^2\right)=\frac{1}{[(k_\pm+q)^2-M^2]},
\end{equation}
depends in addition on the photon momentum $q$ and 
leads to a shift of the Breit--Wigner line shape
to smaller invariant masses.
These effects are especially large without recombination cuts since 
mass singularities remain from FSR \cite{Be99}.
With recombination cuts, the mass singularities disappear and are 
effectively replaced by logarithms depending on these cuts.
A small recombination cut leads to large distortions of the
Breit--Wigner resonance shape, while large recombination cuts
yield relatively small effects.

If the photon is inside the recombination cuts, it is combined with 
the nearest charged fermion and hence is included in the invariant 
mass of the corresponding resonant $\PW$ boson:
\begin{eqnarray}
\label{eq:resombM}
M_+^2&=&(k_++q)^2 \quad \mbox{or} \quad M_-^2=(k_-+q)^2.
\end{eqnarray}
The resonance of the $\PW$-boson propagator,
\begin{equation}
\label{eq:Pproprecomb}
P(k_\pm^2)=\frac{1}{[k_\pm^2-M^2]},
\end{equation}
which do not depend on the photon momentum,
is located at larger invariant masses, \ie $M_{\pm}>\MW$, where $M_{\pm}$ is
defined in \refeq{eq:resombM}.
This leads to a shift of the Breit--Wigner line shape 
to larger invariant masses.

These effects all visible in \reffis{fi:invariantmass184m}, 
\ref{fi:invariantmass500m}, \ref{fi:invariantmass184p}, 
and \ref{fi:invariantmass500p}.
For $M_\pm\lsim \MW$, 
the relative corrections increase for decreasing
invariant masses due to propagators defined in \refeq{eq:Pprop}
and events, which are outside the recombination cuts.
This effect is large for small recombination cuts, as 
expected from the previous discussion.
On the other hand, for $M_\pm\gsim \MW$
the relative corrections increase for increasing invariant mass
owing to the propagators defined in \refeq{eq:Pproprecomb}
and events, where the photon is recombined with the nearest fermion.
This effect is visible for the large recombination cut b.
Both effects result in a shift and a broadening of the Breit--Wigner 
line shape.

Note that the radiative corrections from FSR include the squares
of the charges of the final-state fermions, which emit the bremsstrahlung 
photons. 
Thus, the FSR corrections are proportional to $Q_{l^\pm}^2=1$
for invariant-mass distributions of $\PW$-bosons, 
which decay into leptons, and proportional to $Q_q ^2 +Q_{\bar q}^2=5/9$, 
if the $\PW$-boson decays into a quark--antiquark pair.

\begin{figure}
\centerline{
\setlength{\unitlength}{1cm}
\begin{picture}(7.5,8)
\put(-1,0){\includegraphics{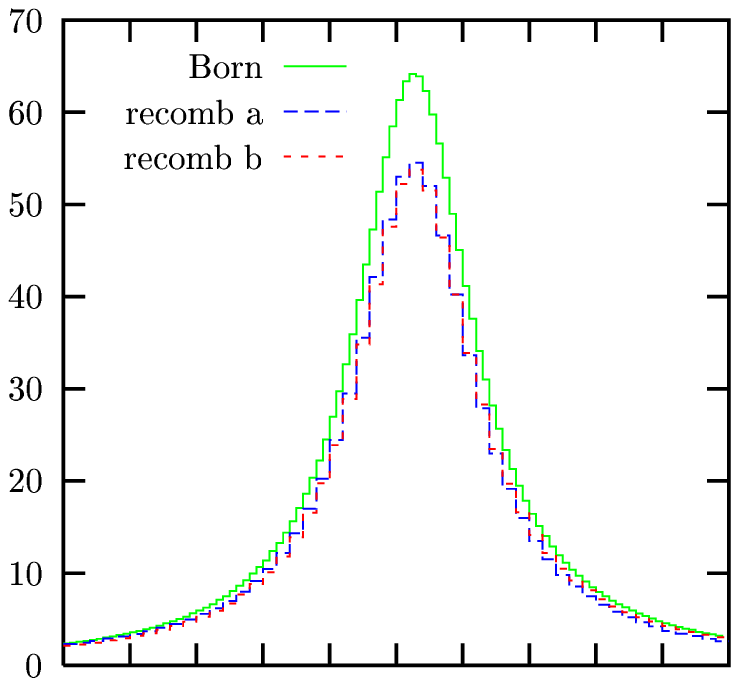}}
\put(0,7.7){\makebox(1,1)[l]{$\frac{\rd \si}{\rd M_{+}}/\frac{\fba}{\GeV}$}}
\end{picture}
\setlength{\unitlength}{1cm}
\begin{picture}(7.5,8)
\put(-1,0){\includegraphics{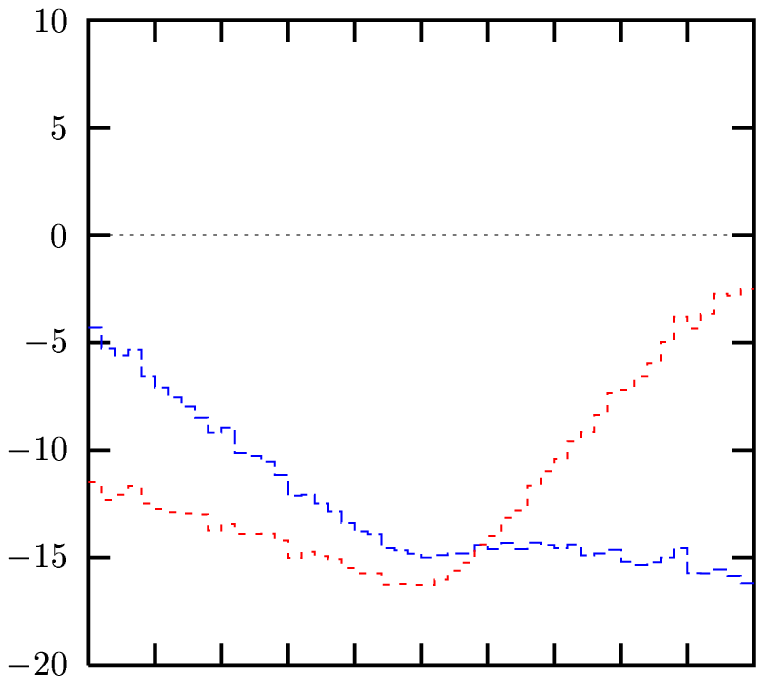}}
\put(2.4,6.6){\makebox(1,1)[l]
{$\Pep \Pem \to \nu_\mu \mu^+ \tau^- \bar{\nu}_\tau$}}
\put(0,7.7){\makebox(1,1)[l]{$\delta/\%$}}
\end{picture}
}
\centerline{
\setlength{\unitlength}{1cm}
\begin{picture}(7.5,6.8)
\put(-1,0){\includegraphics{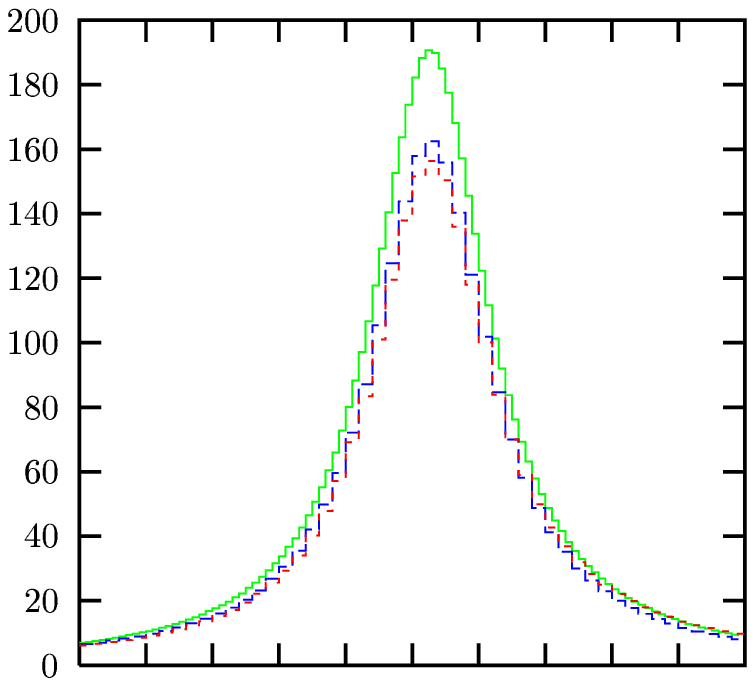}}
\end{picture}
\setlength{\unitlength}{1cm}
\begin{picture}(7.5,7)
\put(-1,0){\includegraphics{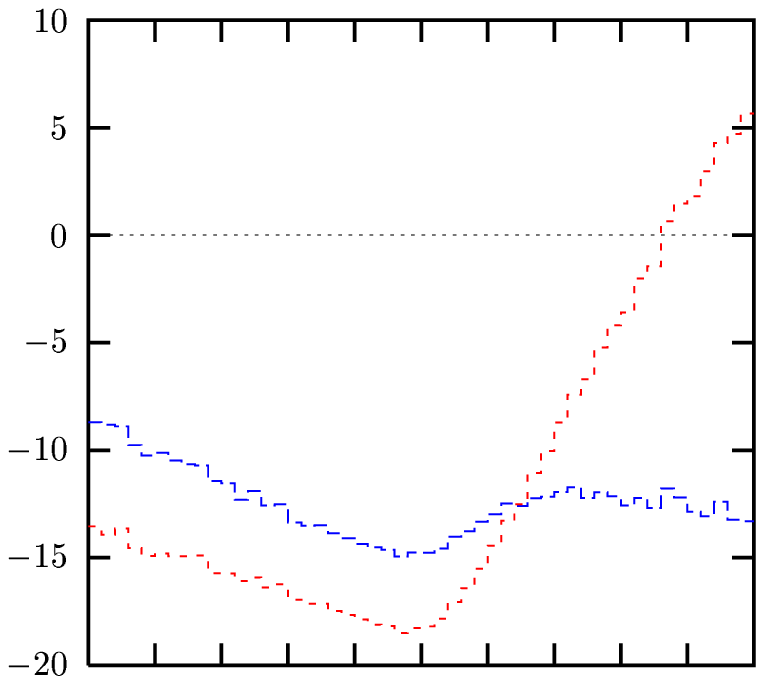}}
\put(2.4,6.6){\makebox(1,1)[l]
{$\Pep \Pem \to \Pu \, \Pdbar \, \mu^- \bar{\nu}_\mu$}}
\end{picture}
}
\centerline{
\setlength{\unitlength}{1cm}
\begin{picture}(7.5,7)
\put(-1,0){\includegraphics{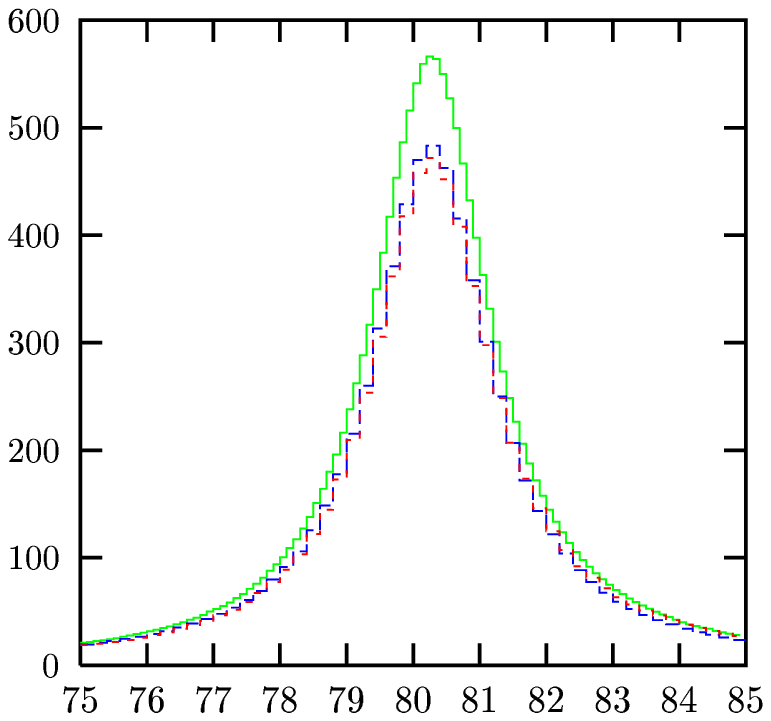}}
\put(4,-0.3){\makebox(1,1)[cc]{{$M_{+}/\GeV$}}}
\end{picture}
\setlength{\unitlength}{1cm}
\begin{picture}(7.5,7)
\put(-1,0){\includegraphics{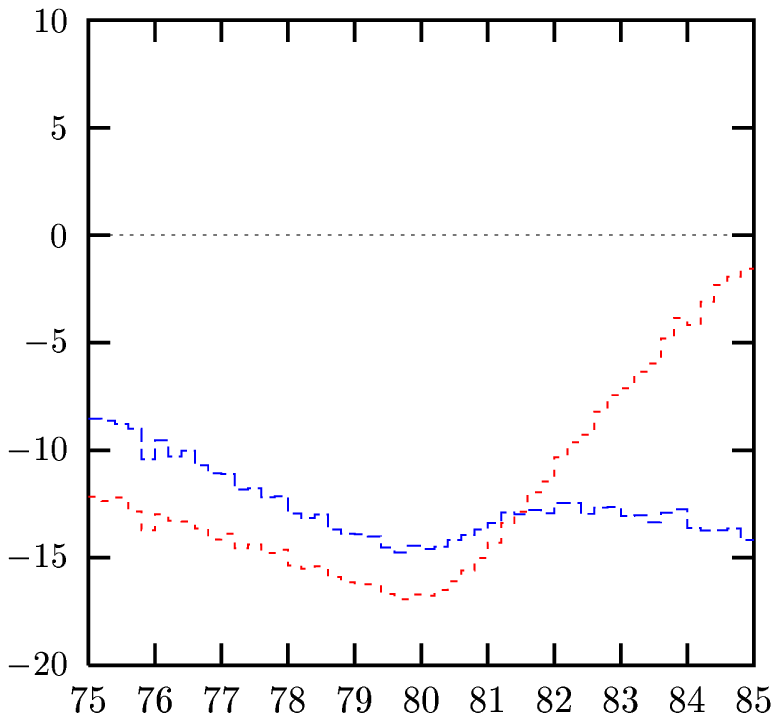}}
\put(2.4,6.6){\makebox(1,1)[l]{$\Pep \Pem \to \Pu \, \Pdbar \, \Ps \, \Pcbar$}}
\put(4,-0.3){\makebox(1,1)[cc]{{$M_{+}/\GeV$}}}
\end{picture}
}
\caption[]{Invariant-mass distributions for different photon-recombination 
cuts, different processes, and energy $\sqrt{s}=184 \GeV$}
\label{fi:invariantmass184p}
\end{figure}

\begin{figure}
\centerline{
\setlength{\unitlength}{1cm}
\begin{picture}(7.5,8)
\put(-1,0){\includegraphics{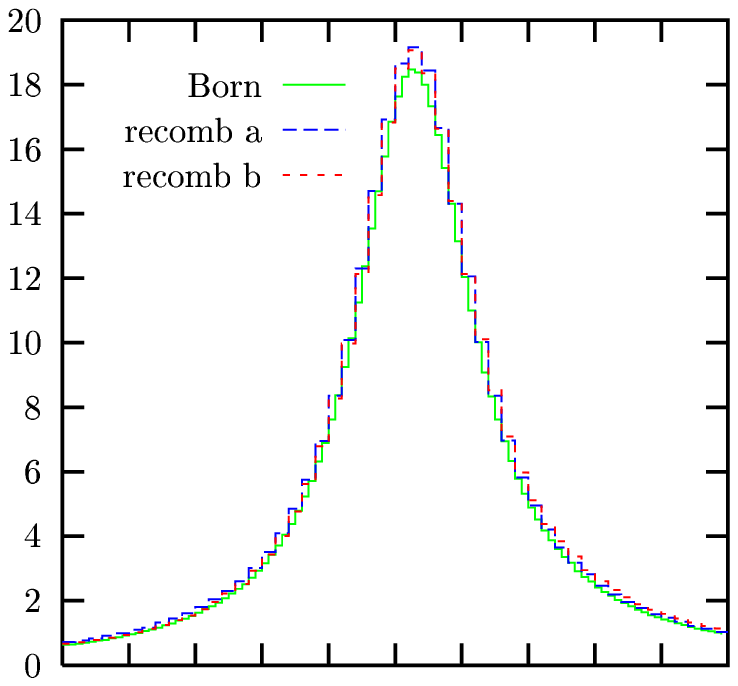}}
\put(0,7.7){\makebox(1,1)[l]{$\frac{\rd \si}{\rd M_{+}}/\frac{\fba}{\GeV}$}}
\end{picture}
\setlength{\unitlength}{1cm}
\begin{picture}(7.5,8)
\put(-1,0){\includegraphics{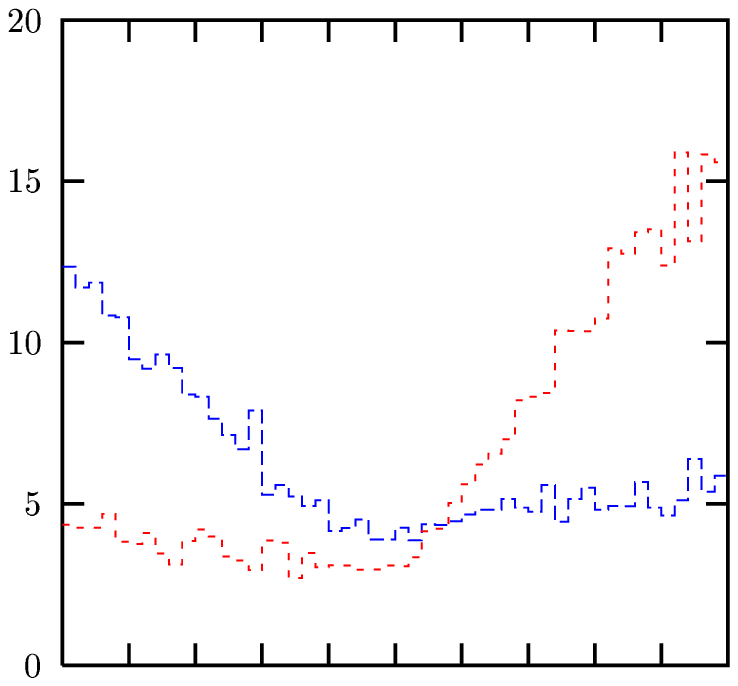}}
\put(2.4,6.6){\makebox(1,1)[l]
{$\Pep \Pem \to \nu_\mu \mu^+ \tau^- \bar{\nu}_\tau$}}
\put(0,7.7){\makebox(1,1)[l]{$\delta/\%$}}
\end{picture}
}
\centerline{
\setlength{\unitlength}{1cm}
\begin{picture}(7.5,6.8)
\put(-1,0){\includegraphics{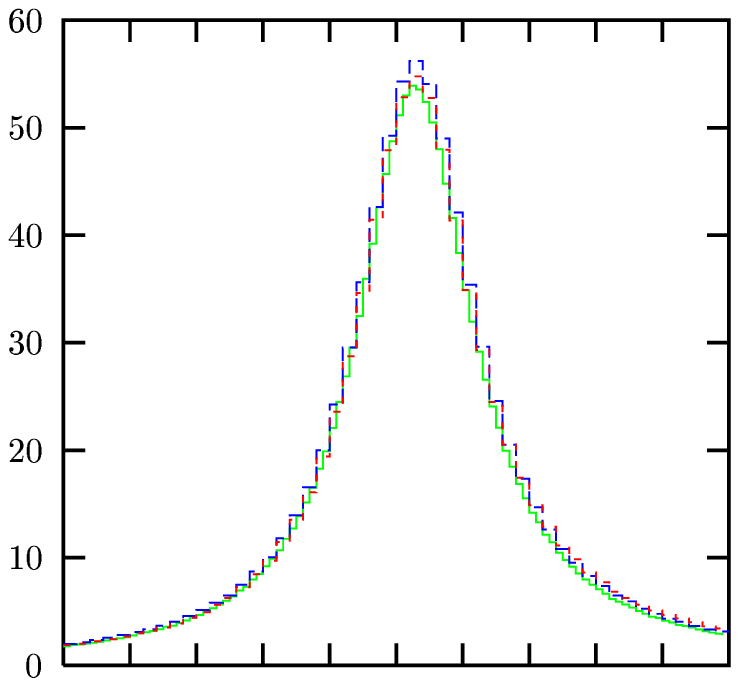}}
\end{picture}
\setlength{\unitlength}{1cm}
\begin{picture}(7.5,7)
\put(-1,0){\includegraphics{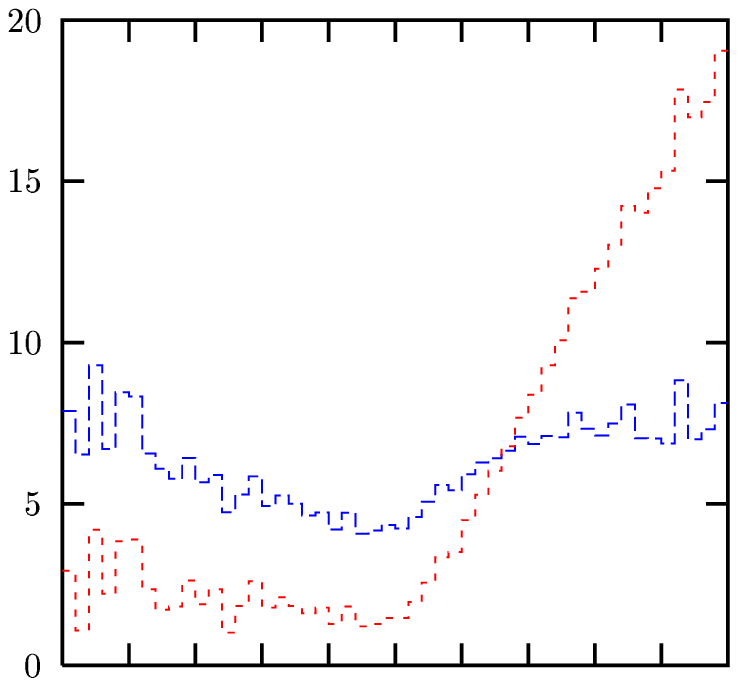}}
\put(2.4,6.6){\makebox(1,1)[l]
{$\Pep \Pem \to \Pu \, \Pdbar \, \mu^- \bar{\nu}_\mu$}}
\end{picture}
}
\centerline{
\setlength{\unitlength}{1cm}
\begin{picture}(7.5,7)
\put(-1,0){\includegraphics{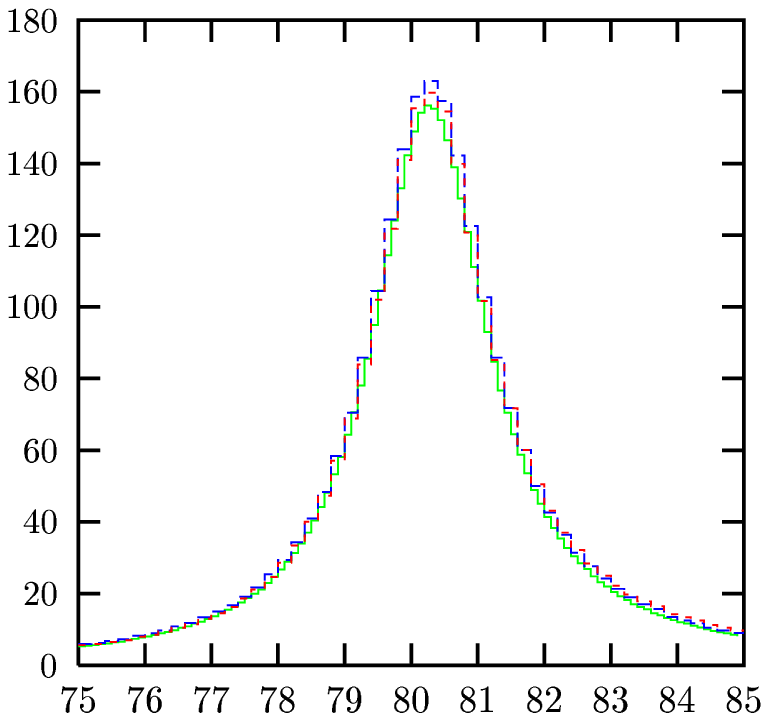}}
\put(4,-0.3){\makebox(1,1)[cc]{{$M_{+}/\GeV$}}}
\end{picture}
\setlength{\unitlength}{1cm}
\begin{picture}(7.5,7)
\put(-1,0){\includegraphics{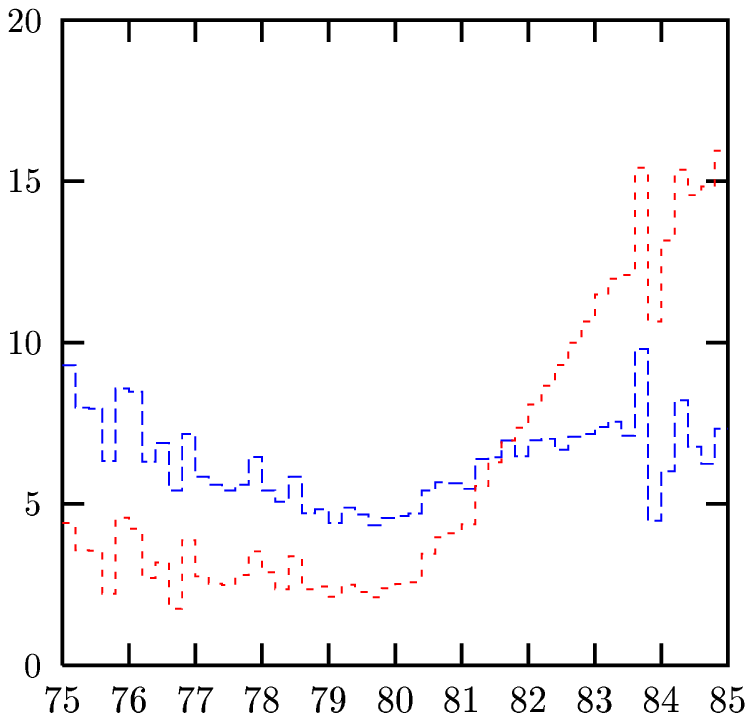}}
\put(2.4,6.6){\makebox(1,1)[l]{$\Pep \Pem \to \Pu \, \Pdbar \, \Ps \, \Pcbar$}}
\put(4,-0.3){\makebox(1,1)[cc]{{$M_{+}/\GeV$}}}
\end{picture}
}
\caption[]{Invariant-mass distributions for different photon-recombination 
cuts, different processes, and energy $\sqrt{s}=500 \GeV$}
\label{fi:invariantmass500p}
\end{figure}

\subsection{Test of the subtraction method}
\label{se:testsub}

\begin{table}
\newdimen\digitwidth
\setbox0=\hbox{0}
\digitwidth=\wd0
\catcode`!=\active
\def!{\kern\digitwidth}
\begin{center}
{\begin{tabular}{|c|c|c|c|c|}
\hline
$\sqrt{s}/\GeV $ & $189$ & $500$ & $2000$ & $10000$ \\
\hline
$\si/\fba$ & $1284.9(5)$ & $570.4(3)$ & $83.4(6)$ & $6.52(1)$ \\
\hline
\end{tabular}}
\end{center}
\caption[]{Comparison with Table 3 of \citere{Je99}:
Total cross section of the process 
$\Pep \Pem \to \Pu \, \Pdbar \, \mu^- \bar\nu_\mu \ga$
with the photon mass $m_\ga=10^{-6}\GeV$
}
\label{ta:Je99}
\end{table}

In Table 3 of \citere{Je99}, the sum of the soft-photon cross section with 
$E_\ga<\omega$ and 
the hard-photon cross section with photon energy $E_\ga>\omega$ of the
process $\Pep \Pem \to \Pu \, \Pdbar \mu^- \bar\nu_\mu \ga$ is given
for several values of the separation cut $\omega$,
in order to show the independence on the parameter $\omega$. 
Since the soft-photon cross section depends on the photon mass
$m_\ga=10^{-6}\GeV$, it is not a physical observable.
However, it is a good test for the implementation of the subtraction method.

In \citere{Je99}, the fixed-width scheme and the following 
input parameters have been used:
$\MW=80.23\GeV$, $\GW=2.085\GeV$, $\MZ=91.1888\GeV$, $\GZ=2.4974\GeV$,
$m_\Pe=0.51099906\MeV$, $m_\Pu=5\MeV$, $m_\Pd=10\MeV$, and $\sw^2=0.22591$.
All couplings have been parameterized by $\al(\MW)=1/128.07$, 
except for the couplings of the bremsstrahlung photon which have been
parameterized by $\al=1/137.0359895$.

Our results for these input parameters are shown in \refta{ta:Je99} and
agree well within the statistical error, apart from the CM energy $10 \TeV$,
where the numerical integration is most complicated.

\begin{appendix}

\chapter*{Appendix}
\addcontentsline{toc}{chapter}{Appendix}

\section{Calculation of real and virtual corrections}
\label{ap:integrals}

\subsection{Useful definitions}
\label{prelim}

In the main text, we have already used the following short-hand expressions,
\beqar\label{abbrev}
\beta_\PW &=& \sqrt{1-\frac{4\MW^2}{s}+\ri\epsilon},
\nl
x_\PW &=& \frac{\beta_\PW-1}{\beta_\PW+1},
\nl
\beta &=&  \sqrt{1-\frac{4M^2}{s}},\nl
\bar\beta &=& \frac{\sqrt{\lambda(s,k_+^2,k_-^2)}}{s},\nl
\kappa_\PW &=& \sqrt{ \lambda(\MW^4,s_{13}s_{24},s_{14}s_{23})
        -\ri\epsilon},\nl
\Delta_M &=& \frac{|k_+^2-k_-^2|}{s}, 
\eeqar
where 
\beq
\label{eq:lambda}
\lambda(x,y,z) = x^2+y^2+z^2-2xy-2xz-2yz.
\eeq
The evaluation of one-loop $n$-point functions naturally leads
to the usual dilogarithm,
\beq
\Li(z) = -\int_0^z\,\frac{\rd t}{t}\,\ln(1-t),
\qquad |\arc(1-z)|<\pi, 
\eeq
and its analytically continued form
\beqar
\cLi(x,y) &=& \Li(1-xy)+[\ln(xy)-\ln(x)-\ln(y)]\ln(1-xy),
\nl && 
|\arc(x)|,|\arc(y)|<\pi. 
\eeqar

\subsection{Calculation of the real bremsstrahlung integrals}
\label{appbrcal}

In this appendix we describe a general method for evaluating the
bremsstrahlung 3- and 4-point integrals defined in \refeq{realCDE0}. 
We make use of the generalized Feynman-parameter representation \cite{Fpar}
\begin{equation}\label{eq:FeynPar}
\frac{1}{\prod_{i=1}^n N'_i} =
\Gamma(n) \int_0^\infty \rd x_1 \cdots \rd x_n  
\frac{\delta (1- \sum_{i=1}^n \alpha_i x_i)}{(\sum_{i=1}^n N'_i x_i)^n},
\end{equation} 
where the real variables $\alpha_i\ge0$ are arbitrary, but not all 
equal to zero simultaneously. The sum $\sum_{i=1}^n N'_i x_i$ must be 
non-zero over the entire integration domain.

We first consider the 3-point integral $\Cbr_0$. We only need the difference 
between the general IR-finite integral and the corresponding
IR-divergent on-shell integral,
\beqar\label{eq:rc0diff}
\lefteqn{\Cbr_0(p_1,p_2,0,m_1,m_2) - 
\Cbr_0\Bigl(p_1,p_2,\la,\sqrt{p_1^2},\sqrt{p_2^2}\Bigr)} \nn\\
&=& \int\!\frac{\rd^3{\bf q}}{\pi q_0}\,
\biggl\{\frac{1}{(2p_1q+p_1^2-m_1^2)(2p_2q+p_2^2-m_2^2)}
-\frac{1}{(2p_1q)(2p_2q)}\biggr\}\bigg|_{q_0=\sqrt{{\bf q}^2+\lambda^2}},
\eeqar
which is UV-finite and Lorentz-invariant. Extracting the signs 
$\sigma_i$ of $p_{i0}$ via definition \refeq{eq:momredef}, and assuming
for the moment $\sigma_i\Re(p_i^2-m_i^2)>0$, the Feynman-parameter
representation \refeq{eq:FeynPar} can be applied to each term in the
integrand of \refeq{eq:rc0diff}. The integration over $\rd^3{\bf q}$ 
can be carried out and yields 
\beqar\label{eq:rc0diffFP}
\lefteqn{\Cbr_0(p_1,p_2,0,m_1,m_2) -
\Cbr_0\Bigl(p_1,p_2,\la,\sqrt{p_1^2},\sqrt{p_2^2}\Bigr)} \\
&=& \sigma_1\sigma_2\int_0^\infty\rd x_1\rd x_2\,
\de\biggl(1-\sum_{i=1}^2\alpha_i x_i\biggr)
\frac{\disp \ln(\lambda^2)
+\ln\biggl[\biggl(\sum_{i=1}^2\tilde p_i x_i\biggr)^2\biggr]
-2\ln\biggl[\sum_{i=1}^2\sigma_i(p_i^2-m_i^2)x_i\biggr]}
{\disp 2\biggl(\sum_{i=1}^2\tilde p_i x_i\biggr)^2}. \nn
\eeqar
Putting for instance $\al_1=0$ and $\al_2=1$, and using
\beq
\int_0^\infty \rd x \left(\frac{1}{x+z_1}-\frac{1}{x+z_2}\right)
\ln(1+z x)=
\cLi(z_1,z)-\cLi(z_2,z), \nl
\eeq
where $|\arc(z_{1,2},z,z\,z_{1,2})|<\pi$,
the remaining one-dimensional integration and the analytical
continuations to arbitrary complex $(p_i^2-m_i^2)$ are straightforward.

Next we consider the IR-finite 4-point function
\beqar
\lefteqn{\Dbr_0(p_1,p_2,p_3,0,m_1,m_2,m_3)} \nn\\
&=& \left.\int\!\frac{\rd^{3}{\bf q}}{\pi q_0}\,
\frac{1}{(2p_1q+p_1^2-m_1^2)(2p_2q+p_2^2-m_2^2)(2p_3q+p_3^2-m_3^2)}
\right|_{q_0=|{\bf q}|}.
\eeqar
For $\sigma_i\Re(p_i^2-m_i^2)>0$ we can proceed as above and find
\beq\label{eq:rd0diffFP}
\Dbr_0(p_1,p_2,p_3,0,m_1,m_2,m_3) 
= \sigma_1\sigma_2\sigma_3\int_0^\infty\rd x_1\rd x_2\rd x_3\,
\frac{\disp \de\biggl(1-\sum_{i=1}^3\alpha_i x_i\biggr)}
{\disp \biggl[\sum_{i=1}^3\sigma_i(p_i^2-m_i^2)x_i\biggr]
\biggl(\sum_{i=1}^3\tilde p_i x_i\biggr)^2}.
\eeq
Again, the two-dimensional integration over the Feynman parameters
and the analytical continuations in $(p_i^2-m_i^2)$ are straightforward.

Finally, we inspect the IR-divergent 4-point function
\beq
\Dbr_0\Bigl(p_1,p_2,p_3,\la,\sqrt{p_1^2},m_2,\sqrt{p_3^2}\Bigr) =
\left.\int\!\frac{\rd^{3}{\bf q}}{\pi q_0}\,
\frac{1}{(2p_1q)(2p_2q+p_2^2-m_2^2)(2p_3q)}
\right|_{q_0=\sqrt{{\bf q}^2+\la^2}}.
\eeq
Instead of applying the Feynman-parameter representation directly, it is
more convenient to extract the infrared singularity by subtracting and adding
an IR-divergent 3-point function containing the same IR structure. Since
the difference between
the 4- and 3-point functions is IR-finite, we can
regularize the IR divergence in these two integrals in a more convenient
way. Therefore, we write
\beqar
\lefteqn{\Dbr_0\Bigl(p_1,p_2,p_3,\la,\sqrt{p_1^2},m_2,\sqrt{p_3^2}\Bigr)
= \lim\limits_{m^2\rightarrow p_1^2}
\bigg[\Dbr_0\Bigl(p_1,p_2,p_3,0,m,m_2,\sqrt{p_3^2}\Bigr)}
\nn\\
&& 
-\frac{1}{p_2^2-m_2^2}\Cbr_0\Bigl(p_1,p_3,0,m,\sqrt{p_3^2}\Bigr)
+\frac{1}{p_2^2-m_2^2}\Cbr_0\Bigl(p_1,p_3,\la,\sqrt{p_1^2},\sqrt{p_3^2}\Bigr)
\bigg],
\eeqar
\ie we regularize the IR divergence in the 4-point function by the
off-shellness $p_1^2-m^2\ne 0$. Both the 4-point function as well as the
difference of 3-point functions can be treated as above, yielding
straightforward Feynman-parameter integrals according to 
\refeq{eq:rc0diffFP} and \refeq{eq:rd0diffFP}, respectively. 
The limit $m^2\to p_1^2$
can easily be taken after the integrations have been performed.

\subsection{Explicit results for scalar integrals}
\label{appscalint}

\subsubsection{Loop integrals in double-pole approximation}
\label{appvirtual}

The following one-loop integrals are  required for the calculation of 
the non-factorizable corrections in \refses{gfcf} and \ref{se:vnfc}.
The 2-point and 3-point functions of the (\mm), (\mf), and (\im) 
corrections read in DPA:
\beqar
\lefteqn{\frac{1}{k_+^2-M^2}\left\{B_0(k_+,0,M)
- \Bigl[B_0\left(k_+,0,M\right)\Bigr]_{k_+^2=M^2} \right\}
-\Bigl[B_0^\prime \left(k_+,\la,\MW\right)\Bigr]_{k_+^2=\MW^2}}
\quad\nn\\ {}
&\sim&\frac{1}{\MW^2}\biggl\{ \ln\biggl(\frac{\lambda\MW}{-K_+}\biggr)+1
\biggr\},
\label{eq:B0}
\\[.5em]
\lefteqn{C_0(k_2,k_+,0,m_2,M)
-\Bigl[C_0(k_2,k_+,\lambda,m_2,\MW)\Bigr]_{k_+^2=\MW^2}}
\quad\nn\\
&\sim&
-\frac{1}{\MW^2} \biggl\{ 
\ln\biggl(\frac{m_2^2}{\MW^2}\biggr)
\ln\biggl(\frac{-K_+}{\lambda\MW}\biggr)
+\ln^2\biggl(\frac{m_2}{\MW}\biggr) + \frac{\pi^2}{6} \biggr\},
\\[.5em] 
\lefteqn{C_0(p_+,k_+,0,\Me,M)
-\Bigl[C_0(p_+,k_+,\lambda,\Me,\MW)\Bigr]_{k_+^2=\MW^2}}
\quad\nn\\
&\sim&
\frac{1}{t-\MW^2} \biggl\{ 
\ln\biggl(\frac{\Me\MW}{\MW^2-t}\biggr)
\biggl[ \ln\biggl(\frac{-K_+}{\MW^2-t}\biggr)
+\ln\biggl(\frac{-K_+}{\lambda^2}\biggr)
+\ln\biggl(\frac{\Me}{\MW}\biggr) \biggr] + \frac{\pi^2}{6} \biggr\}.\quad
\eeqar

In the (\mmp) correction the following
combination of 3-point functions appears in the strict DPA:
\beqar
\label{dC0virt}
\lefteqn{ C_0(k_+,-k_-,0,M,M) -
\Bigl[C_0(k_+,-k_-,\la,\MW,\MW)\Bigr]_{k_\pm^2=\MW^2} }
\quad\nn\\
&\sim& \frac{1}{s\beta_\PW}\biggl\{
-\cLi\biggl(\frac{K_-}{K_+},x_\PW\biggr)
+\cLi\biggl(\frac{K_-}{K_+},x_\PW^{-1}\biggr)
+\Li(1-x_\PW^2)+\pi^2 +\ln^2(-x_\PW)
\nn\\ && {}
+2\ln\biggl(\frac{K_+}{\lambda\MW}\biggr)\ln(x_\PW) -2\pi\ri\ln(1-x_\PW^2)
\biggr\}.
\eeqar
In order to include the full off-shell Coulomb singularity, one has to
add
\beq
 \frac{2\pi\ri}{s \betap} \ln\biggl(
    \frac{\betaM+\Delta_M-\betap}{\betaM+\Delta_M+\betap}\biggr) 
-  \frac{2\pi\ri}{s \betaW} \ln\biggl(
    \frac{K_+ + K_- + s\betaW\Delta_M}{2\betaW^2s}\biggr) 
\eeq
to the r.h.s.\ of \refeq{dC0virt}.

For the (\ifp), (\mfp), and (\ffp) corrections we need the 
following 4-point functions:
\beqar \label{D00}
\nn
\lefteqn{D_0(p_+,k_+,k_2,\lambda,\Me,M,m_2)
\sim\frac{1}{t_{+ 2}K_+} \biggl\{
2\ln\biggl(\frac{\Me m_2}{-t_{+ 2}}\biggr)
\ln\biggl(\frac{\lambda\MW}{-K_+}\biggr)
-\ln^2\biggl(\frac{m_2}{\MW}\biggr)
-\frac{\pi^2}{3}}\\
\lefteqn{\phantom{D_0(p_+,k_+,k_2,\lambda,\Me,M,m_2)
\sim\frac{1}{t_{+ 2}K_+} \biggl\{}
-\ln^2\biggl(\frac{\Me\MW}{\MW^2-t}\biggr)
-\Li\biggl(1-\frac{t-\MW^2}{t_{\pm i}}\biggr)
\biggr\},}\\
\lefteqn{ D_0(0) = D_0(-k_4,k_+ +k_3,k_2+k_3,0,M,M,0) }
\quad\nn\\
&\sim& \frac{1}{\kappa_\PW} \sum_{\si=1,2} (-1)^\si \biggl\{
 \cLi\biggl(-\frac{s_{13}+s_{23}}{\MW^2}-\ri\epsilon,-x_\si\biggr)
+\cLi\biggl(-\frac{\MW^2}{s_{23}+s_{24}}+\ri\epsilon,-x_\si\biggr)
\nn\\ && \qquad {}
-\cLi\biggl(x_\PW,-x_\si\biggr)
-\cLi\biggl(x_\PW^{-1},-x_\si\biggr)
-\ln\biggl(1+\frac{s_{24}}{s_{23}}\biggr)\ln(-x_\si)
\biggr\},
\\[.5em] &&
\mbox{with} \qquad
x_1 = \frac{s_{24}z}{\MW^2}, \quad x_2 = \frac{\MW^2}{s_{13}z}, \quad
z = \frac{\MW^4+s_{13}s_{24}-s_{14}s_{23}+\kappa_\PW}{2s_{13}s_{24}},
\label{eq:xtilde}
\\[1em]
\lefteqn{ D_0(1) = D_0(-k_-,k_+,k_2,0,M,M,m_2) }
\quad\nn\\
&\sim& \frac{1}{K_+(s_{23}+s_{24})+K_-\MW^2} \biggl\{
\sum_{\tau=\pm 1} \biggl[
 \cLi\biggl(\frac{K_+}{K_-},x_\PW^\tau\biggr)
-\cLi\biggl(-\frac{\MW^2}{s_{23}+s_{24}}+\ri\epsilon,x_\PW^\tau\biggr)
\biggr]
\\ && {}
-2\cLi\biggl(\frac{K_+}{K_-},-\frac{s_{23}+s_{24}}{\MW^2}-\ri\epsilon\biggr)
-\ln\biggl(\frac{m_2^2}{\MW^2}\biggr) \biggl[
\ln\biggl(\frac{K_+}{K_-}\biggr)
+\ln\biggl(-\frac{s_{23}+s_{24}}{\MW^2}-\ri\epsilon\biggr) \biggr]
\biggr\},
\hspace{1.8em}
\nn\\[1em]
\lefteqn{ D_0(2) = D_0(-k_3,k_+,k_2,\la,m_3,M,m_2) }\quad
\nn\\*
&\sim& \frac{1}{K_+s_{23}} \biggl\{
-\Li\biggl(-\frac{s_{13}}{s_{23}}\biggr)
-\frac{\pi^2}{3}
+2\ln\biggl(-\frac{s_{23}}{m_2 m_3}-\ri\epsilon\biggr)
\ln\biggl(\frac{-K_+}{\la\MW}\biggr)
-\ln^2\biggl(\frac{m_2}{\MW}\biggr)
\nn\\ && {}
-\ln^2\biggl(-\frac{s_{13}+s_{23}}{m_3\MW}-\ri\epsilon\biggr)
\biggr\},
\\[1em]
\lefteqn{D_0(3) = D_0(-k_3,-k_-,k_2,\la,m_3,M,m_2) 
= D_0(2) \Big|_{K_+\leftrightarrow K_-,m_2\leftrightarrow m_3,
s_{13}\leftrightarrow s_{24}},
}\quad
\\[1em]
\lefteqn{ D_0(4) = D_0(-k_3,-k_-,k_+,0,m_3,M,M) 
= D_0(1) \Bigr|_{K_+\leftrightarrow K_-,m_2\leftrightarrow m_3,
s_{13}\leftrightarrow s_{24}}.
}\quad
\eeqar

\subsubsection{Bremsstrahlung integrals in double-pole approximation}
\label{appreal}

In the (\mmp) interference corrections the following
combination of 3-point functions appears:
\beqar\label{C0br}
\lefteqn{ \Cbr_0(k_+,k_-,0,M,M^*) -
\Bigl[\Cbr_0(k_+,k_-,\la,\MW,\MW)\Bigr]_{k_\pm^2=\MW^2} }
\quad\nn\\
&\sim& 
\biggl\{C_0(k_+,-k_-,0,M,M)- \Bigl[
C_0(k_+,-k_-,\la,\MW,\MW)\Bigr]_{k_\pm^2=\MW^2}\biggr\}\bigg|_{K_-\to -K_-^*}
\nn\\ && {}
- \frac{2\pi\ri}{s\beta_\PW}
\ln\biggl[\frac{K_+ +K_-^*x_\PW}{\ri\lambda\MW(1-x_\PW^2)}\biggr].
\eeqar

For the (\mfp) and (\ffp) interference corrections the
following 4-point functions are required:
\beqar\label{Dbr00}
\lefteqn{\tilde\Dbr_0(0) = -D_0(k_4,k_+-k_3,k_2-k_3,0,M_-,M_+,0)  } \quad&&
\nn\\
&\sim& -D_0(0)
+\frac{2\pi\ri}{\kappa_\PW}\biggl\{
\ln\biggl(-x_\PW\frac{s_{23}}{\MW^2}\biggr)
+\ln\biggl[1+\frac{s_{13}}{s_{23}}(1-z)\biggr]
+\ln\biggl[1+\frac{s_{24}}{s_{23}}(1-z)\biggr] 
\nn\\ && {} 
\qquad\qquad
-\ln\biggl(1+\frac{s_{13}}{\MW^2}z x_\PW\biggr)
-\ln\biggl(1+\frac{s_{24}}{\MW^2}z x_\PW\biggr) \biggr\},
\\[.5em] &&
\mbox{with $z$ from \refeq{eq:xtilde},} 
\nn\\[1em]\label{D0br1}
\lefteqn{ \Dbr_0(1) = \Dbr_0(k_-,k_+,k_2,0,M^*,M,m_2) }\quad
\nn\\
&\sim& D_0(1)\Big|_{K_-\to -K_-^*}
+\frac{2\pi\ri}{K_+(s_{23}+s_{24})-K_-^*\MW^2} \biggl[
2\ln\biggl(1-\frac{K_+}{K_-^*}\frac{s_{23}+s_{24}}{\MW^2}\biggr)
\nn\\ && {}
\qquad\qquad\qquad\qquad
-\ln\biggl(1+\frac{K_+}{K_-^* x_\PW}\biggr)
-\ln\biggl(1+\frac{x_\PW \MW^2}{s_{23}+s_{24}}\biggr)
-\ln\biggl(\frac{m_2^2}{\MW^2}\biggr) 
\biggr],
\hspace{2em}
\\[1em]
\lefteqn{ \Dbr_0(2) = \Dbr_0(k_3,k_+,k_2,\la,m_3,M,m_2) }\quad
\nn\\
&\sim& D_0(2)
+\frac{2\pi\ri}{K_+s_{23}} 
\ln\biggl[\frac{K_+s_{23}}{\ri\lambda m_2(s_{13}+s_{23})}\biggr],
\\[1em]
\lefteqn{\Dbr_0(3) = \Dbr_0(k_3,k_-,k_2,\la,m_3,M^*,m_2) 
}\quad
\nn\\
&\sim& -D_0(3)\Big|_{K_-\to -K_-^*}
+\frac{2\pi\ri}{K_-^*s_{23}}
\ln\biggl[\frac{\ri K_-^*m_2}{\lambda(s_{23}+s_{24})}\biggr],
\\[1em]
\lefteqn{ \Dbr_0(4) = \Dbr_0(k_3,k_-,k_+,0,m_3,M^*,M) 
}\quad
\nn\\
&\sim& -D_0(4)\Big|_{K_-\to -K_-^*}
+\frac{2\pi\ri}{K_-^*(s_{13}+s_{23})-K_+\MW^2}
\biggl[\ln\biggl(1+\frac{K_-^* x_\PW}{K_+}\biggr)
-\ln\biggl(1+\frac{x_\PW \MW^2}{s_{13}+s_{23}}\biggr)\biggr].
\nn\\
\eeqar
We note that the logarithmic terms on the r.h.s.\ of \refeq{Dbr00}
yield purely imaginary contributions if
$(s_{13}+s_{23})>-\MW^2 x_\PW$, $(s_{23}+s_{24})>-\MW^2 x_\PW$, 
and $\kappa_\PW$ is imaginary. These conditions are fulfilled on 
resonance. Off resonance, this is no longer true
near the boundary of phase space.
But since these conditions are only violated in a fraction of
phase space of order $|k_\pm^2-\MW^2|/\MW^2$, which is irrelevant in DPA
(compare the discussion of the relevance of the Landau singularities in
\refse{se:ambig}), it would
even be allowed to replace $\tilde\Dbr_0(0)$, which is real,
by $-\Re\{D_0(0)\}$, as it was done in \citeres{Be97a,Be97b}.

\section[Four-fermion momenta for on-shell $\mathrm W$ bosons]
{\boldmath Four-fermion momenta for on-shell $\mathrm W$ bosons}
\label{ap:onshell}

In order to define a DPA for the virtual corrections of 
four-fermion production, one has to project the four-fermion momenta 
for off-shell intermediate $\PW$ bosons to the ones, where the $\PW$ bosons 
are on shell.
Therefore, eight independent parameters have to be chosen,
which determine the four-fermion phase space uniquely and
include the invariant masses of the resonant $\PW$ bosons $k_\pm^2$.
One possible choice is to fix the directions 
of the particles \PWp, $f_1$, and $f_3$ in the CM frame of the 
initial-state $\Pep \Pem$ pair while taking the limit $k_\pm^2 \to \MW^2$.
This results in the on-shell projection:
\begin{eqnarray}
\nn
k_{+0}^{\mathrm {on}}&=&\frac{\sqrt{s}}{2},\quad 
{\mathbf k}_{+}^{\mathrm {on}}=\frac{{\mathbf k}_+}{|{\mathbf k}_+|}
\frac{\sqrt{s}}{2}\beta_\PW, \quad
k_-^{\mathrm {on}}=p_+ + p_- -k_+^{\mathrm {on}},\\
k_1^{\mathrm {on}}&=&
k_1 \frac{\MW^2}{2 k_1 k_+^{\mathrm {on}}},\quad
k_2^{\mathrm {on}}=k_+^{\mathrm {on}}-k_1^{\mathrm {on}},\quad
\label{eq:onshell}
k_3^{\mathrm {on}}=
k_3 \frac{\MW^2}{2 k_3 k_-^{\mathrm {on}}},\quad
k_4^{\mathrm {on}}=k_-^{\mathrm {on}}-k_3^{\mathrm {on}}
\end{eqnarray}
with $k_+=k_1+k_2$ and $k_-=k_3+k_4$.
The momenta of the initial-state $\Pe^\pm$ are denoted by $p_\pm$ 
and the momenta of the final-state fermions by $k_i, i=1,\dots, 4$.
The on-shell momenta are marked by the superscript 'on'.

\section{Construction of phase-space generators}
\label{ap:phasespace}
\newcommand{\Rot}{{\cal R}}
\newcommand{\Dec}{{\cal D}}
\newcommand{\Boost}{{\cal B}}

The scattering amplitude of a given process 
includes usually a lot of different propagators
corresponding to the different diagrams.
The different propagators peak in different phase-space regions and, 
therefore, slow down the convergence of the numerical integration or 
cause numerical instabilities.
If the differential cross section varies too much, the result and,
in particular, the statistical error of the Monte Carlo integration
can not be trusted.

As described in \refse{se:MC}, the multichannel approach is applied 
to obtain accurate predictions for the processes $\eeffff(\,+\,\ga)$, 
where each channel is responsible for a diagram or a class of diagrams
with the same propagators.
A channel is defined by the mappings from random
numbers to the momenta of the event.
For each diagram, a phase-space generator is constructed 
in such a way that the corresponding propagators 
are smoothed by the Jacobian of the specific mapping.
To do so, the inverse of the Jacobian of the channel
has to simulate the cross section in those regions
of phase space where the cross section becomes large.
The construction of the phase-space generators is explained in the 
following (see in addition \citeres{Multichannel,By73,Ba87}).

\subsection{Smoothing propagators depending on time-like momenta}

Diagrams usually involve several propagators 
depending on time-like or space-like momenta.
In this section, diagrams which depend only on propagators with 
time-like momenta are considered.
It is appropriate to decompose the integral over the phase space into
several integrals over the invariants of the propagators and 
the remaining integrals over the angles in appropriate rest frames.
Each invariant is mapped to a random number $r$
such that the integral takes the following form:
\begin{eqnarray}
\int_{s_{\min} }^{s_{\max} } \rd s &=&
\int_0^1 \frac{\rd r}{g_s(s(r),m^2,\nu,s_{\min} ,s_{\max} )}
\end{eqnarray} 
with $0\le r \le 1$ and the density $g_s$ defined by
\begin{eqnarray}
\frac{1}{g_s(s(r),m^2,\nu,s_{\min} ,s_{\max} )}&=&
\frac{\rd h(r,m^2,\nu,s_{\min} ,s_{\max} )}{\rd r}.
\end{eqnarray} 
The mapping of the random number $r$ into the invariant $s$
\begin{eqnarray}
s(r)&=&h(r,m^2,\nu,s_{\min} ,s_{\max} )
\end{eqnarray}
has to be chosen such that the density $g_s$ 
simulates the behaviour of the integrand in the region where the
integrand is large.
This is called {\it importance sampling}, 
because more events are sampled in the important integration region of $s$
in which the integrand becomes large.
Thus, it is a reasonable approach to require that the density has to include
the inverse of the propagator.   

The mappings belonging to the different propagator types read:
\begin{itemize}
\item {\bf Propagator with vanishing width} \cite{Multichannel} {\bf :}
\qquad$\si\propto 1/|s-m^2|^2$
\begin{eqnarray}
\nn
h(r,m^2,\nu,s_{\min} ,s_{\max} )&=& 
\Big[r (s_{\max}-m^2 )^{1-\nu}+
(1-r) (s_{\min}-m^2 )^{1-\nu}\Big]^{\frac{1}{1-\nu}}+m^2,\qquad \\
\label{eq:mapne1}
g_s(s,m^2,\nu,s_{\min} ,s_{\max} )&=&
\frac{1-\nu}{\Big[(s_{\max}-m^2 )^{1-\nu}-
(s_{\min}-m^2 )^{1-\nu}\Big](s-m^2)^\nu }
\end{eqnarray}
for $\nu\ne 1$ and
\begin{eqnarray}
\nn
h(r,m^2,1,s_{\min} ,s_{\max} )&=& \exp \Big[r \ln (s_{\max}-m^2 )-
(1-r) \ln (s_{\min}-m^2 )\Big]+m^2,\qquad\\
\label{eq:mapeq1}
g_s(s,m^2,1,s_{\min} ,s_{\max} )&=&\frac{1}
{\Big[\ln (s_{\max}-m^2 )-\ln (s_{\min}-m^2 )\Big](s-m^2)}
\end{eqnarray}
for $\nu=1$. 
\item {\bf Breit--Wigner propagator:}
\qquad$\si\propto 1/[(s-\MV^2)^2+\MV^2\GV^2]$
\begin{eqnarray}
\label{eq:BreitWigner}
\nn
h(r ,\MV^2-\ri \MV \GV,2,s_{\min} ,s_{\max} )&=&
\MV \Gamma \tan \Big[y_1 +(y_2 - y_1 ) r \Big]+\MV^2,\\
g_s(s,\MV^2-\ri \MV \Gamma,2,s_{\min} ,s_{\max} ) &=&
\frac{\MV \GV}{(y_2 - y_1 ) [(s-\MV^2)^2+\MV^2 \GV^2]}
\end{eqnarray} 
with
\begin{eqnarray}
y_{1/2}&=&{\mathrm {arctan}}
\left(\frac{s_{{\min} /{\max }}-\MV^2}{\MV \GV}\right).
\end{eqnarray}
\end{itemize}
These mappings are applicable not only for the Mandelstam variable $s$, 
but also for the absolute value of the
Mandelstam variables $|t|$ and $|u|$.

The parameter $\nu $ can be tuned to optimise the 
Monte Carlo integration and should be chosen $\gsim 1$.
The naive expectation $\nu =2$ is not necessarily the best
choice, because the propagator poles of the differential cross section
are partly cancelled in the collinear limit.  
Note that the mappings \refeq{eq:mapne1} and \refeq{eq:mapeq1} 
are undefined for $\nu \ge 1$ and $s_{\min}\le m^2$.

Since all fermion masses are neglected in our calculations for 
four-fermion production, 
$m$ can be omitted in \refeq{eq:mapne1} and \refeq{eq:mapeq1}.
However, I found it convenient to use a small negative mass parameter 
$m^2=-a$ to avoid numerical problems for $s_{\min }=0$.

\subsubsection{Isotropic decay into two particles} 
\label{se:decay}

Phase-space generators of diagrams, which include only propagators depending 
on time-like momenta, can be composed by isotopic particle decays.
Therefore, the isotopic decay is described in the following.

A particle with momentum $p$ decays into 
two particles with momenta $k_1$ and $k_2$.
The momenta of the initial-state particle and the masses of the
final-state particles are fixed.
The polar angle $\phi^*$ and azimuthal angle $\theta^*$ 
in the rest frame of the decaying particle 
are suitable integration variables:
\begin{eqnarray}
\nn
\int \rd \Phi_d(p,m_1^2,m_2^2)&=&
\int \rd^4 k_1 \rd^4 k_2 \delta(k_1^2-m_1^2) \theta(k_{10})
\delta(k_2^2-m_2^2) \theta(k_{20})
\delta^{(4)} (p-k_1-k_2 )\\
&=&\frac{\lambda^\frac{1}{2}(p^2,m_1^2,m_2^2)}{8 p^2}
\int_0^{2 \pi} \rd \phi^*
\int_{-1}^1 \rd \cos \theta^* \qquad\qquad
\label{eq:decay}
\end{eqnarray}
with the function $\la$ defined in \refeq{eq:lambda}.
For the numerical integration using the Monte Carlo technique, 
the angles have to be expressed by random numbers $r_1$ and $r_2$
with $0 \le r_1, r_2 \le 1$:
\begin{eqnarray}
\int \rd \Phi_d(p,m_1^2,m_2^2)
&=&\frac{1}{g_d(p^2,m_1^2,m_2^2)}
\int_0^1 \rd r_1 \rd r_2 \qquad
\end{eqnarray}
with $\phi =2 \pi r_1$, $\cos \theta^*=2 r_2-1$, and the density
\begin{eqnarray}
g_d(p^2,m_1^2,m_2^2)&=&\frac{2 p^2}{\lambda^\frac{1}{2}(p^2,m_1^2,m_2^2) \pi}.
\end{eqnarray}

Since the laboratory frame usually does not coincide with the 
rest frame of the 
decaying particle, the proper Lorentz transformation is introduced.
The Lorentz transformation of momentum $k$ into the rest frame of the 
particle with momentum $p$ is defined by
\begin{eqnarray}
k^\prime &=&\Boost (p_0,{\mathbf p})k
\end{eqnarray}
with the explicit form (see \eg \citere{By73})
\begin{eqnarray}
\label{eq:boost}
k_0^\prime&=&\gamma k_0+{\mathbf b}{\mathbf k},\qquad
{\mathbf k}^{\,\prime} = 
{\mathbf k}+{\mathbf b}\frac{{\mathbf b}{\mathbf k}}{1-\gamma}+{\mathbf b} k_0,
\end{eqnarray}
where ${\mathbf b}=-{\mathbf p}/m$, $\gamma=p_0/m$ and $m=\sqrt{p^2}$.
The inverse Lorentz transform is obtained by substituting 
${\mathbf p}$ into $-{\mathbf p}$.
For instance, if the momentum $k$ is identical to $p$, 
the particle is transformed into its rest frame and vice versa:
\begin{eqnarray}
p^* &=&\Boost (p_0,{\mathbf p})p, \qquad
p=\Boost (p_0,-{\mathbf p}) p^*
\end{eqnarray}
with $p^*=(m,0,0,0)^{\mathrm T}$.

Since the decay is isotropic, the orientation of the coordinate system
can be arbitrarily chosen and
the momenta of the outgoing particles read
\begin{eqnarray}
k_1&=&\Boost (p_0,-{\mathbf p}) \Rot (\phi^*,\cos \theta^*) 
\left(
\begin{array}{c}
\frac{p^2+k_1^2-k_2^2}{2 \sqrt{p^2} } \\       
0 \\
0 \\
\frac{\lambda^\frac{1}{2}(p^2,k_1^2,k_2^2)}{2 \sqrt{p^2} }
\end{array}
\right), \qquad
k_2=p-k_1
\end{eqnarray}
with the explicit rotation 
\begin{eqnarray}
\Rot (\phi^*,\cos \theta^*)=
\left(
\begin{array}{cccc}
1 & 0            & 0            & 0 \\       
0 & \cos \phi^*  & \sin \phi^*  & 0 \\
0 & -\sin \phi^* & \cos \phi^*  & 0 \\
0 & 0            & 0            & 1 
\end{array}
\right)
\left(
\begin{array}{cccc}
1 & 0              & 0 & 0             \\
0 & \cos \theta^*  & 0 & \sin \theta^* \\
0 & 0              & 1 & 0             \\
0 & -\sin \theta^* & 0 & \cos \theta^*   
\end{array}
\right).
\end{eqnarray}

\subsubsection{Example}

\begin{figure}
\begin{center}
\begin{picture}(150,110)(0,0)
\ArrowLine(0,30)(25,55)
\ArrowLine(0,80)(25,55)
\Photon(25,55)(70,55){2}{4}
\Photon(70,55)(100,85){2}{4}
\Photon(70,55)(100,25){2}{4}
\ArrowLine(100,85)(125,105)
\ArrowLine(125,65)(100,85)
\ArrowLine(100,25)(125,45)
\ArrowLine(125,5)(100,25)
\Vertex(25,55){2}
\Vertex(70,55){2}
\Vertex(100,85){2}
\Vertex(100,25){2}
\put(-40,75){\makebox(0,0)[lb]{$\Pe^+(p_+)$}}
\put(-40,25){\makebox(0,0)[lb]{$\Pe^-(p_-)$}}
\put(70,75){\makebox(0,0)[lb]{$\PW$}}
\put(70,25){\makebox(0,0)[lb]{$\PW$}}
\put(40,60){\makebox(0,0)[lb]{$\ga/ \PZ$}}
\put(130,100){\makebox(0,0)[lb]{$\nu_\mu(k_1)$}}
\put(130,60){\makebox(0,0)[lb]{$\mu^+(k_2)$}}
\put(130,40){\makebox(0,0)[lb]{$\tau^-(k_3)$}}
\put(130,0){\makebox(0,0)[lb]{$\bar{\nu}_\tau(k_4)$}}
\end{picture}
\end{center}
\caption{
Diagram with two resonant W bosons of the 
process $\Pep\Pem\to \nu_\mu \mu^+ \tau^- \bar{\nu}_\tau$}
\label{fi:example1}
\end{figure}
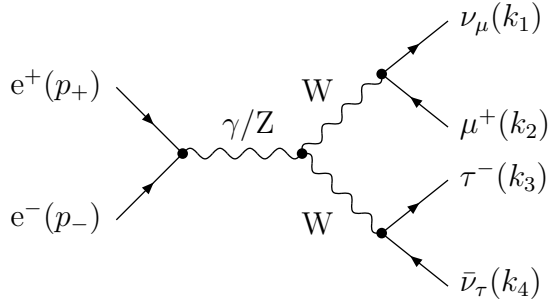

As an example, 
a phase-space generator for the diagram of \reffi{fi:example1}
is constructed.
For both intermediate particles, photon or $\PZ$ boson, only
one generator is required, since the corresponding propagator is fixed 
by the momenta of the initial-state particles and, hence, no mapping
is needed.
The phase-space integral can be decomposed of the 
integrals over the invariants of the intermediate $\PW$ bosons,
the \PW-pair production, and the two $\PW$ decays:
\begin{eqnarray}
\nn
\int \rd \Phi&=&\int_0^{p^2} \rd k_+^2 
\int_0^{\left(\sqrt{p^2}-\sqrt{k_+^2}\right)^2} \rd k_-^2
\int \rd \Phi_d(p,k_+^2,k_-^2)\\
&&\times \int \rd \Phi_d(k_+,k_1^2,k_2^2)
\int \rd \Phi_d(k_-,k_3^2,k_4^2)
\end{eqnarray}
with $p=p_++p_-$, $k_+=k_1+k_2$, and $k_-=k_3+k_4$.
The invariant masses of the \PW-boson propagators are determined by
\begin{eqnarray}
k_+^2&=&h(r_1,\MW^2-\ri \MW \GW,2, 0, p^2),\\
k_-^2&=&h\Bigl(r_2,\MW^2-\ri \MW \GW,2, 0, 
\Bigl(\sqrt{p^2}-\sqrt{k_+^2}\Bigr)^2\Bigr)
\end{eqnarray}
with the function $h$ defined in \refeq{eq:BreitWigner}. 
The total density reads
\begin{eqnarray}
\nn
g_{\mathrm {tot}}&=&g_s(k_+^2,\MW^2-\ri \MW \GW,2, 0, p^2)
g_s\Bigl(k_-^2,\MW^2-\ri \MW \GW,2, 0, \Bigl(\sqrt{p^2}-\sqrt{k_+^2}\Bigr)^2
\Bigr)\\
&&\times
g_d(p^2,k_+^2,k_-^2) g_d(k_+^2,k_1^2,k_2^2) g_d(k_-^2,k_3^2,k_4^2).
\end{eqnarray}
including the two \PW-boson propagators.

\subsection[The $t$-channel diagram of a $2\to 2\,\mathrm{particle}$ process]
{\boldmath The $t$-channel diagram of a $2\to 2\,\mathrm{particle}$ process}
\label{se:process}

So far, only phase-space generators for diagrams with 
propagators depending on time-like momenta are considered.
The simplest diagram including virtual particles with space-like momenta 
is the $t$-channel diagram of a $2\to 2\,\mathrm{particle}$ scattering process
which is outlined in the following.

Two particles with momenta $p_+$ and $p_-$ 
transform into two particles with momenta $k_1$ and $k_2$,
where the momenta of the initial-state particles and
the masses of the final-state particles are fixed. 
Furthermore, the diagram includes a propagator depending on the 
Mandelstam variable $t=(p_+-k_1)^2$ which has to be smoothed.

The calculation is performed 
in the rest frame of $p=p_++p_-$ with the 3-axis in ${\mathbf p_+}$ direction.
The momenta of the external particles take the form
\begin{eqnarray}
p^*_\pm=
\left(
\begin{array}{c}
\frac{p^2+p_\pm^2-p_\mp^2}{2 \sqrt{p^2}} \\ 0 \\ 0 \\
\pm \frac{\lambda^\frac{1}{2}(p^2,p_+^2,p_-^2)}{2 \sqrt{p^2}}
\end{array}\right),\qquad
k^*_1&=&
\Rot(\phi^*,\cos \theta^*)
\left(
\begin{array}{c}
\frac{p^2+k_1^2-k_2^2}{2 \sqrt{p^2}} \\ 0 \\ 0 \\
\frac{\lambda^\frac{1}{2}(p^2,k_1^2,k_2^2)}{2 \sqrt{p^2}}
\end{array}
\right), 
\end{eqnarray}
and $k_2^*=p_+^*+p_-^*-k_1^*$.
Since the invariant mass of the propagator $t$ depends 
exclusively on the azimuthal angle $\cos \theta^*$,
\begin{eqnarray}
\label{eq:reltcos}
t&=&k_1^2+p_+^2-\frac{(p^2+k_1^2-k^2_2)(p^2+p_+^2-p_-^2)-
\lambda^\frac{1}{2}(p^2,k_1^2,k_2^2) \lambda^\frac{1}{2}(p^2,p_+^2,p_-^2)
\cos \theta^*}{2 p^2},\qquad
\end{eqnarray} 
the phase-space integral can be converted into an integral over the polar 
angle $\phi^*$ and over the absolute value of the 
Mandelstam variable $t$:
\begin{eqnarray}
\nn
\lefteqn{\int \rd \Phi_p (p_+,p_-,m_1^2,m_2^2)}\\
\nn
&=&
 \int \rd^4 k_1  \rd^4 k_2 \delta^{(4)} (p-p_1-p_2 ) 
\delta\left(k_1^2-m_1^2\right)\theta (k_{10})
\delta\left(k_2^2-m_2^2\right)\theta (k_{20})\\
&=&\frac{1}{4 \lambda^\frac{1}{2}(p^2,p_+^2,p_-^2)}
\int_0^{2 \pi} \rd \phi^*
\int_{-t_{\max}}^{-t_{\min}} \rd |t|,
\end{eqnarray}
where the integration boundaries can be calculated from 
\refeq{eq:reltcos} with $-1\le \cos \theta^* \le 1$.
Like in the previous section, the polar angle $\phi^*$ and 
the invariant $t$ of the propagator
are determined by 
\begin{eqnarray}
\phi^*&=&2 \pi r_1,\qquad |t|=h(r_2,m^2,\nu,-t_{\max},-t_{\min})
\end{eqnarray}
resulting in
\begin{eqnarray}
\int \rd \Phi_p (p_+,p_-,k_1^2,k_2^2)=\int_0^1 \rd r_1 \int_0^1 \rd r_2  
\frac{1}{g_p (p^2,p_+^2,p_-^2,t,m^2,\nu,t_{\min},t_{\max})}
\end{eqnarray}
with the density
\begin{eqnarray}
g_p (p^2,p_+^2,p_-^2,t,m^2,\nu,t_{\min},t_{\max})&=&
\frac{2}{\pi} \lambda^\frac{1}{2}(p^2,p_+^2,p_-^2)
g_s(-t,m^2,\nu,-t_{\max},-t_{\min}).\quad
\end{eqnarray} 

To obtain the momenta of the final-state particles
in the laboratory frame, a rotation and a proper Lorentz transformation 
are performed:
\begin{eqnarray}
k_i&=&\Boost(p_0,-{\mathbf p}) \Rot (-\hat{\phi},\cos \hat{\theta}) k^*_i
\end{eqnarray} 
with $i=1,2$ and the angles
\begin{eqnarray}
\hat{\phi}&=&
\left\{\begin{array}{cl}
\mathrm{arctan}\left(\frac{\hat{p}_{+2}}{\hat{p}_{+1}}\right)&
,\qquad \hat{p}_{+1}>0\\
\mathrm{arctan}\left(\frac{\hat{p}_{+2}}{\hat{p}_{+1}}\right)+\pi&
,\qquad \hat{p}_{+1}<0
\end{array}\right. 
,\qquad \cos \hat{\theta}=\frac{\hat{p}_{+3}}{|{\mathbf{\hat{p}_+}}|}
\end{eqnarray}
with $\hat{p}_+=\Boost(p_0,{\mathbf p})  p_+$.

In order to make the numerical integration more efficient, 
cuts can already be introduced in the generation of the 
event, \ie an upper limit on $t$ or on $\cos \theta^*$.
Since the Mandelstam variable $t$ is 
Lorentz invariant the cut on $t$ is valid in arbitrary frames.
Angular cuts have to be transformed into the rest frame of the 
incoming particles.  

\subsection{Phase-space generator for a multi-peripheral diagram}

\begin{figure}
\begin{center}
\begin{picture}(150,120)
\ArrowLine(50,105)(0,105)
\ArrowLine(0,5)(50,5)
\ArrowLine(150,105)(50,105)
\ArrowLine(50,5)(150,5)
\Photon(50,105)(50,55){2}{4}
\Photon(50,5)(50,55){2}{4}
\Photon(100,55)(50,55){2}{5}
\ArrowLine(100,55)(125,70)
\ArrowLine(125,70)(150,85)
\ArrowLine(150,25)(100,55)
\Photon(125,70)(150,55){-2}{3}
\Vertex(50,105){2}
\Vertex(50,5){2}
\Vertex(50,55){2}
\Vertex(100,55){2}
\Vertex(125,70){2}
\put(-40,100){\makebox(0,0)[lb]{$\Pe^+(p_+)$}}
\put(-40,0){\makebox(0,0)[lb]{$\Pe^-(p_-)$}}
\put(30,75){\makebox(0,0)[lb]{$\PW$}}
\put(30,25){\makebox(0,0)[lb]{$\PW$}}
\put(70,60){\makebox(0,0)[lb]{$\ga$}}
\put(105,65){\makebox(0,0)[lb]{$\mu$}}
\put(160,100){\makebox(0,0)[lb]{$\bar\nu_\Pe(k_1)$}}
\put(160,0){\makebox(0,0)[lb]{$\nu_\Pe(k_5)$}}
\put(160,80){\makebox(0,0)[lb]{$\mu^-(k_2)$}}
\put(160,20){\makebox(0,0)[lb]{$\mu^+(k_4)$}}
\put(160,50){\makebox(0,0)[lb]{$\gamma(k_3)$}}
\end{picture}
\end{center}
\caption[]{An example for a multi-peripheral diagram contributing to the
process $\Pep\Pem\to \Pne\Pnebar \mu^- \mu^+ \gamma$}
\label{fi:example2}
\end{figure}

The multi-peripheral diagram 
of \reffi{fi:example2} is investigated as an example, because it 
involves all types of propagators.
In the following, all fermion masses are neglected and the 
following definitions are used:
\beq
\begin{array}[b]{rlrlrl}
k_{23} =&k_2+k_3, & \qquad
k_{234} =&k_2+k_3+k_4,& \qquad
k_{1234}=&k_1+k_2+k_3+k_4,\\
q_+=&p_+-k_1,& \qquad
q_-=&p_--k_5,& \qquad
p=&p_++p_-.
\end{array}
\eeq
The infrared and collinear singularities are excluded by a lower limit 
on $k_{23} ^2>k^2_{{23}, \min}$.

The phase-space integral
\begin{eqnarray}
\nn
\int \rd \Phi &=&
\int_{k_{{23}, \min}^2}^{p^2} \rd k_{23}^2
\int_{k_{23}^2}^{p^2} \rd k_{234}^2
\int_{k_{234}^2}^{p^2} \rd k_{1234}^2
\int \rd \Phi_p (p_+,p_-,k_{1234}^2,k_5^2)
\\&&  {} \times
\int \rd \Phi_p (p_+,q_-,k_1^2,k_{234}^2)
\int \rd \Phi_d (k_{234},k_{23}^2,k_4^2)
\int \rd \Phi_d (k_{23},k_2^2,k_3^2)\qquad
\end{eqnarray}
is decomposed of two scattering processes
and two particle decays.
The intermediate particles of this decomposition are the
virtual particles and an additional fictitious particle
with momentum $k_{1234}$.
The invariant masses of the external particles of the 
scattering processes and particle decays 
have to be determined first:
\begin{eqnarray}
k_{{23} }^2&=&h(r_1,0,\nu _1,k^2_{{23}, \min },p^2),\\
k_{{234} }^2&=&h(r_2,0,\nu _2,k_{23} ^2 ,p^2),\\
\label{eq:detinvariants}
k_{1234}^2&=&h(r_3,0,0,k_{234} ^2 ,p^2).
\end{eqnarray}

As a second step, the momenta of the final-state particles are calculated:

\subsubsection{\fbox{$p_+ + p_- \to k_{1234} + k_5$}}
The initial-state electron and positron transforms into 
the final-state $\nu_\Pe$
and a fictitious particle with momentum $k_{1234}$.
The invariant mass of the $\PW$-boson propagator is determined by
\begin{eqnarray}
|q_-^2|&=&h(r_4,\MW^2-\ri \MW \GW,2, 0, p^2-k_{1234}^2),
\end{eqnarray}
where the boundaries are taken from \refeq{eq:reltcos}.
The momenta can be calculated according to \refse{se:process}.

\subsubsection{\fbox{$p_+ + q_- \to k_1 + k_{{234}}$}}
The incoming particles are the initial-state positron and the 
incoming virtual $\PW$ boson which couples to the initial-state electron.
Note, that the momentum of the incoming $\PW$ boson is fixed 
by the momentum of the 
initial-state electron and the already calculated momentum of the 
final-state $\nu_\Pe$.
The invariant mass of the $\PW$-boson propagator reads
\begin{eqnarray}
|q_+^2|&=&h(r_5,\MW^2-\ri \MW \GW,2,0, 
(k_{1234}^2-k_{234}^2)(k_{1234}^2-q_-^2)/k_{1234}^2)
\end{eqnarray}
with the boundaries calculated from \refeq{eq:reltcos}.
In this way, the momenta of the final-state $\bar\nu_\Pe$ and
the virtual photon are determined.

\subsubsection{\fbox{$k_{{234} }\to k_{{23} } + k_4$}}
The virtual photon decays isotropically into the final-state $\mu^+$
and the virtual $\mu$ according to \refse{se:decay}.

\subsubsection{\fbox{$k_{{23} }\to k_2 + k_3$}}
Finally, the virtual $\mu$ decays isotropically into the final-state
$\mu^-$ and the final-state $\ga$.
\bigskip\\
The total density reads
\begin{eqnarray}
\nn
g_{\mathrm {tot}}&=&
g_s(k_{23}^2,0,\nu _1,k^2_{{23}, \min },p^2)
g_s(k_{234}^2,0,\nu _2,k_{23} ^2 ,p^2)
g_s(k_{1234}^2,0,0,k_{234} ^2 ,p^2)\\
\nn
&&\times
g_p(p^2,0,0,q_-^2,\MW^2-\ri \MW \GW,2, 0, p^2-k_{1234}^2)\\
\nn
&&\times
g_p(k_{1234}^2,0,q_-^2,q_+^2,\MW^2-\ri \MW \GW,2,0, 
(k_{1234}^2-k_{234}^2)(k_{1234}^2-q_-^2)/k_{1234}^2)\\
&&\times
g_d(k_{234}^2,k_{23}^2,0)
g_d(k_{23}^2,0,0)
\end{eqnarray}
and includes all propagators of this diagram.

\subsection{Remarks on four-fermion production}

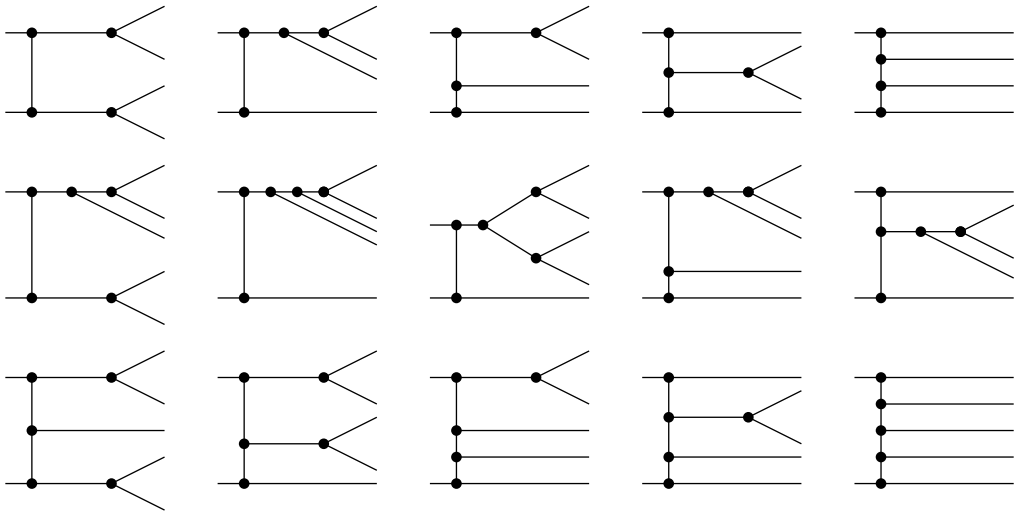
\begin{figure}
\begin{center}
\begin{picture}(400,150)(0,0)
\Line(10,160)(50,160)
\Line(10,130)(50,130)
\Line(20,130)(20,160)
\Line(50,130)(70,140)
\Line(50,130)(70,120)
\Line(50,160)(70,170)
\Line(50,160)(70,150)
\Vertex(50,160){2}
\Vertex(50,130){2}
\Vertex(20,160){2}
\Vertex(20,130){2}
\Line(90,160)(130,160)
\Line(90,130)(150,130)
\Line(100,130)(100,160)
\Line(115,160)(150,142.5)
\Line(130,160)(150,170)
\Line(130,160)(150,150)
\Vertex(130,160){2}
\Vertex(100,130){2}
\Vertex(100,160){2}
\Vertex(115,160){2}
\Line(170,160)(210,160)
\Line(170,130)(230,130)
\Line(180,130)(180,160)
\Line(180,140)(230,140)
\Line(210,160)(230,170)
\Line(210,160)(230,150)
\Vertex(210,160){2}
\Vertex(180,130){2}
\Vertex(180,160){2}
\Vertex(180,140){2}
\Line(250,160)(310,160)
\Line(250,130)(310,130)
\Line(260,130)(260,160)
\Line(260,145)(290,145)
\Line(290,145)(310,155)
\Line(290,145)(310,135)
\Vertex(260,130){2}
\Vertex(260,145){2}
\Vertex(260,160){2}
\Vertex(290,145){2}
\Line(330,160)(390,160)
\Line(330,130)(390,130)
\Line(340,130)(340,160)
\Line(340,150)(390,150)
\Line(340,140)(390,140)
\Vertex(340,130){2}
\Vertex(340,140){2}
\Vertex(340,150){2}
\Vertex(340,160){2}
\Line(10,100)(50,100)
\Line(10,60)(50,60)
\Line(20,60)(20,100)
\Line(50,60)(70,70)
\Line(50,60)(70,50)
\Line(50,100)(70,110)
\Line(50,100)(70,90)
\Line(35,100)(70,82.5)
\Vertex(50,100){2}
\Vertex(50,60){2}
\Vertex(20,100){2}
\Vertex(20,60){2}
\Vertex(35,100){2}
\Line(90,100)(130,100)
\Line(90,60)(150,60)
\Line(100,60)(100,100)
\Line(130,100)(150,110)
\Line(130,100)(150,90)
\Line(120,100)(150,85)
\Line(110,100)(150,80)
\Vertex(130,100){2}
\Vertex(100,60){2}
\Vertex(130,100){2}
\Vertex(100,100){2}
\Vertex(120,100){2}
\Vertex(110,100){2}
\Line(170,87.5)(190,87.5)
\Line(170,60)(230,60)
\Line(180,60)(180,87.5)
\Line(190,87.5)(210,100)
\Line(190,87.5)(210,75)
\Line(210,100)(230,110)
\Line(210,100)(230,90)
\Line(210,75)(230,85)
\Line(210,75)(230,65)
\Vertex(180,60){2}
\Vertex(180,87.5){2}
\Vertex(190,87.5){2}
\Vertex(210,100){2}
\Vertex(210,75){2}
\Line(250,100)(290,100)
\Line(250,60)(310,60)
\Line(260,60)(260,100)
\Line(290,100)(310,110)
\Line(290,100)(310,90)
\Line(275,100)(310,82.5)
\Line(260,70)(310,70)
\Vertex(290,100){2}
\Vertex(260,60){2}
\Vertex(290,100){2}
\Vertex(260,100){2}
\Vertex(275,100){2}
\Vertex(260,70){2}
\Line(330,100)(390,100)
\Line(330,60)(390,60)
\Line(340,60)(340,100)
\Line(370,85)(390,95)
\Line(370,85)(390,75)
\Line(355,85)(390,67.5)
\Line(340,85)(370,85)
\Vertex(370,85){2}
\Vertex(340,60){2}
\Vertex(340,100){2}
\Vertex(370,85){2}
\Vertex(340,85){2}
\Vertex(355,85){2}
\Line(10,30)(50,30)
\Line(10,-10)(50,-10)
\Line(20,-10)(20,30)
\Line(50,-10)(70,0)
\Line(50,-10)(70,-20)
\Line(50,30)(70,40)
\Line(50,30)(70,20)
\Line(20,10)(70,10)
\Vertex(50,30){2}
\Vertex(50,-10){2}
\Vertex(20,30){2}
\Vertex(20,-10){2}
\Vertex(20,10){2}
\Line(90,30)(130,30)
\Line(90,-10)(150,-10)
\Line(100,-10)(100,30)
\Line(130,30)(150,40)
\Line(130,30)(150,20)
\Line(130,5)(150,15)
\Line(130,5)(150,-5)
\Line(100,5)(130,5)
\Vertex(130,30){2}
\Vertex(100,-10){2}
\Vertex(100,30){2}
\Vertex(100,5){2}
\Vertex(130,5){2}
\Line(170,30)(210,30)
\Line(170,-10)(230,-10)
\Line(180,-10)(180,30)
\Line(210,30)(230,40)
\Line(210,30)(230,20)
\Line(180,10)(230,10)
\Line(180,0)(230,0)
\Vertex(210,30){2}
\Vertex(180,-10){2}
\Vertex(180,30){2}
\Vertex(180,0){2}
\Vertex(180,10){2}
\Line(250,30)(310,30)
\Line(250,-10)(310,-10)
\Line(260,-10)(260,30)
\Line(290,15)(310,25)
\Line(290,15)(310,5)
\Line(260,15)(290,15)
\Line(260,0)(310,0)
\Vertex(290,15){2}
\Vertex(260,-10){2}
\Vertex(260,30){2}
\Vertex(260,0){2}
\Vertex(260,15){2}
\Line(330,30)(390,30)
\Line(330,-10)(390,-10)
\Line(340,-10)(340,30)
\Line(340,20)(390,20)
\Line(340,0)(390,0)
\Line(340,10)(390,10)
\Vertex(340,20){2}
\Vertex(340,-10){2}
\Vertex(340,30){2}
\Vertex(340,0){2}
\Vertex(340,10){2}
\end{picture}
\end{center}
\caption{
The generic phase-space generators symbolized by the topologies 
of the corresponding diagrams}
\label{fi:topologies}
\end{figure}

In this way, five generic phase-space generators
for $\eeffff$ and ten for $\eeffffg$ are constructed.
The topologies corresponding to the phase-space generators for the 
processes $\eeffff$ are shown in the 
first row of \reffi{fi:topologies} and
the topologies corresponding to $\eeffffg$ 
can be found in the second and third row of \reffi{fi:topologies}.
For each generic generator the order of the external 
particles and the mappings for each propagator can be chosen.
In particular, flat mappings of a subset of propagators 
correspond to topologies where these propagators are contracted. 
With the help of these generators all propagators of the 
cross section are smoothed.
Special phase-space generators for interference contributions to the 
cross section are not constructed.

For the process $\Pep \Pem \to \Pep \Pem \Pep \Pem \ga$ with
1008 diagrams 928 different channels are included in the
calculation. 
The difference is due to diagrams where the initial-state
$\Pep \Pem$ pair couples to a virtual photon or $\PZ$ boson.
Since the invariant masses of the virtual photon and \PZ-boson 
are fixed by the CM energy,
only one phase-space generator is required for both diagrams.
For instance, the diagram of \reffi{fi:example1} 
corresponds to the first topology of \reffi{fi:topologies},
where the mapping of the propagator which couples to 
both incoming particles is flat, \ie $\nu_i=0$.

Phase-space generators for diagrams with a quartic-gauge-boson vertex 
are also included.
This diagrams have one propagator less than the other diagrams.
Therefore, one propagator in \reffi{fi:topologies}
is omitted and 
the mapping of the invariant of this propagator is flat.

The order of the particle decays, the $2 \to 2\,\mathrm{particle}$ processes,
and the determination of the invariants 
influence the convergence behaviour of the numerical integration,
since the integration domains of the single mappings depend on the
already fixed invariants and momenta.
In particular, the invariant masses of decaying particles
in a decay chain are not independent of each other.
To improve the numerical stability, the invariants of the 
virtual massless particles that couple to external particles
should be calculated first,
applying already the cuts in the generation of the momenta.
In the here discussed Monte Carlo program, invariant-mass cuts are 
used for the propagators of the virtual particles with time-like momenta.
For the virtual particles with space-like momenta
only one angular cut is applied to the first $2\to 2\,\mathrm{particle}$ 
process.
Note, that the angular cut does not influence the boundaries of
the integrations over the invariants of the virtual or fictitious particles
with time-like momenta, which are calculated at the beginning. 

For the numerical integration of the subtraction terms,
additional phase-space generators are constructed,
which take into account the propagators of the Born-cross section
included in the subtraction terms.
Since the Born cross section depends on the four-particle phase space, 
the four-particle momenta have to be generated first with the phase-space 
generators for the processes $\eeffff$.
These momenta are then mapped into the five-particle phase space
according to the different mappings defined \refse{se:subtractionterm}.

Besides momentum conservation and the requirements on 
the momenta of external particles, \ie $k_i^2=m_i^2$ and $k_{i0}\ge 0$,
the phase-space volume is a very helpful variable to test the 
phase-space generators.
For massless external particles, the phase-space volume can be calculated 
analytically as follows
\begin{eqnarray}
\int \prod\limits_{i=1,n} \rd^4 k_i \delta(k_i^2)\theta(k_{i0}) 
\delta^{(4)} ( p-\mbox{$\sum_{i=1}^n$} k_i) 
&=&\left(\frac{\pi}{2}\right)^{n-1}\frac{p^{2 n-4}}{\Gamma (n) \Gamma(n-1)},
\end{eqnarray}
where $n$ is the number of final-state 
particles (see \eg \citere{Kl86}).

\end{appendix}

\newpage
\chapter*{Curriculum Vitae}
\noindent{\Large \bf Personal data}\\
\begin{tabbing}
{\bf Name:} \hspace*{2cm}\= Roth, Markus \\
{\bf Date of birth: }    \> March 8, 1969 \\
{\bf Place of birth:}    \> Werneck, Germany \\
{\bf Nationality:}       \> German
\end{tabbing}
\vspace*{0.6cm}
\noindent{\Large \bf Education}\\
\begin{description}
\item[{\bf Since 1996:}]
Ph.~D.~student under the supervision of Dr.~Ansgar Denner and 
Prof.~Zoltan Kunszt at the ETH Z\"urich and 
the Paul Scherrer Institute, Switzerland.
\item[{\bf 1996:}] 
Ph.~D.~student under the supervision of Dr.~Ansgar Denner 
at the University of W\"urzburg.
\item[{\bf 1995 -- 1996:}] 
Compulsory military service
\item[{\bf 1989 -- 1995:}]
Diploma student in physics at the University of W\"urzburg, Germany.\\
Thesis title: {\em ``High-energy approximation of massive Feynman-integrals''}
\item[{\bf 1988:}] 
Abitur at the Alexander--von--Humboldt-Gymnasium, Schweinfurt, Germany.
\end{description}
\vspace*{0.6cm}
\noindent{\Large \bf Publications}
\vspace*{0.3cm}
\begin{itemize}
\item
{\boldmath \bf Predictions for all processes 
$\mathrm e^+ \mathrm e^- \to 4\,{\mathrm {fermions}} + \gamma$},\\
A.~Denner, S.~Dittmaier, M.~Roth and D.~Wackeroth, 
{\sl Nucl.~Phys.} {\bf B560} (1999) 33.
\item
{\boldmath \bf Further numerical results on non-factorizable corrections to \\
$ \mathrm e^+ \mathrm e^- \to 4\,{\mathrm {fermions}}$},\\
A.~Denner, S.~Dittmaier and M.~Roth, {\sl Phys.~Lett.} {\bf B429} (1998) 145.
\item
{\boldmath \bf Non-factorizable photonic corrections to 
$ \mathrm e^+ \mathrm e^- \to \PW \PW \to 4\,{\mathrm {fermions}}$},\\
A.~Denner, S.~Dittmaier and M.~Roth, {\sl Nucl.~Phys.} {\bf B519} (1998) 39.
\item
{\bf High-energy approximation of one-loop Feynman integrals},\\
M.~Roth and A.~Denner, {\sl Nucl.~Phys.} {\bf B479} (1996) 495.
\end{itemize}

\newpage
\chapter*{Acknowledgements}

First of all, I am very much grateful to my advisor A.~Denner 
for the interesting topic of my thesis and for 
supporting me actively during my work on this thesis.
I profited a lot from his great experience and knowledge in 
electroweak physics and especially enjoyed the very friendly atmosphere. 
\vspace*{0.4em}\\
I would like to thank Prof.~Z.~Kunszt 
for the official supervision of my thesis and the interest in my work.
\vspace*{0.4em}\\
Furthermore, I am grateful to Prof.~J.~Fr\"ohlich for the coreferat 
of my thesis.
\vspace*{0.4em}\\
I enjoyed a lot the collaboration with S.~Dittmaier and D.~Wackeroth.
I am grateful to them for reading the manuscript of my thesis.
\vspace*{0.4em}\\
I would like to thank also D.~Graudenz for useful discussions about 
the subtraction method,
R.~Pittau for the assistance in implementing
the non-factorizable corrections in EXCALIBUR, and my brother S.~Roth
for reading the manuscript of this thesis.
\vspace*{0.4em}\\
Special thanks to D.~Bodmer, S.~W\"uger, M.~Fosco, and M.~Beckers 
for the wonderful time that we spend together in the Schartenstrasse 11a.
\vspace*{0.4em}\\
I thank my parents for enabling and encouraging all my activities.

\end{document}